\documentclass[useAMS,usenatbib]{./aa}
\usepackage{txfonts}
\usepackage{natbib}
\bibpunct{(}{)}{;}{a}{}{,}
\usepackage[dvips]{graphicx}

\def\ls{\mathrel{\hbox{\rlap{\hbox{\lower4pt\hbox{$\sim$}}}\hbox{$<$}}}}
\def\gs{\mathrel{\hbox{\rlap{\hbox{\lower4pt\hbox{$\sim$}}}\hbox{$>$}}}}

\begin{document}

\title{LoCuSS: comparison of observed X-ray and lensing galaxy cluster
scaling relations with simulations \thanks{This work is based on
observations made with the \emph{XMM-Newton}, an ESA science mission
with instruments and contributions directly funded by ESA member
states and the USA (NASA).}}

\author{Y.-Y. Zhang\inst{1,2}, 
A. Finoguenov\inst{1,3},
H. B\"ohringer\inst{1}, 
J.-P. Kneib\inst{4},
G. P. Smith\inst{5},
R. Kneissl\inst{6},
N. Okabe\inst{7}, 
\and
H. Dahle\inst{4}
}

\institute{Max-Planck-Institut f\"ur extraterrestrische Physik, Giessenbachstra\ss e, 85748 Garching, Germany
\and Argelander-Institut f\"ur Astronomie, Universit\"at Bonn, Auf dem H\"ugel 71, 53121 Bonn, Germany
\and University of Maryland, Baltimore County, 1000 Hilltop Circle, Baltimore, MD 21250, USA
\and OAMP, Laboratoire d'Astrophysique de Marseille, traverse du Siphon, 13012 Marseille, France
\and School of Physics and Astronomy, University of Birmingham, Edgbaston, Birmingham, B152TT, UK
\and Max-Planck-Institut f\"ur Radioastronomie, Auf dem H\"ugel 69, 53121 Bonn, Germany
\and Astronomical institute, Tohoku University, Aramaki, Aoba-ku, Sendai, 980-8578, Japan 
}

\authorrunning{Zhang et al.}

\titlerunning{LoCuSS: comparison of observed X-ray and lensing galaxy
cluster scaling relations with simulations}

\date{Received 20/11/07 / Accepted 06/02/08}

\offprints{Y.-Y. Zhang}

\abstract{The Local Cluster Substructure Survey (LoCuSS, Smith et al.)
  is a systematic multi-wavelength survey of more than 100 X-ray
  luminous galaxy clusters in the redshift range 0.14-0.3 selected
  from the ROSAT All Sky Survey. We used data on 37 LoCuSS clusters
  from the \emph{XMM-Newton} archive to investigate the global scaling
  relations of galaxy clusters. The scaling relations based solely on
  the X-ray data ($S$--$T$, $S$--$Y_{\rm X}$, $P$--$Y_{\rm X}$,
  $M$--$T$, $M$--$Y_{\rm X}$, $M$--$M_{\rm gas}$, $M_{\rm gas}$--$T$,
  $L$--$T$, $L$--$Y_{\rm X}$, and $L$--$M$) obey empirical
  self-similarity and reveal no additional evolution beyond the
  large-scale structure growth.  They also reveal up to 17 per cent
  segregation between all 37 clusters and non-cool core clusters. Weak
  lensing mass measurements are also available in the literature for
  19 of the clusters with \emph{XMM-Newton} data. The average of the
  weak lensing mass to X-ray based mass ratio is $1.09\pm 0.08$,
  setting the limit of the non-thermal pressure support to $9 \pm 8$
  per cent. The mean of the weak lensing mass to X-ray based mass
  ratio of these clusters is $\sim 1$, indicating good agreement
  between X-ray and weak lensing masses for most clusters, although
  with 31--51 per cent scatter. The scatter in the mass--observable
  relations ($M$--$Y_{\rm X}$, $M$--$M_{\rm gas}$, and $M$--$T$) is
  smaller using X-ray based masses than using weak lensing masses by a
  factor of 2. With the scaled radius defined by the $Y_{\rm X}$
  profile -- $r_{500}^{\rm Y_{\rm X},X}$, $r_{500}^{\rm Y_{\rm
      X},wl}$, and $r_{500}^{\rm Y_{\rm X},si}$, we obtain lower
  scatter in the weak lensing mass based mass--observable relations,
  which means the origin of the scatter is $M^{\rm wl}$ and $M^{\rm
    X}$ instead of $Y_{\rm X}$. The normalization of the $M$--$Y_{\rm
    X}$ relation using X-ray mass estimates is lower than the one from
  simulations by up to 18--24 per cent at $3\sigma$ significance.
  This agrees with the $M$--$Y_{\rm X}$ relation based on weak lensing
  masses, the normalization of the latter being $\sim 20$ per cent
  lower than the one from simulations at $\sim 2\sigma$ significance.
  This difference between observations and simulations is also
  indicated in the $M$--$M_{\rm gas}$ and $M$--$T$ relations.  Despite
  the large scatter in the comparison of X-ray to lensing, the
  agreement between these two completely independent observational
  methods is an important step towards controlling astrophysical and
  measurement systematics in cosmological scaling relations.

}

\maketitle

\section{Introduction}
\label{s:intro}

The gravitational growth of fluctuations in the matter density
distribution can be traced by the evolution of the galaxy cluster mass
function (e.g. Schuecker et al. 2003). The mass function of luminous
galaxy clusters probes the cosmic evolution of large-scale structure
(LSS) and is an extremely effective test of cosmological models. It is
sensitive to the matter density, $\Omega_{\rm m}$, and the amplitude
of the cosmic power spectrum on cluster scales, $\sigma_8$
(e.g. Schuecker et al. 2003).

The construction of the mass function of galaxy clusters for large
cosmological cluster samples is based on robust calibration of cluster
mass--observable scaling relations for representative galaxy cluster
samples using reliable mass measurements (e.g. Reiprich \& B\"ohringer
2002; Voit 2005; Stanek et al. 2006). 

Deep X-ray observations using \emph{XMM-Newton} and \emph{Chandra} can
recover precisely the intra-cluster medium (ICM) density and
temperature distributions.  \emph{XMM-Newton} (also \emph{Chandra})
allows us for the first time to measure mass distributions for relaxed
nearby clusters with statistical uncertainties below 15 per cent up to
$r_{500}$, which yields a tight $M_{500}$--$T$ relation with only 8
per cent scatter (Arnaud et al.  2005).  Using \emph{Chandra} data,
Vikhlinin et al. (2006a) measured precise mass distributions up to
radii of $r_{2500}$, which provides a tight $M_{2500}$--$T$ relation
for relaxed nearby clusters with its scatter also within 8 per cent.
However, relaxed clusters only represent a small fraction of the
cluster population in the local Universe and an even smaller fraction
at high redshifts (Jeltema et al. 2005; Smith et al. 2005; Vikhlinin
et al.  2006b) and the cluster mass estimates from X-ray observations
are limited by additional physical processes in the ICM.  Recent
simulations (e.g.  Rasia et al. 2004, 2006; Nagai et al.  2007a) show
that the X-ray total mass measured via hydrostatic equilibrium can be
biased low by up to 20 per cent for nearby relaxed clusters and
therefore the normalization of the mass--temperature relation can be
biased low by up to 30 per cent. The bias is more significant (up to
$\sim$30 per cent) for un-relaxed clusters.

Gravitational lensing provides a direct probe of the total cluster
mass without needing to invoke the symmetry and equilibrium
assumptions required in the X-ray analysis. However lensing yields
direct constraints on the mass projected along the line-of-sight
through the cluster and may therefore be prone to projection effects.
It is therefore very powerful to use both X-ray and lensing methods to
measure cluster mass and thus to investigate systematic uncertainties in
mass measurements and their impact on cosmological parameter
estimation.  Early X-ray versus strong lensing mass comparisons
revealed large discrepancies between the methods (Miralda-Escude \&
Babul 1995), which were subsequently attributed to the absence of a
cool core in, and thus presumed more disturbed dynamical state of, the
discrepant clusters (Allen 1998; Wu 2000).  More recent analysis of
\emph{Chandra} and \emph{Hubble Space Telescope} data suggest a
picture that is broadly consistent with Allen (1998) in that
disturbed, likely merging, clusters dominate the scatter in the
mass-temperature relation for cluster cores ($R{\le}250\,{\rm kpc}/h$)
when using lensing to infer cluster mass (Smith et al.\ 2003, 2005).
Wide-field weak lensing studies of cluster scaling relations have to
date generally lacked the statistical precision to investigate
systematic differences between disturbed and undisturbed clusters
(Bardeau et al., 2007; Pedersen \& Dahle 2007; Zhang et al., 2007;
Hoekstra 2007).  However extending the mass range of lensing-based
scaling relation studies down to groups has been shown to help
overcome the large scatter when trying to measure the slope of, for
example, the mass-temperature relation (Berg\'e et al. 2007).

In summary, credible cluster cosmology experiments require well
calibrated measurements of the shape, scatter and evolution of the
mass-observable scaling relations based on large statistical samples
of clusters that are unbiased with respect to cluster morphology.
Elimination of systematic uncertainties from this calibration demands
that the cluster mass measurements are cross-checked between
independent mass measurement techniques, including X-ray and lensing
methods.  The challenges for precision cluster cosmology on the per
cent level are therefore as follows: (1) construction of large
($n_{\rm clus}{\gs}100$) representative cluster samples, and (2)
detailed high quality X-ray and lensing mass measurements of that
sample as a basis for calibrating the mass-observable scaling
relations.  Despite recent progress on weak lensing searches for
galaxy clusters (Wittman et al., 2006; Massey et al., 2007; Miyazaki
et al., 2007), the most efficient method to construct a homogeneous
sample of massive clusters is to use all sky X-ray surveys.  The main
attractions of this approach are the well-defined sensitivity limit of
such surveys and the minimal biases towards different cluster
morphologies. 

The Local Cluster Substructure Survey (LoCuSS, PI: G.\ P.\ Smith;
Smith et al.\ in prep.)  is a systematic multi-wavelength survey of
galaxy clusters designed to tackle the probems outlined above.  LoCuSS
provides a morphology-unbiased sample of ${\sim}100$ X-ray luminous
galaxy clusters in the redshift range $0.14{\le}z{\le}0.3$ selected
from the ROSAT All Sky Survey catalogs (Ebeling et al. 1998, 2000;
B\"ohringer et al.\ 2004).  Studies of the LoCuSS sample are in
progress on (1) gravitational lens modeling of the mass distribution
in the cluster cores, using data from our ongoing \emph{Hubble Space
  Telescope} (HST) observing program (SNAP:10881, GO:11312;
Hamilton-Morris et al., in prep.); (2) the scaling relation between
cluster mass inferred from lensing and the Sunyaev-Zeldovich (SZ)
effect Y-parameter (Marrone et al., in prep.); (3) the distribution of
the luminosity gap statistic in $10^{15}M_\odot$ clusters (Khosroshahi
et al., in prep.); (4) wide-field weak-lensing analysis of clusters
observed with Suprime-CAM on Subaru (Okabe et al., in prep.). A wide
variety of follow-up observations are also underway or imminent,
including recent awards of 300~ks on \emph{Chandra} (PID:9800372),
129~ks at 24~$\mu$m with \emph{Spitzer} (PID:40872), 60~ks on GALEX
(Cycle~4, Proposal \# 90), and a 0.5~Msec Key Program on ESA's
Herschel.  These space-based observations are supported by ongoing
ground-based programs at Keck, VLT, Subaru, Palomar, NOAO and UKIRT.

The goal of this paper is to compare X-ray and lensing cluster mass
measurements of LoCuSS clusters for which archival X-ray data are
available from XMM-Newton, and weak lensing masses are available in
the literature. As at the end of 2006, 37 LoCuSS clusters had
sufficiently high quality archival \emph{XMM-Newton} data to derive
precise temperature distributions for statistically reliable X-ray
mass estimates. Weak lensing mass measurements are also available in
the literature for 19 of these clusters (including 7 duplicates): 10
clusters from CFH12k data (Bardeau et al., 2005; 2007); 15 clusters
from observations at the Nordic Optical Telescope and UH 88in (Dahle
2006); one cluster, Abell~1689, has also been studied using both CFHT and
Subaru data (Limousin et al., 2007; Halkola et al. 2006; Broadhurst et
al., 2005).  Weak lensing measurements for an enlarged sample of
LoCuSS clusters are coming, based on high quality HST and Subaru data
(Okabe et al., in prep.).

This paper is organised as follows.  The \emph{XMM-Newton} data
quality and corresponding optimized reduction method are described in
\S~\ref{s:method}. \S~\ref{s:scale} shows an empirical similarity
of the X-ray scaling relations for our sample and the difference in
the normalization between mass--observable relations from X-ray
observations and simulations. In \S~\ref{s:simuyx}, we used a new
method to determine the radial scale ($r^{\rm Y_{\rm X},X}_{500}$) to
study the scaling relations, in which the scatter can be reduced. In
\S~\ref{s:mpro}, we present X-ray versus weak lensing mass
calibration using different mass definitions to look for the one which
gives least scatter in the X-ray versus weak lensing mass comparison.
\S~\ref{s:ylmwlen} describes the weak lensing mass based
mass--observable relations and the difference in the normalization
between the weak lensing mass based mass--observable relations for 19
LoCuSS clusters and the relations from simulations.  In
\S~\ref{s:dis}, we discuss the scatter and bias in weak lensing
mass estimates, the cluster mass comparison between observations and
simulations, and the variation of the amplitude of cosmic power
spectrum with the mass calibration. We draw our conclusions in
\S~\ref{s:conclusion}.  Unless explicitly stated otherwise, we
adopt a flat $\Lambda$CDM cosmology with the density parameter
$\Omega_{\rm m}=0.3$ and the Hubble constant $H_{\rm
  0}=70$~km~s$^{-1}$~Mpc$^{-1}$. We adopt the solar abundance table of
Anders \& Grevesse (1989). Confidence intervals correspond to the 68
per cent confidence level. We apply the Orthogonal Distance Regression
package (ODRPACK~2.01\footnote{http://www.netlib.org/odrpack/ and
  references therein}, e.g. Boggs et al. 1987) taking into account
measurement errors on both variables to determine the parameters and
their errors of the fitting throughout this paper. We use Monte Carlo
simulations for the uncertainty propagation on all quantities of
interest.

\section{Data description and X-ray mass modeling}
\label{s:method}

Achieving the goals of this paper (\S~\ref{s:intro}), specifically
identifying systematic differences between mass measurement methods,
rests on being able to measure statistically reliable cluster masses
using X-ray data.  For this, a precise temperature distribution
measurement is required. In order to obtain temperature measurements
with uncertainties of $\sim 15$ per cent for X-ray luminous galaxy
clusters, we used the criterion of $\sim 2000$ net counts in the
2--7~keV band per bin (Zhang et al.  2006, 2007). We therefore require
that the screened \emph{XMM-Newton} data of each cluster have more
than 4000 net counts in the 2--7~keV band after the subtraction of the
renormalized blank sky background to ensure at least two bins to
measure the temperature profile.  Up to the end of 2006, 37 LoCuSS
clusters in the \emph{XMM-Newton} archives fulfill this criterion
(Fig.~\ref{f:lxz} and Table~\ref{t:obs}), of which 9 clusters are part
of the REFLEX-DXL sample (Zhang et al. 2004, 2006; Finoguenov et al.
2005), 12 clusters are from the LoCuSS Pilot Study sample (Zhang et
al. 2007), and additionally 6 clusters are from the REXCESS sample
(B\"ohringer et al.  2007; Pratt et al.  2007; Croston et al. 2007)
respectively. Since the REFLEX-DXL, LoCuSS Pilot Study, and REXCESS
samples are constructed as morphology-unbiased samples, our sample, as
a collective sample of them only with a redshift cut, provides an
approximately unbiased cluster population. This allows us to
investigate (i) the systematics of both X-ray and weak lensing cluster
mass estimates, and (ii) the morphology dependence of the segregation
and scatter in the scaling relations.

All 37 LoCuSS clusters in the \emph{XMM-Newton} archives were uniformly
analyzed using the optimized reduction method developed for \emph{XMM-Newton}
observations for galaxy clusters at redshifts higher than 0.14 (so that the
cluster X-ray emission covers less than half of the field of view, FOV) with
more than 4000 net photons emitted from the cluster in the 2--7~keV band. More
details about this method can be found in Zhang et al. (2007). The outermost
annulus outside the truncation radius (8--8.33$^{\prime}$ or 9.2-10$^{\prime}$
depending on the truncation radius) was used to model the residual background.
The XMMSAS v6.5.0 software was used for data reduction.  We present the X-ray
determined cluster center and observational information of the sample of all
37 LoCuSS clusters in Table~\ref{t:obs}.

In our X-ray mass modeling, we used the de-projected spectra to
measure the spatially resolved radial temperature distribution. A
double-$\beta$ model is adopted to fit the the ICM density
distribution as follows.  Firstly, the slope parameter $\beta$ is
derived by the local fit of the density profile at outer radii
($>0.5r_{500}$) by a single-$\beta$ model. As the inner component of
the ICM density distribution is insensitive to the slope, we fixed the
slope parameter $\beta$ and fit the ICM density distribution to derive
the inner and outer core radii of a double-$\beta$ model. As the slope
is locally fitted using the outer radii, there should not be a
significant bias due to the $\beta$ steepening in our X-ray mass
modeling. We therefore only show the mass distribution derived from
the spatially resolved temperature and density distributions with no
extrapolation to larger radii.  With deeper observations, the slope of
the temperature profile can be better determined, which will improve
the precision of the mass estimate. The Monte-Carlo error estimation
method used for the error propagation in our X-ray mass modeling
accounts for this. More details about the X-ray mass modeling can be
found in Appendix~\ref{s:result}.

\section{X-ray scaling relations: an empirical similarity}
\label{s:scale}

This section describes the detailed modeling and analysis of the X-ray
data.  Readers interested solely in the X-ray/lensing comparison may
wish to move straight to the summary in \S~\ref{s:scasummary}, and
continue from there.

For this sample the cluster masses are uniformly determined using
\emph{XMM-Newton} data. This guarantees minimum scatter due to the
analysis method. The scaling relations are generally parameterized by
a power law ($Y=Y_{0} X^{\gamma}$).  For each relation, we performed
the fitting with both normalization and slope free. Scatter describes
the dispersion between observational data and the best-fit values. The
scatter in the scaling relations directly measures the degree of
variation of ICM properties from the self-similar behavior. We list
the best-fit power law and the scatter in Table~\ref{t:mtx_lite}. The
ODRPACK package weights the errors of the data, and estimates the
uncertainties of the normalization parameter.

To compare the normalization between our sample and the samples from
other observations and simulations, we also performed the fitting with
the fixed slopes to the values as used for their samples. The
uncertainty in the difference of the normalization takes into account
the errors of both normalization of the two scaling relations for
comparison. Using a fixed slope might enlarge the segregation when the
mass range is not the same for two samples.

Based on the least scattered relations, $M_{\rm gas}$--$T$ and $L_{\rm
  bol}$--$Y_{\rm X}$, for our sample we checked the self-consistency
of the shapes (slopes) of the remaining scaling relations. The
resulting slopes generally appear consistent with the inferred slopes
for our sample. This allows us to fix the slopes to the
self-consistent values inferred from our sample and to study the
segregation and scatter in the scaling relations due to different
cluster populations.

\subsection{Cool cores and X-ray scaling relations}
\label{s:lxcoolcore}

We investigated the structural similarity of the scaled profiles of
the X-ray properties in Appendix~\ref{s:prof} in order to optimize the
definition of the cluster global temperature and luminosity to obtain
least scattered X-ray scaling relations with a brief description as
follows. The global temperature is derived by the volume average of
the radial temperature profile limited to the radial range of
0.2--0.5$r_{500}$ ($T_{0.2-0.5 r_{500}}=\frac{\Sigma_{k=1,n}
  (T_{k}/V_{k})}{\Sigma_{k=1,n} (1/V_{k})}$) as listed in
Table~\ref{t:global}, where $T_{k}$ and $V_{k}$ are the temperature
and volume of the $k$th radial shell of all the shells (from $1$st to
$n$th) between $0.2-0.5 r_{500}$. We used the values in Table~4 in
Zhang et al.  (2007), and found that the volume average of the radial
temperature profile gives nearly the same value as the spectroscopic
temperature with a spectroscopic temperature versus volume averaged
temperature ratio of 0.98.  According to simulations from Rasia et al.
(2005), our temperature definition will be lower than the emission
measured temperature by 20-30 per cent. The core radii populate a
broad range of values up to 0.2~$r_{500}$.  With the bolometric X-ray
luminosity including and excluding the $< 0.2 r_{500}$ region
(Table~\ref{t:global}), we show the normalized cumulative cluster
number count as a function of the fraction of the luminosity
attributed to the cluster core ($< 0.2 r_{500}$) in
Fig.~\ref{f:fraclcc}. Up to 80 per cent of the bolometric X-ray
luminosity is contributed by the cluster core ($< 0.2 r_{500}$). We
therefore use the X-ray luminosity corrected for the cool core for the
$R<0.2 r_{500}$ region ($L_{\rm bol}^{\rm corr}$ in
Table~\ref{t:global}).

\subsection{Mass versus temperature}

Hereafter $M_{500}^{\rm X}$ denotes the total cluster mass derived
from X-ray data at a density contrast of 500. In Fig.~\ref{f:my_2}, we
present the $M_{500}^{\rm X}$--$T$ relation for our sample and its
best fit, which gives a slope of $1.65\pm 0.26$.

We also performed the fitting with the slope fixed to the values 1.49 and
1.489 (consistent with our fit $1.65\pm 0.26$) as used in Arnaud et al. (2005)
and Vikhlinin et al (2006a). Our $M_{500}^{\rm X}$--$T$ relation agrees ($1
\pm 7$ per cent) with the $M$--$T$ relations for 13 low-redshift clusters
($z<0.23$, 0.7--9~keV) using \emph{Chandra} observations (Vikhlinin et al
2006a), and for 6 relaxed nearby clusters ($z<0.15$, $T>3.5$~keV) using
\emph{XMM-Newton} data (Arnaud et al.  2005), respectively. No evident
evolution is found for the $M$--$T$ relation after the redshift evolution
correction due to LSS growth (defined in Appendix~\ref{s:prof}).

Under the assumption that the temperature is constant within the cluster, the
$M_{500}^{\rm X}$--$T$ relation based on ROSAT observations and the
temperature measurements from literature related to the HIFLUGCS sample
(Reiprich \& B\"ohringer 2002; Popesso et al. 2005; Chen et al. 2007) gives
higher normalization by $\sim (40$--$50)\pm 20$ per cent than the
$M_{500}^{\rm X}$--$T$ relation based on \emph{Chandra}/\emph{XMM-Newton}
observations.  As the temperature distribution is used in the mass modeling to
obtain the \emph{XMM-Newton}/\emph{Chandra} results, the high normalization in
the ROSAT results is most probably due to the isothermal assumption in the
mass modeling. We fixed the slope to 1.48 as used for the hot clusters
($>3$~keV) with ASCA measured temperature profiles in Finoguenov et al.
(2001), and found the normalization is higher by $19 \pm 21$ per cent than the
value in Finoguenov et al.  (2001).  It indicates the importance of well
defined global temperature based on a spatially resolved temperature profile
on the scaling relation studies.

We also fixed the slope to 1.59 (consistent with our fit $1.65\pm
0.26$), and found that the normalization of our X-ray $M$--$T$
relation is lower by $19 \pm 6$ per cent than the one from simulations
(SPH code) including radiative cooling, star formation, and SN
feedback in Borgani et al. (2004). This is similar to the finding in
Nagai et al. (2007a) that the X-ray measured cluster masses within
$r_{500}$ are lower than the cluster masses from simulations.

The $M$--$T$ relation of our observations uses the temperature
comparable to the spectral-like temperature. The $M$--$T$ relation
from simulations in Borgani et al. (2004) uses the emission-weighted
temperature. As Rasia et al. (2005) mentioned that the normalization
of the $M$--$T$ relation from simulations in Borgani et al. (2004) is
higher by $\sim 30$ per cent using the spectral-like temperature
instead of the emission-weighted temperature, $19 \pm 6$ per cent
gives the minimum difference between our observations and simulations
in Borgani et al. (2004).

Evrard et al. (1996, emission weighted temperature) simulated ROSAT
observations of 58 nearby clusters ($z \sim 0.04$, 1--10~keV), for
which the normalization of the $M$--$T$ relation is higher by $40 \pm
16$ per cent than ours (with the slope fixed to 1.5). It might be due
to the neglect of complex gas physics (e.g. radiative cooling, star
formation and SN feedback) in their simulations.

\subsection{Mass versus $Y_{\rm X}$}

Simulations (Kravtsov et al. 2006) indicate that the X-ray analog of the
integrated SZ flux ($Y_{\rm X}= M_{\rm gas} \cdot T$) can be used as a robust
low scatter cluster mass indicator with less scatter compared to other X-ray
observables. Kravtsov et al. (2006, Fig.~7 in their paper; also see Table~4 in
Nagai et al. 2007b) found the $M$--$Y_{\rm X}$ relation for the sample in
Vikhlinin et al. (2006a) confirms the low scatter and shows an empirical
similarity. Maughan et al. (2007) investigated the $M$--$Y_{\rm X}$ relation
using \emph{Chandra} observations and found that the scatter varies with the
data quality. It is therefore important to calibrate the $M$--$Y_{\rm X}$
relation and to check whether $Y_{\rm X}$ does show the least scatter as a
mass proxy using high quality observations from \emph{XMM-Newton}.

We present the $M$--$Y_{\rm X}$ relation for the sample in
Fig.~\ref{f:my_1}, in which the best fit gives a slope of $0.624\pm
0.061$, consistent with $0.526 \pm 0.038$ derived from \emph{Chandra}
data (see Fig.~7 in Kravtsov et al.  2006 and Table~4 in Nagai et al.
2007b). We fixed the slope to $0.526$, and found an agreement ($2\pm12
$ per cent) of the normalization between $M$--$Y_{\rm X}$ relations
for our sample and theirs. With the slope fixed to 0.548, the
normalization for our sample is higher than for the sample in Arnaud
et al.  (2007) by 6 per cent, which is also based on \emph{XMM-Newton}
observations. There is no obvious evolution for the $M$--$Y_{\rm X}$
relation after the redshift evolution correction due to LSS growth. We
found that the scatter in $Y_{\rm X}$ for $M$--$Y_{\rm X}$ is similar
to the scatter in $T$ for $M$--$T$. The comparison between our sample
and the samples in Vikhlinin et al.  (2006a), Maughan et al. (2007)
and Arnaud et al.  (2007) might indicate that the slope of the
$M$--$Y_{\rm X}$ relation varies as a function of the mass range of
the sample.

As $Y_{\rm X}$ appears to be a good mass proxy, the cluster mass can
be estimated from $Y_{\rm X}$ via the $M$--$Y_{\rm X}$ relation.
Hereafter such a mass estimate is called the $Y_{\rm X}$ inferred
cluster mass, $M^{\rm Y_{\rm X}}$.

We also fixed the slope to 0.568 (consistent with our fit $0.624\pm
0.061$) as obtained in simulations (Nagai et al.\ 2007b), and found
the normalization of our X-ray $M$--$Y_{\rm X}$ relation is lower than
the one in simulations from Nagai et al. (2007b, the mass--observable
relation using the spectroscopic temperature and true mass; also see
Kravtsov et al. 2006) including cooling and galaxy feedback by $18 \pm
4$ per cent. From simulations in Nagai et al.\ (2007b), the
hydrostatic mass estimate is about 20\% lower than the true mass,
similar to the differences we observe. Here we compare our mass
measurements with ``the truth'' from simulations rather than the
hydrostatic masses derived from the simulations in Nagai et al.
(2007b). In part this is because it gives a consistent basis upon
which to compare lensing and X-ray masses with the simulations.

\subsection{Mass versus gas mass}

The total mass versus gas mass relation ($M$--$M_{\rm gas}$) for our
sample is shown in Fig.~\ref{f:my_1}, giving a slope of $0.906 \pm
0.078$, consistent with $0.811 \pm 0.067$ derived from \emph{Chandra}
data (Kravtsov et al. 2006, Nagai et al. 2007b). We fixed the slope to
0.811 as derived for the sample in Kravtsov et al. (2006), and found
that the normalization of the $M$--$M_{\rm gas}$ relation for our
sample agrees ($2 \pm 11$ per cent) with theirs. No evident evolution
is found for the $M$--$M_{\rm gas}$ relation after the redshift
evolution correction due to LSS growth.

We also fixed the slope to 0.894 (consistent with our fit $0.906 \pm
0.078$) as derived in Nagai et al. (2007b), and found the
normalization of our X-ray $M$--$M_{\rm gas}$ relation is lower than
the one in simulations in Nagai et al. (2007b) by $13 \pm 4$ per cent.

\subsection{Gas mass versus temperature}

Fig.~\ref{f:mgt} shows the $M_{gas, 500}$--$T$ relation for our sample
and its best fit giving a slope of $1.859\pm 0.187$.

We obtained a slope of 1.84 for the $M_{\rm gas,500}$--$T$ relation for the
sample in Vikhlinin et al. (2006a) using the scaling relations in Kravtsov et
al. (2006). We then fixed the slope to 1.84 (consistent with our fit $1.859\pm
0.187$), and found that the $M_{\rm gas,500}$--$T$ relation for our sample
agrees ($3 \pm 11$ per cent) with their nearby relaxed cluster sample based on
\emph{Chandra} observations. No evident evolution is found for the $M_{\rm
  gas}$--$T$ relation after the redshift evolution correction due to LSS
growth.

Castillo-Morales \& Schindler (2003) investigated a sample of 10 nearby
clusters ($0.03<z<0.09$, 4.7--9.4~keV) observed by ROSAT and ASCA. The
normalization for their samples is higher by $19 \pm 3$ per cent (no errors
available for the normalization in Castillo-Morales \& Schindler 2003) than
ours with the slope fixed to 1.8 as used in their work. The $M_{\rm
  gas,500}$--$T$ relation in Mohr et al. (1999) is based on a flux-limited
sample of 45 nearby clusters spanning a temperature range of 2--10~keV also
observed by ROSAT and ASCA. Their normalization is higher than ours by $81\pm
7$ per cent using their slope of 1.98. Chen et al. (2007) present the $M_{\rm
  gas,500}$--$T$ relation based on the non-cool core clusters (non-CCCs) with
$T>3$~keV in HIFLUGCS, in which the normalization is higher than ours by a
factor of 2 with the slope fixed to 1.8. The difference in the normalization
might be due to the difficulty of performing spatially resolved temperature
measurements with ROSAT/ASCA data. At that time, one had to use the $r_{500}$
defined from simulations (e.g. Evrard et al.  1996) which is larger than the
value obtained from observations based on the radial temperature profile using
\emph{XMM-Newton} data.

We also fixed the slope to 1.8 (consistent with our fit $1.859\pm
0.187$) as shown in Borgani et al. (2004), and found the normalization
of our X-ray $M_{\rm gas}$--$T$ relation is lower than the one from
simulations in Borgani et al. (2004, emission weighted temperature) by
$7 \pm 11$ per cent. If only considering the massive systems, we
observe a slightly smaller segregation between our observations and
their simulations as shown in Fig.~\ref{f:mgt}.

\subsection{Luminosity versus $Y_{\rm X}$}

The detailed calculation of the X-ray luminosity is described in
Appendix\ref{s:lumi}. The $L_{\rm bol}$--$Y_{\rm X}$ relation for this
sample is shown in Fig.~\ref{f:mgt}, with a best-fit slope of $0.951
\pm 0.081$.

We fixed the slope to 0.94 (consistent with our fit $0.951 \pm 0.081$) as used
in Maughan (2007), and found that the normalization for our sample is higher
than for the sample of 116 massive clusters ($0.1<z<1.3$) in their work by $10
\pm 4$ per cent. Compared to the observational errors, there is no obvious
evolution in the $L_{\rm bol}$--$Y_{\rm X}$ relation after the redshift
evolution correction due to LSS growth.

\subsection{Luminosity versus temperature}

The $L$--$T$ relations and their best fits, giving slopes of $2.127
\pm 0.323$ for $L_{\rm 0.1-2.4}^{\rm corr}$--$T$ and $2.614 \pm 0.319$
for $L_{\rm bol}^{\rm corr}$--$T$, are shown in Fig.~\ref{f:l0124t}.

The $L$--$T$ relation has been thoroughly investigated for nearby cluster
samples based on ROSAT and ASCA observations, for example, in (i) Markevitch
(1998, 30 clusters, $0.04<z<0.09$, $T>3.5$~keV), (ii) Arnaud \& Evrard (1999,
24 clusters, $z<0.37$, $T>2$~keV), (iii) Reiprich \& B\"ohringer (2002, in
which the flux-limited morphology-unbiased sample HIFLUGCS was initially
constructed and investigated. The fits used here are for the whole HIFLUGCS
sample including groups), (iv) Ikebe et al. (2002, a flux-limited sample of 62
clusters, $z<0.16$, 1--10~keV), and (v) Chen et al. (2007, HIFLUGCS).

The best-fit slope $2.614 \pm 0.319$ for the $L_{\rm bol}^{\rm corr}$--$T$
relation for our sample rules out the standard self-similar prediction $L
\propto T^2$ at $2 \sigma$ significance. The normalization for our sample
agrees with the normalization for the nearby samples using representative
non-CCCs in Markevitch (1998) on $6 \pm 11$ per cent level with the slope
fixed to 2.64, in Arnaud \& Evrard (1999) on $4 \pm 9$ per cent level with the
slope fixed to 2.88, and in Chen et al. (2007) on $13\pm 22$ per cent level
with the slope fixed to 2.78, respectively.

The normalization of the $L_{\rm 0.1-2.4keV}^{\rm corr}$--$T$ relation for our
sample is slightly lower than for the nearby samples in Markevitch (1998) by
$13\pm 11$ per cent with the slope fixed to 2.10, but shows an agreement with
the sample in Ikebe et al. (2002) on $5 \pm 38$ per cent level with the slope
fixed to 2.44.

The observed evolution in the $L$--$T$ relation is consistent with the
prediction of the LSS growth and is therefore not seen on the
corrected relations. 

There is no significant difference between the normalization of the
$L$--$T$ relations using ROSAT/ASCA data and
\emph{XMM-Newton}/\emph{Chandra} data, which is not the case for the
$M$--$T$ and $M_{\rm gas}$--$T$ relations. The spatial resolution is
of prime importance to get precise temperature gradient for a good
X-ray mass measurements.  However, $L$ and $T$ can be derived reliably
without spatially resolved spectroscopy. Theforefore the $L$--$T$
relation from ASCA/ROSAT is more likely to be in agreement with
Chandra/XMM than mass--observable relations.

The normalization is higher by $41 \pm 24$ per cent using a fixed slope of
2.598 for $L_{\rm 0.1-2.4keV}^{\rm corr}$--$T$ and by $36 \pm 23$ per cent
with the slope fixed to 2.98 for $L_{\rm bol}^{\rm corr}$--$T$ for the
HIFLUGCS sample in Reiprich \& B\"ohringer (2002). The temperature can be
biased low and the X-ray luminosity can be biased high for CCCs in the
HIFLUGCS sample. At the same time both temperature and luminosity can be
boosted high for merging systems. It is difficult to predict the combined
effects without looking into the detailed temperature and density profiles for
the HIFLUGCS sample.

\subsection{Luminosity versus mass}

The $L_{\rm bol}^{\rm corr}$--$M$ relation is shown in
Fig.~\ref{f:my_2} giving a best-fit slope of $2.325 \pm 0.701$. The
scatter in the $L$--$M$ relation is much larger than the scatter in
the other X-ray scaling relations for our sample.

The normalization of the $L$--$M$ relations for our sample is higher
than for the HIFLUGCS sample based on ROSAT/ASCA observations
(Reiprich \& B\"ohringer 2002; Popesso et al. 2005; Chen et al.  2007)
by a factor of $\sim 2$. The difference in the normalization might be
due to the isothermal assumption in their mass modeling and 
boosting effects due to CCCs in HIFLUGCS.

We also fixed the slope to 1.572 (smaller than our fit $2.325 \pm
0.701$) as derived from the $L_{\rm bol}$--$T$ and $M$--$T$ relations
in Borgani et al.  (2004, emission weighted temperature), and found
the normalization of our X-ray $L_{\rm bol}$--$M$ relation is lower
than the one from simulations in Borgani et al. (2004) by $18 \pm 32$
per cent.

\subsection{$r_{500}$ versus global temperature}

The best fit for our sample gives $r_{500}=10^{-0.38 \pm 0.07}
T_{0.2-0.5r_{500}}^{0.60\pm 0.08} E(z)^{-1} $~Mpc. Similar to the finding in
Arnaud et al. (2005) observing $r_{500} \propto T^{0.50 \pm 0.05}$ for 6
relaxed nearby clusters ($z<0.15$, 2--9~keV, $T>3.5$~keV) using
\emph{XMM-Newton} data. Pacaud et al. (2007) found $r_{500}=0.375
(T/keV)^{0.63} h_{73}(z)^{-1}$~Mpc based on the measurements in Finoguenov et
al. (2001), which is in agreement with ours within the uncertainty.

\subsection{Self-consistent scaling relations}
\label{s:empi}

As shown in Table~\ref{t:mtx_lite}, both $M_{\rm gas}$--$T$ and
$L_{\rm bol}$--$Y_{\rm X}$ relations show smaller scatter than the
remaining scaling relations. The errors (10 per cent) of the slope
parameters are also the smallest for these 2 relations compared to the
other relations. We therefore checked the self-consistency of the
empirical scaling relations based on these 2 relations ($M_{\rm gas}
\propto T^{1.859 \pm 0.187}$ and $L_{\rm bol} \propto Y_{\rm X}^{0.951
  \pm 0.081}$) for our sample as follows.

The definition of the X-ray bolometric luminosity gives $L_{\rm bol}
\propto M_{\rm gas}^2 T^{0.5} M^{-1}$. Combining with $L_{\rm bol}
\propto Y_{\rm X}^{0.951}$, we derive $ M_{\rm gas}^2 T^{0.5}
M^{-1}=(M_{\rm gas} \cdot T)^{0.951}$. Given $M_{\rm gas} \propto
T^{1.859}$, the $M$--$T$ relation is $M \propto T^{1.499}$, and the
$M$--$M_{\rm gas}$ relation is $M \propto M_{\rm gas}^{0.806}$,
respectively.

Combining $M_{\rm gas}
\propto T^{1.859}$ and $M \propto T^{1.499}$ yields 
$M \propto Y_{\rm X}^{0.524} $. 

We obtain $L_{\rm bol} \propto T^{2.719}$ on the basis of $L_{\rm bol}
\propto Y_{\rm X}^{0.951}$ and $M_{\rm gas}\propto T^{1.859}$, and 
$L_{\rm bol} \propto M^{1.814}$ on the basis of $L_{\rm bol} \propto
T^{2.719}$ and $M \propto T^{1.499}$, respectively. 

The slopes of the empirical scaling relations for our sample inferred from
$M_{\rm gas}\propto T^{1.859 \pm 0.256}$ and $L_{\rm bol} \propto Y_{\rm
  X}^{0.951\pm 0.081}$ are in agreement with the best-fit slope values in
Table~\ref{t:mtx_lite}. This indicates that our sample shows a
self-consistent, empirical similarity in the scaling relations.

\subsection{Cool core dependence of the normalization and scatter}

Different cluster populations might show different normalization and
scatter in the scaling relations. Most CCCs show simpler morphologies
than non-CCCs, though this is not always the case, e.g. Abell~2390
(Allen et al., 2001; Swinbank et al., 2006). We checked the cool core
dependence of the normalization and scatter using our sample. There is
no significant bimodal distributions in the X-ray scaling relations.
However, the CCCs and non-CCCs defined in Appendix~\ref{s:result} are
still used to divide the sample into two subsets in order to quantify
the segregation in the scaling relations due to cool cores. We fixed
the slope parameters to the prediction in \S~\ref{s:empi} on the basis
of $M_{\rm gas}\propto T^{1.859 \pm 0.256}$ and $L_{\rm bol} \propto
Y_{\rm X}^{0.951 \pm 0.081}$, and compared the normalization and
scatter in the scaling relations between the whole sample and
non-CCCs\footnote{The CCC subsample only consists of 9 clusters as
  shown in Table~\ref{t:catalog2}, which is statistically less
  sufficient compared to the non-CCC subsample to investigate the
  scatter, specially for the mean determined using a Gaussian model
  fit to the scatter histogram.} (Table~\ref{t:mtx_lite}). The
normalization agrees to better than 17 per cent.  The difference in
scatter is most marked in the $L$--$M$ relations (also cf. Chen et al.
2007). This may be due to the use of the hydrostatic equilibrium
assumption for the determination of the X-ray cluster mass.

\subsection{Mass--observable relations from
  X-ray observations and simulations}
\label{s:xmykinematic}

In the section we summarized the comparison between our sample and the
samples from numerical simulations. Given a wide variety for sources
of the simulated data, we provide here a comparison between the
different methods and also comment on the sensitivity of various
predictions in the adopted schemes.  In simulations the measurements,
such as the total mass and gas mass, are known precisely, while some
measurements, such as the X-ray luminosity, are calculated in a
straightforward manner, and some measurements, such as the observed
temperature, are method-dependent.  For example, Rasia et al.  (2005)
mention that the normalization of the $M$--$T$ relation from
simulations in Borgani et al.  (2004) is higher by $\sim 30$ per cent
using the spectral-like temperature instead of the emission weighted
temperature. We used the following simulations in the comparison:
Evrard et al. (1996, emission weighted temperature), Borgani et al.
(2004, emission weighted temperature), and Nagai et al.  (2007b,
spectroscopic temperature).  Further details regarding the numerical
resolution, cosmological parameters, and codes (grid versus particle)
of simulations can be found in the papers cited above.

The normalization of the X-ray mass--observable relations is lower
than the one from simulations ($M$--$Y_{\rm X}$ and $M$--$M_{\rm gas}$
in Nagai et al.  2007b; $M$--$T$ and $L_{\rm bol}$--$M$ in Borgani et
al. 2004, emission weighted temperature) by (1) $18 \pm 4$ per cent
for $M$--$Y_{\rm X}$, (2) $13 \pm 4$ per cent for $M$--$M_{\rm gas}$,
(3) $19 \pm 6$ per cent for $M$--$T$, and (4) $18 \pm 32$ per cent for
$L_{\rm bol}$--$M$, respectively.

The upward correction for the X-ray masses was suggested to be due to
the steepening of the ICM density and temperature distributions
assuming hydrostatic equilibrium by the comparison between
\emph{Chandra} observations (Kravtsov et al. 2006) and simulations
(Nagai et al. 2007b). For our sample, the density profiles are well
resolved spatially ($>100$ bins) while the temperature profiles are
poorly resolved spatially ($\ge 2$ bins).  Luckily, the universal
temperature profile in Vikhlinin et al.  (2005) is based on the
temperature profiles well resolved spatially for 13 low redshift
galaxy clusters. We therefore investigated the steepening of the ICM
density and temperature profiles in the outer region using our sample
for the density distribution, and the sample in Vikhlinin et al.
(2005) for the temperature distribution. We found that the mass can be
higher by up to 25 per cent at $r_{500}$ taking into account our ICM
density steepening and the temperature steepening using the universal
temperature profile in Vikhlinin et al.  (2005) in the hydrostatic
equilibrium assumption. One should be cautious with this point since
we have almost no resolution on temperature profiles between
$0.5r_{500}$ and $r_{500}$, where the temperature profile steepening
can not be verified. One requires deep X-ray observations with 8--10
radial bins for the temperature measurements on the per cent precision
level covering the area of more than $0.75 r_{500}$ for a
representative sample of more than 30 clusters to sufficiently verify
the steepening of the ICM temperature profiles to investigate the
bias.

\subsection{Summary of the X-ray scaling relations}
\label{s:scasummary}

The above comparison of the X-ray scaling relations between this sample and
published nearby and more distant samples shows that the evolution of the
scaling relations can be accounted for by the redshift evolution due to LSS
growth as given in Appendix~\ref{s:prof}.

Based on our sample, the scatter in the X-ray mass-observable scaling
relations is 7--40\% (0.03-0.15dex).

The normalization agrees to better than 17 per cent for the scaling
relations between the sample and the non-CCC subsample, with by far
the worst case being the $L_{\rm bol}$--$M$ relation. In general, the
slopes of the scaling relations (e.g.  $1.859\pm 0.187$ for $M_{\rm
  gas}$--$T$) indicate the need for non-gravitational processes. This
fits a general opinion that galaxy clusters show an empirical
similarity.

The normalization of the X-ray mass--observable relations (i.e. $M$--$T$,
$M$--$Y_{\rm X}$ and $M$--$M_{\rm gas}$) for our sample is lower by 13--19 per
cent with less than 6 per cent uncertainties than the relations from
simulations (Borgani et al. 2004; Kravtsov et al.  2006; Nagai et al. 2007b).

\section{$r^{\rm Y_{\rm X},X}_{500}$ and the reduced scatter in the
  mass--observable relations and $L_{\rm bol}$--$Y_{\rm X}$ relation}
\label{s:simuyx}

As shown in the left panels in Figs.~\ref{f:my_1}--\ref{f:my_2}, a few
merging systems show significant deviation from the best-fit
mass--observable relations. Since simulations (Nagai et al. 2007b)
have shown that $Y_{\rm X}$ is claimed to be a good mass proxy even
for merging clusters in which hydrodynamical equilibrium may be
disturbed, we may improve the radial scaling by using a mass profile
based on $Y_{\rm x}$ as a mass estimator. We used the following method
to reduce the deviation in mass--observable relations for merging
clusters by using the measured masses and X-ray observables at a new
radial scale, $r^{\rm Y_{\rm X},X}_{500}$.

The radial scale $r^{\rm Y_{\rm X},X}_{500}$ is derived as follows.
The procedure is initiated by deriving the best-fit $M$--$Y_{\rm X}$
relation using the X-ray masses $M_{500}$ and $Y_{\rm X}$ parameters
at the X-ray determined $r_{500}$ for the 37 LoCuSS clusters with the
slope fixed to 0.568, the prediction from simulations in Nagai et al.
(2007). This best fit is used to calculate the mass profile $M^{\rm
  Y_{\rm X}}(r)$ from the profile $Y_{\rm X}(r)=M_{\rm gas}(r) *
T_{\rm 0.2-0.5r_{500}}$. The radial scale, within which the density
contrast derived from the mass profile $M^{\rm Y_{\rm X}}(r)$ is 500,
is called $r^{\rm test}$ .

We re-fit the $M$--$Y_{\rm X}$ relation but using the X-ray masses and
$Y_{\rm X}$ parameters at $r^{\rm test}$ for the 37 LoCuSS clusters
with the fixed slope of 0.568. This best fit is used to re-derive the
mass profile $M^{\rm Y_{\rm X}}(r)$ from the profile $Y_{\rm X}(r)$,
and $M^{\rm Y_{\rm X}}(r)$ can then be used to re-determine the radial
scale $r^{\rm test}$, within which the density contrast, derived from
the mass profile $M^{\rm Y_{\rm X}}(r)$, is 500. We iteratively re-fit
the $M$--$Y_{\rm X}$ relation using the X-ray masses and $Y_{\rm X}$
parameters at $r^{\rm test}$, and re-determine the radial scale
$r^{\rm test}$ using $M^{\rm Y_{\rm X}}(r)$ till the best fit of the
$M$--$Y_{\rm X}$ relation does not vary any more for our sample. The
iteration is less than 10 times to fulfill the goal. The radial scale,
$r^{\rm Y_{\rm X},X}_{500}$, is the final $r^{\rm test}$. The radial
scale $r^{\rm Y_{\rm X},X}_{500}$ and the X-ray mass and $Y_{\rm X}$
parameter at $r^{\rm Y_{\rm X},X}_{500}$ for each cluster are listed
in Table~\ref{t:yx500}.

In Figs.~\ref{f:my_simuyx_1}--\ref{f:my_simuyx_2}, we show the
mass--observable scaling relations using the X-ray masses and
observables at $r^{\rm Y_{\rm X},X}_{500}$ for our sample. The
deviation from the best fit is indeed reduced for merging systems.

The normalization of the mass--observable relations using X-ray masses
and observables measured at $r^{\rm Y_{\rm X},X}_{500}$ is lower than
the one from simulations by, (1) $24 \pm 3$ per cent for $M$--$Y_{\rm
X}$, (2) $17 \pm 4$ per cent for $M$--$M_{\rm gas}$, (3) $17 \pm 5$ per
cent for $M$--$T$, and (4) $17 \pm 31$ per cent for $L_{\rm
bol}$--$M$, respectively. The difference in the normalization from
X-ray observations and simulations does not vary significantly
compared to the investigation using the X-ray masses and observables
measured at $r_{500}$. This difference in the normalization is
relatively independent of the scatter in the X-ray scaling relations.

\section{Weak lensing versus X-ray mass comparison}
\label{s:mpro}

Weak lensing and X-ray approaches provide independent cluster mass
estimates.  Comparison of X-ray and weak lensing masses is a powerful
test of systematic errors in cluster mass determination. In total,
weak lensing mass estimates are available in the literature for 19 of
the total sample of 37 clusters.  Ten of the 19 were observed over a
wide-field (i.e.\ out to $R>r_{200}$) with CFHT by Bardeau et al.,
(2005; 2007 -- hereafter the B07 sample).

Fifteen of the 19 were studied by Dahle (2006 -- hereafter the D06
subsample) using the shear estimator based on the method in Kaiser
(2000) but applied for small-field (2k chip, typically to
$0.4r_{500}$) data from NOT and UH88.  For this shear estimator, the
polarizability includes calculating various derivatives with respect
to galaxy flux, size and ellipticity. In practice, this is done
numerically by binning the observed galaxies in a 3-D flux-size-shape
space.  For small-field data from NOT and UH88, there are only $\sim
10^3$ suitable background galaxies available for each cluster, which
gives sparsely populated bins and thus introduces some additional
uncertainties to the calculation of the polarizability. This shear
estimator overcomes the problem by including galaxy information from a
number of different fields before the binning and polarizability
calculation, and produces more stable results.  The overlap between
the D06 and B07 samples is 6 clusters, giving a net literature-based
weak lensing sample of 19 clusters.

The following sections describe the mass comparison using $M_{500}$
(\S~\ref{s:m500}), and the mass measured at $r^{\rm Y_{\rm
    X},X}_{500}$ (\S~\ref{s:myx500}).

\subsection{$M_{500}$ comparison}
\label{s:m500}

To avoid the uncertainties caused by the extrapolation of the X-ray
measured mass profile beyond $r_{500}$ in the projection, we present
the weak lensing versus X-ray mass calibration using spherical masses
within radii of a density contrast of 500 ($M_{500}$). Hereafter
$M_{500}^{\rm wl}$ denotes the weak lensing mass within weak lensing
measured $r_{500}$. The average ratio of $M_{500}^{\rm wl}$ to
$M_{500}^{\rm X}$ gives $1.7 \pm 0.2$ for the D06 subsample, and $1.1
\pm 0.2$ for the B07 subsample, respectively. The better agreement
obtained from B07 subsample arises from the large field of view of the
CFHT data used by B07 (see above), in contrast the D06 data do not
reach $r_{500}$, and thus the D06 masses require extrapolation. This
shows the advantage of the LoCuSS project using uniformly observed and
reduced data with large FOV to calibrate the weak lensing and X-ray
masses and to investigate the scaling relations and their scatter.

The parameter $Y_{\rm X}$ shows less scatter than the X-ray mass as
shown in the $M$--$Y_{\rm X}$ relation. To check whether the scatter
in the weak lensing versus X-ray mass comparison is due to the scatter
in X-ray masses, we replace the X-ray mass estimates by the mass given
by $Y_{\rm X}$ at $r_{500}$ and the X-ray $M_{500}$--$Y_{\rm X}$
relation, $M^{Y_{\rm X}}_{500} = M (Y_{\rm X}(r_{500}))$.

In the upper panels in Fig.~\ref{f:szwlen}, we show $M_{500}^{\rm wl}$
against $M_{500}^{\rm X}$ and $M_{500}^{\rm Y_{\rm X}}$, respectively,
for the D06 and B07 subsamples.  The scatter is reduced using
$M_{500}^{\rm Y_{\rm X}}$ compared to the scatter using $M_{500}^{\rm
  X}$. The slope of the best-fit power law is $1.61 \pm 0.39$
($\chi^2=0.06/23$) for the $M_{500}^{\rm wl}$ versus $M_{500}^{\rm X}$
relation, and $1.57 \pm 0.38$ ($\chi^2=21.2/23$) for the $M_{500}^{\rm
  wl}$ versus $M_{500}^{\rm Y_{\rm X}}$ relation. The reduced $\chi^2$
for the $M_{500}^{\rm wl}$ versus $M_{500}^{\rm X}$ relation is much
smaller than 1, which might indicate the over-estimation of the errors
of $M_{500}^{\rm X}$. When the slope is fixed to 1, the normalization
of the best-fit power law is $1.28 \pm 0.14$ for the $M_{500}^{\rm
  wl}$ versus $M_{500}^{\rm X}$ relation, and $1.21 \pm 0.13$ for the
$M_{500}^{\rm wl}$ versus $M_{500}^{\rm Y_{\rm X}}$ relation,
respectively. On average, the weak lensing masses ($M^{\rm wl}_{500}$)
are higher than the X-ray masses ($M^{\rm X}_{500}$) by $28\pm 14$ per
cent according to the normalization. Systematically higher weak
lensing masses than X-ray masses at $r_{500}$ have also been reported
by Mahdavi et al. (2007).

The weak lensing to X-ray mass ratio derived above takes into account
the error weighting in both X-ray and weak lensing masses. The error
in X-ray masses does not vary much from cluster to cluster. However,
for a few clusters, the error in the weak lensing mass is a few times
as large as for the remaining clusters.  This may introduce a bias in
the derived weak lensing to X-ray mass ratio using the above error
weighted method. To avoid such a bias, we checked the mean of the
$M_{500}^{\rm wl}$ to $M_{500}^{\rm X}$ ratios and $M_{500}^{\rm wl}$
to $M_{500}^{\rm Y_{\rm X}}$ ratios as follows.  In the lower panels
in Fig.~\ref{f:szwlen}, we show the normalized differential cluster
number count as a function of the $M_{500}$ ratio. It can be fitted by
a Gaussian model $Y=Y_0\; exp^{-((X-X_0)/W)^2/2}$. The histogram of
$M_{500}^{\rm wl}/M_{500}^{\rm X}$ gives $Y_0=0.15 \pm 0.01$,
$X_0=0.99 \pm 0.05$, and $W=0.51 \pm 0.05$. And the histogram of
$M_{500}^{\rm wl}/M_{500}^{\rm Y_{\rm X}}$ gives $Y_0=0.15 \pm 0.01$,
$X_0=1.00 \pm 0.05$, and $W=0.45 \pm 0.05$. In logarithmic space, the
best Gaussian fit provides mean values lower than the values obtained
in linear space by $\sim 15$ per cent for both cases.  In linear
space, the mean is 0.99 with 51 per cent scatter in $M_{500}^{\rm
  wl}/M_{500}^{\rm X}$, and 1.00 with 45 per cent scatter in
$M_{500}^{\rm wl}/M_{500}^{\rm Y_{\rm X}}$, respectively. As shown in
the lower part of the lower panels in Fig.~\ref{f:szwlen}, the
discrepancy between $M_{500}^{\rm wl}$ and $M_{500}^{\rm X}$ (or
$M_{500}^{\rm Y_{\rm X}}$) can be up to 5 times the errors of
$M_{500}^{\rm wl}$ for individual clusters. To our surprise, the
clusters in the tails of the Gaussian distribution show both
pronounced mass discrepancies and very small errors in weak lensing
masses. It indicates that the clusters showing large mass
discrepancies may be over-weighted in the determination of the weak
lensing to X-ray mass ratio using the error weighted method. As a
consequence, the deviation of the weak lensing to X-ray mass ratio
from 1 can be biased high using the error weighted method. On the
contrary, the Gaussian fit to the histogram of the weak lensing to
X-ray mass ratio gives the value of the dominant population.

In the above test, we used the best-fit X-ray $M$--$Y_{\rm X}$
relation for our sample with the slope fixed to 0.568 predicted from
simulations (Nagai et al. 2007b) to derive $M_{500}^{\rm Y_{\rm X}}$
from $Y_{\rm X,500}$. We made an additional attempt to repeat such
tests by using the best-fit X-ray $M$--$Y_{\rm X}$ relations for our
sample with (i) a free slope, and (ii) the slope fixed to 0.524 as
predicted from the empirical self-similarity of our sample described
in \S~\ref{s:empi}, respectively. The mean of the $M_{500}^{\rm wl}$
to $M_{500}^{\rm Y_{\rm X}}$ ratios varies within 14 per cent, and the
scatter varies within 7 per cent in the tests via different X-ray
$M$--$Y_{\rm X}$ relation. Within the scatter, it holds that the mean
of the weak lensing to X-ray $M_{500}$ ratio is about 1.

\subsection{Mass comparison at  $r_{500}^{\rm Y_{\rm X},X}$}
\label{s:myx500}

In the above mass comparison, we used $M_{500}$ for X-rays taken at the X-ray
determined $r_{500}$ and $M_{500}$ for weak lensing taken from the weak
lensing determined $r_{500}$. This can introduce two sources of difference,
(1) the normalization of the mass profile, and (2) the radius up to which the
mass is integrated. To reduce such differences, we investigated the mass
comparison using the masses measured at the same radius as follows.

Despite the mass comparison using both weak lensing masses and X-ray
masses measured at the radial scale of $r_{500}^{\rm Y_{\rm X},X}$, we
also made a comparison between the weak lensing mass estimates and the
mass given by $Y_{\rm X}$ at $r_{500}^{\rm Y_{\rm X},X}$ and the X-ray
$M_{500}$--$Y_{\rm X}$ relation, $M^{Y_{\rm X}}(r_{500}^{\rm Y_{\rm
    X},X}) = M (Y_{\rm X}(r_{500}^{\rm Y_{\rm X},X}))$, as shown in
Fig.~\ref{f:szwlen_simuyx}.  When the slope is fixed to 1, the
normalization of the best-fit power law for the $M^{\rm
  wl}(r_{500}^{\rm Y_{\rm X},X})$ versus $M^{\rm X}(r_{500}^{\rm
  Y_{\rm X},X})$ relation is $1.09 \pm 0.08$, and for the $M^{\rm
  wl}(r_{500}^{\rm Y_{\rm X},X})$ versus $M^{\rm Y_{\rm
    X}}(r_{500}^{\rm Y_{\rm X},X})$ relation is $1.04 \pm 0.08$,
respectively.  On average, the weak lensing masses, $M^{\rm
  wl}(r_{500}^{\rm Y_{\rm X},X})$, are higher than the X-ray masses,
$M^{\rm X}(r_{500}^{\rm Y_{\rm X},X})$, by $9\pm 8$ per cent according
to the normalization.  Following the arguments presented in Churazov
et al.  (2007), the current comparison limits the non-thermal pressure
support to $9 \pm 8$ per cent according to the ratio between the weak
lensing and X-ray mass estimates.

With a Gaussian fit $Y=Y_0\; exp^{-((X-X_0)/W)^2/2}$, the histogram of
$M^{\rm wl}(r_{500}^{\rm Y_{\rm X},X})/M^{\rm X}(r_{500}^{\rm Y_{\rm
    X},X})$ gives $Y_0=0.19 \pm 0.02$, $X_0=1.16 \pm 0.04$, and
$W=0.42 \pm 0.04$.  And the histogram of $M^{\rm wl}(r_{500}^{\rm
  Y_{\rm X},X})/M^{\rm Y_{\rm X}}(r_{500}^{\rm Y_{\rm X},X})$ gives
$Y_0=0.22 \pm 0.02$, $X_0=1.01 \pm 0.03$, and $W=0.33 \pm 0.03$. The
scatter is therefore reduced to 33--42 per cent compared to 45--51 per
cent for $M_{500}^{\rm wl}/M_{500}^{\rm Y_{\rm X}}$ and $M_{500}^{\rm
  wl}/M_{500}^{\rm X}$.

\section{Weak lensing mass $M^{\rm wl}_{500}$ based mass--observable relations}
\label{s:ylmwlen}

We now investigate the scaling relations between weak lensing mass and
four representative X-ray observable proxies of the cluster mass: (1)
$Y_{\rm X}$, which has been suggested as the mass proxy with least
scatter on the basis of simulations (e.g.  Kravtsov et al. 2006), (2)
gas mass, which is an observable almost independent of X-ray mass, (3)
global temperature, and (4) X-ray bolometric luminosity. We present
the mass--observable relations using weak lensing masses for the
subsample of the 19 clusters on the right column of panels in
Figs.~\ref{f:my_1}--\ref{f:my_2}.

\subsection{Scatter and systematics}

The scatter using weak lensing masses is larger than the scatter using
X-ray masses by about a factor of 2 (Table~\ref{t:myl_wlen}).  This
difference is probably caused by a combination of (i) larger
measurement errors on the weak lensing mass measurements than on the
X-ray mass estimates, and (ii) possible under-estimation of the
scatter in the X-ray scaling relation due to intrinsic correlation
between the axes when both are based on the same dataset.  

We also seek a physical origin of the scatter in the lensing relation,
by dividing the clusters into single and merger X-ray morphologies
(defined in Appendix~\ref{s:kt}).  The single clusters only show
significantly less scatter in mass at fixed observable than the merger
clusters in the luminosity--mass relation, the reduction in scatter
being ${\sim}50\%$.  This indicates that, in general, merging systems
introduce insignificant scatter in the weak lensing mass based
mass--observable relations at $r_{500}$.  The larger scatter in the
luminosity--mass relation of merger clusters may indicate shock
heating of gas in merger clusters, although it is odd that this effect
does not also affect the mass--temperature relation. There is also no
evidence for a difference in normalization between single and merger
clusters in any of the scaling relations examined above except for the
$L_{\rm bol}$--$M$ relation, showing 32 per cent difference. This
might indicate that boost effects are significant on the X-ray
luminosity than on the temperature for merging systems.

\begin{figure*}
\begin{center}
\includegraphics[angle=270,width=8.5cm]{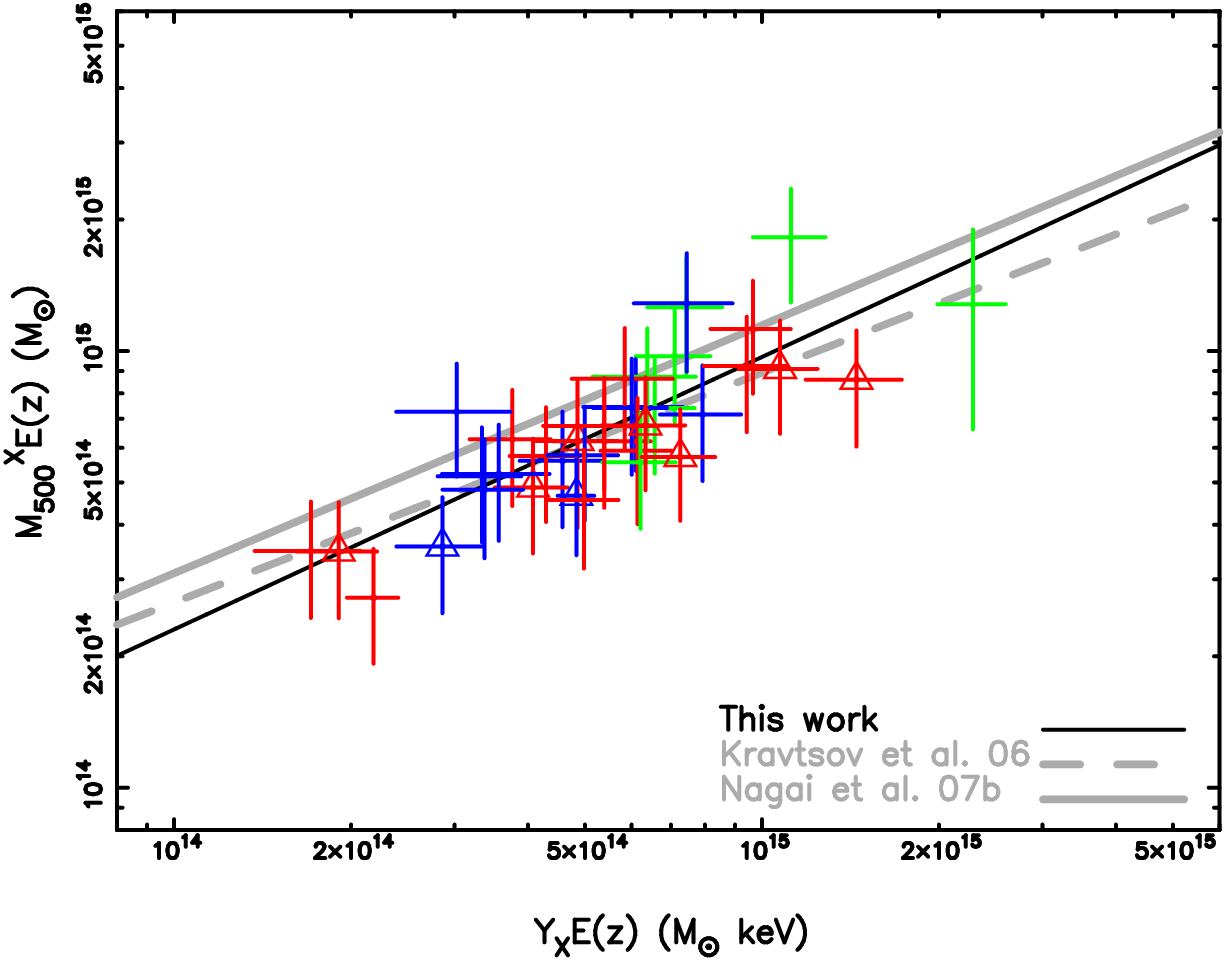}
\includegraphics[angle=270,width=8.5cm]{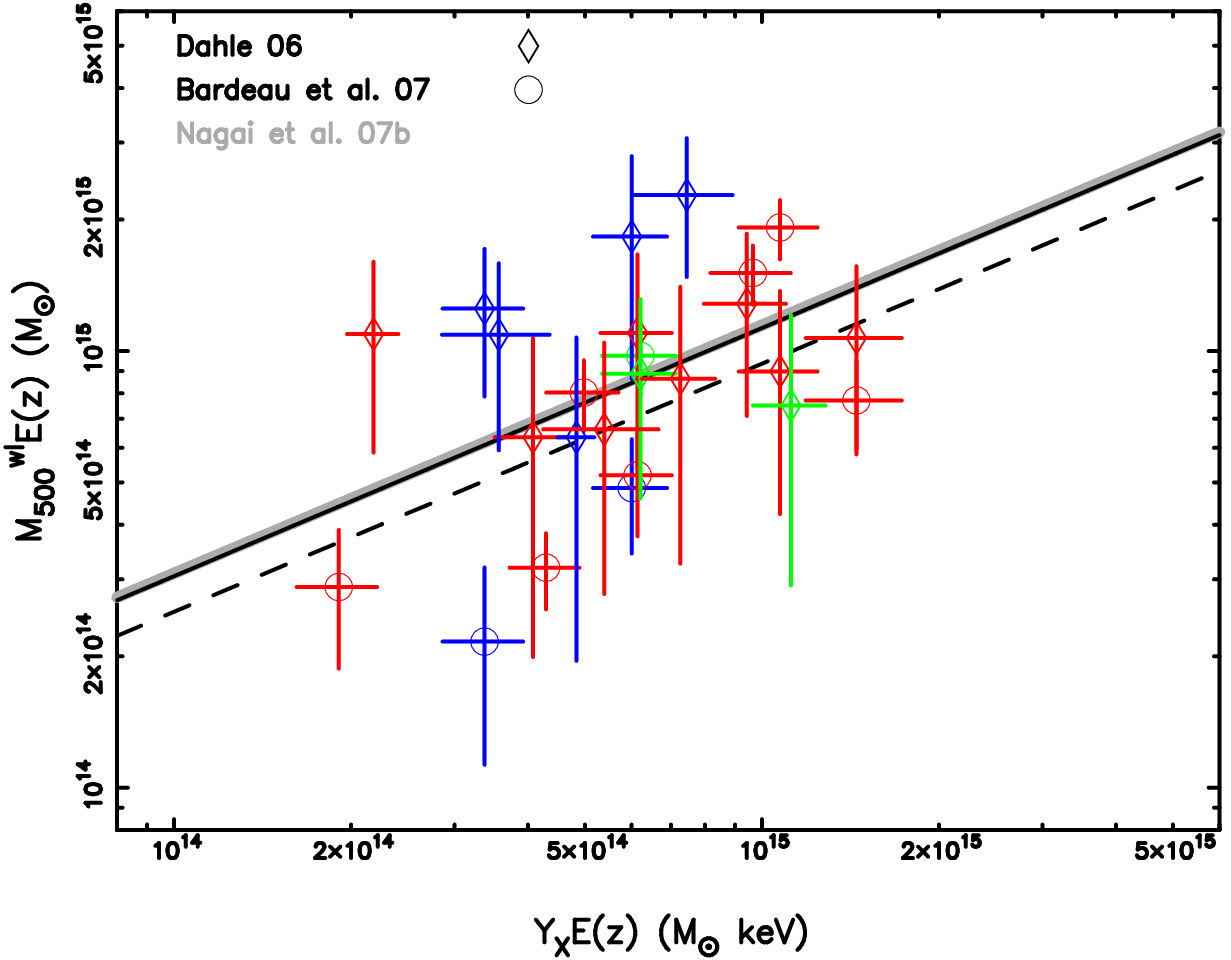}

\includegraphics[angle=270,width=8.5cm]{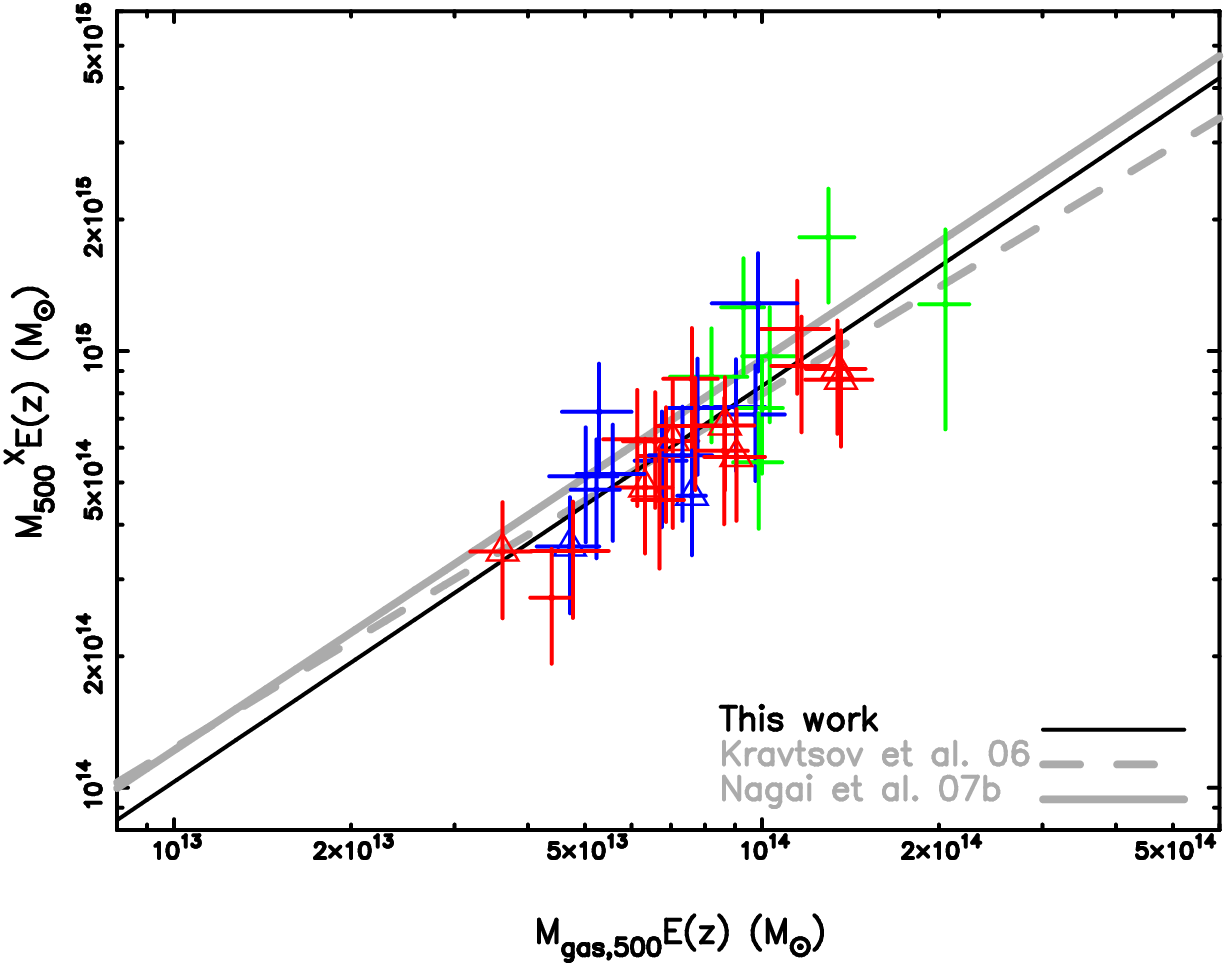}
\includegraphics[angle=270,width=8.5cm]{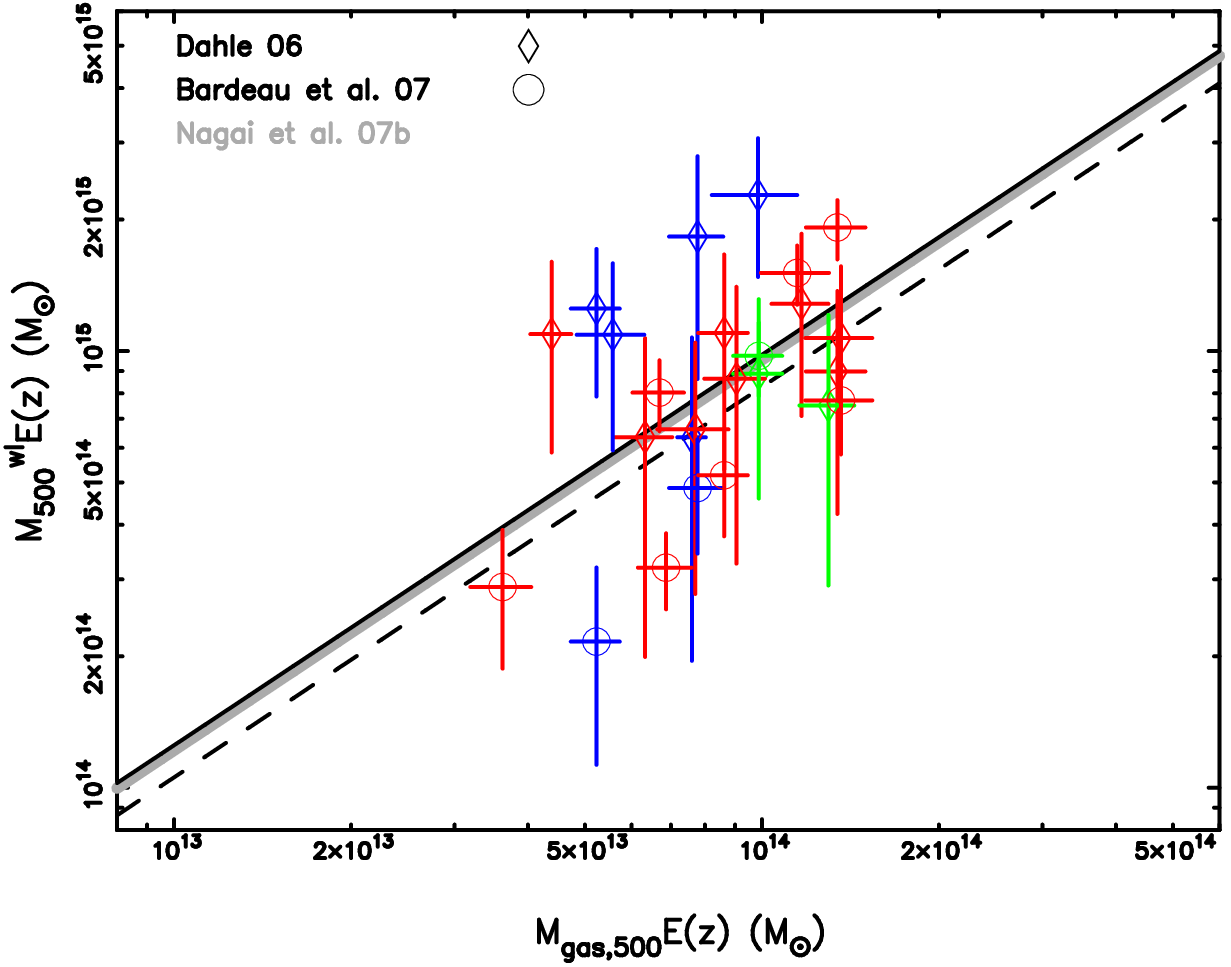}
\end{center}
\caption{Mass--$Y_{\rm X}$ (upper panels) and mass--gas mass (lower
  panels) relations using X-ray masses (left panels) and weak lensing
  masses (right panels). The CCCs are marked by triangles in the left
  panels. The best-fit power law in the right panels (solid, using
  weak lensing masses; dashed, using X-ray masses) with their slopes
  fixed to the values as found in simulations in order to compare the
  normalization (Table~\ref{t:myl_wlen}). The clusters appearing
  ``primary with small secondary'' or ``off-center'' morphology are in
  green, appearing ``elliptical'' or ``complex'' morphology are in
  blue, and appearing ``single'' morphology are in red using the
  classification in Jones \& Forman (1992).
  \label{f:my_1}}
\end{figure*}

\begin{figure*}
\begin{center}
\includegraphics[angle=270,width=8.5cm]{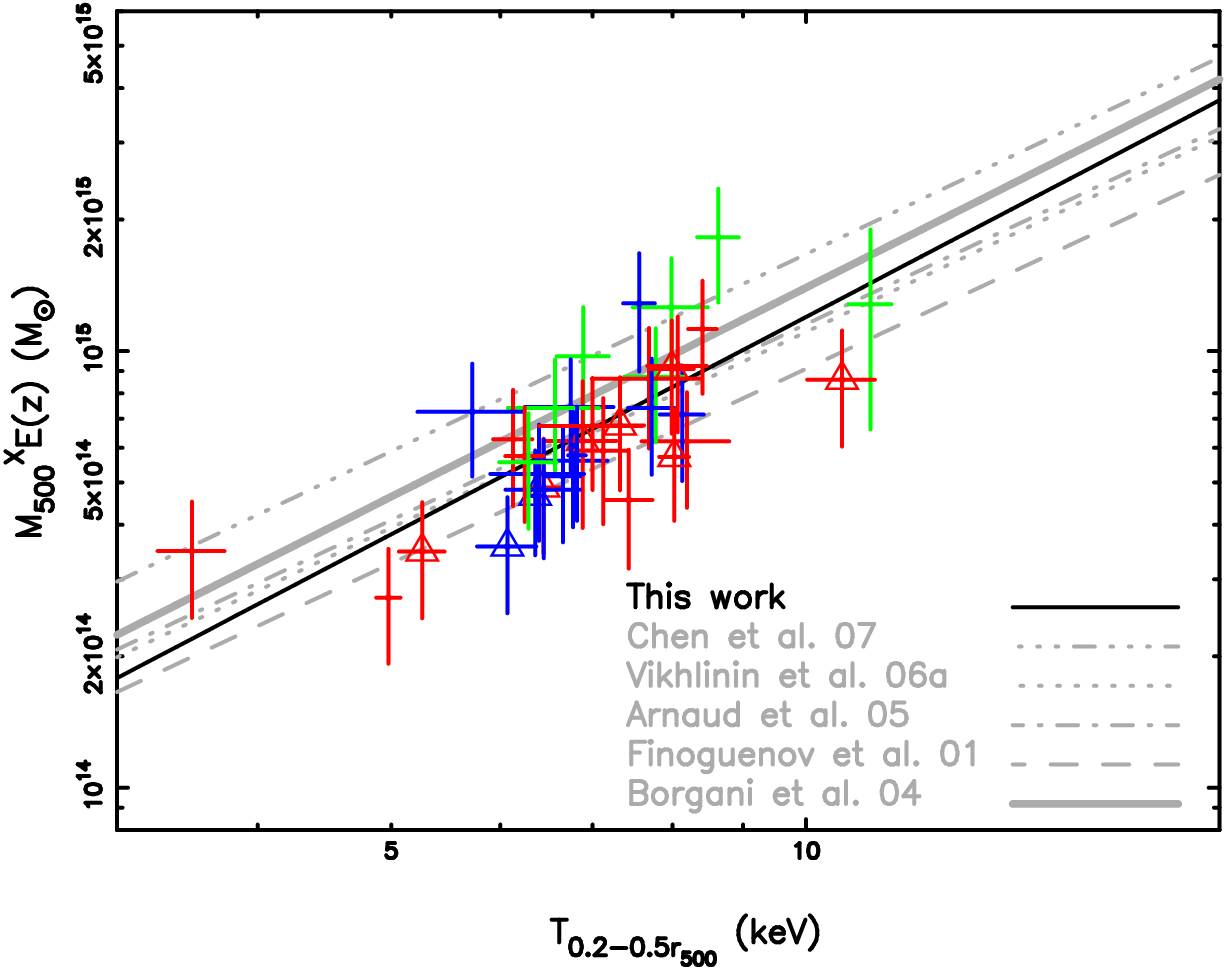}
\includegraphics[angle=270,width=8.5cm]{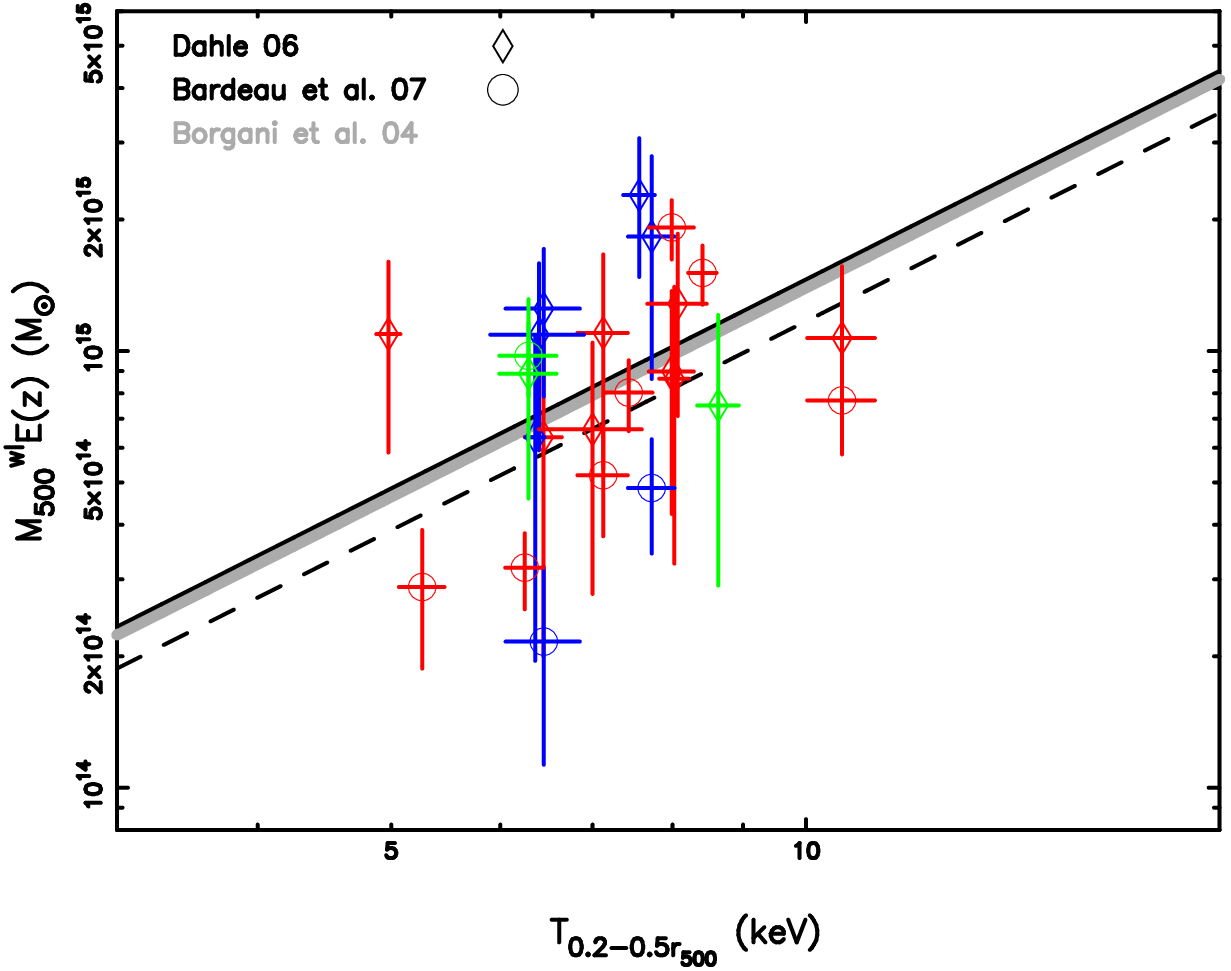}

\includegraphics[angle=270,width=8.5cm]{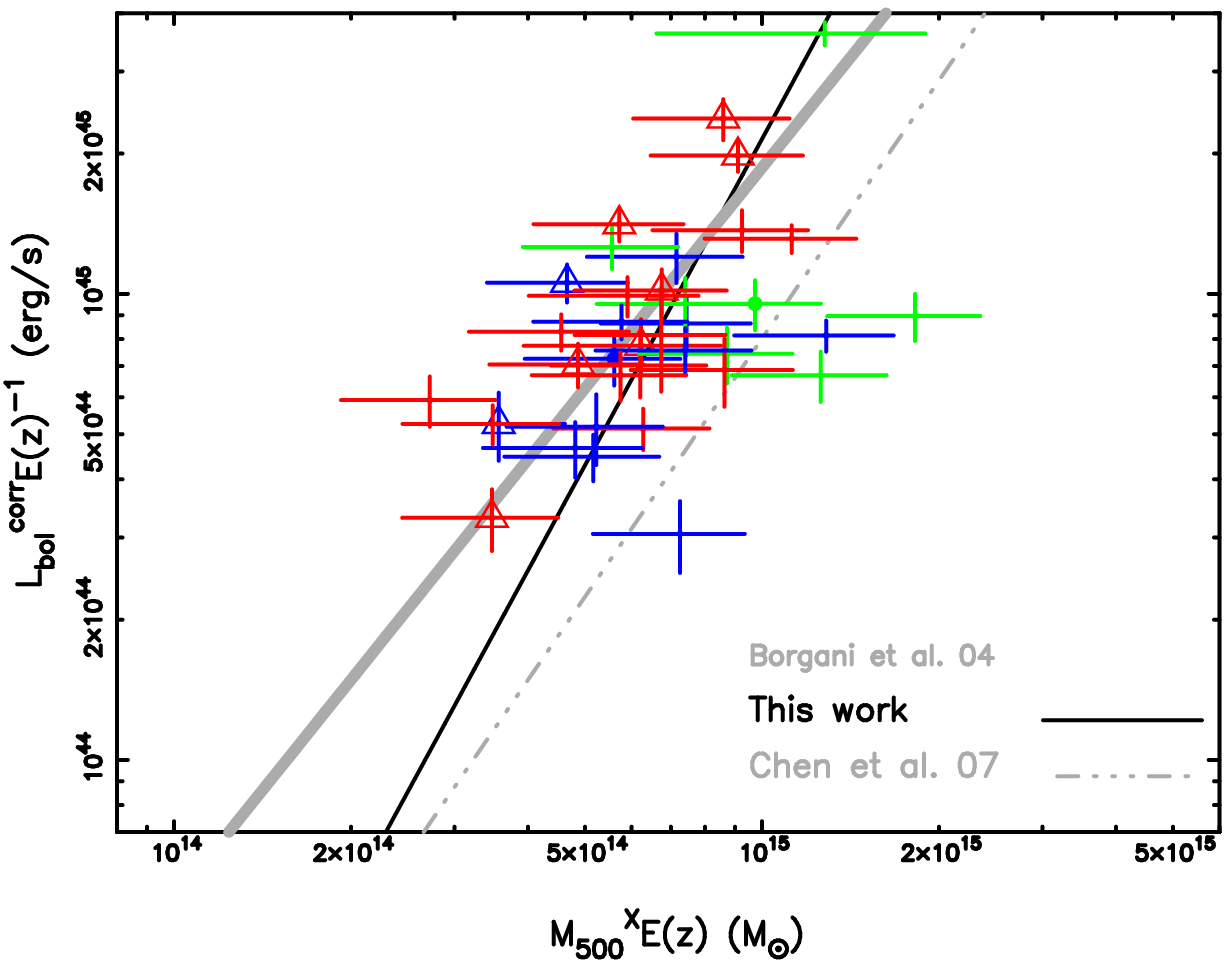}
\includegraphics[angle=270,width=8.5cm]{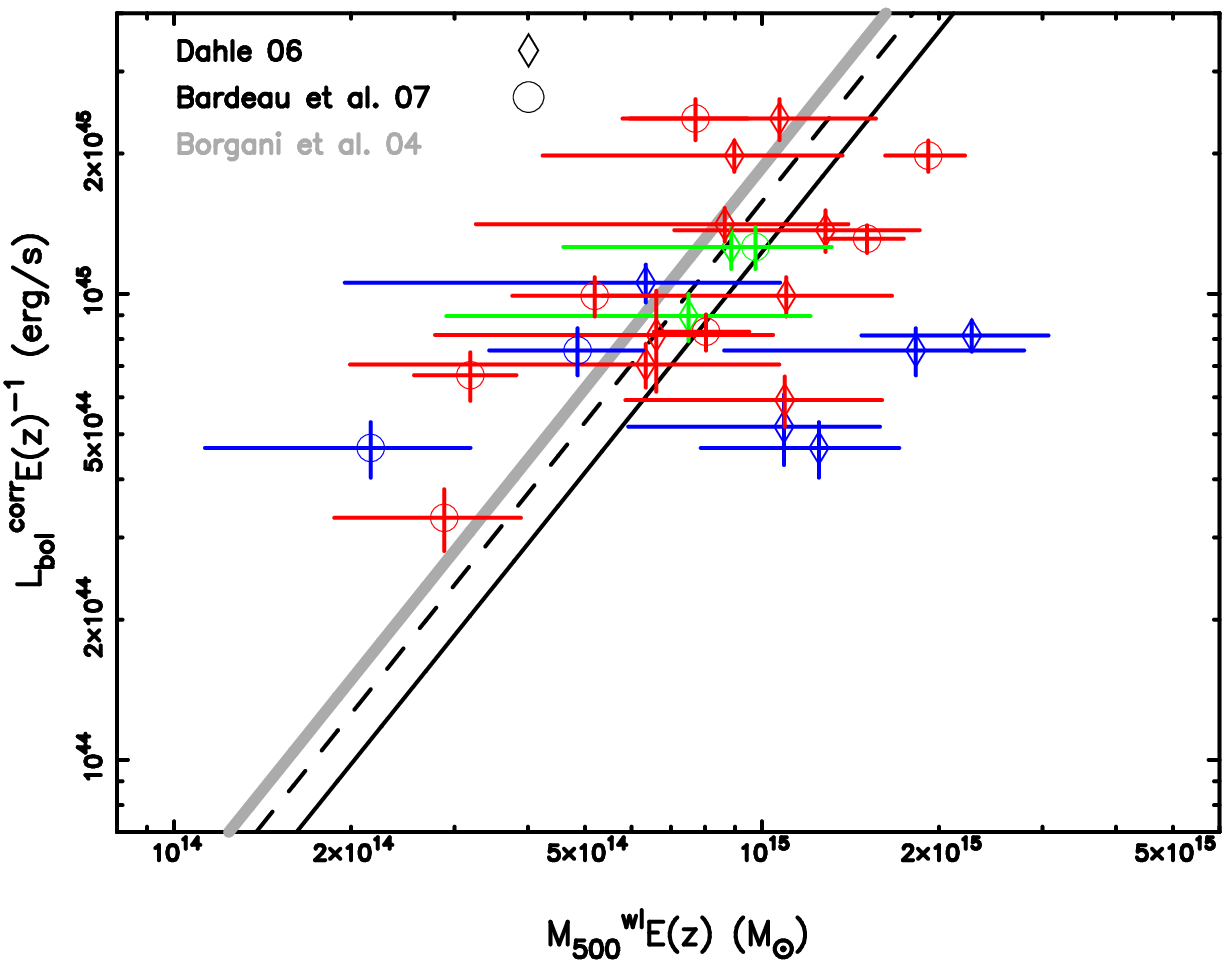}
\end{center}
\caption{Mass--temperature (upper panels) and luminosity--mass (lower 
panels) relations using X-ray masses (left panels) and weak lensing 
masses (right panels). The CCCs are marked by triangles in the left
panels. The best-fit power law in the right panels (solid, using
weak lensing masses; dashed, using X-ray masses) with their slopes
fixed to the values as found in simulations in order to compare the
normalization (Table~\ref{t:myl_wlen}).  The colors have the same
meaning as those in Fig.~\ref{f:my_1}.
\label{f:my_2}}
\end{figure*}

\begin{figure*}
\begin{center}
\includegraphics[angle=270,width=8.5cm]{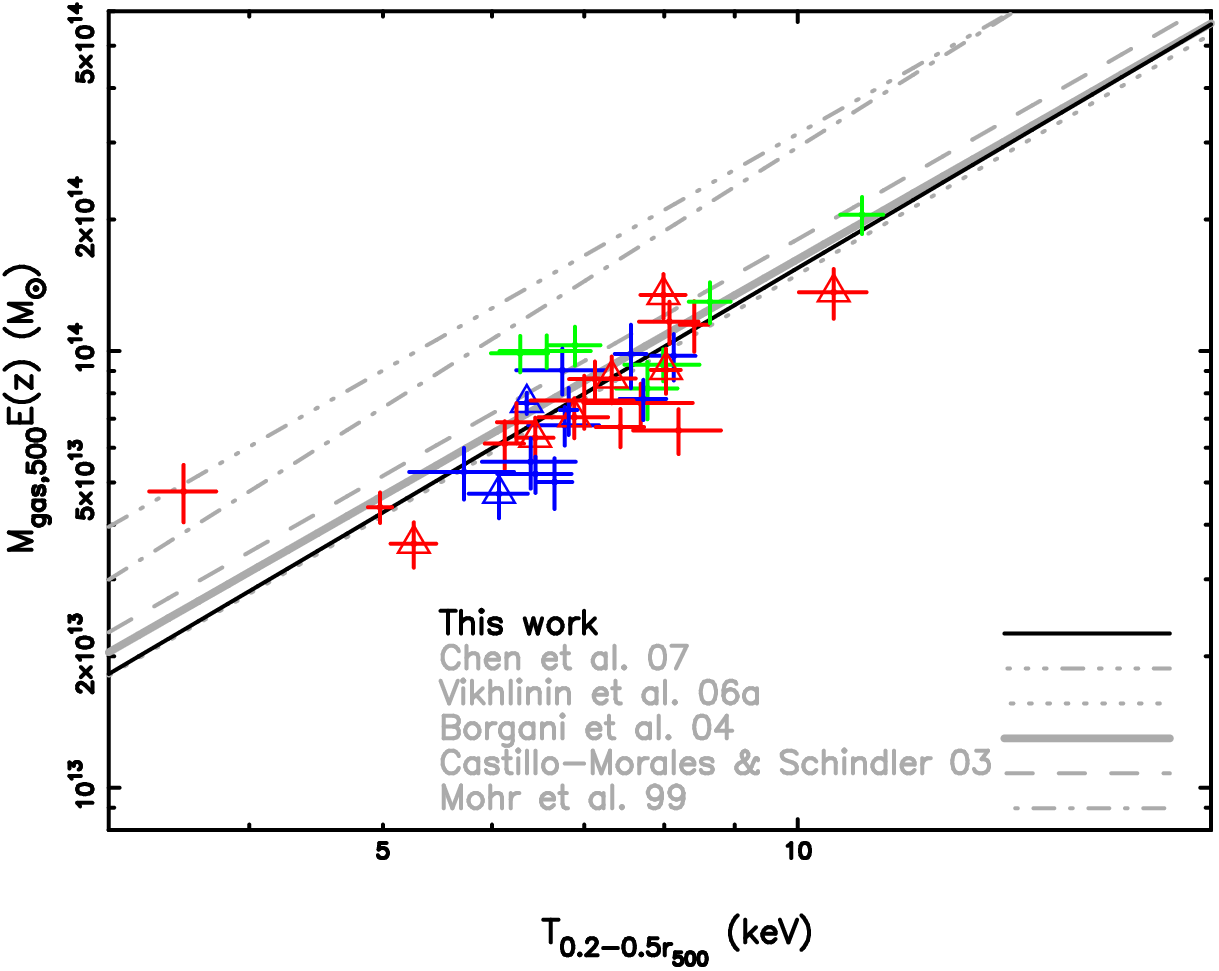}
\includegraphics[angle=270,width=8.5cm]{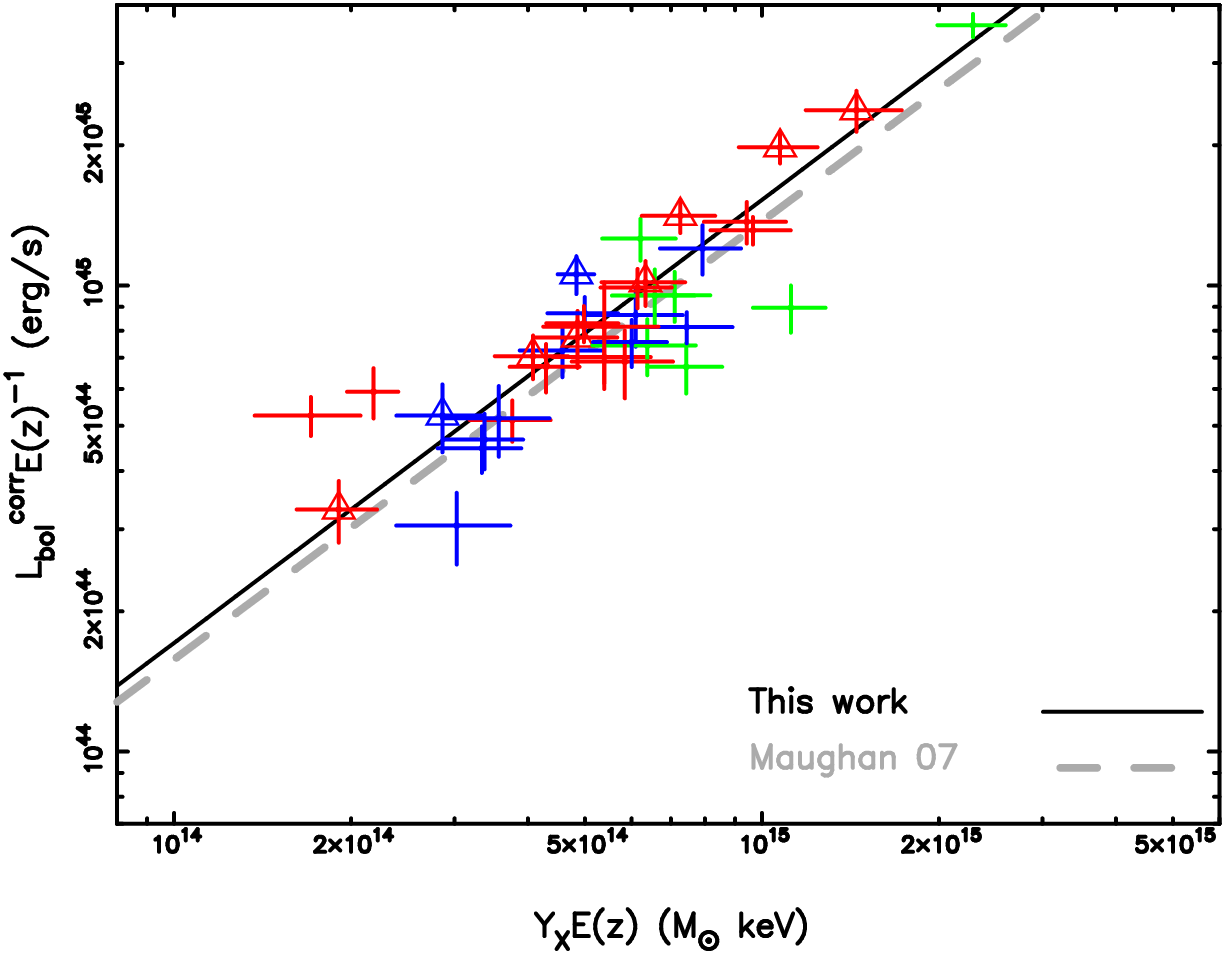}
\end{center}
\caption{{\it Left:} Gas mass--temperature
relation. {\it Right:} Luminosity--$Y_{\rm X}$ relation.  The CCCs are
marked by triangles. The colors have the same meaning as
those in Fig.~\ref{f:my_1}.
\label{f:mgt}}
\end{figure*}

\begin{figure*}
\begin{center}
\includegraphics[angle=270,width=8.5cm]{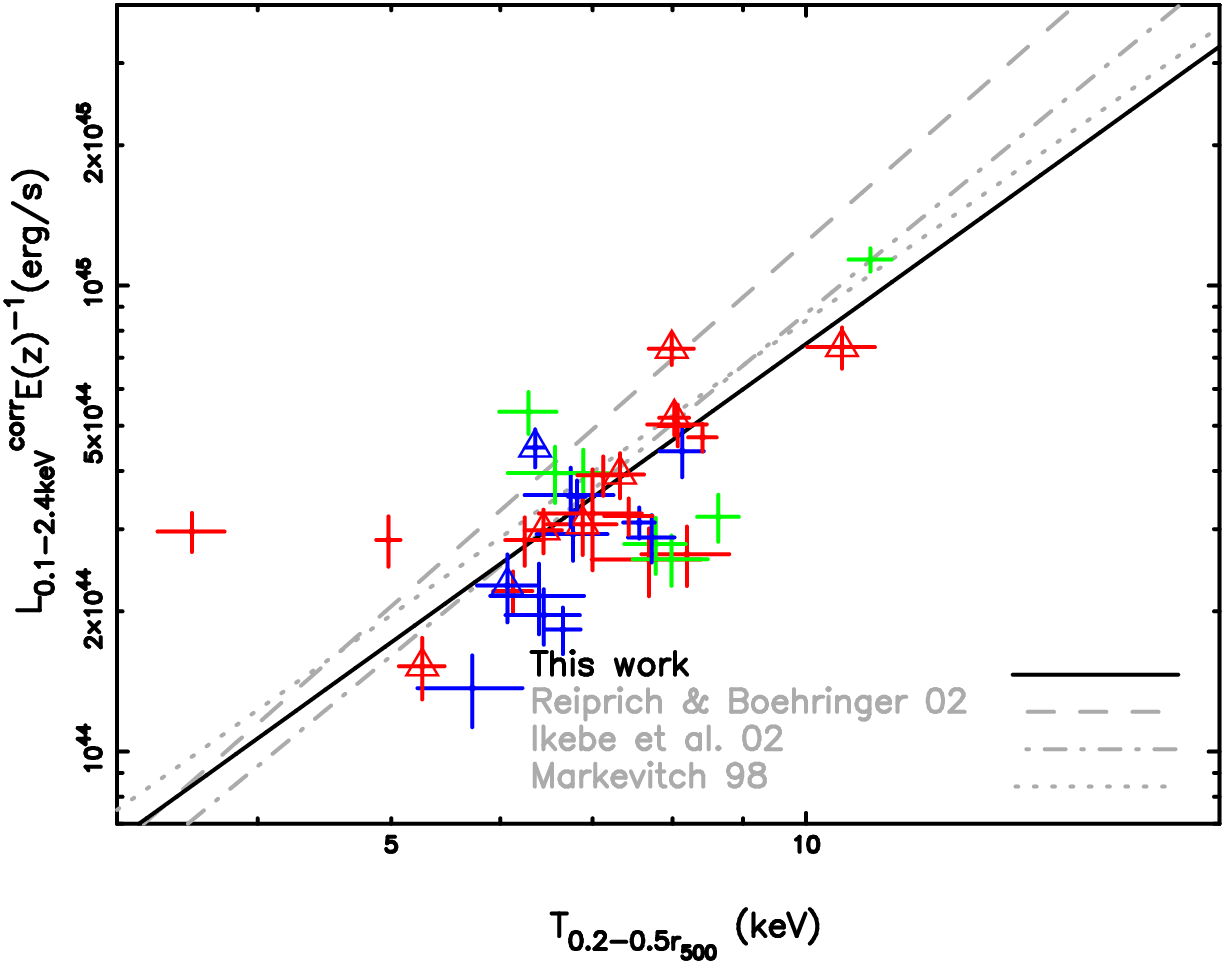}
\includegraphics[angle=270,width=8.5cm]{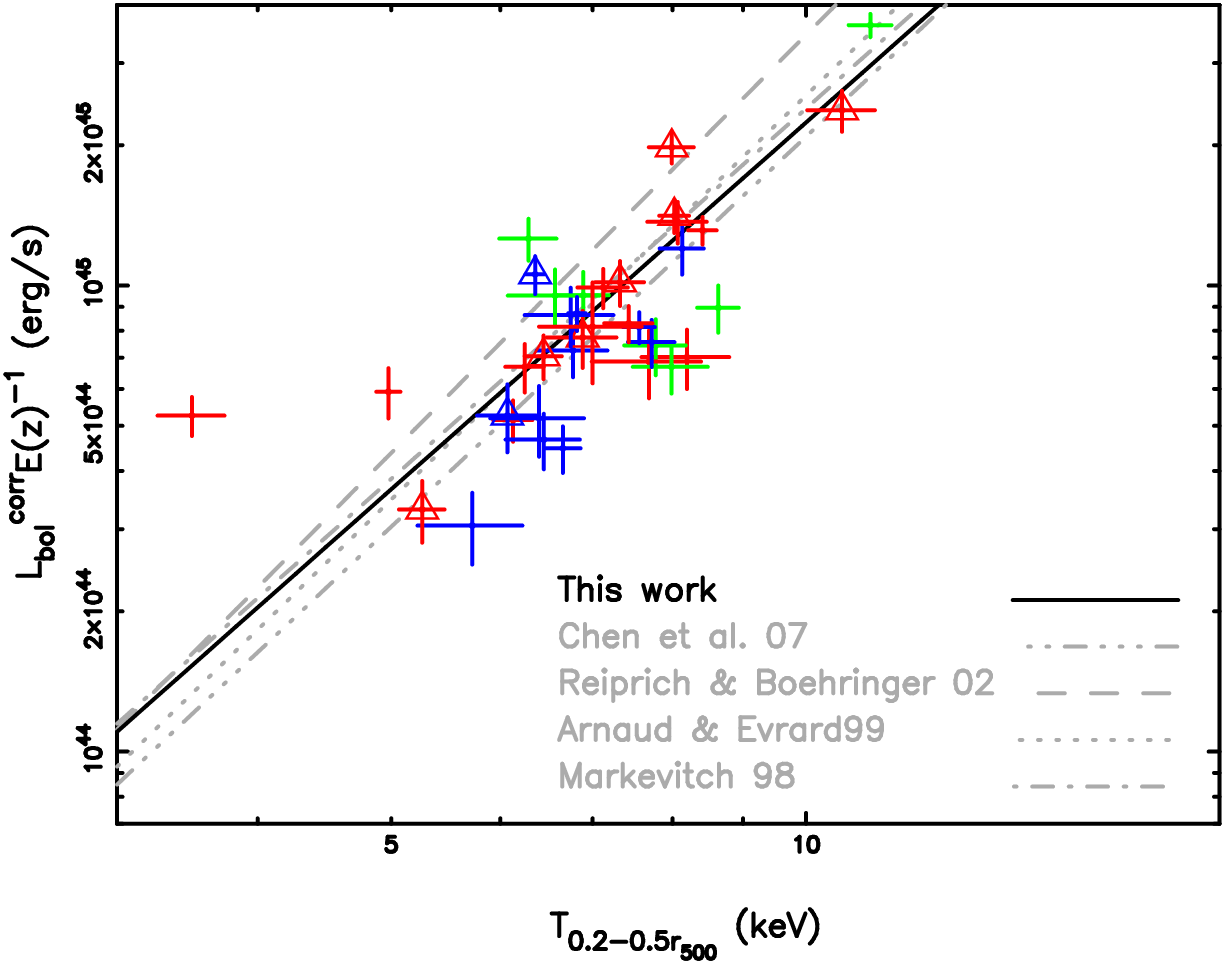}
\end{center}
\caption{X-ray luminosity in the 0.1--2.4~keV band versus temperature
(left) and bolometric X-ray luminosity versus temperature (right). The
CCCs are marked by triangles. The colors have the same meaning as
those in Fig.~\ref{f:my_1}.
\label{f:l0124t} }
\end{figure*}

The scatter in the scaling relations is nearly symmetric using weak
lensing masses irrespective of being classified as CCCs or non-CCCs
(Figs.~\ref{f:my_1}--\ref{f:my_2}). The normalization is nearly the
same using weak lensing masses between all 19 clusters and the non-CCC
subsample, giving up to 7 per cent difference (i.e. 7 per cent
difference in the $L_{\rm bol}$--$M$ relation). This indicates low
systematics in the weak lensing mass based mass--observable scaling
relations concerning being CCCs and non-CCCs.

\subsection{Observations versus simulations}
\label{s:ylmwlen_simuyxall}

To study the difference in the normalization between the
mass--observable relations using weak lensing masses and the one from
simulations, we fixed the slope to the value derived from simulations
and fitted the normalization for the 19 LoCuSS clusters.

\subsubsection{First Attempt}
\label{s:first}

First we use weak lensing masses computed within $r_{500}$ derived
from the lensing data with X-ray observables measured at the X-ray
determined $r_{500}$.

We fixed the slope to 0.568 as obtained from simulations (Nagai et al. 2007b)
for the $M$--$Y_{\rm X}$ relation, and found that the normalization using weak
lensing masses is marginally lower than the one from simulations (AMR code) in
Nagai et al. (2007b, also see Kravtsov et al. 2006) by $1 \pm 10$ per cent.

With the slope fixed to 0.894 predicted from simulations (Nagai et al.  2007b)
for the $M$--$M_{\rm gas}$ relation, we found a marginally higher
normalization of the $M$--$M_{\rm gas}$ relation using weak lensing masses
than the one from simulations (AMR code) in Nagai et al. (2007b) by $3\pm 11$
per cent.

Using the slope fixed to 1.59, we found that the normalization of the weak
lensing mass based $M$--$T$ relation is marginally higher by $4 \pm 12$ per
cent than the one from simulations (SPH code) including radiative cooling,
star formation, and SN feedback in Borgani et al. (2004).

The $L_{\rm bol}$--$M$ relation from simulations can be derived from the
$M$--$T$ and $L_{\rm bol}$--$T$ relations from simulations (SPH code) in
Borgani et al.  (2004), which shows a slope of 1.572. With the slope fixed to
1.572, we found the normalization of the $L_{\rm bol}$--$M$ relation using
weak lensing masses is lower than the one from simulations by $52 \pm 34$.

In summary, there is no statistically significant difference between
the normalization of the observed and simulated relations when using
lensing mass and X-ray observables derived from within $r_{500}$
obtained from each respective dataset.

\subsubsection{All measurements within $r_{500}^{\rm Y_X,X}$}
\label{s:myl_wlen_simuyx}

In \S~\ref{s:first}, the scatter in the weak lensing mass based
scaling is so much larger than the difference in the normalization
between observations and simulation. It is therefore not clear whether
there is consistency between weak lensing mass based mass--observable
relations and mass--observable relations from simulations. In
Table~\ref{t:myl_wlen_simuyx}, we present the best-fit normalization
of weak lensing mass based mass--observable relations using the
measurements at $r_{500}^{\rm Y_{\rm X},X}$ listed in
Table~\ref{t:yx500}. As shown in
Figs.~\ref{f:my_simuyx0_1}--\ref{f:my_simuyx0_2}, the scatter is
greatly reduced compared to the relations using the weak lensing
masses $M^{\rm wl}_{500}$ at the weak lensing determined $r_{500}$ and
X-ray observables measured at the X-ray determined $r_{500}$.

As derived in \S~\ref{s:simuyx}, the normalization of the mass--observable
relations using X-ray masses and observables measured at $r^{\rm Y_{\rm
    X},X}_{500}$ is lower than the one from simulations by, (1) $24 \pm 3$ per
cent for $M$--$Y_{\rm X}$, (2) $17 \pm 4$ per cent for $M$--$M_{\rm gas}$, (3)
$17 \pm 5$ per cent for $M$--$T$, and (4) $17 \pm 31$ per cent for $L_{\rm
  bol}$--$M$, respectively. The normalization of the mass--observable
relations using the weak lensing masses and X-ray observables measured at
$r^{\rm Y_{\rm X},X}_{500}$ is lower than the one from simulations by, i.e.
(1) $18 \pm 8$ per cent for $M$--$Y_{\rm X}$, (2) $17 \pm 8$ per cent for
$M$--$M_{\rm gas}$, (3) $6 \pm 10$ per cent for $M$--$T$, and (4) $25 \pm 32$
per cent for $L_{\rm bol}$--$M$, respectively. At $r^{\rm Y_{\rm X},X}_{500}$,
we found a similar difference in the normalization between simulations and
observations using either X-ray masses or weak lensing masses.  

We also derived the weak lensing mass based and X-ray mass based
mass--observable relations at $r^{\rm Y_{\rm X},X}_{500}$ without the
LSS evolution correction
(Figs.~\ref{f:my_simuyx0_1}--\ref{f:my_simuyx0_2}). For $M$--$Y_{\rm
  X}$, $M$--$M_{\rm gas}$, the conclusions are relatively similar as
for those with the LSS evolution correction as the LSS evolution
correction has the same impact on both variables. However, the LSS
evolution correction gives an opposite impact on $L_{\rm bol}$ and
$Y_{\rm X}$ (also $M$). Therefore the $L_{\rm bol}$--$Y_{\rm X}$
relation comparison between our sample and the sample in Maughan et
al. (2007) shows that the LSS evolution correction is required to
account for the evolutionary effect of the scaling relations and the
scatter is significantly reduced using $Y_{\rm X}$ measured at
$r_{500}^{\rm Y_{\rm X},X}$ (Fig.~\ref{f:lboly_simuyx2}).

\subsubsection{All measurements within $r_{500}^{\rm Y_X,wl}$}
\label{s:ylmwlen_simuyx1}

We followed a similar procedure outlined in \S~\ref{s:simuyx}, to
derive iteratively the radius $r_{500}$ for each cluster, but using
the mass derived from the weak lensing data. Hereafter we call it
$r^{\rm Y_{\rm X,wl}}_{500}$. 

The procedure is initiated by deriving the mass profile $M^{\rm Y_{\rm
    X}}(r)$ from the profile $Y_{\rm X}(r)=M_{\rm gas}(r) * T_{\rm
  0.2-0.5r_{500}}$ using the $M^{\rm wl}_{500}$--$Y_{\rm X}$ relation
(see Table~\ref{t:myl_wlen}). The radial scale, within which the
density contrast derived from the mass profile $M^{\rm Y_{\rm X}}(r)$
is 500, is called $r^{\rm test}$. We re-fit the $M^{\rm wl}$--$Y_{\rm
  X}$ relation but using the weak lensing masses and $Y_{\rm X}$
parameters at $r^{\rm test}$ for the 19 LoCuSS clusters with the slope
fixed to 0.568. This best fit is used to re-derive the mass profile
$M^{\rm Y_{\rm X}}(r)$ from the profile $Y_{\rm X}(r)$, and $M^{\rm
  Y_{\rm X}}(r)$ can then be used to re-determine the radial scale
$r^{\rm test}$, within which the density contrast, derived from the
mass profile $M^{\rm Y_{\rm X}}(r)$, is 500. We iteratively re-fit the
$M^{\rm wl}$--$Y_{\rm X}$ relation using the weak lensing masses and
$Y_{\rm X}$ parameters at $r^{\rm test}$, and re-determine the radial
scale $r^{\rm test}$ using $M^{\rm Y_{\rm X}}(r)$ till the best fit of
the $M^{\rm wl}$--$Y_{\rm X}$ relation does not vary any more for our
sample.  The radial scale, $r^{\rm Y_{\rm X},wl}_{500}$, is the final
$r^{\rm test}$. The X-ray and weak lensing masses at $r^{\rm Y_{\rm
    X},wl}_{500}$ are listed in Table.~\ref{t:ywl500}. The X-ray mass
and weak lensing mass based mass--observable relations at $r^{\rm
  Y_{\rm X},wl}_{500}$ are shown in
Figs.~\ref{f:my_simuyx1_1}--\ref{f:my_simuyx1_2} and
Table~\ref{t:myl_wlen_simuyx1}. The normalization of the
mass--observable relations using the X-ray masses and X-ray
observables measured at $r^{\rm Y_{\rm X},wl}_{500}$ is lower than the
one from simulations by, i.e. (1) $20 \pm 3$ per cent for $M$--$Y_{\rm
  X}$, (2) $16 \pm 4$ per cent for $M$--$M_{\rm gas}$, (3) $18 \pm 5$
per cent for $M$--$T$, and (4) $13 \pm 31$ per cent for $L_{\rm
  bol}$--$M$, respectively.  The normalization of the mass--observable
relations using the weak lensing masses and X-ray observables measured
at $r^{\rm Y_{\rm X},wl}_{500}$ is lower than the one from simulations
by, i.e. (1) $17 \pm 8$ per cent for $M$--$Y_{\rm X}$, (2) $14 \pm 8$
per cent for $M$--$M_{\rm gas}$, (3) $14 \pm 10$ per cent for
$M$--$T$, and (4) $9 \pm 32$ per cent for $L_{\rm bol}$--$M$,
respectively.

\subsubsection{All measurements within $r_{500}^{\rm Y_X,si}$}
\label{s:ylmwlen_simuyx2}

We also derived $r_{500}$, the radius by combing the $Y_{\rm
  X}(r)$ profile with the $M$--$Y_{\rm X}$ relation from simulations
in Nagai et al.  (2007b). The $M_{500}$--$Y_{\rm X}$ relation in Nagai
et al. (2007b) can be used to derive the mass profile $M^{\rm Y_{\rm
    X}}(r)$ from the profile $Y_{\rm X}(r)=M_{\rm gas}(r) * T_{\rm
  0.2-0.5r_{500}}$. The radius, within which the density contrast
derived from the mass profile $M^{\rm Y_{\rm X}}(r)$ is 500, is the
radial scale. Hereafter we call it $r^{\rm Y_{\rm X},si}_{500}$. The
X-ray and weak lensing masses at $r^{\rm Y_{\rm X},wl}_{500}$ are
listed in Table.~\ref{t:ywl500si}. The X-ray mass and weak lensing
mass based mass--observable relations at $r^{\rm Y_{\rm X},si}_{500}$
are shown in Figs.~\ref{f:my_simuyx2_1}--\ref{f:my_simuyx2_2} and
Table~\ref{t:myl_wlen_simuyx2}. The normalization of the
mass--observable relations using the X-ray masses and X-ray
observables measured at $r^{\rm Y_{\rm X},si}_{500}$ is lower than the
one from simulations by, i.e. (1) $15 \pm 3$ per cent for $M$--$Y_{\rm
  X}$, (2) $16 \pm 5$ per cent for $M$--$M_{\rm gas}$, (3) $8 \pm 5$
per cent for $M$--$T$, and (4) $26 \pm 31$ per cent for $L_{\rm
  bol}$--$M$, respectively.  The normalization of the mass--observable
relations using the weak lensing masses and X-ray observables measured
at $r^{\rm Y_{\rm X},si}_{500}$ is lower than the one from simulations
by, i.e. (1) $15 \pm 8$ per cent for $M$--$Y_{\rm X}$, (2) $21 \pm 8$
per cent for $M$--$M_{\rm gas}$, (3) $6 \pm 10$ per cent for $M$--$T$,
and (4) $25 \pm 32$ per cent for $L_{\rm bol}$--$M$, respectively.

\bigskip

We thus draw similar conclusions concerning the scatter and
normalization in the mass--observable relations using the weak lensing
masses at $r^{\rm Y_{\rm X,wl}}_{500}$ and $r^{\rm Y_{\rm
X},si}_{500}$, respectively, as for those measured at $r^{\rm Y_{\rm
X},X}_{500}$. As precisely shown above, the normalization of the
$M$--$Y_{\rm X}$, $M$--$M_{\rm gas}$, and $M$--$T$ relations using
X-ray mass estimates is lower than the values from simulations by
$\sim 20$ per cent at $3\sigma$ significance. In good agreement, these
relations based on weak lensing masses also show lower normalization
by $\sim 20$ per cent than the values from simulations but at $\sim
2\sigma$ significance.

In Fig.~\ref{f:lboly_simuyx2}, we also summarize the
luminosity--$Y_{\rm X}$ relations at $r_{500}^{\rm Y_{\rm X},X}$,
$r_{500}^{\rm Y_{\rm X},wl}$ and $r_{500}^{\rm Y_{\rm X},si}$. The
upper panels show that the LSS evolution correction accounts for the
segregation of the Luminosity versus $Y_{\rm X}$ relations between
cluster samples at different redshifts.

\section{Discussion} 
\label{s:dis}

\subsection{Scatter and bias in weak lensing mass estimates}
\label{s:scatter}

Since we observed a large scatter in weak lensing masses for the
sample, we summarize the possible sources which can enlarge scatter in
weak lensing mass estimates as follows.

(1) Accuracy of the faint galaxy shape measurements.  Residuals in the
removal of the point spread function (PSF) can masquerade as spurious
shear signal, and/or introduce strong systematic errors, often
referred to as B-models.  Future LoCuSS weak lensing studies will use
multiple shape measurement techniques to validate the fidelity of the
PSF removal.

(2) Size of FOV. Ideally the FOV should extend beyond the virialized
region of each cluster, which for a $10^{15}M_\odot$ cluster at
$z{\simeq}0.2$ corresponds to $r_{200}{\simeq}2.0\,{\rm Mpc}$, or
${\gs}10^{\prime}$. Small FOV weak lensing observations typically
incur systematic errors in cluster mass determination due to the
mass-sheet degeneracy and the need to extrapolate the observed shear
signal out to larger radii.  Dahle's (2006) data are a good example of
this; the NOT and UH 88in observations reach a maximum radius of just
${\sim}7^{\prime}$. In contrast, the LoCuSS pilot study used wide
field data from CFHT (Bardeau et al.\ 2007), and LoCuSS is using
Suprime-CAM on Subaru to make weak lensing measurements (Okabe et al.,
in prep.) -- both reach out to $r{\gs}15^{\prime}$.  This will
therefore not be a major concern for future LoCuSS papers.

(3) Accuracy of the background galaxy redshifts. The weak lensing mass
estimate depends on the assumed redshift distribution of background
galaxies, including the extent to which the background galaxy catalog
is contaminated by faint cluster members and foreground galaxies.
Several color-selection techniques have been developed recently
(Broadhurst et al. 2005; Limousin et al. 2007) to deal with this
issue.  We will apply these methods within LoCuSS weak lensing
analysis and explore refined techniques to measure reliable photmetric
redshifts to minimize contamination of the shear catalogs.

(4) Projection of LSS.  The scatter in the weak lensing masses could
be slightly boosted by effects of projection due to LSS.  For example,
Hoekstra (2003) investigated effects of uncorrelated LSS on weak
lensing mass measurements, showing that the accuracy of the measured
mass of an NFW model does not improve once the outer radius for the
fitting reaches the angular scale $\sim 15^{\prime}$.  Fortunately,
the LoCuSS clusters are at redshift around 0.2, which places their
virial radii at $r{\ls}13^{\prime}$.  A similar investigation of the
impact of correlated LSS is also required.

(5) Mass models. The weak lensing mass measurements using a tangential
shear fitting are dependent on the assumed mass model, e.g.\ NFW model
or SIS model.  Usually an NFW model is considered to be more
representative for the mass distribution in clusters than an SIS
model. However, for some clusters, especially merging clusters, a NFW
model is not favored statistically, and the SIS model has only been
ruled out at large significance in a small number of cases (e.g.\
Kneib et al., 2003).  LoCuSS weak lens modeling will test for the
presence of substructures and attempt to fit multi-component models.
The superb image quality of Subaru, and consequently higher number
density of background galaxies will also increase the probability that
we can discriminate between NFW and SIS models.

(6) Effects of triaxiality and interior substructures. Clowe et al.\
(2004) investigated effects of cluster triaxiality and substructures
on the mass measurements and found that the triaxiality is more
dominant. The value of $r_{200}$ obtained by fitting the shear profile
with an NFW model can vary by 10--15 per cent due to different
projection, which corresponds to a mass ($M_{200}$) change of $\sim40$
per cent. King et al. (2001) show that the mass measurement can be
affected by interior substructures by up to 10 per cent.

\subsection{Cluster mass comparison between observations and simulations }
\label{s:simobs}

The cluster mass estimates using different approaches can be different
as e.g.  boost effects caused by merging systems might be different on
the X-ray and weak lensing masses. For the purpose of precise cluster
cosmology, one requires accurate knowledge of the systematics in X-ray
cluster mass estimates and weak lensing cluster mass estimates and the
difference between cluster mass estimates from simulations and
observations.

Despite the larger individual measurement uncertainties in weak
lensing masses, the normalization agrees between mass--observable
relations using X-ray mass estimates and weak lensing mass estimates.
The mean of the weak lensing to X-ray mass ratios is about 1 with up
to 51 per cent scatter (Fig.~\ref{f:szwlen} and
Fig.~\ref{f:szwlen_simuyx}).

Therefore the X-ray and weak lensing approaches provide a very
valuable tool to cross calibrate the normalization and to constrain
the systematics of the mass--observable relations. The systematics in
the calibration of cluster mass estimates can be reduced by using a
sufficiently large sample of galaxy clusters with good statistic. The
average of the weak lensing versus X-ray mass discrepancy is about two
times the errors of the weak lensing masses (Fig.~\ref{f:szwlen}).
This indicates either that the true errors in the weak lensing masses
might be two times as large as the value quoted or that the difference
is a real discrepancy. Therefore, one really needs a sufficiently
large sample of galaxy clusters with both X-ray and weak lensing mass
measured by large FOV observations to reduce the scatter in weak
lensing mass estimates. This point will be more obvious with the
progress of the LoCuSS project -- the completion of the X-ray
(\emph{XMM-Newton}) and weak lensing (HST and Subaru) observations.

\begin{figure*}
\begin{center}
\includegraphics[angle=270,width=8.5cm]{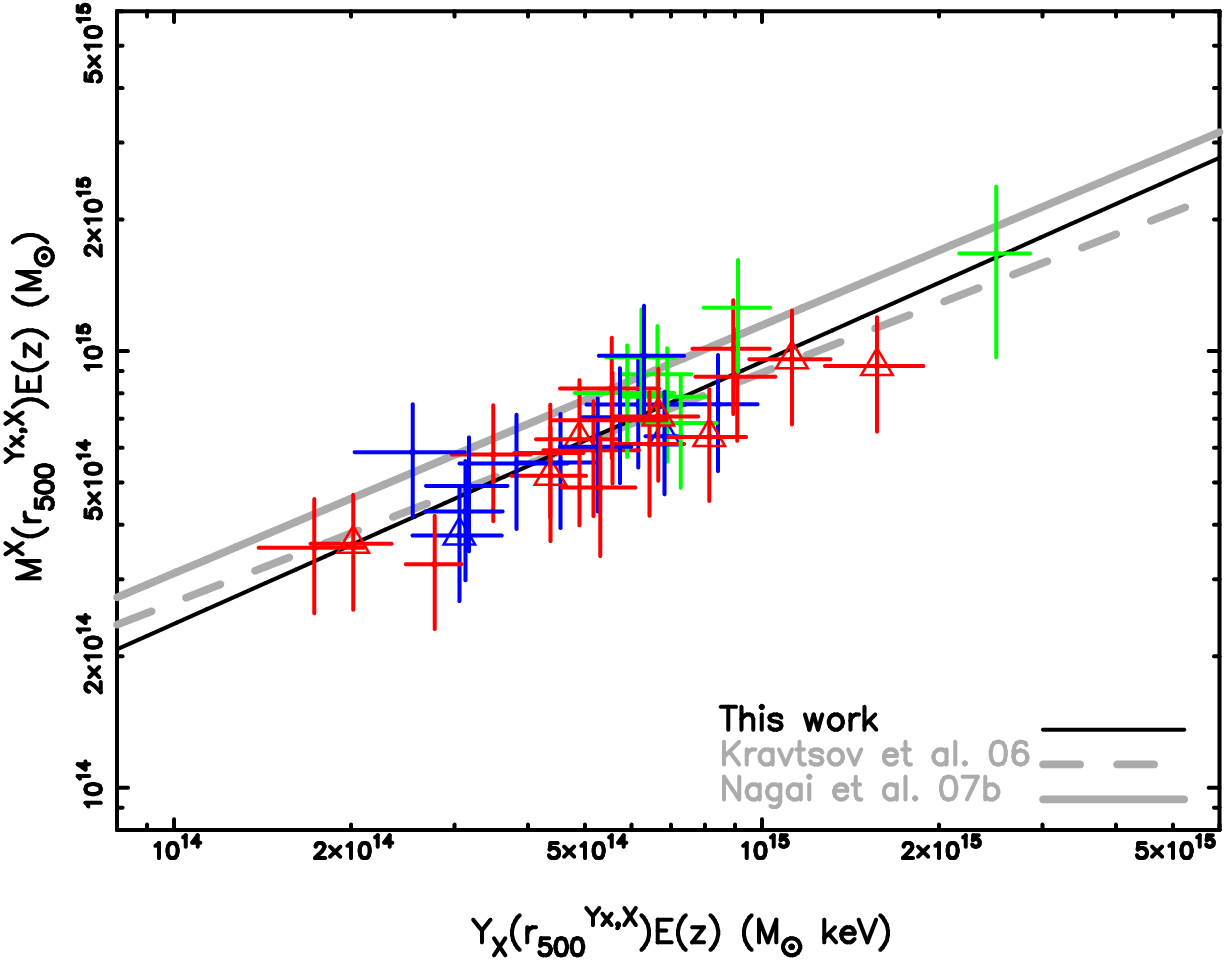}
\includegraphics[angle=270,width=8.5cm]{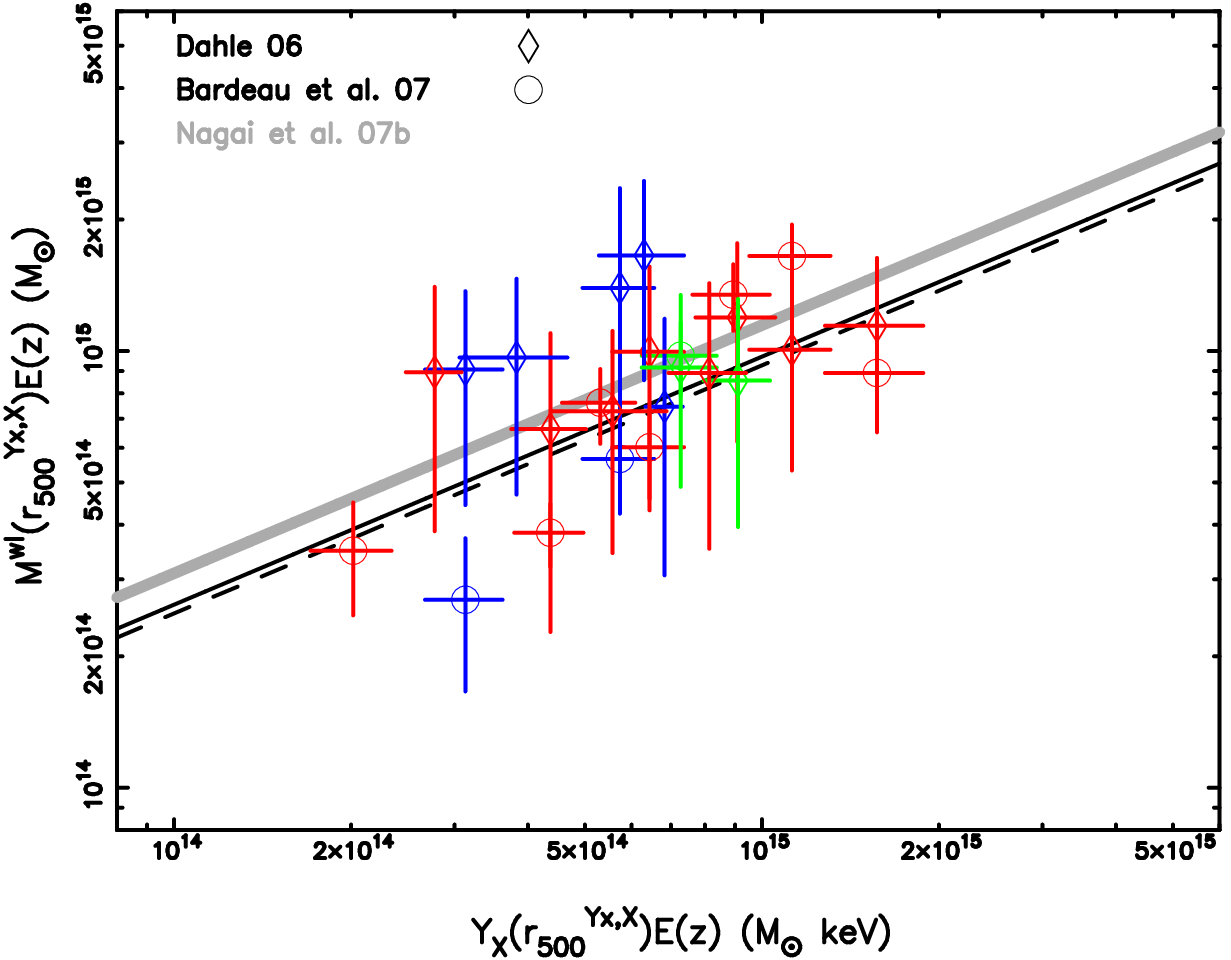}

\includegraphics[angle=270,width=8.5cm]{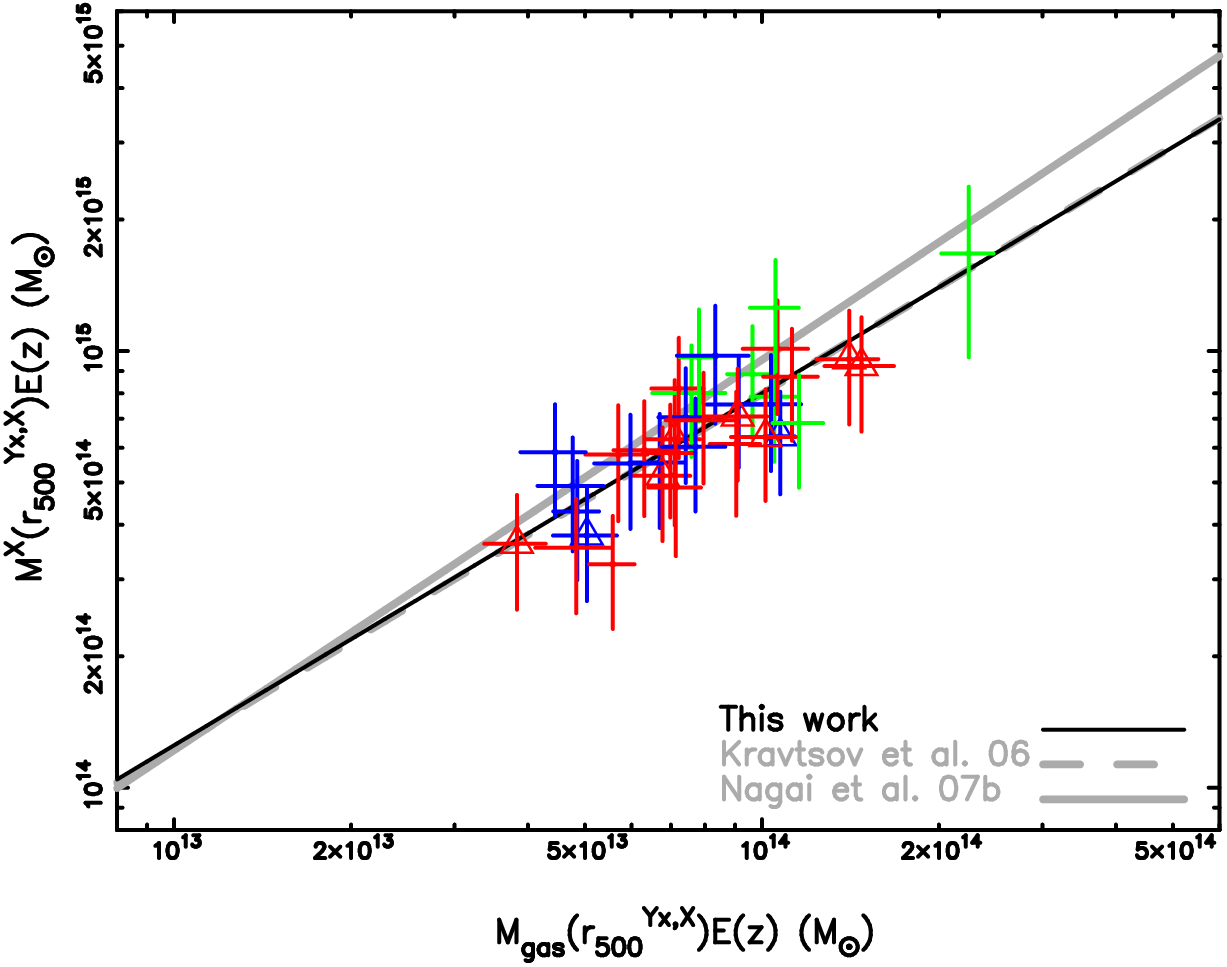}
\includegraphics[angle=270,width=8.5cm]{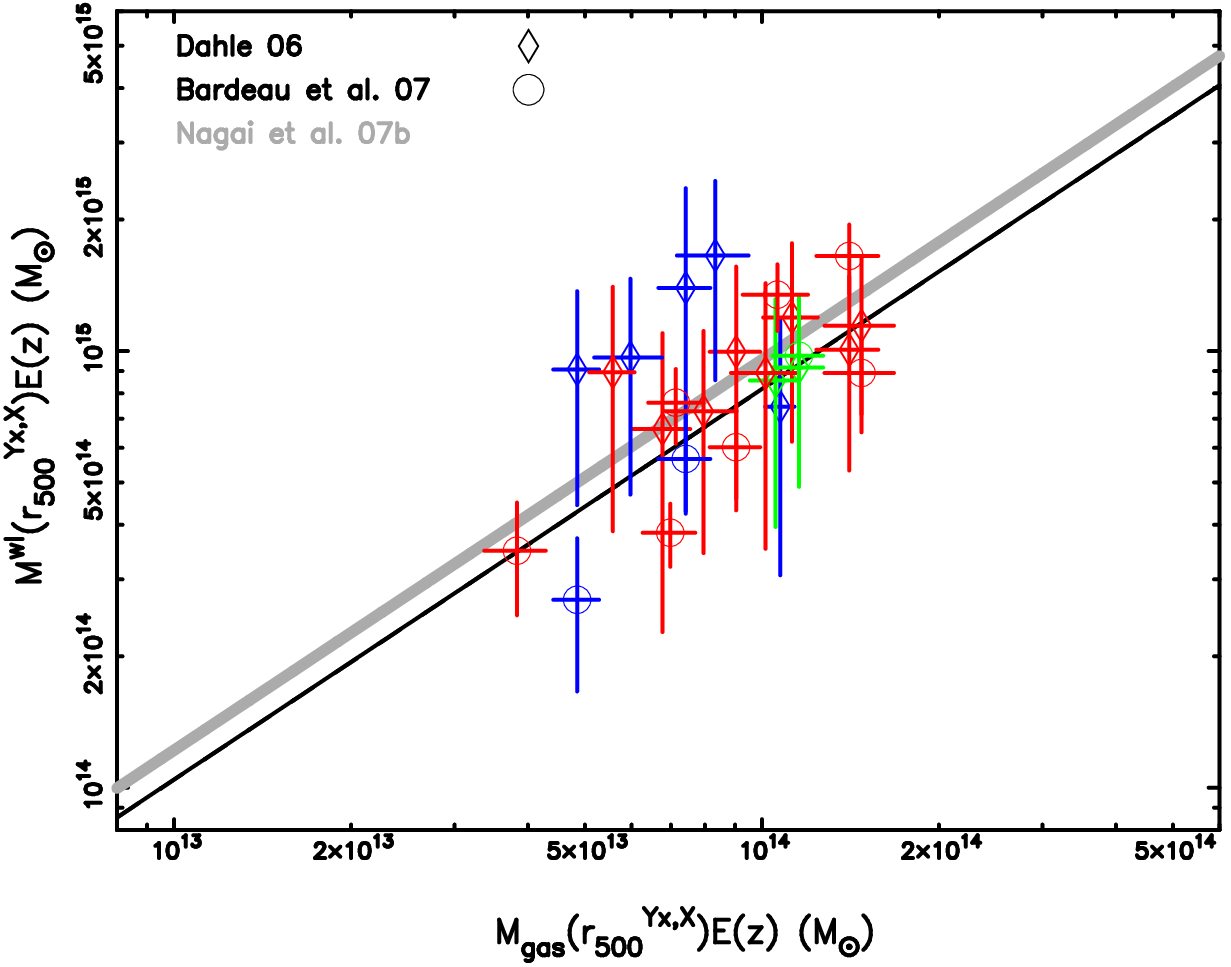}
\end{center}
\caption{See caption in Fig.~\ref{f:my_1} except that the weak lensing
  masses, X-ray masses and $Y_{\rm X}$ parameters are determined
  within the radii ($r_{500}^{\rm Y_{\rm X},X}$) determined by the
  method described in \S~\ref{s:simuyx} (also see
  Table~\ref{t:myl_wlen_simuyx}).
          \label{f:my_simuyx_1}}
\end{figure*}

\begin{figure*}
\begin{center}
\includegraphics[angle=270,width=8.5cm]{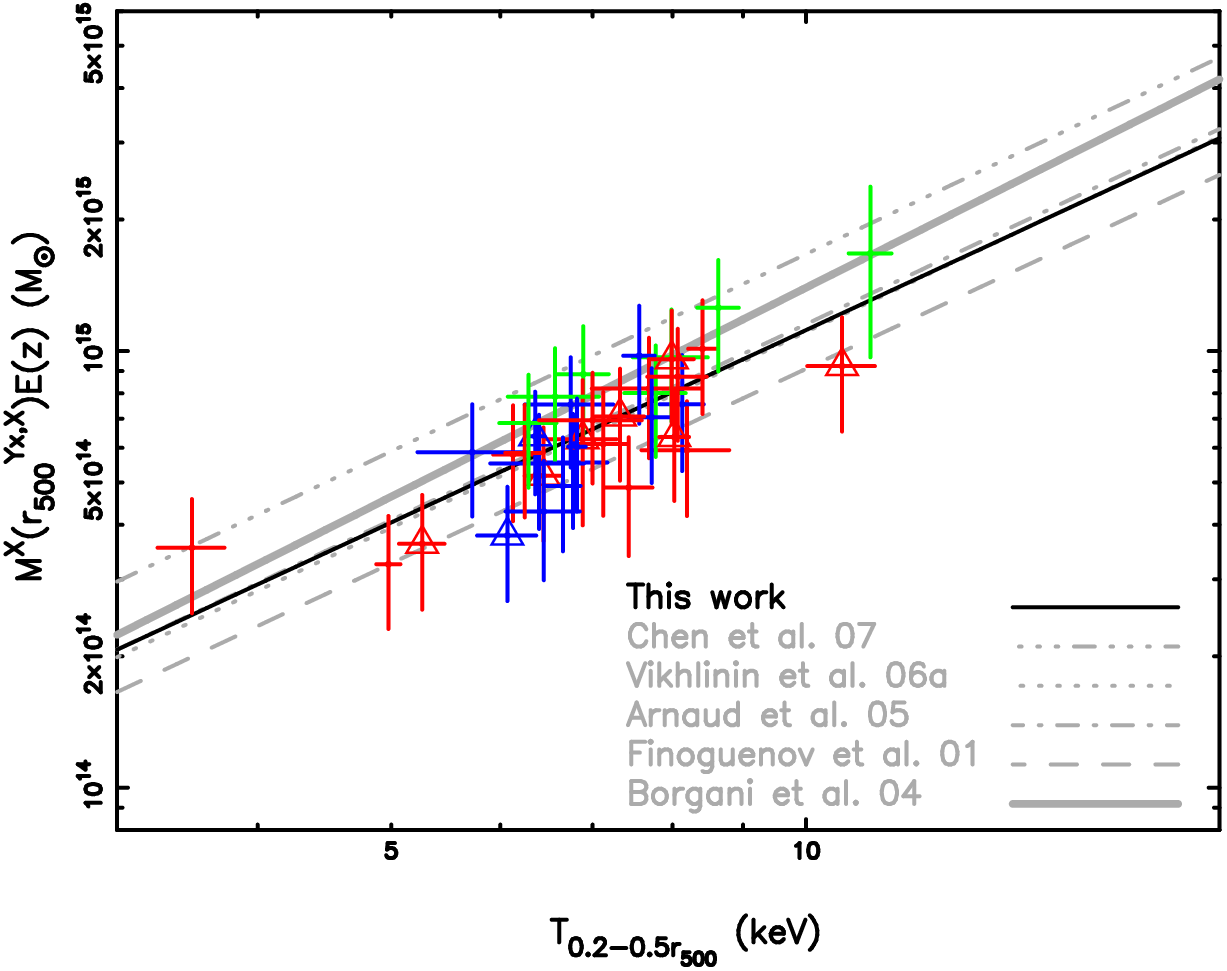}
\includegraphics[angle=270,width=8.5cm]{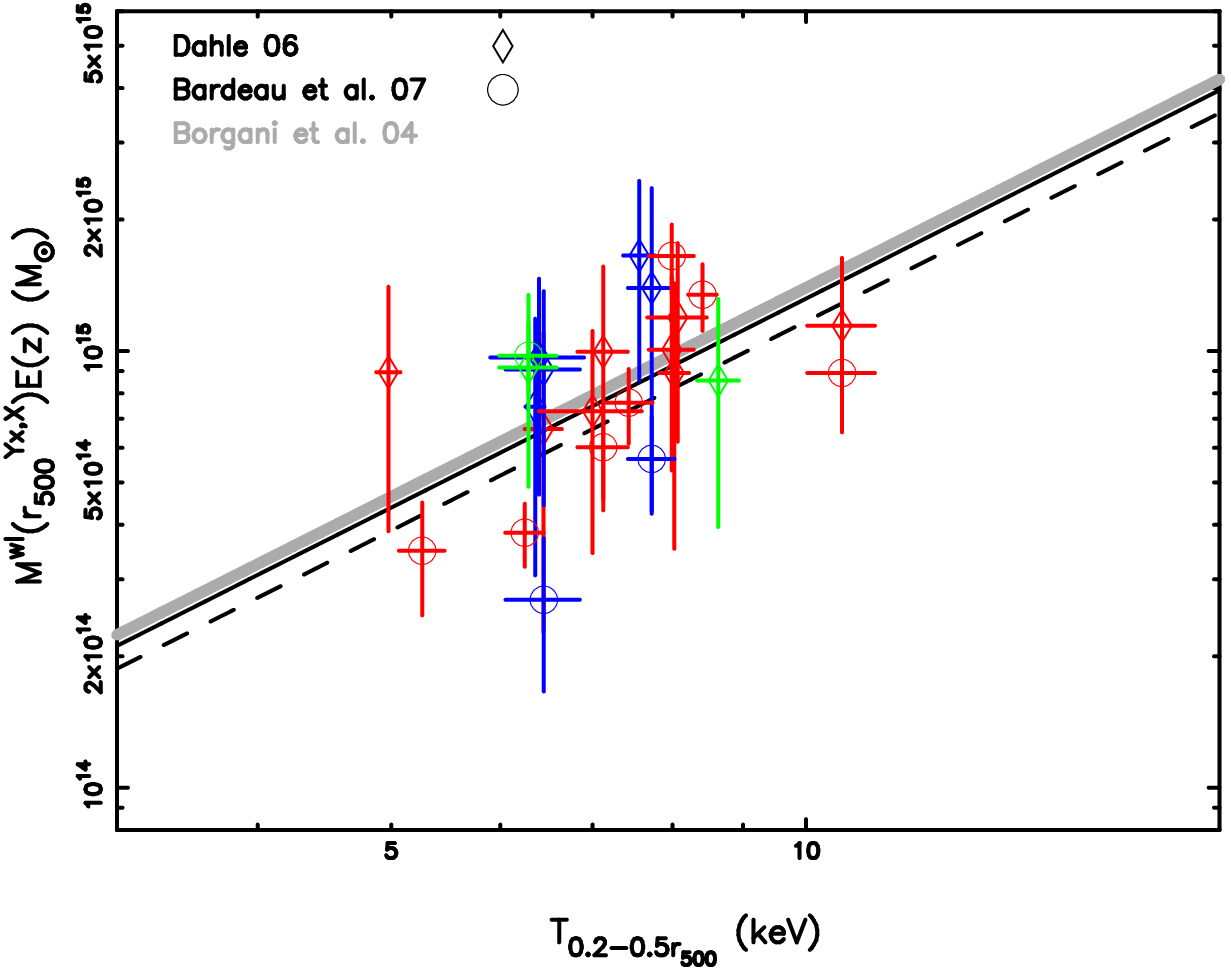}

\includegraphics[angle=270,width=8.5cm]{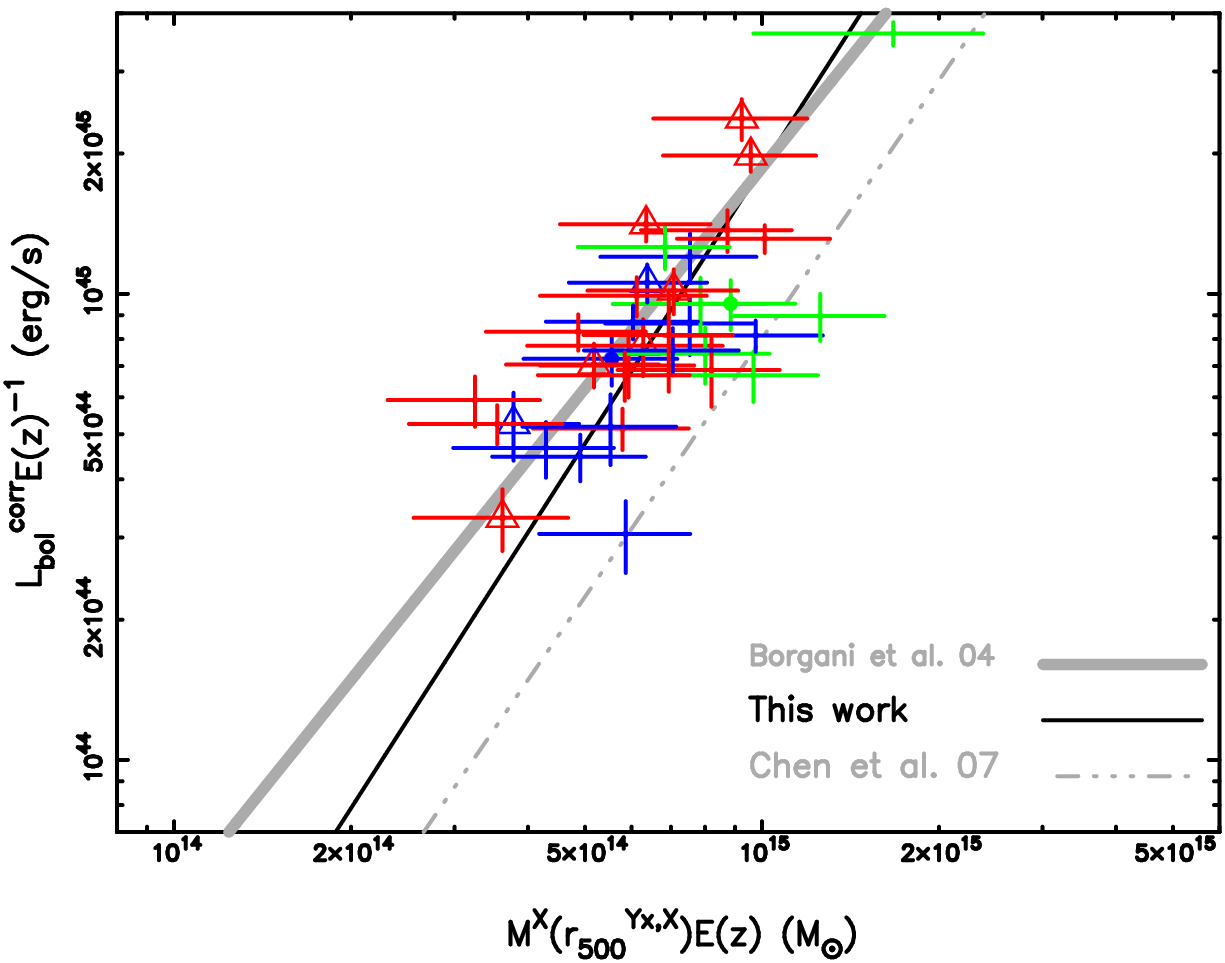}
\includegraphics[angle=270,width=8.5cm]{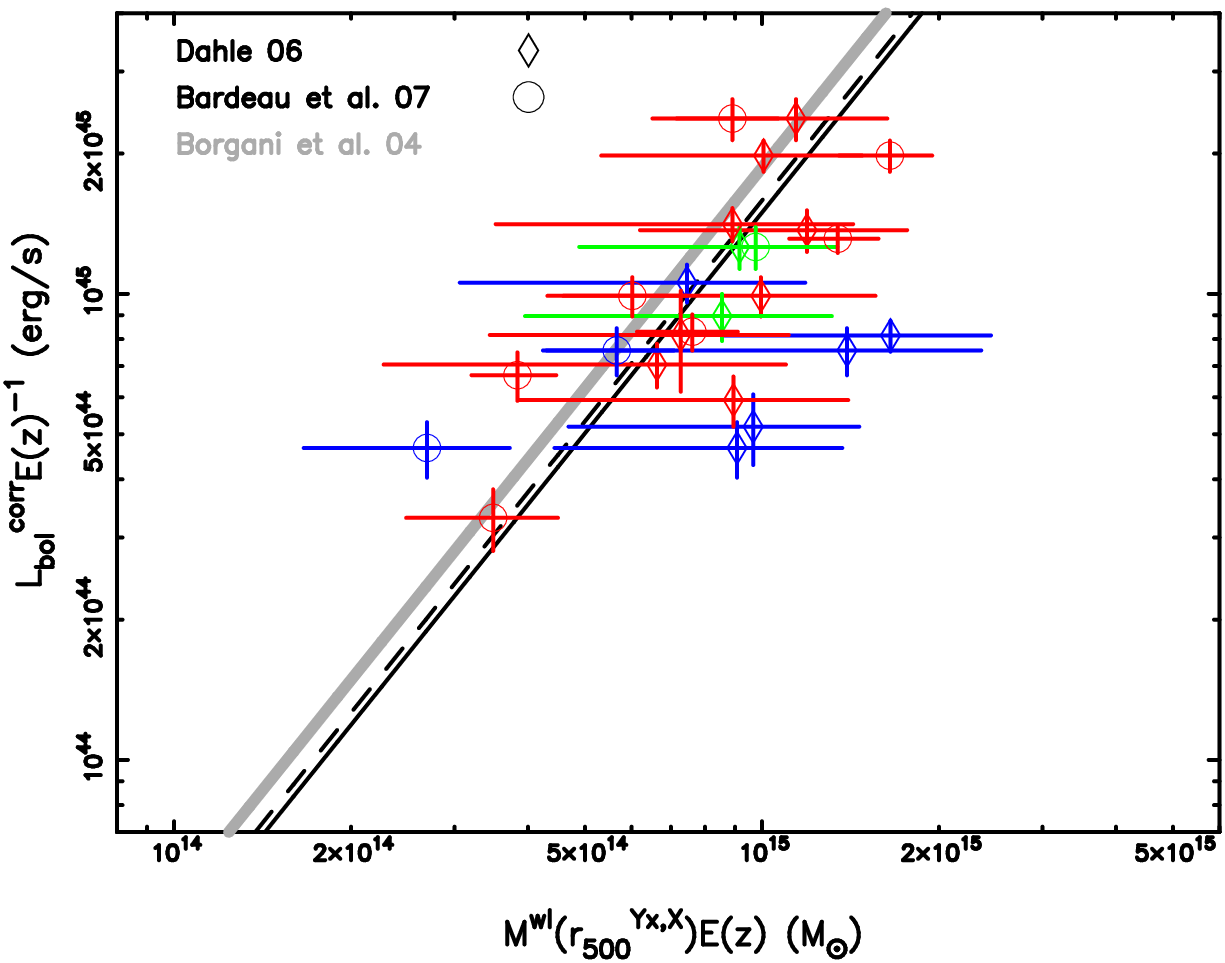}
\end{center}
\caption{See caption in Fig.~\ref{f:my_2} except that the weak lensing
  masses, X-ray masses and $Y_{\rm X}$ parameters are determined
  within the radii ($r_{500}^{\rm Y_{\rm X},X}$) determined by the
  method described in \S~\ref{s:simuyx} (also see
  Table~\ref{t:myl_wlen_simuyx}).
          \label{f:my_simuyx_2}}
\end{figure*}

\begin{figure*}
\begin{center}
\includegraphics[angle=270,width=8.5cm]{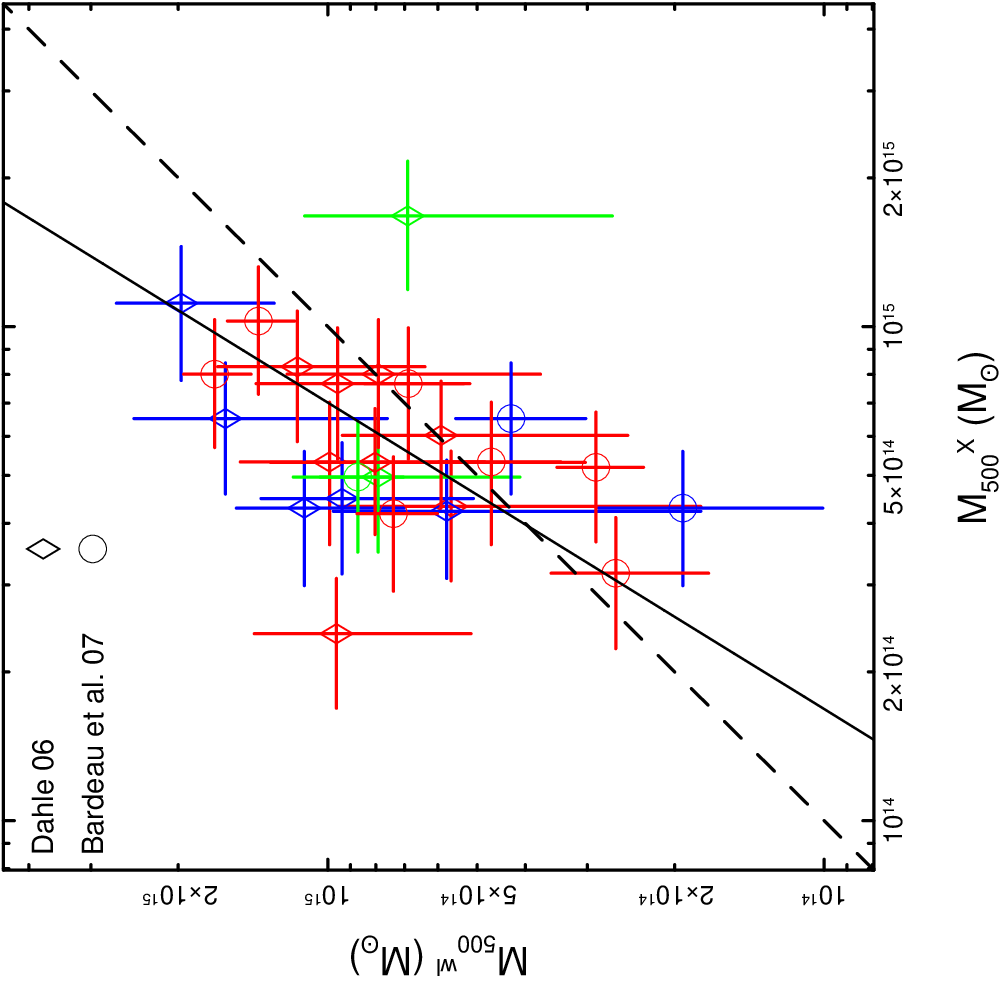}
\includegraphics[angle=270,width=8.5cm]{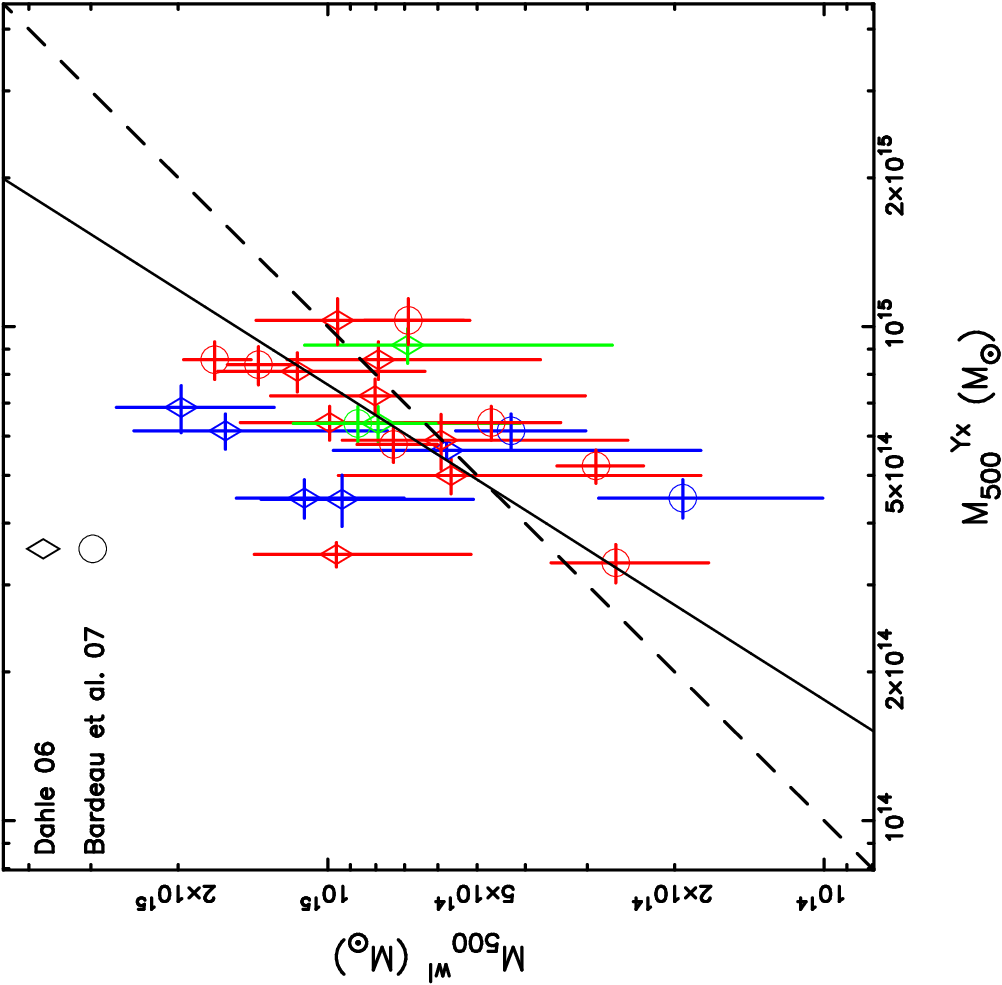}

\includegraphics[angle=270,width=8.5cm]{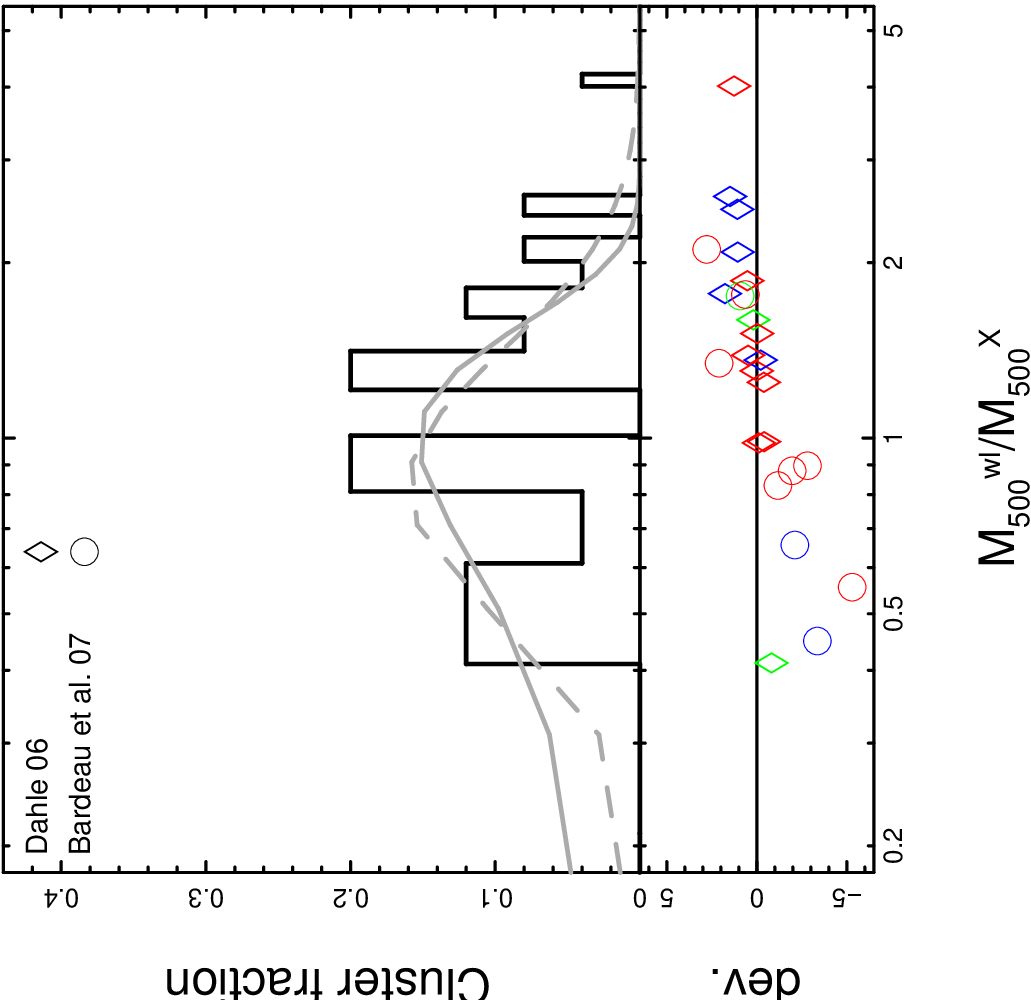}
\includegraphics[angle=270,width=8.5cm]{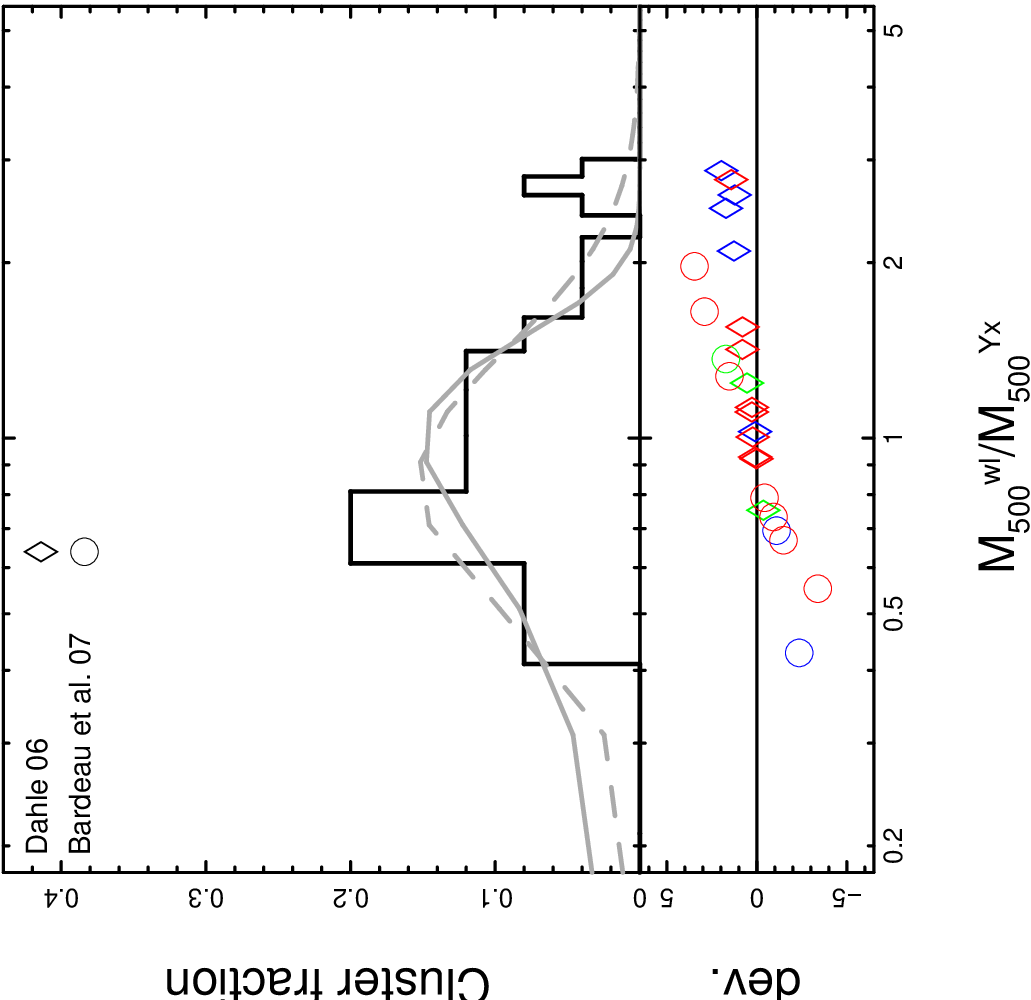}
\end{center}
\caption{{\it Upper left: } $M_{500}^{\rm wl}$ versus $M_{500}^{\rm
    X}$ (a best-fit slope of $1.61 \pm 0.39$). {\it Upper right: }
  $M_{500}^{\rm wl}$ versus $M_{500}^{\rm Y_{\rm X}}$ (a best-fit
  slope of $1.57 \pm 0.38$).  $M_{500}^{\rm Y_{\rm X}}$ is derived
  from $Y_{\rm X}$ with the slope fixed to 0.568 (the prediction from
  simulations in Nagai et al. 2007b) fitting $M$--$Y_{\rm X}$ relation
  for our sample. {\it Lower panels: } Normalized differential cluster
  number count as a function of the $M_{500}^{\rm wl}$ to
  $M_{500}^{\rm X}$ ratio and $M_{500}^{\rm wl}$ to $M_{500}^{\rm
    Y_{\rm X}}$ ratio, respectively, with their Gaussian fits (gray,
  solid curves denote the fits with x-axis in linear space, and dashed
  curves in logarithmic space).  The lower parts of the lower panels
  show the $M_{500}^{\rm wl}$ versus $M_{500}^{\rm X}$ discrepancy and
  $M_{500}^{\rm wl}$ versus $M_{500}^{\rm Y_{\rm X}}$ discrepancy,
  respectively, normalized to the error bar of $M_{500}^{\rm wl}$.
  The colors have the same meaning as those in Fig.~\ref{f:my_1}.
  \label{f:szwlen}}
\end{figure*}

\begin{figure*}
\begin{center}
\includegraphics[angle=270,width=8.5cm]{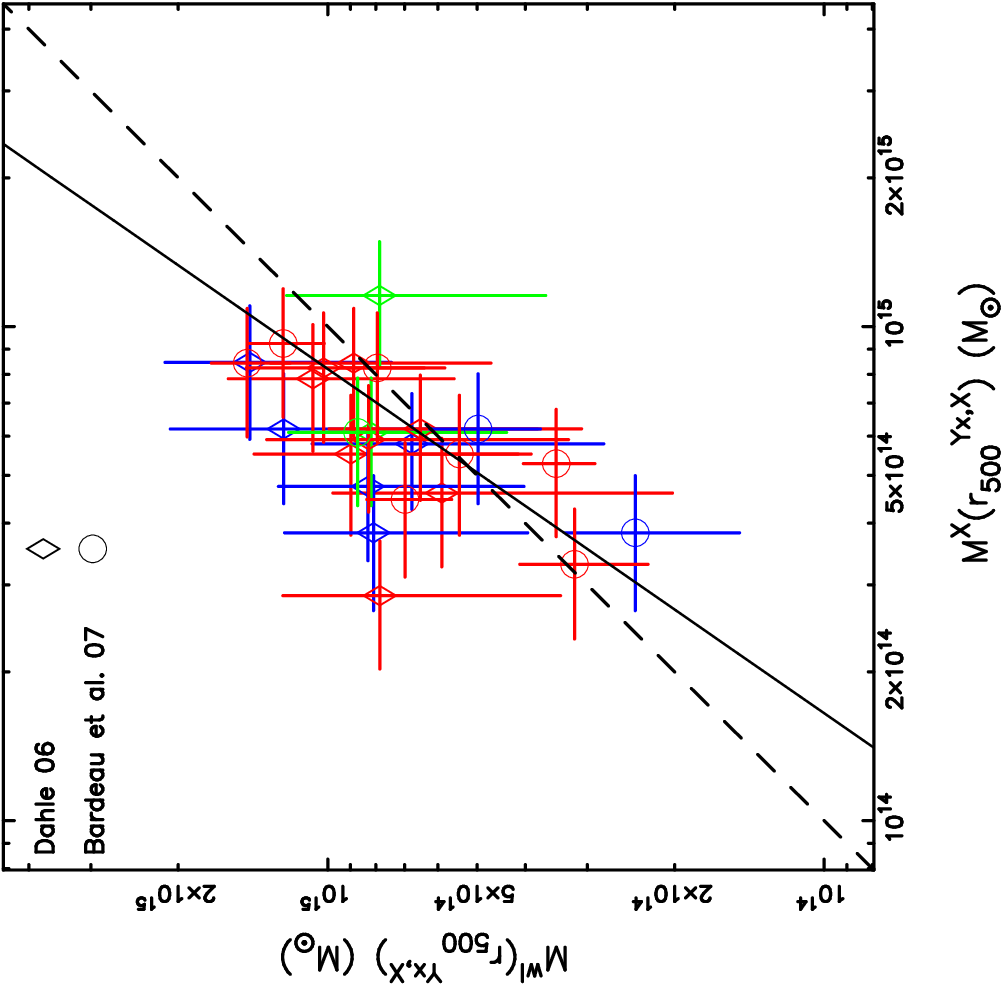}
\includegraphics[angle=270,width=8.5cm]{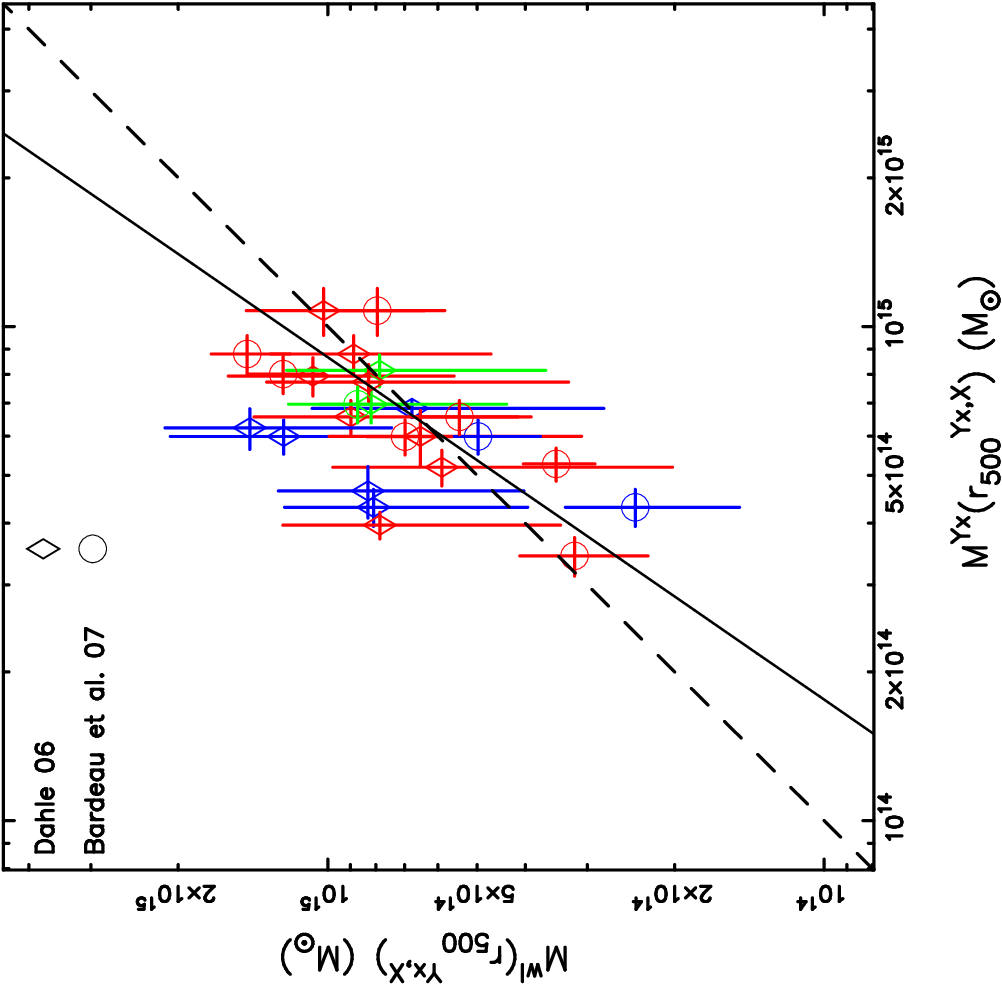}

\includegraphics[angle=270,width=8.5cm]{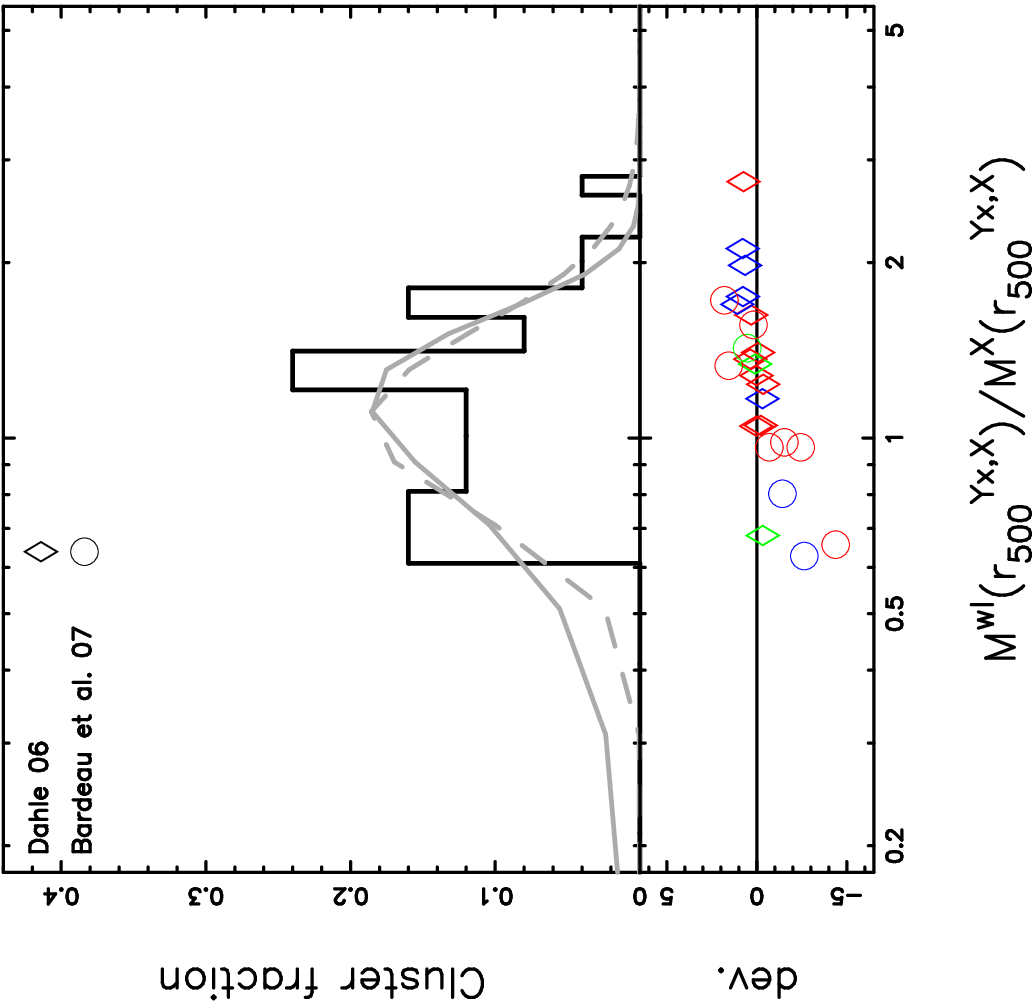}
\includegraphics[angle=270,width=8.5cm]{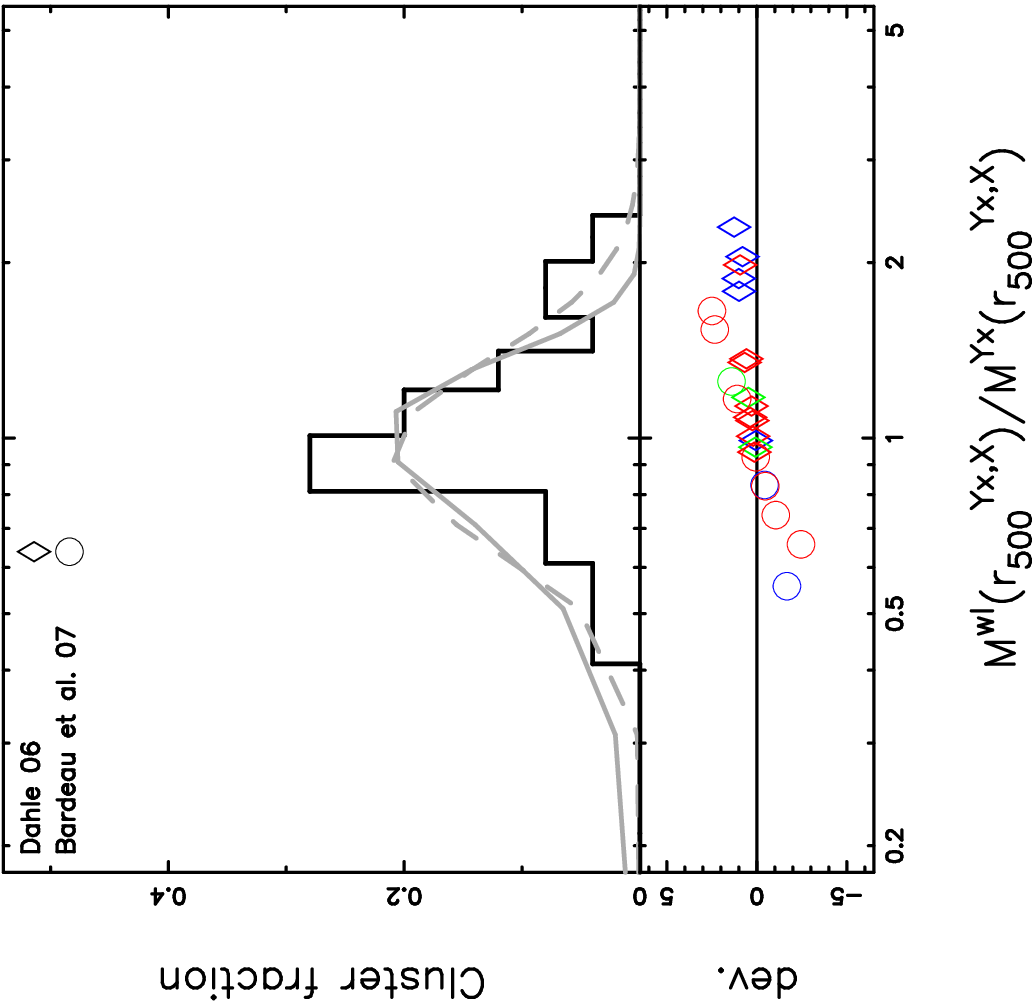}
\end{center}
\caption{See caption in Fig.~\ref{f:szwlen} except that the weak
  lensing masses and X-ray masses are determined within the radii
  ($r_{500}^{\rm Y_{\rm X},X}$) determined by the method described in
  \S~\ref{s:simuyx}.
          \label{f:szwlen_simuyx}}
\end{figure*}

\begin{figure*}
\begin{center}
\includegraphics[angle=270,width=8.5cm]{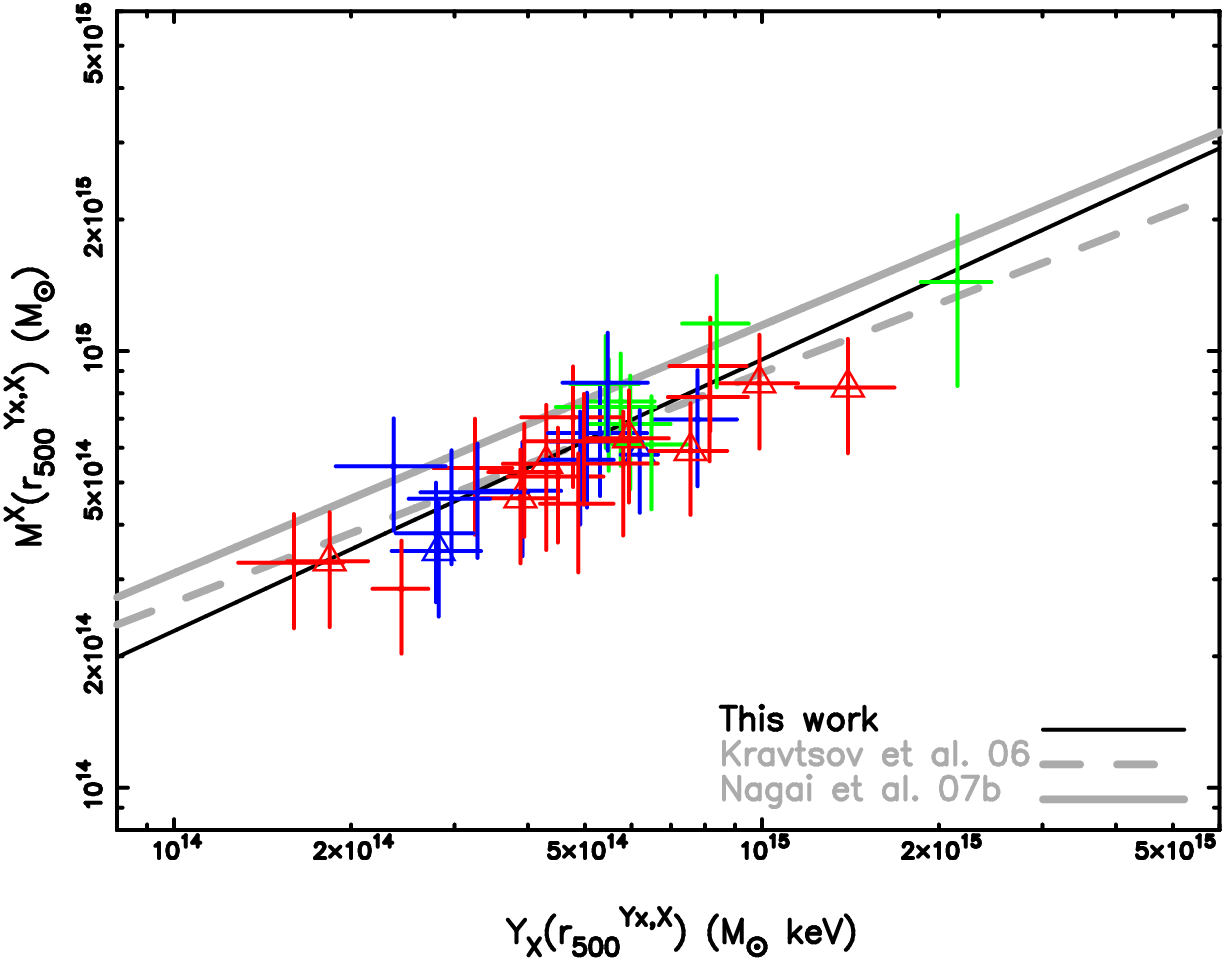}
\includegraphics[angle=270,width=8.5cm]{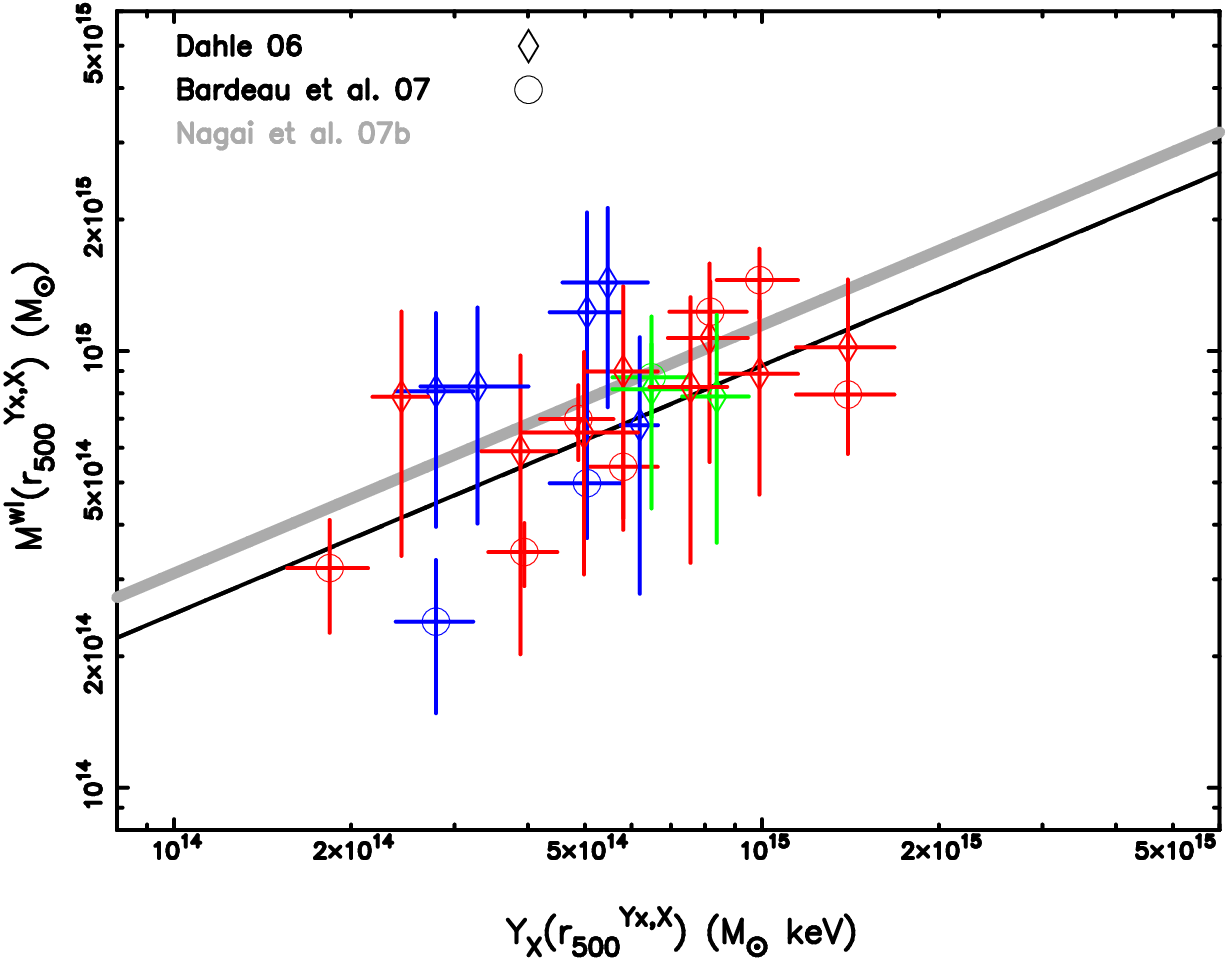}

\includegraphics[angle=270,width=8.5cm]{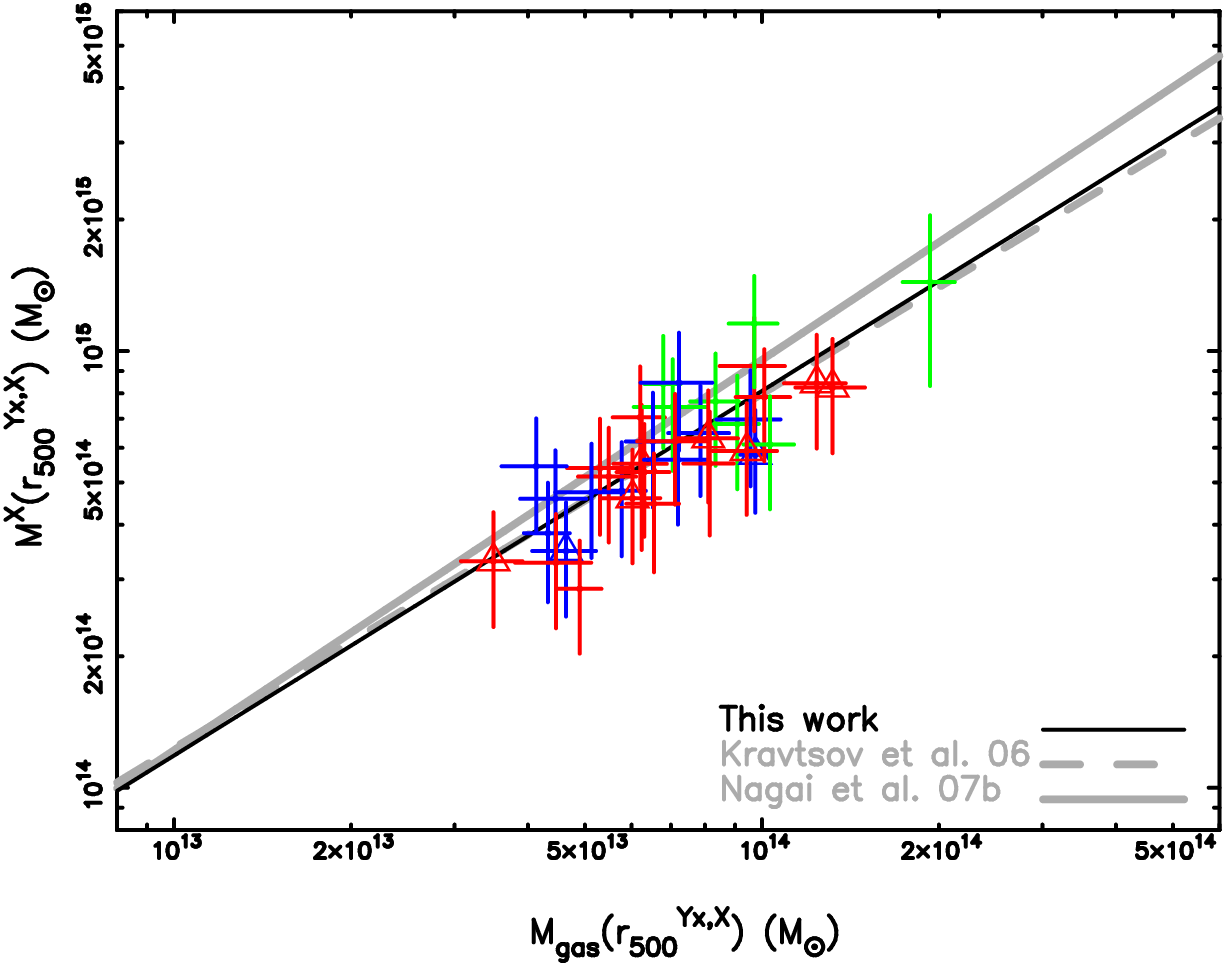}
\includegraphics[angle=270,width=8.5cm]{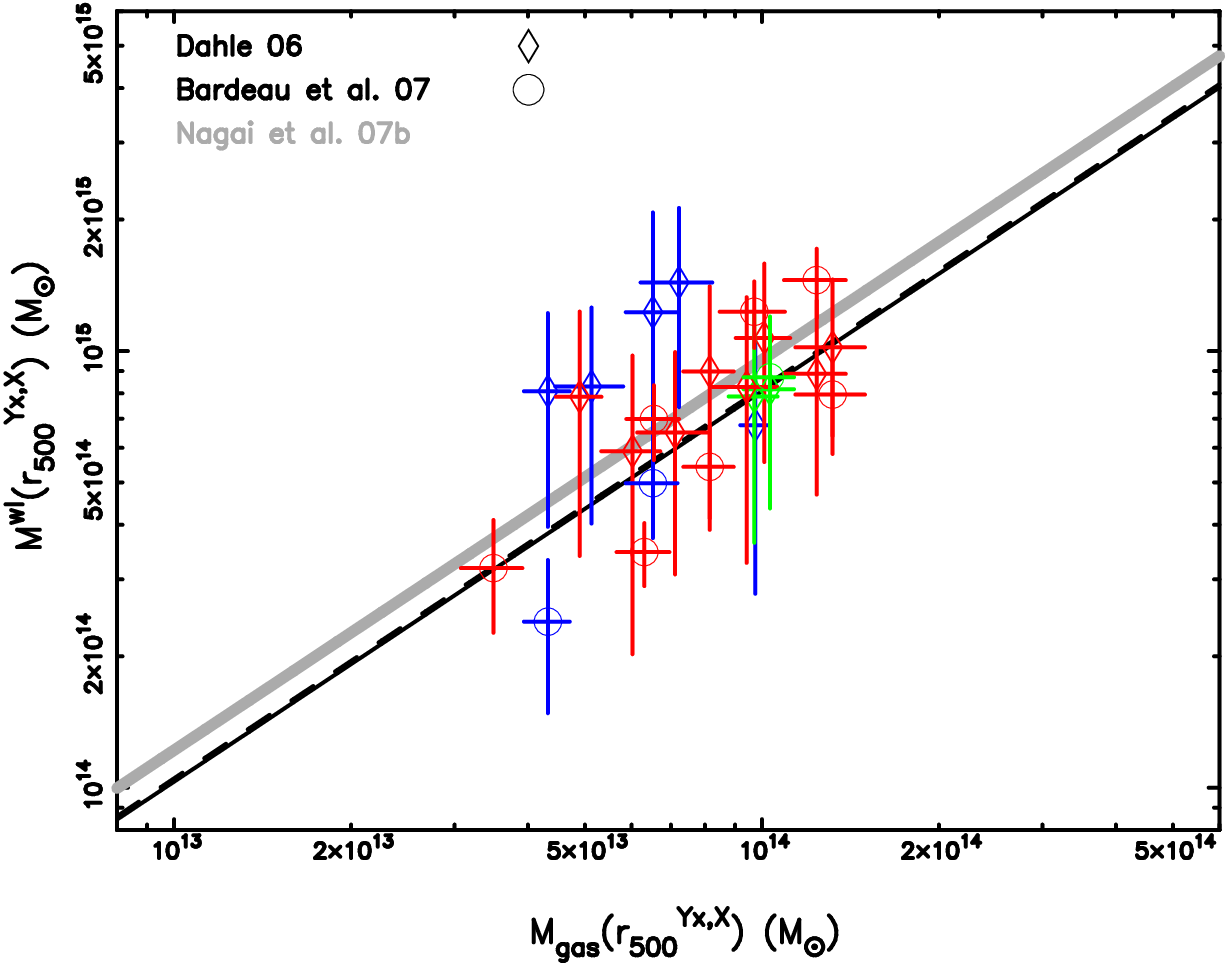}
\end{center}
\caption{See caption in Fig.~\ref{f:my_simuyx_1} except that no LSS
  evolution correction is applied in the mass--observable relations.
  \label{f:my_simuyx0_1}}
\end{figure*}

\begin{figure*}
\begin{center}
\includegraphics[angle=270,width=8.5cm]{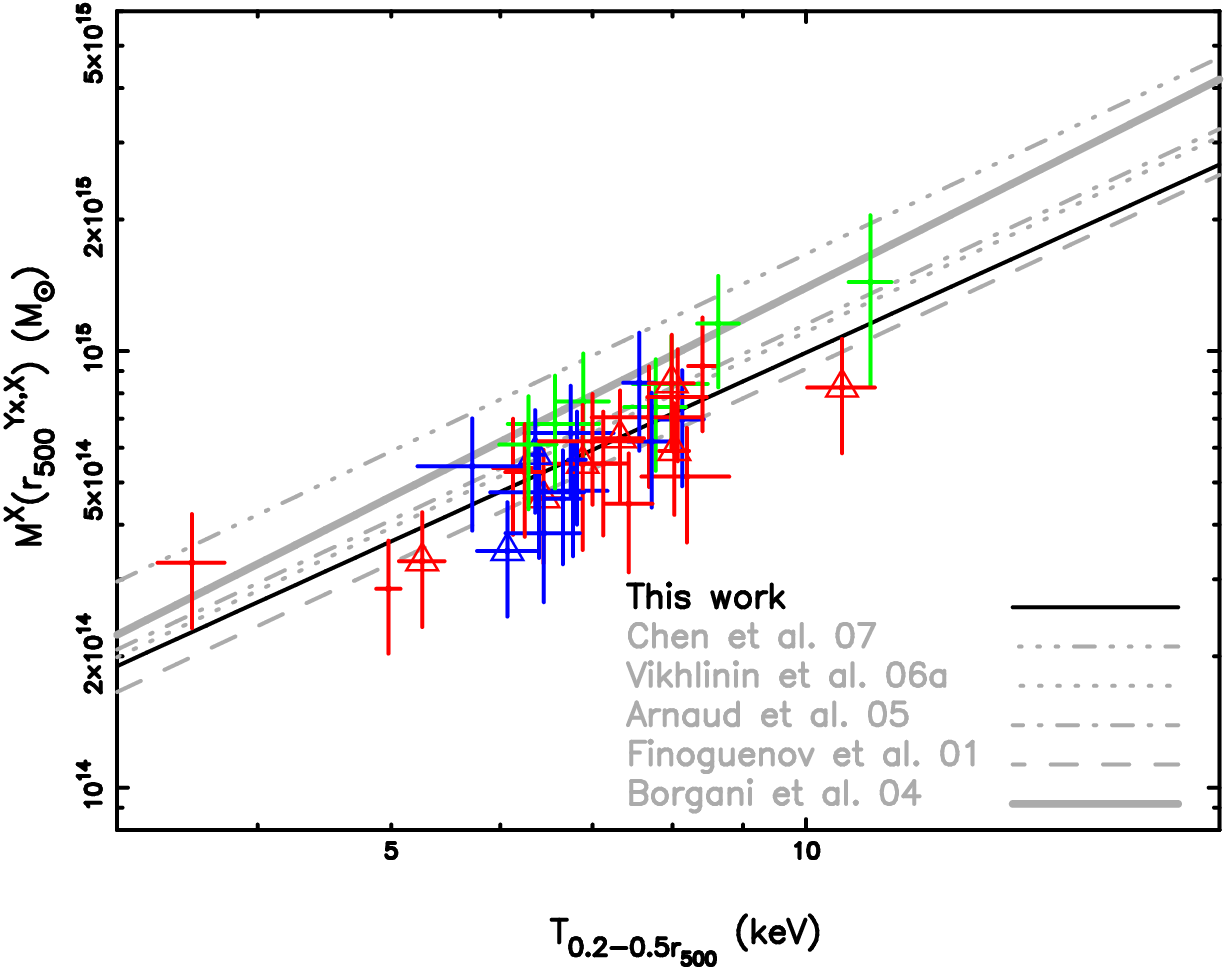}
\includegraphics[angle=270,width=8.5cm]{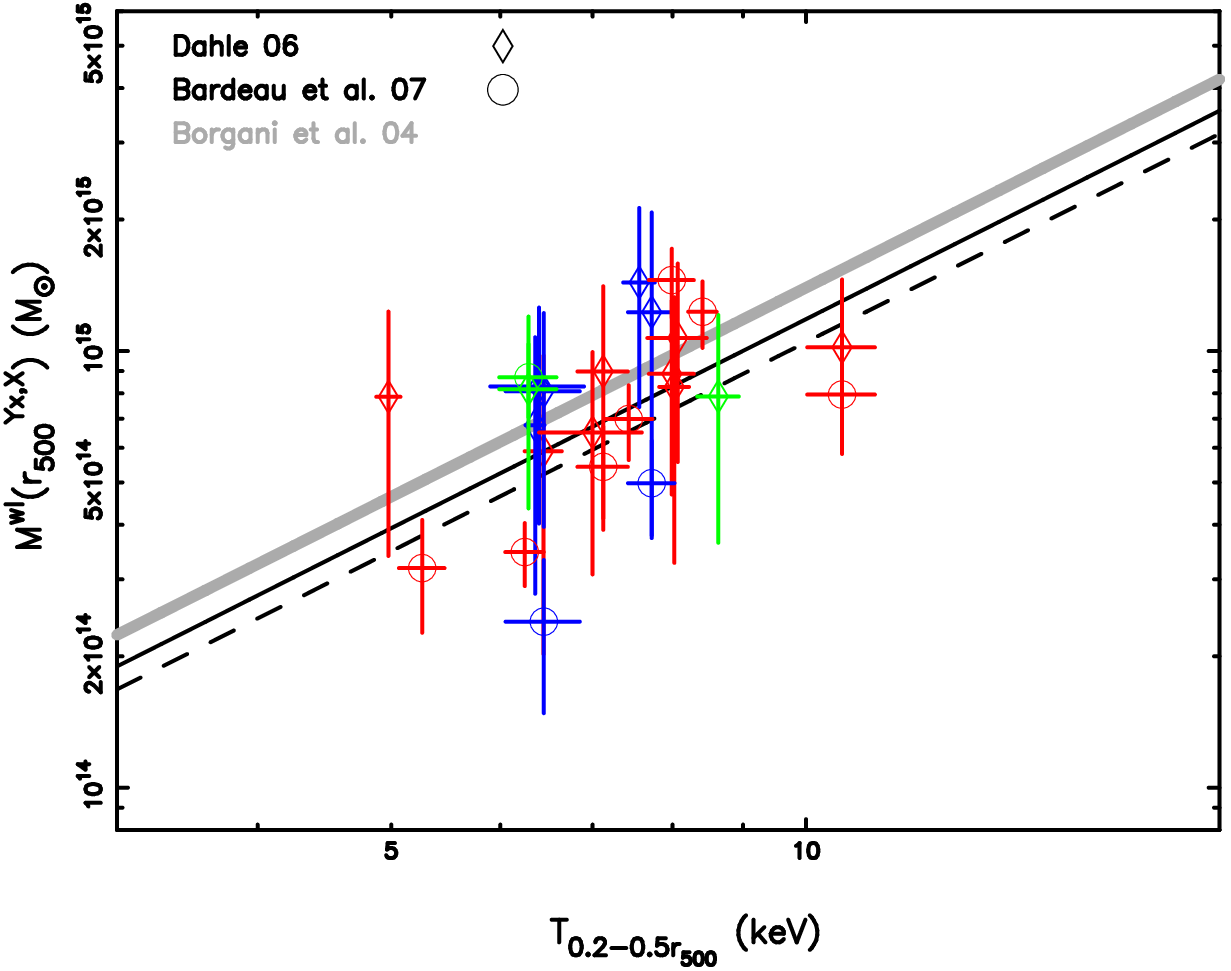}

\includegraphics[angle=270,width=8.5cm]{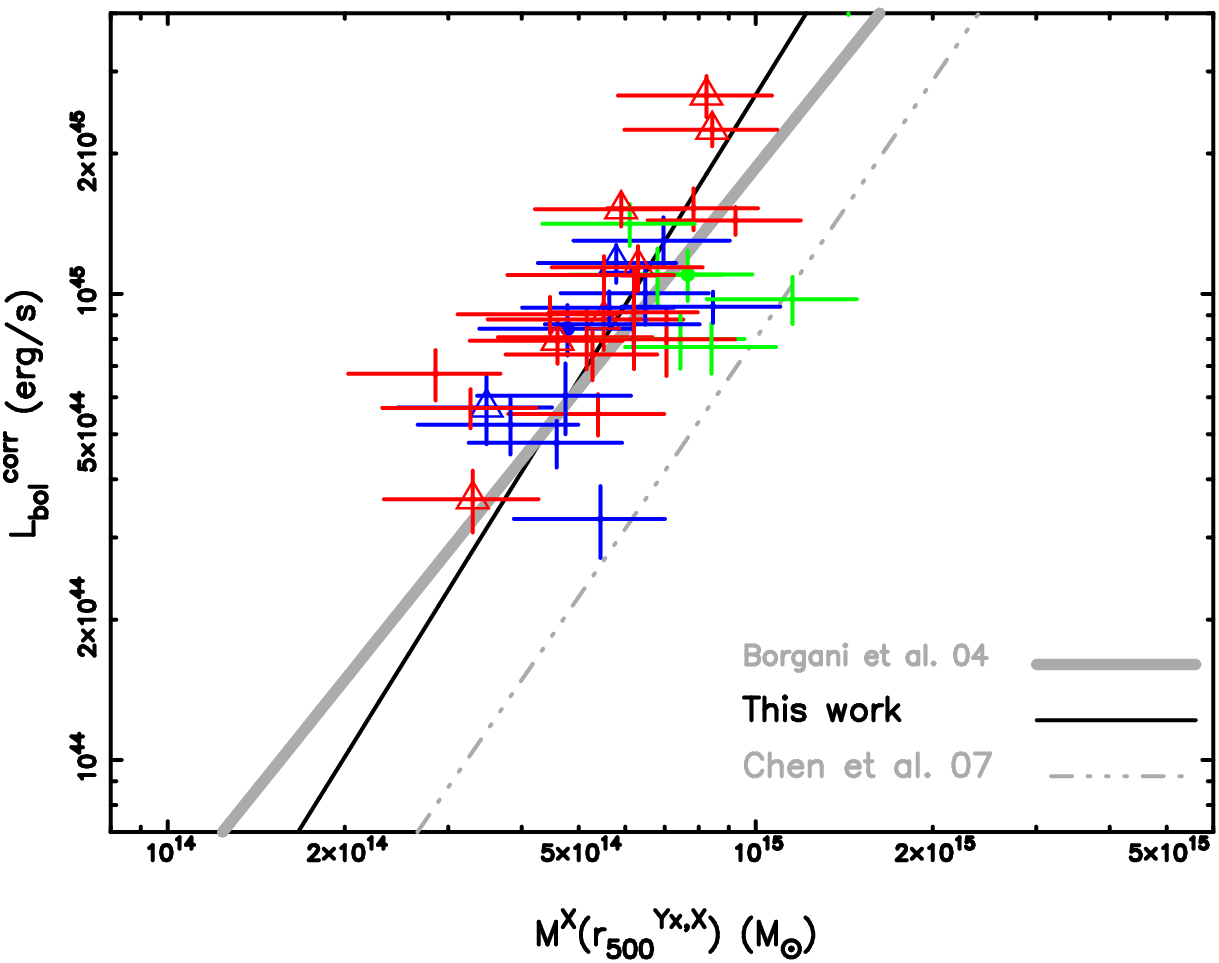}
\includegraphics[angle=270,width=8.5cm]{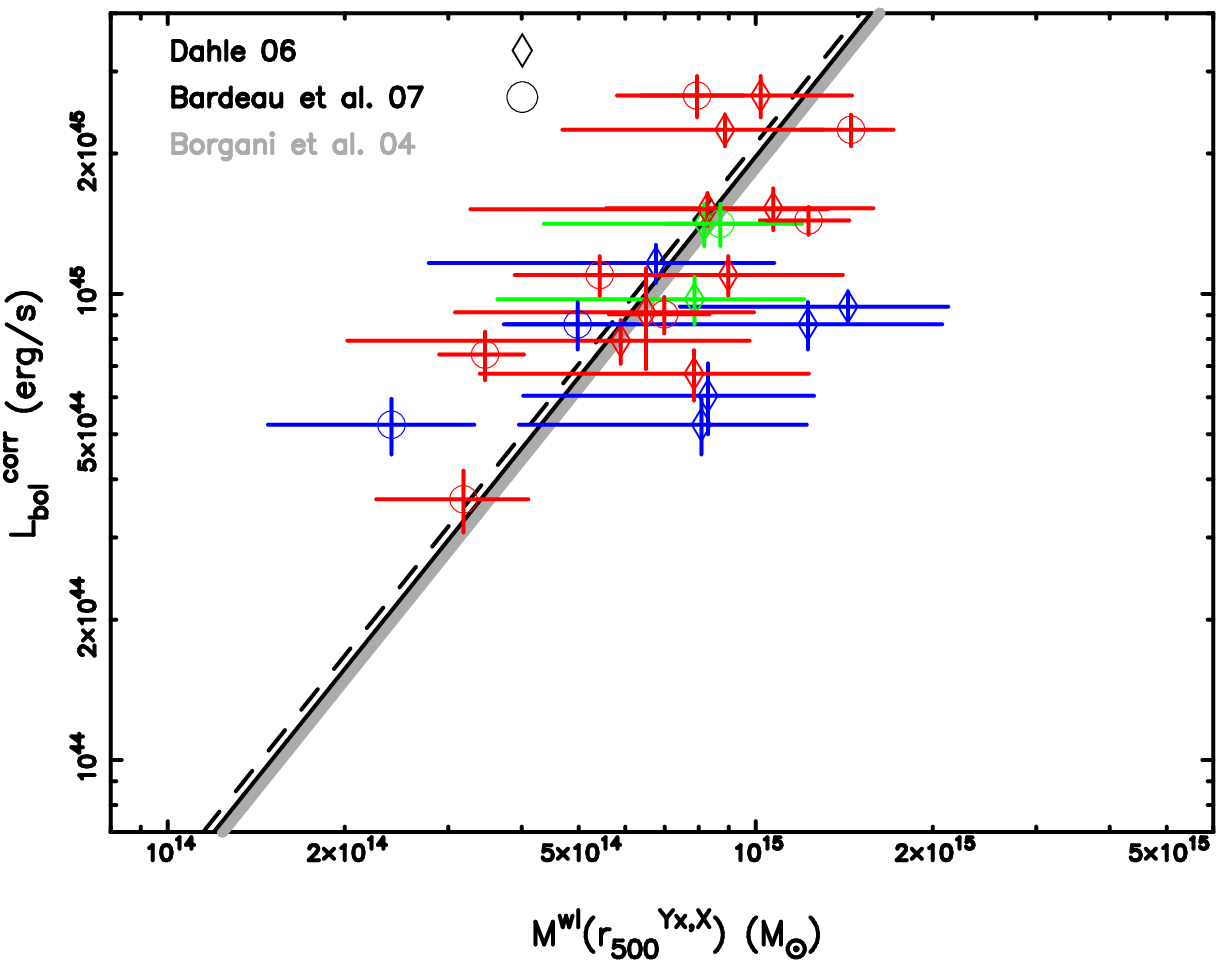}
\end{center}
\caption{See caption in Fig.~\ref{f:my_simuyx_2} except that no LSS
  evolution correction is applied in the mass--observable relations.
  \label{f:my_simuyx0_2}}
\end{figure*}

At the radial scale of $r^{\rm Y_{\rm X},X}_{500}$, both the X-ray
determined masses and the weak lensing determined masses
systematically lie below the prediction from the mass--observable
relations (i.e. $M$--$Y_{\rm X}$, $M$--$M_{\rm gas}$ and $M$--$T$)
from simulations (Borgani et al. 2004; Kravtsov et al. 2006; Nagai et
al.  2007b) by up to $\sim 20$ per cent with $\sim 3 \sigma$
significance for the X-ray masses and $\sim 2 \sigma$ significance for
the weak lensing masses (Figs.~\ref{f:my_simuyx_1}--\ref{f:my_simuyx_2},
Figs.~\ref{f:my_simuyx1_1}--\ref{f:my_simuyx2_2}). If one believes that
the two independent approaches, weak lensing and X-ray, give reliable
mass estimates, more real physics has to be included in simulations to
reproduce the real galaxy clusters. This can be further verified by
comparing the mass--observable relations between simulations and
observations using a sufficiently large sample of galaxy clusters with
both X-ray and weak lensing mass measurements. Since the difference in
the normalization is only up to 20 per cent between mass--observable
relations from observations and simulations, it is important to pursue
a few per cent level systematics in the mass estimates. Therefore one
has to improve the statistic for both simulation and observational
samples by using a large sample (e.g. $\sim 100$ galaxy clusters).
And one also has to reduce the errors in observations by using both
(1) deep X-ray observations (e.g. an accuracy of the ICM temperature
measurements on the per cent level precision for 8--10 radial bins)
performed with a large FOV (e.g. covering the area of more than
$r_{500}$), and (2) weak lensing observations performed with an
instrument with a good PSF and a large FOV (e.g. covering the area of
more than $r_{200}$) together with a photo-z database to characterize
the background source galaxy redshift distribution.

\subsection{Mass calibration and $\sigma_8$}
\label{s:sigma8}

Based on (1) the number density of clusters with $T>6$~keV (Henry
2004, Ikebe et al. 2002) and (2) three bins of the cluster density
with certain velocity dispersion from Rines et al (2007), Evrard et
al. (2007) found a high amplitude of cosmic power spectrum, $S_8 =
\sigma_8 (\Omega_{\rm m}/0.3)^{0.35}=0.8$ using the mass--temperature
relation ($M^{\prime}$--$T^{\prime}$) from simulations combined with
the gas mass--temperature relation ($M^{\prime}_{\rm
  gas}$--$T^{\prime}$) from ROSAT and ASCA observations in Mohr et al.
(1999) also regarding massive clusters in a similar temperature
range.  Our comparison shows that the normalization of the
mass--temperature relation in Evrard et al. (1996) is higher than ours
($M^{X}_{500}$--$T$) by 40 per cent, and the normalization of the gas
mass--temperature relation in Mohr et al.  (1999) is higher than ours
($M_{\rm gas,500}$--$T$) by 81 per cent. The gas mass to mass ratio in
Evrard et al.  (2007) is $M^{\prime}_{\rm
  gas}/M^{\prime}=(1.81/1.4)\times M_{\rm gas}/M$.  According to the
gas mass to mass ratio, $M^{\prime}_{\rm
  gas}/M^{\prime}=0.083\;h_{70}^{-1.5}S_8^{\prime\;-5/3} $, in Evrard
et al.  (2007), we obtained $M_{\rm gas}/M=0.083\times
(1.4/1.81)\;h_{70}^{-1.5}\;S_8^{-5/3}$. For our sample, we then found
$S_8=0.857174 S_8^{\prime}$, where $S_8^{\prime}$ is 0.8 for the
sample in Evrard et al. (2007). We thus obtained $S_8=0.69$ for our
sample. The X-ray mass calibration using $M^{\rm X}_{500}$ for our
sample supports a lower amplitude of cosmic power spectrum on the
basis of the comparison between observations and simulations.

To be conservative, we repeated the above calculation using the
mass--temperature and gas mass-temperature relations at the radius $r^{\rm
  Y_{\rm X,si}}_{500}$, which is determined by our $Y_{\rm X}$ profile and the
mass--$Y_{\rm X}$ relation from simulations in Nagai et al. (2007b). In this
case, the normalization of the mass--temperature relation in Evrard et al.
(1996) is higher than ours by 28 per cent, and the normalization of the gas
mass--temperature relation in Mohr et al. (1999) is higher than ours by 58 per
cent. Therefore the gas mass to mass ratio in Evrard et al. (2007) is
$M^{\prime}_{\rm gas}/M^{\prime}=(1.58/1.28)\; M_{\rm gas}/M$. According to
the gas mass to mass ratio, $M^{\prime}_{\rm
  gas}/M^{\prime}=0.083\;h_{70}^{-1.5}S_8^{\prime\;-5/3} $, in Evrard et al.
(2007), we obtained $M_{\rm gas}/M=0.083\times
(1.28/1.58)\;h_{70}^{-1.5}\;S_8^{-5/3}$. For our sample, we then found
$S_8=0.88 S_8^{\prime}=0.70$.

Assuming weak lensing masses agree better with cluster masses from
simulations, we also derived the $S_8$ parameter using the weak
lensing mass based mass--temperature relation ($M_{500}^{\rm
  wl}$--$T_{0.2-0.5r_{500}}$) with the slope fixed to 1.5, which
normalization is $4.49 \times 10^{13} M_{\odot}$. It gives $S_8=0.77$,
similar to the value obtained from simulations in Evrard et al.
(2007).

The above three experiments show the importance of the calibration of
the scaling relations for the cluster cosmology. Depending on the mass
calibration (e.g. via X-ray or weak lensing approach) the derived
value of the amplitude of cosmic power spectrum varies by 10 per cent
between 0.69 and 0.77 for our sample.
 
\section{Summary and conclusions}
\label{s:conclusion}

We performed a systematic analysis to measure the X-ray quantities based on
\emph{XMM-Newton} observations for a sample of 37 LoCuSS clusters. We
investigated (1) the empirical similarity of the X-ray scaling relations, (2)
the scatter in X-ray and weak lensing mass estimates, (3) the difference in
the normalization between mass--observable relations from observations (using
X-ray masses and weak lensing masses, respectively) and simulations, and (4)
the impact of the mass calibration on the determination of the amplitude of
cosmic power spectrum. We summarize the main conclusions as follows.

\bigskip

\noindent
(i) The self-consistent X-ray scaling relations for our sample support
an empirical similarity for massive galaxy clusters. The X-ray scaling
relations of our sample show no evident evolution compared to the
nearby and more distant samples after the redshift evolution
correction due to LSS growth.  Specially, the X-ray $M$--$T$ relations
agree on 1 per cent level between our sample and the samples in Arnaud
et al. (2005) and Vikhlinin et al. (2006a).

The cluster cores ($<0.2 r_{500}$) contribute up to 80 per cent to the
bolometric X-ray luminosity. Using X-ray luminosity corrected for the
cool core and temperature excluding the cool core, leads both to less
scatter (7--40 per cent) in the X-ray scaling relations and to better
than 17 per cent agreement of the normalization between the sample and
the non-CCC subsample.

\bigskip

\noindent
(ii) Using a Gaussian fit to the histogram of the weak lensing to
X-ray $M_{500}$ ratios, the mean weak lensing to X-ray $M_{500}$
ratios is about 1, indicating a good agreement between X-ray and weak
lensing mass estimates for most clusters. Therefore the weak lensing
approach can be a very valuable tool to cross calibrate the
systematics in cluster mass estimates and the normalization of the
mass--observable relations together with the X-ray approach. The
comparison between weak lensing and X-ray mass estimates shows a large
scatter of the individual results. The most robust way to characterize
this scatter is by the width of the Gaussian fit to the distribution
of deviations which is 31--51 per cent. The large scatter is mostly
due to uncertainties in the weak lensing mass caused in part by the
LSS into which the clusters are embedded. For the purpose of precise
cosmology, one requires a sufficiently large sample with both X-ray
and weak lensing measurements to verify (1) the scatter in X-ray and
weak lensing mass estimates on a few per cent level precision, and (2)
the difference in the normalization between mass--observable relations
between observations and simulations on per cent level accuracy.

\bigskip 

\noindent
(iii) The scatter in the mass--observable relations is larger using
weak lensing masses than using X-ray masses by a factor of 2.
However, the scatter using weak lensing masses neither depends on
being classified as CCCs and non-CCCs nor depends on being classified
as ``single''s and ``merger''s.  Using our method, we can define a
radial scale $r^{\rm Y_{\rm X},X}_{500}$, which can not only reduce
the scatter in the mass--observable scaling relations, but also
provide consistency between weak lensing mass based and X-ray mass
based mass--observable scaling relations.

\bigskip 

\noindent
(iv) The X-ray mass estimates are about 20 per cent lower than the
prediction of the mass--observable relations (i.e. $M$--$Y_{\rm X}$,
$M$--$M_{\rm gas}$ and $M$-$T$) from simulations with $\sim 3 \sigma$
significance. The comparison between the observed mass-observable
relations (i.e.  $M$--$Y_{\rm X}$, $M$--$M_{\rm gas}$ and $M$-$T$ at
$r^{\rm Y_{\rm X},X}_{500}$, $r^{\rm Y_{\rm X},wl}_{500}$ or $r^{\rm
  Y_{\rm X},si}_{500}$ ) and the relations from simulations (Borgani
et al. 2004; Kravtsov et al. 2006; Nagai et al.  2007b) shows that
weak lensing mass estimates, similar to X-ray mass estimates, are
about 20 per cent lower than the cluster masses predicted by the
mass--observable relations from simulations, but with $\sim 2 \sigma$
significance. Since the X-ray and weak lensing approaches are
independent, this might indicate that more complex physics is required
in simulations to reproduce the real physics in galaxy cluster.

\bigskip

\noindent
(v) With a fixed slope to 1, the normalization of the best-fit power
law for the $M^{\rm wl}(r_{500}^{\rm Y_{\rm X},X})$ versus $M^{\rm
  X}(r_{500}^{\rm Y_{\rm X},X})$ relation is $1.09 \pm 0.08$. On
average, the weak lensing masses, $M^{\rm wl}(r_{500}^{\rm Y_{\rm
    X},X})$, are higher than the X-ray masses, $M^{\rm X}(r_{500}^{\rm
  Y_{\rm X},X})$, by $9\pm 8$ per cent according to the normalization.
Following the arguments presented in Churazov et al.  (2007), the
current comparison limits the non-thermal pressure support to $9 \pm
8$ per cent according to the ratio between the weak lensing and X-ray
mass estimates.

\bigskip

\noindent
(vi) Depending on the mass calibration (e.g. via X-ray or weak
lensing approaches), the derived value of the amplitude of cosmic
power spectrum varies by 10 per cent between 0.69 and 0.77 for our
sample.

\bigskip

Since the weak lensing mass determination is expected to be unbiased
on average for a large sample of clusters, it is also very important
to compare the agreement between lensing and X-ray mass for the sample
average. We find that the results are consistent with an agreement of
both methods. But different approaches to characterize the
distribution of the deviation gives slightly different results. Taking
these differences as a very conservative measure of systematic
uncertainties we conclude that the two methods of mass determination
agree within an uncertainty of less than 20 per cent for the sample
average.

It is clear from this result that the latter conclusion, which is very
important for the cosmological application of clusters, will benefit
very much from an increase of the sample size, in addition to the
currently happening improvements of lensing mass determinations.  The
present paper therefore highlights the crucial importance of the
completion of the entire LoCuSS sample, which is planned to comprise
in the end about 100 clusters.

\begin{acknowledgements}

  The \emph{XMM-Newton} project is an ESA Science Mission with instruments and
  contributions directly funded by ESA Member States and the USA (NASA). The
  \emph{XMM-Newton} project is supported by the Bundesministerium f\"ur
  Wirtschaft und Technologie/Deutsches Zentrum f\"ur Luft- und Raumfahrt
  (BMWI/DLR, FKZ 50 OX 0001) and the Max-Planck Society. We acknowledge the
  anonymous referee for very useful suggestions led to improvement in our
  understanding.  YYZ acknowledges G. W. Pratt, A.  Vikhlinin, T.  J.  Ponman,
  A. E. Evrard, T. H.  Reiprich, A. Babul, O. Czoske, P. Mazzotta, G.
  Hasinger, and D.  Pierini for useful discussions. YYZ acknowledges support
  from the German BMBF through the Verbundforschung under grant
  No.\,50\,OR\,0601 and MPG. AF acknowledges support from BMBF/DLR under grant
  No.\,50\,OR\,0207 and MPG.  HB and AF acknowledge support through the
  funding of the DfG for the Excellence Cluster Universe EXC153. GPS
  acknowledges support from a Royal Society University Research Fellowship. NO
  acknowledges Grants-in-Aid for the 21st Century COE Program ``Exploring New
  Science by Bridging Particle-Matter Hierarchy'' at Tohoku University, funded
  by the Ministry of Education, Science, Sports and Culture of Japan.

\end{acknowledgements}


\clearpage

\begin{table*} { \begin{center} \footnotesize
      {\renewcommand{\arraystretch}{1.3} \caption[]{ Power law,
          $Y=Y_0\;X^{\gamma}$, parameterized X-ray scaling relations.
          The mean of the scatter is determined by a Gaussian model
          fit to the histogram of the scatter.}
  \label{t:mtx_lite}}
\begin{tabular}{llllccccl}
\hline
\hline
$X$ & $Y$ & $Y_0$ & $\gamma$ & \multicolumn{4}{c}{Scatter (dex)} & Sample \\
\hline
    &     &       &           &  \multicolumn{2}{c}{mean} & \multicolumn{2}{c}{average} & \\
    &     &       &           &  $X$ & $Y$ &  $X$ & $Y$ & \\
\hline
$\frac{Y_{\rm X}}{M_{\odot}\;{\rm keV}} \; E(z) $
   & $\frac{M_{500}^{\rm X}}{M_{\odot}}\;E(z) $
   & $10^{7.10 \pm 0.02}h_{70}^{-1}$ & 0.524 (fixed)  & 0.06 & 0.12 & 0.15 & 0.14   & whole sample  \\
&  & $10^{7.12 \pm 0.02}h_{70}^{-1}$ & 0.524 (fixed)  & 0.07 & 0.13 & 0.15 & 0.14   & non-CCC subsample  \\
&  & $10^{5.63 \pm 0.90}h_{70}^{-1}$ & $0.624 \pm 0.061$ & 0.08 & 0.13 & 0.13 & 0.13   & whole sample  \\
&  & $10^{5.03 \pm 1.07}h_{70}^{-1}$ & $0.666 \pm 0.072$ & 0.05 & 0.12 & 0.12 & 0.13   & non-CCC subsample  \\
\hline
$\frac{M_{\rm gas,500}}{M_{\odot}} \; E(z) $
   & $\frac{M_{500}^{\rm X}}{M_{\odot}}\;E(z) $
   & $10^{3.62 \pm 0.02}h_{70}^{-1}$ & 0.806 (fixed)  & 0.03 & 0.13 & 0.10 & 0.14   & whole sample  \\
&  & $10^{3.65 \pm 0.02}h_{70}^{-1}$ & 0.806 (fixed)  & 0.03 & 0.13 & 0.10 & 0.14   & non-CCC subsample  \\
&  & $10^{2.23 \pm 1.08}h_{70}^{-1}$ & $0.906 \pm 0.078$ & 0.03 & 0.13 & 0.09 & 0.14   & whole sample  \\
&  & $10^{1.44 \pm 1.32}h_{70}^{-1}$ & $0.965 \pm 0.095$ & 0.03 & 0.13 & 0.08 & 0.13   & non-CCC subsample  \\
\hline
$\frac{T_{0.2-0.5r_{500}}}{\rm keV}$
   & $\frac{M_{500}^{\rm X}}{M_{\odot}} \; E(z) $ 
   & $10^{13.55\pm0.02}h_{70}^{-1}$ & 1.499 (fixed) & 0.03 & 0.13 & 0.06 & 0.13 & whole sample \\
&  & $10^{13.58\pm0.02}h_{70}^{-1}$ & 1.499 (fixed) & 0.03 & 0.13 & 0.06 & 0.13 & non-CCC subsample \\
&  & $10^{13.42\pm0.22}h_{70}^{-1}$ & $1.654\pm0.256$ & 0.03 & 0.13 & 0.06 & 0.14 & whole sample \\
&  & $10^{13.40\pm0.25}h_{70}^{-1}$ & $1.718\pm0.298$ & 0.03 & 0.13 & 0.06 & 0.13 & non-CCC subsample \\
\hline
$\frac{T_{0.2-0.5r_{500}}}{\rm keV}$
   & $\frac{M_{\rm gas,500}}{M_{\odot}} \; E(z) $ 
   & $10^{12.33\pm0.01}h_{70}^{-1}$ & 1.859 (fixed) & 0.03 & 0.03 & 0.04 & 0.08 & whole sample \\
&  & $10^{12.33\pm0.02}h_{70}^{-1}$ & 1.859 (fixed) & 0.03 & 0.03 & 0.04 & 0.08 & non-CCC subsample \\
&  & $10^{12.33\pm0.16}h_{70}^{-1}$ & $1.859\pm 0.187$ & 0.03 & 0.11 & 0.04 & 0.08 & whole sample \\
&  & $10^{12.39\pm0.18}h_{70}^{-1}$ & $1.792\pm 0.215$ & 0.07 & 0.13 & 0.04 & 0.08 & non-CCC subsample \\
\hline
$\frac{T_{0.2-0.5r_{500}}}{\rm keV}$
   &
$\frac{L_{\rm 0.1-2.4keV}^{\rm corr}}{\rm erg\;s^{-1}}\;E(z)^{-1} $
   & $10^{42.67\pm0.03}h_{70}^{-1}$ & 2.219 (fixed) & 0.03 & (0.03, 0.20) & 0.06 & 0.13 & whole sample  \\
&  & $10^{42.65\pm0.04}h_{70}^{-1}$ & 2.219 (fixed) & 0.04 & 0.18 & 0.06 & 0.13 & non-CCC subsample  \\
&  & $10^{42.75\pm0.28}h_{70}^{-1}$ & $2.127\pm 0.323$ & 0.03 & (0.03, 0.14) & 0.06 & 0.13 & whole sample  \\
&  & $10^{42.84\pm0.31}h_{70}^{-1}$ & $1.994\pm 0.362$ & 0.04 & 0.15 & 0.07 & 0.13 & non-CCC subsample  \\
\hline
$\frac{T_{0.2-0.5r_{500}}}{\rm keV}$
   & $\frac{L_{\rm bol}^{\rm corr}}{\rm erg\;s^{-1}}\;E(z)^{-1} $ 
   & $10^{42.65\pm0.03}h_{70}^{-1}$ & 2.719 (fixed) & 0.05 & (0.03, 0.18) & 0.05 & 0.13 & whole sample \\
&  & $10^{42.62\pm0.03}h_{70}^{-1}$ & 2.719 (fixed) & 0.03 & 0.16 & 0.05 & 0.13 & non-CCC subsample \\
&  & $10^{42.74\pm0.27}h_{70}^{-1}$ & $2.614 \pm 0.319$ & 0.05 & (0.03, 0.15) & 0.05 & 0.13 & whole sample \\
&  & $10^{42.85\pm0.31}h_{70}^{-1}$ & $2.457 \pm 0.354$ & 0.04 & 0.15 & 0.05 & 0.13 & non-CCC subsample \\
\hline
$\frac{Y_{\rm X}}{M_{\odot}\;{\rm keV}} \; E(z) $
   & $\frac{L_{\rm bol}^{\rm corr}}{\rm erg\;s^{-1}}\;E(z)^{-1} $
   & $10^{30.92 \pm 0.02}h_{70}^{-1}$ & 0.951 (fixed)  & 0.07 & 0.03 & 0.09 & 0.09   & whole sample  \\
&  & $10^{30.89 \pm 0.02}h_{70}^{-1}$ & 0.951 (fixed)  & 0.05 & 0.03 & 0.09 & 0.09   & non-CCC subsample  \\
&  & $10^{30.92 \pm 1.19}h_{70}^{-1}$ & $0.951 \pm 0.081$ & 0.07 & 0.03 & 0.09 & 0.09   & whole sample  \\
&  & $10^{31.42 \pm 1.38}h_{70}^{-1}$ & $0.915 \pm 0.093$ & 0.06 & 0.03 & 0.10 & 0.09   & non-CCC subsample  \\
\hline
$\frac{M_{500}^{\rm X}}{M_{\odot}} \; E(z) $
   & $\frac{L_{\rm bol}^{\rm corr}}{\rm erg\;s^{-1}}\;E(z)^{-1} $
   & $10^{18.03 \pm 0.05}h_{70}^{-1}$ & 1.814 (fixed)  & 0.13 & (0.06, 0.25) & 0.15 & 0.21   & whole sample  \\
&  & $10^{17.96 \pm 0.06}h_{70}^{-1}$ & 1.814 (fixed)  & 0.13 & (0.03, 0.18) & 0.16 & 0.22   & non-CCC subsample  \\
&  & $10^{10.46 \pm 10.39}h_{70}^{-1}$ & $2.325 \pm 0.701$ & 0.13 & 0.09 & 0.15 & 0.26   & whole sample  \\
&  & $10^{15.12 \pm 10.91}h_{70}^{-1}$ & $2.005 \pm 0.735$ & 0.13 & (0.03, 0.19) & 0.16 & 0.24   & non-CCC subsample  \\
\hline
\hline
  \end{tabular}
  \end{center}
  \hspace*{0.3cm}{\footnotesize Here we provide the mean scatter derived by 
    a Gaussian fit to the histogram, in which double values are from the
    histogram with a bi-modal Gaussian distribution. In Zhang et
    al. (2006, 2007), we only presented the average scatter due to the
    limited data points.} }
\end{table*}

\begin{table*} { \begin{center} \footnotesize
      {\renewcommand{\arraystretch}{1.3} \caption[]{ Power law,
          $Y=Y_0\;X^{\gamma}$, parameterized mass--observable scaling
          relations using weak lensing masses ($M_{500}^{\rm wl}$) for
          the 19 LoCuSS clusters having weak lensing masses.}
        \label{t:myl_wlen}}
\begin{tabular}{lllllllll}
\hline
\hline
$X$ & $Y$ & $Y_0$ & $\gamma$ & \multicolumn{4}{c}{Scatter (dex)} & Sample \\
\hline
    &     &       &           &  \multicolumn{2}{c}{mean}  & \multicolumn{2}{c}{average} &                  \\
    &     &       &           &  $X$  & $Y$ &  $X$  & $Y$ &                  \\
\hline
$\frac{Y_{\rm X}}{M_{\odot}\;{\rm keV}} \; E(z) $
   & $\frac{M_{500}^{\rm wl}}{M_{\odot}}\;E(z) $
   & $10^{6.53 \pm 0.04}h_{70}^{-1}$ & 0.568 (fixed)  & 0.09 & 0.24 & 0.32 & 0.35   & all \\
&  & $10^{6.54 \pm 0.06}h_{70}^{-1}$ & 0.568 (fixed)  & 0.25 & 0.23 & 0.37 & 0.32   & non-CCC \\
&  & $10^{6.51 \pm 0.05}h_{70}^{-1}$ & 0.568 (fixed)  & 0.32 & 0.24 & 0.26 & 0.35   & single \\
&  & $10^{6.58 \pm 0.08}h_{70}^{-1}$ & 0.568 (fixed)  & 0.30 & 0.24 & 0.40 & 0.38   & merger \\
\hline
$\frac{M_{\rm gas,500}}{M_{\odot}} \; E(z) $
   & $\frac{M_{500}^{\rm wl}}{M_{\odot}}\;E(z) $
   & $10^{2.48 \pm 0.04}h_{70}^{-1}$ & 0.894 (fixed)  & 0.11 & 0.24 & 0.20 & 0.35   & all \\
&  & $10^{2.48 \pm 0.06}h_{70}^{-1}$ & 0.894 (fixed)  & 0.34 & 0.22 & 0.23 & 0.32   & non-CCC \\
&  & $10^{2.46 \pm 0.05}h_{70}^{-1}$ & 0.894 (fixed)  & 0.08 & 0.23 & 0.16 & 0.34   & single \\
&  & $10^{2.52 \pm 0.08}h_{70}^{-1}$ & 0.894 (fixed)  & 0.31 & 0.23 & 0.26 & 0.36   & merger \\
\hline
$\frac{T_{0.2-0.5r_{500}}}{{\rm keV}} $
   & $\frac{M_{500}^{\rm wl}}{M_{\odot}}\;E(z) $
   & $10^{13.57 \pm 0.05}h_{70}^{-1}$ & 1.59 (fixed)  & 0.06 & 0.26 & 0.12 & 0.37   & all \\
&  & $10^{13.57 \pm 0.06}h_{70}^{-1}$ & 1.59 (fixed)  & 0.09 & 0.23 & 0.14 & 0.33   & non-CCC \\
&  & $10^{13.55 \pm 0.06}h_{70}^{-1}$ & 1.59 (fixed)  & 0.04 & 0.27 & 0.10 & 0.35   & single \\
&  & $10^{13.62 \pm 0.09}h_{70}^{-1}$ & 1.59 (fixed)  & 0.10 & 0.33 & 0.14 & 0.39   & merger \\
\hline
$\frac{M_{500}^{\rm wl}}{M_{\odot}} \; E(z) $
   & $\frac{L_{\rm bol}^{\rm corr}}{\rm erg\;s^{-1}}\;E(z)^{-1} $
   & $10^{21.51 \pm 0.07}h_{70}^{-1}$ & 1.572 (fixed)  & 0.24 & 0.26 & 0.35 & 0.29   & all  \\
&  & $10^{21.48 \pm 0.09}h_{70}^{-1}$ & 1.572 (fixed)  & 0.21 & 0.34 & 0.32 & 0.31   & non-CCC  \\
&  & $10^{21.58 \pm 0.09}h_{70}^{-1}$ & 1.572 (fixed)  & 0.22 & 0.19 & 0.33 & 0.24   & single  \\
&  & $10^{21.41 \pm 0.13}h_{70}^{-1}$ & 1.572 (fixed)  & 0.17 & 0.31 & 0.37 & 0.39   & merger  \\
\hline
\hline
  \end{tabular}
  \end{center}
  \hspace*{0.3cm}{\footnotesize In Col.~Sample, ``merger'' 
    (which denotes ``primary with small secondary'', ``elliptical'',
    ``off-center'' and ``complex'') and ``single'' are classified as
    described in Jones \& Forman (1992). Here we provide the mean
    scatter derived by a Gaussian fit to the histogram, in which double
    values are from the histogram with a bi-modal Gaussian distribution.}
}
\end{table*}

\begin{table*} { \begin{center} \footnotesize
      {\renewcommand{\arraystretch}{1.3} \caption[]{See the caption in
          Table~\ref{t:myl_wlen} except that the weak lensing masses,
          X-ray masses and $Y_{\rm X}$ parameters are determined
          within the radii ($r_{500}^{\rm Y_{\rm X},X}$) determined by
          the method described in \S~\ref{s:simuyx}.}
        \label{t:myl_wlen_simuyx}}
\begin{tabular}{lllllllll}
\hline
\hline
$X$ & $Y$ & $Y_0$ & $\gamma$ & \multicolumn{4}{c}{Scatter (dex)} & Sample \\
\hline
    &     &       &           &  \multicolumn{2}{c}{mean}  & \multicolumn{2}{c}{average} &                  \\
    &     &       &           &  $X$  & $Y$ &  $X$  & $Y$ &                  \\
\hline
$\frac{Y_{\rm X}(r^{\rm Y_{\rm X},X}_{500})}{M_{\odot}\;{\rm keV}} \; E(z) $
   & $\frac{M^{\rm X}(r^{\rm Y_{\rm X},X}_{500})}{M_{\odot}}\;E(z) $
   & $10^{6.45 \pm 0.01}h_{70}^{-1}$ & 0.568 (fixed)  & 0.07 & 0.13 & 0.14 & 0.20   & all \\
&  & $10^{6.47 \pm 0.01}h_{70}^{-1}$ & 0.568 (fixed)  & 0.07 & 0.13 & 0.14 & 0.20   & non-CCC  \\
$\frac{Y_{\rm X}(r^{\rm Y_{\rm X},X}_{500})}{M_{\odot}\;{\rm keV}} \; E(z) $
   & $\frac{M^{\rm wl}(r^{\rm Y_{\rm X},X}_{500})}{M_{\odot}}\;E(z) $
   & $10^{6.47 \pm 0.03}h_{70}^{-1}$ & 0.568 (fixed)  & 0.07 & 0.22 & 0.32 & 0.35   & all \\
&  & $10^{6.47 \pm 0.04}h_{70}^{-1}$ & 0.568 (fixed)  & 0.09 & 0.20 & 0.32 & 0.31   & non-CCC  \\
\hline
$\frac{M_{\rm gas}(r^{\rm Y_{\rm X},X}_{500})}{M_{\odot}} \; E(z) $
   & $\frac{M^{\rm X}(r^{\rm Y_{\rm X},X}_{500})}{M_{\odot}}\;E(z) $
   & $10^{2.40 \pm 0.01}h_{70}^{-1}$ & 0.894 (fixed)  & 0.03 & 0.13 & 0.10 & 0.20   & all \\
&  & $10^{2.42 \pm 0.02}h_{70}^{-1}$ & 0.894 (fixed)  & 0.03 & 0.13 & 0.09 & 0.20   & non-CCC  \\
$\frac{M_{\rm gas}(r^{\rm Y_{\rm X},X}_{500})}{M_{\odot}} \; E(z) $
   & $\frac{M^{\rm wl}(r^{\rm Y_{\rm X},X}_{500})}{M_{\odot}}\;E(z) $
   & $10^{2.40 \pm 0.03}h_{70}^{-1}$ & 0.894 (fixed)  & 0.03 & 0.18 & 0.20 & 0.35   & all \\
&  & $10^{2.39 \pm 0.04}h_{70}^{-1}$    & 0.894 (fixed)  & 0.03 & 0.03 & 0.32 & 0.31   & non-CCC  \\
\hline
$\frac{T_{0.2-0.5r_{500}}}{{\rm keV}} $
   & $\frac{M^{\rm X}(r^{\rm Y_{\rm X},X}_{500})}{M_{\odot}}\;E(z) $
   & $10^{13.48 \pm 0.02}h_{70}^{-1}$ & 1.59 (fixed)  & 0.03 & 0.13 & 0.05 & 0.20   & all \\
&  & $10^{13.49 \pm 0.02}h_{70}^{-1}$ & 1.59 (fixed)  & 0.01 & 0.12 & 0.05 & 0.20   & non-CCC  \\
$\frac{T_{0.2-0.5r_{500}}}{{\rm keV}} $
   & $\frac{M^{\rm wl}(r^{\rm Y_{\rm X},X}_{500})}{M_{\odot}}\;E(z) $
   & $10^{13.53 \pm 0.04}h_{70}^{-1}$ & 1.59 (fixed)  & 0.09 & 0.24 & 0.12 & 0.37   & all \\
&  & $10^{13.53 \pm 0.04}h_{70}^{-1}$ & 1.59 (fixed)  & 0.10 & 0.22 & 0.32 & 0.31   & non-CCC  \\
\hline
$\frac{M^{\rm X}(r^{\rm Y_{\rm X},X}_{500})}{M_{\odot}} \; E(z) $
   & $\frac{L_{\rm bol}^{\rm corr}}{\rm erg\;s^{-1}}\;E(z)^{-1} $
   & $10^{21.62 \pm 0.03}h_{70}^{-1}$ & 1.572 (fixed)  & 0.13 & 0.03 & 0.20 & 0.14   & all  \\
&  & $10^{21.58 \pm 0.03}h_{70}^{-1}$ & 1.572 (fixed)  & 0.13 & 0.03 & 0.20 & 0.15   & non-CCC  \\
$\frac{M^{\rm wl}(r^{\rm Y_{\rm X},X}_{500})}{M_{\odot}} \; E(z) $
   & $\frac{L_{\rm bol}^{\rm corr}}{\rm erg\;s^{-1}}\;E(z)^{-1} $
   & $10^{21.59 \pm 0.05}h_{70}^{-1}$ & 1.572 (fixed)  & 0.13 & 0.03 & 0.35 & 0.29   & all  \\
&  & $10^{21.56 \pm 0.06}h_{70}^{-1}$ & 1.572 (fixed)  & 0.21 & 0.19 & 0.32 & 0.31   & non-CCC  \\
\hline
\hline
  \end{tabular}
  \end{center}
  \hspace*{0.3cm}{\footnotesize }
}
\end{table*}

\begin{table*} { \begin{center} \footnotesize
      {\renewcommand{\arraystretch}{1.3} \caption[]{See the caption in
          Table~\ref{t:myl_wlen} except that the weak lensing masses, X-ray
          masses and $Y_{\rm X}$ parameters are determined within the radii
          ($r_{500}^{\rm Y_{\rm X},wl}$) determined by the method described in
          \S~\ref{s:ylmwlen_simuyx1}.}
        \label{t:myl_wlen_simuyx1}}
\begin{tabular}{lllllllll}
\hline
\hline
$X$ & $Y$ & $Y_0$ & $\gamma$ & \multicolumn{4}{c}{Scatter (dex)} & Sample \\
\hline
    &     &       &           &  \multicolumn{2}{c}{mean}  & \multicolumn{2}{c}{average} &                  \\
    &     &       &           &  $X$  & $Y$ &  $X$  & $Y$ &                  \\
\hline
$\frac{Y_{\rm X}(r^{\rm Y_{\rm X},wl}_{500})}{M_{\odot}\;{\rm keV}} \; E(z) $
   & $\frac{M^{\rm X}(r^{\rm Y_{\rm X},wl}_{500})}{M_{\odot}}\;E(z) $
   & $10^{6.44 \pm 0.01}h_{70}^{-1}$ & 0.568 (fixed)  & 0.08 & 0.13 & 0.13 & 0.20   & all \\
&  & $10^{6.46 \pm 0.01}h_{70}^{-1}$ & 0.568 (fixed)  & 0.07 & 0.12 & 0.13 & 0.20   & non-CCC  \\
$\frac{Y_{\rm X}(r^{\rm Y_{\rm X},wl}_{500})}{M_{\odot}\;{\rm keV}} \; E(z) $
   & $\frac{M^{\rm wl}(r^{\rm Y_{\rm X},wl}_{500})}{M_{\odot}}\;E(z) $
   & $10^{6.46 \pm 0.03}h_{70}^{-1}$ & 0.568 (fixed)  & 0.05 & 0.19 & 0.22 & 0.38   & all \\
&  & $10^{6.46 \pm 0.04}h_{70}^{-1}$ & 0.568 (fixed)  & 0.14 & 0.14 & 0.26 & 0.36   & non-CCC  \\
\hline
$\frac{M_{\rm gas}(r^{\rm Y_{\rm X},wl}_{500})}{M_{\odot}} \; E(z) $
   & $\frac{M^{\rm X}(r^{\rm Y_{\rm X},wl}_{500})}{M_{\odot}}\;E(z) $
   & $10^{2.39 \pm 0.01}h_{70}^{-1}$ & 0.894 (fixed)  & 0.03 & 0.12 & 0.09 & 0.20   & all \\
&  & $10^{2.41 \pm 0.02}h_{70}^{-1}$ & 0.894 (fixed)  & 0.03 & 0.12 & 0.09 & 0.20   & non-CCC  \\
$\frac{M_{\rm gas}(r^{\rm Y_{\rm X},wl}_{500})}{M_{\odot}} \; E(z) $
   & $\frac{M^{\rm wl}(r^{\rm Y_{\rm X},wl}_{500})}{M_{\odot}}\;E(z) $
   & $10^{2.40 \pm 0.03}h_{70}^{-1}$ & 0.894 (fixed)  & 0.02 & 0.15 & 0.15 & 0.38   & all \\
&  & $10^{2.40 \pm 0.04}h_{70}^{-1}$ & 0.894 (fixed)  & 0.12 & 0.06 & 0.17 & 0.36   & non-CCC  \\
\hline
$\frac{T_{0.2-0.5r_{500}}}{{\rm keV}} $
   & $\frac{M^{\rm X}(r^{\rm Y_{\rm X},wl}_{500})}{M_{\odot}}\;E(z) $
   & $10^{13.47 \pm 0.02}h_{70}^{-1}$ & 1.59 (fixed)  & 0.03 & 0.13 & 0.05 & 0.20   & all \\
&  & $10^{13.49 \pm 0.02}h_{70}^{-1}$ & 1.59 (fixed)  & 0.01 & 0.12 & 0.05 & 0.20   & non-CCC  \\
$\frac{T_{0.2-0.5r_{500}}}{{\rm keV}} $
   & $\frac{M^{\rm wl}(r^{\rm Y_{\rm X},wl}_{500})}{M_{\odot}}\;E(z) $
   & $10^{13.49 \pm 0.04}h_{70}^{-1}$ & 1.59 (fixed)  & 0.08 & 0.24 & 0.09 & 0.39   & all \\
&  & $10^{13.49 \pm 0.04}h_{70}^{-1}$ & 1.59 (fixed)  & 0.10 & 0.23 & 0.11 & 0.38   & non-CCC  \\
\hline
$\frac{M^{\rm X}(r^{\rm Y_{\rm X},wl}_{500})}{M_{\odot}} \; E(z) $
   & $\frac{L_{\rm bol}^{\rm corr}}{\rm erg\;s^{-1}}\;E(z)^{-1} $
   & $10^{21.63 \pm 0.03}h_{70}^{-1}$ & 1.572 (fixed)  & 0.13 & 0.03 & 0.20 & 0.14   & all  \\
&  & $10^{21.59 \pm 0.03}h_{70}^{-1}$ & 1.572 (fixed)  & 0.13 & 0.07 & 0.20 & 0.15   & non-CCC  \\
$\frac{M^{\rm wl}(r^{\rm Y_{\rm X},wl}_{500})}{M_{\odot}} \; E(z) $
   & $\frac{L_{\rm bol}^{\rm corr}}{\rm erg\;s^{-1}}\;E(z)^{-1} $
   & $10^{21.65 \pm 0.05}h_{70}^{-1}$ & 1.572 (fixed)  & 0.21 & 0.03 & 0.38 & 0.20   & all  \\
&  & $10^{21.61 \pm 0.06}h_{70}^{-1}$ & 1.572 (fixed)  & 0.19 & 0.14 & 0.37 & 0.24   & non-CCC  \\
\hline
\hline
  \end{tabular}
  \end{center}
  \hspace*{0.3cm}{\footnotesize }
}
\end{table*}

\begin{table*} { \begin{center} \footnotesize
      {\renewcommand{\arraystretch}{1.3} \caption[]{See the caption in
          Table~\ref{t:myl_wlen} except that the weak lensing masses, X-ray
          masses and $Y_{\rm X}$ parameters are determined within the radii
          ($r_{500}^{\rm Y_{\rm X},si}$) determined by the method described in
          \S~\ref{s:ylmwlen_simuyx2}.}
        \label{t:myl_wlen_simuyx2}}
\begin{tabular}{lllllllll}
\hline
\hline
$X$ & $Y$ & $Y_0$ & $\gamma$ & \multicolumn{4}{c}{Scatter (dex)} & Sample \\
\hline
    &     &       &           &  \multicolumn{2}{c}{mean}  & \multicolumn{2}{c}{average} &                  \\
    &     &       &           &  $X$  & $Y$ &  $X$  & $Y$ &                  \\
\hline
$\frac{Y_{\rm X}(r^{\rm Y_{\rm X},si}_{500})}{M_{\odot}\;{\rm keV}} \; E(z) $
   & $\frac{M^{\rm X}(r^{\rm Y_{\rm X},si}_{500})}{M_{\odot}}\;E(z) $
   & $10^{6.47 \pm 0.01}h_{70}^{-1}$ & 0.568 (fixed)  & 0.08 & 0.13 & 0.14 & 0.20   & all \\
&  & $10^{6.49 \pm 0.02}h_{70}^{-1}$ & 0.568 (fixed)  & 0.08 & 0.13 & 0.13 & 0.20   & non-CCC  \\
$\frac{Y_{\rm X}(r^{\rm Y_{\rm X},si}_{500})}{M_{\odot}\;{\rm keV}} \; E(z) $
   & $\frac{M^{\rm wl}(r^{\rm Y_{\rm X},si}_{500})}{M_{\odot}}\;E(z) $
   & $10^{6.47 \pm 0.03}h_{70}^{-1}$ & 0.568 (fixed)  & 0.07 & 0.22 & 0.23 & 0.34   & all \\
&  & $10^{6.47 \pm 0.04}h_{70}^{-1}$ & 0.568 (fixed)  & 0.09 & 0.20 & 0.27 & 0.32   & non-CCC  \\
\hline
$\frac{M_{\rm gas}(r^{\rm Y_{\rm X},si}_{500})}{M_{\odot}} \; E(z) $
   & $\frac{M^{\rm X}(r^{\rm Y_{\rm X},si}_{500})}{M_{\odot}}\;E(z) $
   & $10^{2.40 \pm 0.02}h_{70}^{-1}$ & 0.894 (fixed)  & 0.03 & 0.13 & 0.09 & 0.20   & all \\
&  & $10^{2.43 \pm 0.02}h_{70}^{-1}$ & 0.894 (fixed)  & 0.03 & 0.13 & 0.10 & 0.20   & non-CCC  \\
$\frac{M_{\rm gas}(r^{\rm Y_{\rm X},si}_{500})}{M_{\odot}} \; E(z) $
   & $\frac{M^{\rm wl}(r^{\rm Y_{\rm X},si}_{500})}{M_{\odot}}\;E(z) $
   & $10^{2.40 \pm 0.03}h_{70}^{-1}$ & 0.894 (fixed)  & 0.03 & 0.18 & 0.15 & 0.34   & all \\
&  & $10^{2.40 \pm 0.04}h_{70}^{-1}$ & 0.894 (fixed)  & 0.15 & 0.09 & 0.17 & 0.32   & non-CCC  \\
\hline
$\frac{T_{0.2-0.5r_{500}}}{{\rm keV}} $
   & $\frac{M^{\rm X}(r^{\rm Y_{\rm X},si}_{500})}{M_{\odot}}\;E(z) $
   & $10^{13.52 \pm 0.02}h_{70}^{-1}$ & 1.59 (fixed)  & 0.05 & 0.12 & 0.05 & 0.19   & all \\
&  & $10^{13.53 \pm 0.02}h_{70}^{-1}$ & 1.59 (fixed)  & 0.02 & 0.13 & 0.06 & 0.20   & non-CCC  \\
$\frac{T_{0.2-0.5r_{500}}}{{\rm keV}} $
   & $\frac{M^{\rm wl}(r^{\rm Y_{\rm X},si}_{500})}{M_{\odot}}\;E(z) $
   & $10^{13.53 \pm 0.04}h_{70}^{-1}$ & 1.59 (fixed)  & 0.09 & 0.24 & 0.09 & 0.35   & all \\
&  & $10^{13.53 \pm 0.04}h_{70}^{-1}$ & 1.59 (fixed)  & 0.10 & 0.22 & 0.11 & 0.33   & non-CCC  \\
\hline
$\frac{M^{\rm X}(r^{\rm Y_{\rm X},si}_{500})}{M_{\odot}} \; E(z) $
   & $\frac{L_{\rm bol}^{\rm corr}}{\rm erg\;s^{-1}}\;E(z)^{-1} $
   & $10^{21.56 \pm 0.03}h_{70}^{-1}$ & 1.572 (fixed)  & 0.13 & 0.14 & 0.20 & 0.15   & all  \\
&  & $10^{21.51 \pm 0.03}h_{70}^{-1}$ & 1.572 (fixed)  & 0.13 & 0.14 & 0.20 & 0.18   & non-CCC  \\
$\frac{M^{\rm wl}(r^{\rm Y_{\rm X},si}_{500})}{M_{\odot}} \; E(z) $
   & $\frac{L_{\rm bol}^{\rm corr}}{\rm erg\;s^{-1}}\;E(z)^{-1} $
   & $10^{21.59 \pm 0.05}h_{70}^{-1}$ & 1.572 (fixed)  & 0.18 & 0.03 & 0.34 & 0.21   & all  \\
&  & $10^{21.56 \pm 0.06}h_{70}^{-1}$ & 1.572 (fixed)  & 0.21 & 0.17 & 0.33 & 0.24   & non-CCC  \\
\hline
\hline
  \end{tabular}
  \end{center}
  \hspace*{0.3cm}{\footnotesize }
}
\end{table*}


\clearpage



\clearpage

\appendix

\clearpage 

\section{X-ray mass modeling}
\label{s:result}

The X-ray mass modeling described here helps to understand the measurement
uncertainties of the X-ray quantities used for studies of the scaling
relations.

\begin{figure}
\begin{center}
\includegraphics[angle=270,width=8.5cm]{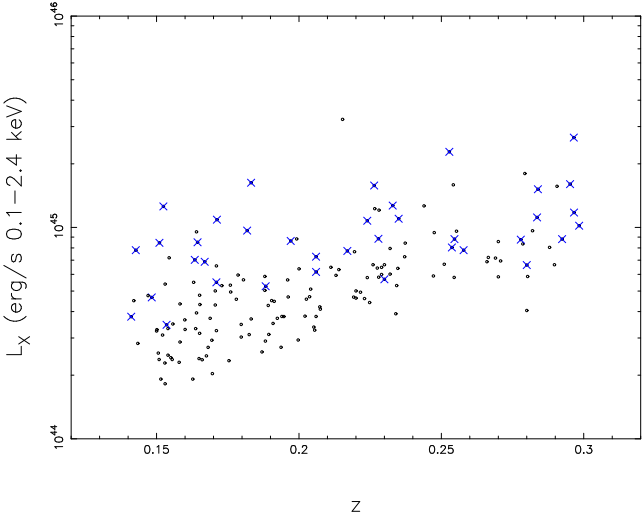}
\end{center}
\caption{ROSAT luminosity versus redshift for the LoCuSS sample. 
The crosses denote the 37 LoCuSS clusters observed by \emph{XMM-Newton}.
\label{f:lxz}}
\end{figure}

We show the ROSAT luminosity versus redshift distribution for the LoCuSS
sample in Fig.~\ref{f:lxz} and the observational information of the
\emph{XMM-Newton} data for the 37 LocuSS cluters in Table~\ref{t:obs}. The
primary parameters of all 37 LoCuSS galaxy clusters are given in
Table~\ref{t:global}.

\subsection{Density contrast}

To determine the global cluster parameters, one needs a fiducial outer
radius that was defined as follows. The mean cluster density
contrast is the average density with respect to the critical density,
\begin{equation}
\Delta(<r)= \frac{3 M(<r)}{4 \pi r^3 \rho_{\rm c}(z)} \;.
\label{e:del}
\end{equation}
The critical density at redshift $z$ is $\rho_{\rm c}(z)=\rho_{\rm c0}
E^2(z)$, where $E^2(z)=\Omega_{\rm
m}(1+z)^3+\Omega_{\Lambda}+(1-\Omega_{\rm
m}-\Omega_{\Lambda})(1+z)^2$. $r_{\Delta}$ is the radius within which
the density contrast is $\Delta$. $M_{\Delta}$ is the total mass
within $r_{\Delta}$. For $\Delta=500$, $r_{500}$ is the radius within
which the density contrast is 500 and $M_{500}$ is the total mass
within $r_{500}$. 

\subsection{Temperature profiles}
\label{s:kt}

\begin{figure*}
\begin{center}

\includegraphics[angle=270,width=8.5cm]{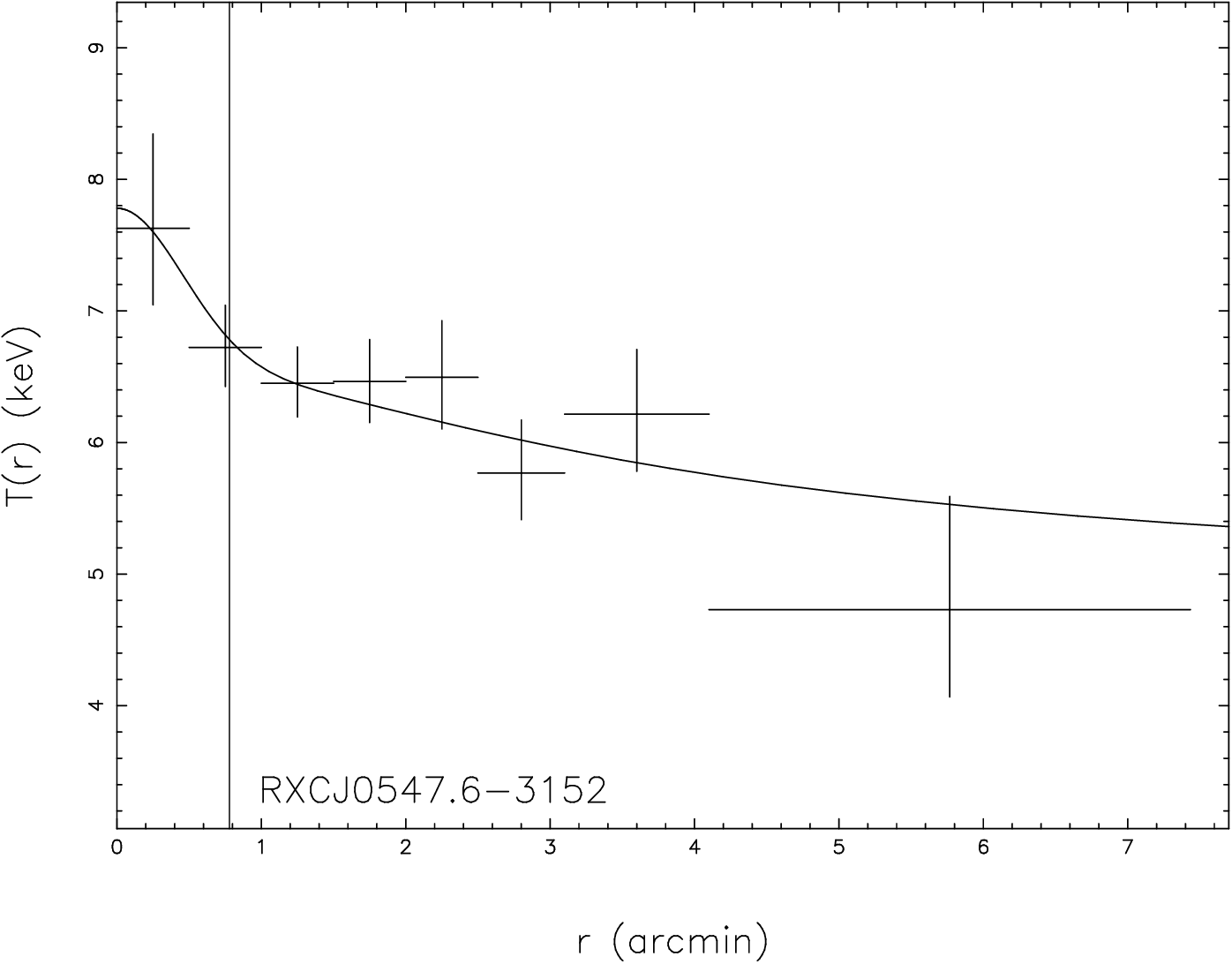}
\includegraphics[angle=270,width=8.5cm]{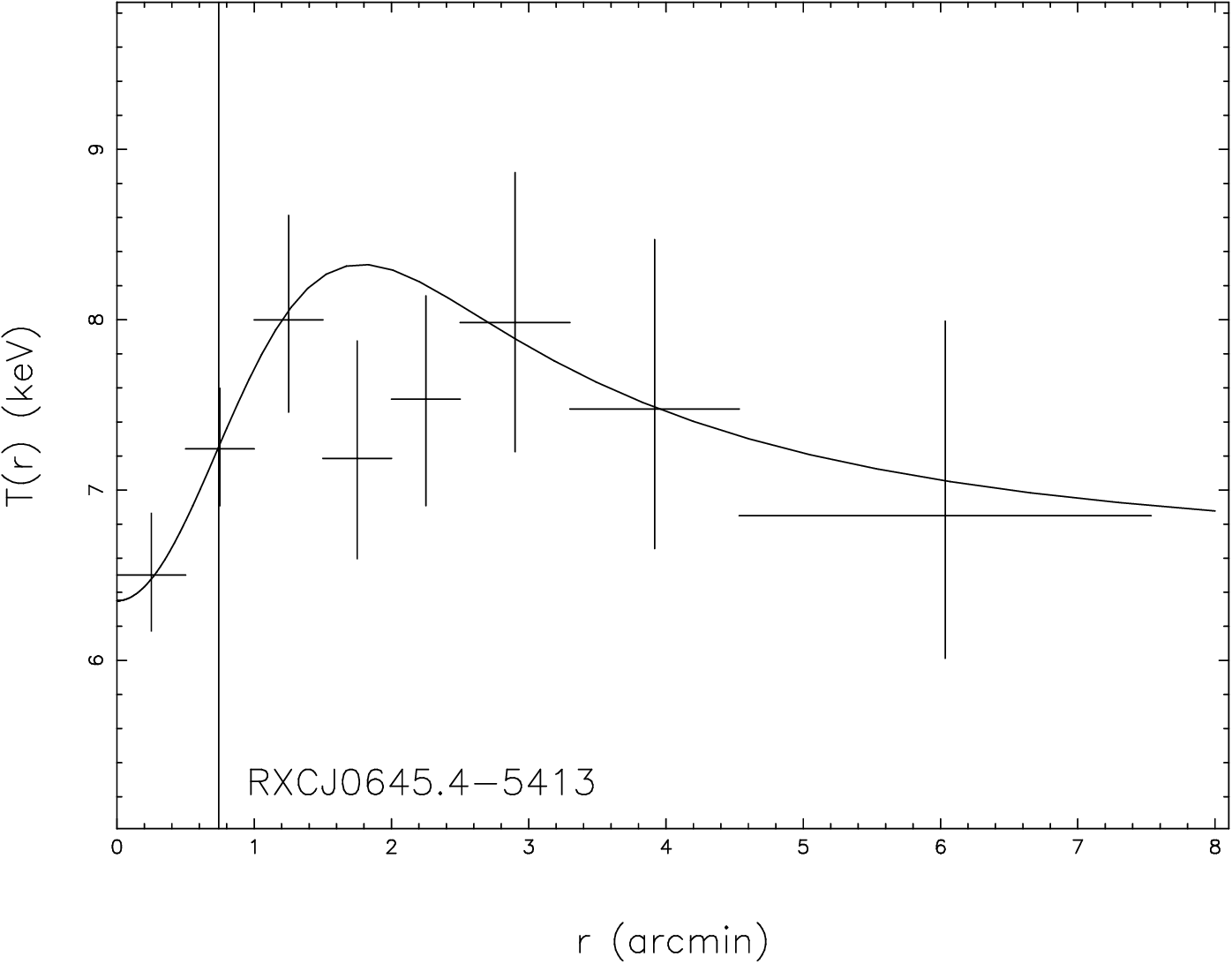}

\includegraphics[angle=270,width=8.5cm]{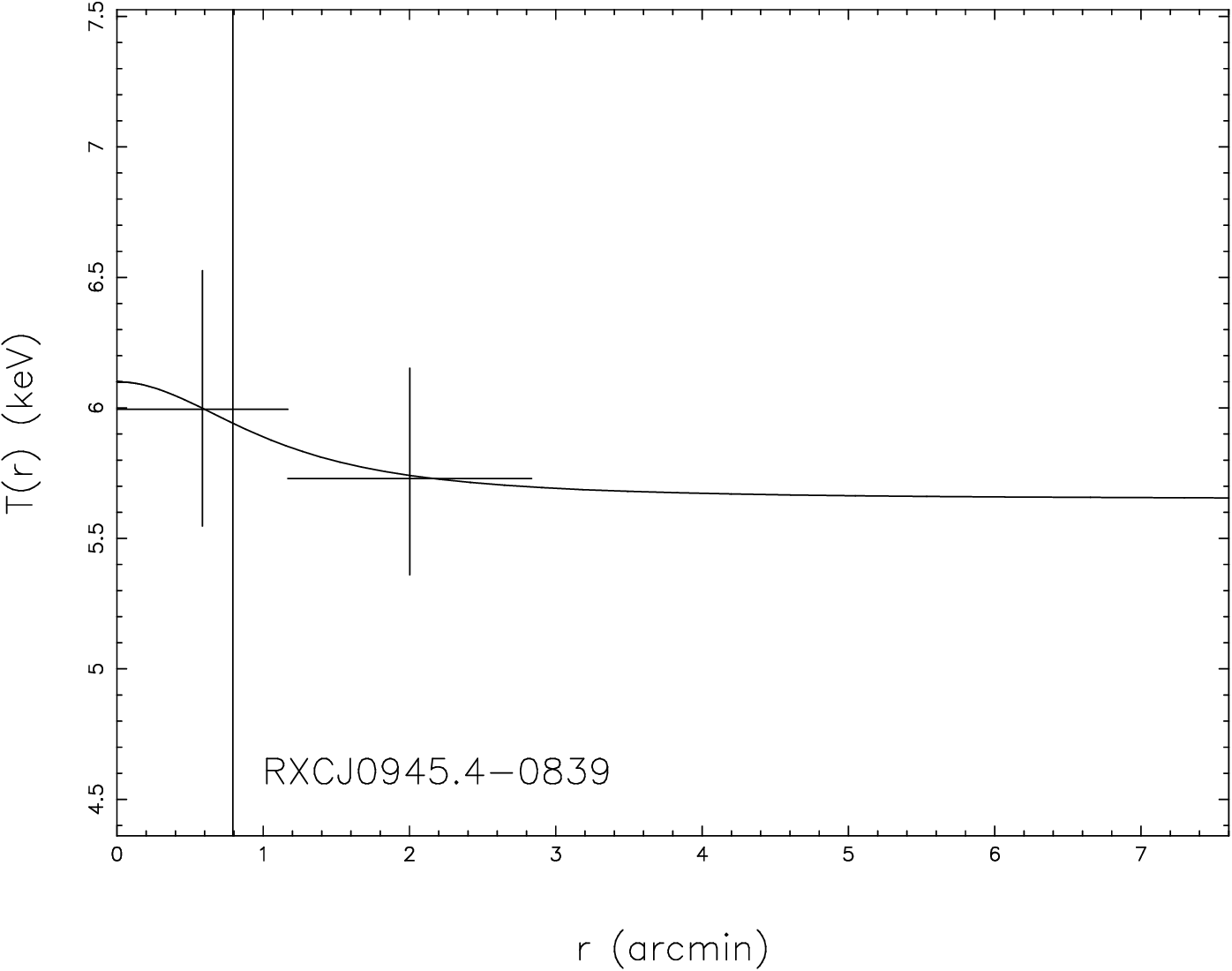}
\includegraphics[angle=270,width=8.5cm]{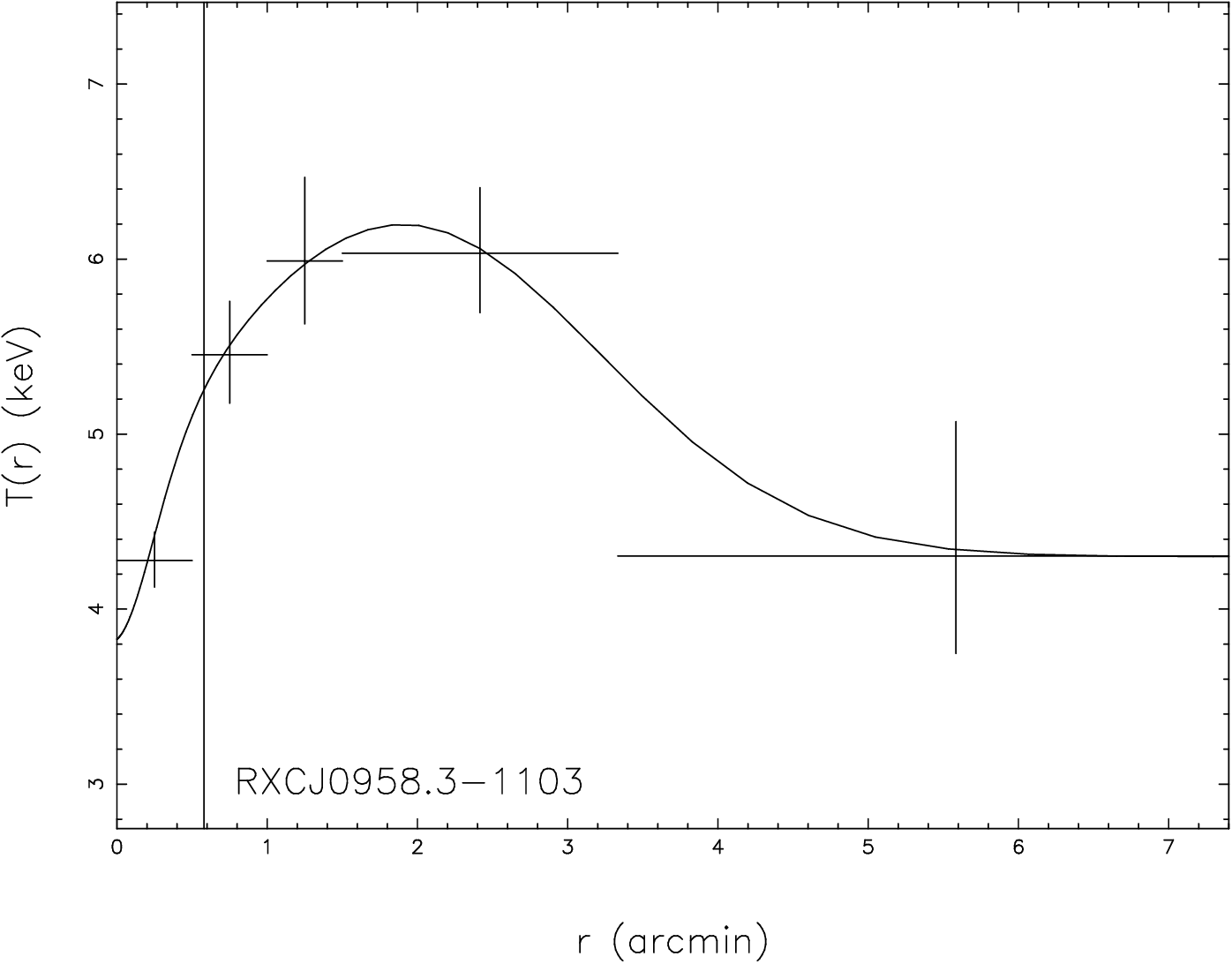}

\includegraphics[angle=270,width=8.5cm]{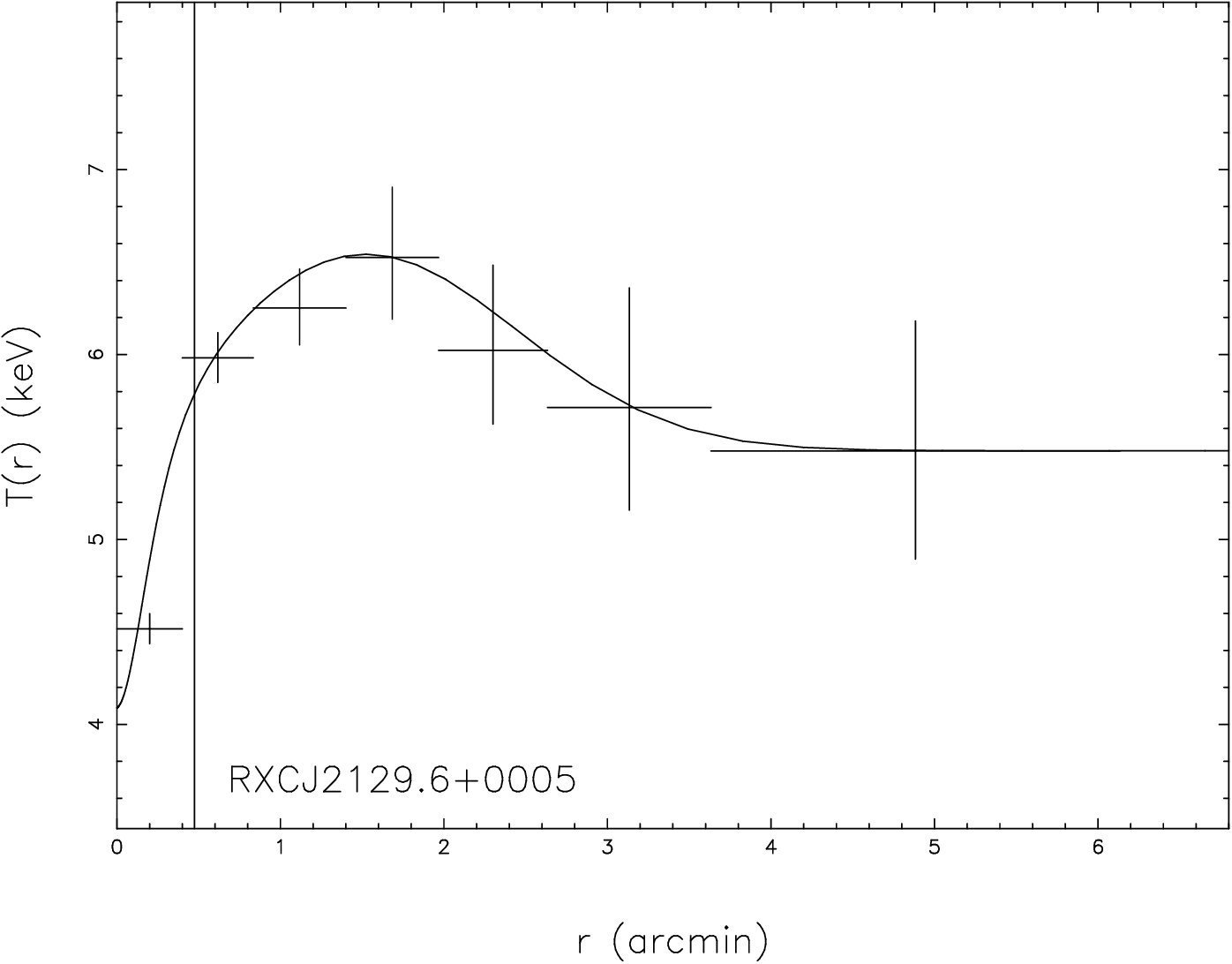}
\includegraphics[angle=270,width=8.5cm]{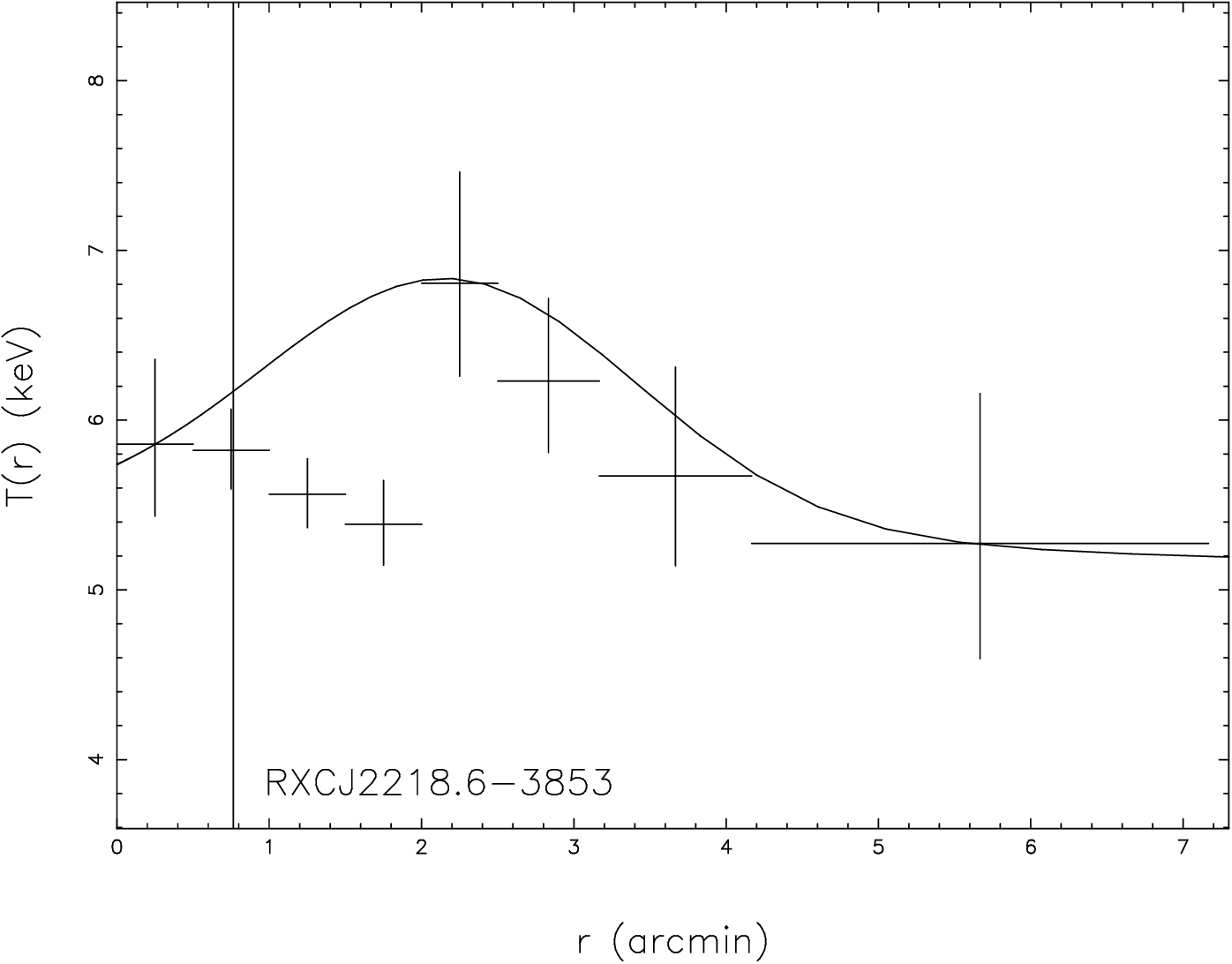}

\end{center}
\caption{De-projected temperature profiles approximated by the
parameterization $T(r)=P_3
\exp[-(r-P_1)^2/P_2]+P_6(1+r^2/P_4^2)^{-P_5}+P_7$ crossing all the
data points (solid). The vertical line denotes $0.1 r_{500}$.
\label{f:ktcom1}}
\end{figure*}

\begin{figure*}
\begin{center}

\includegraphics[angle=270,width=8.5cm]{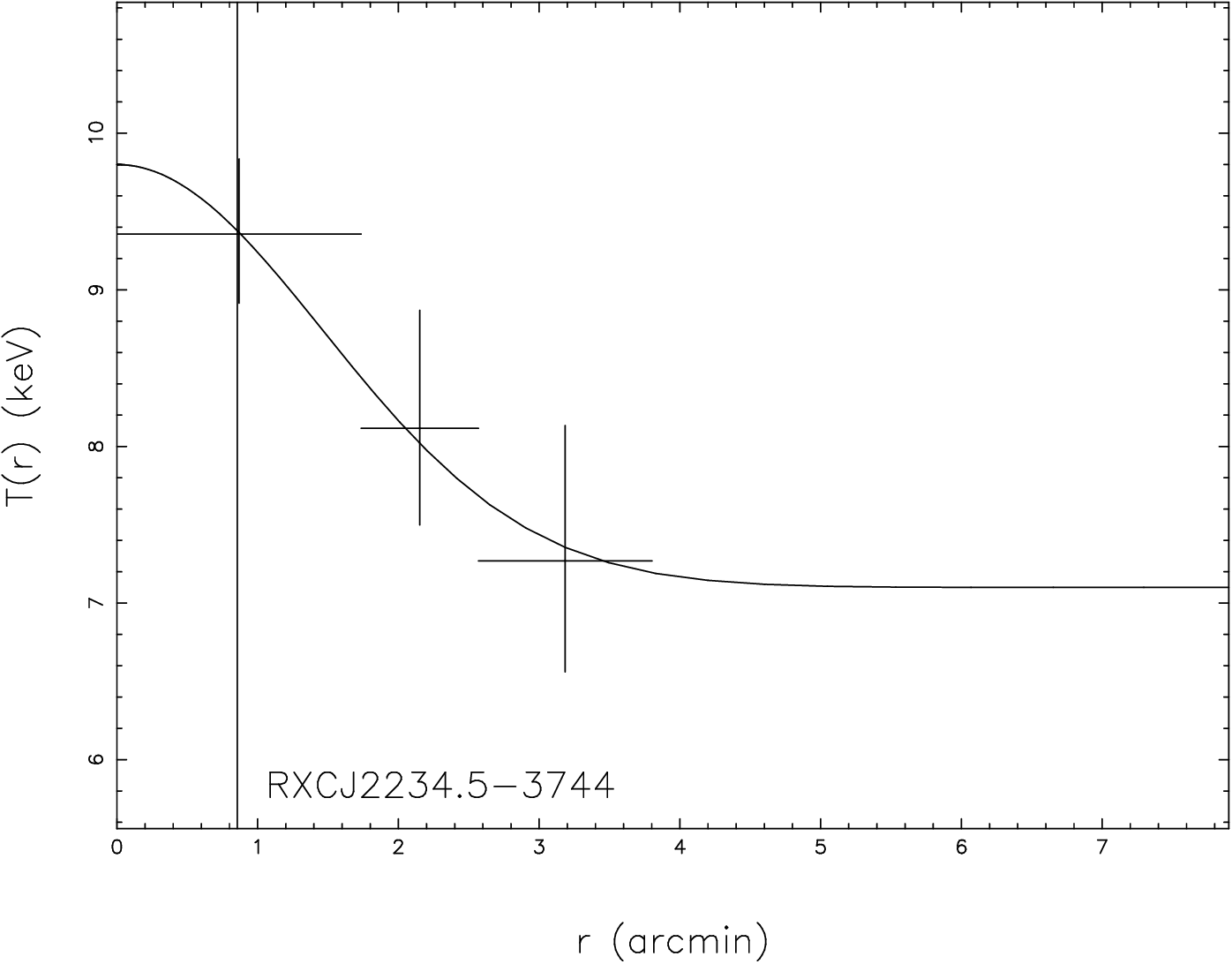}
\includegraphics[angle=270,width=8.5cm]{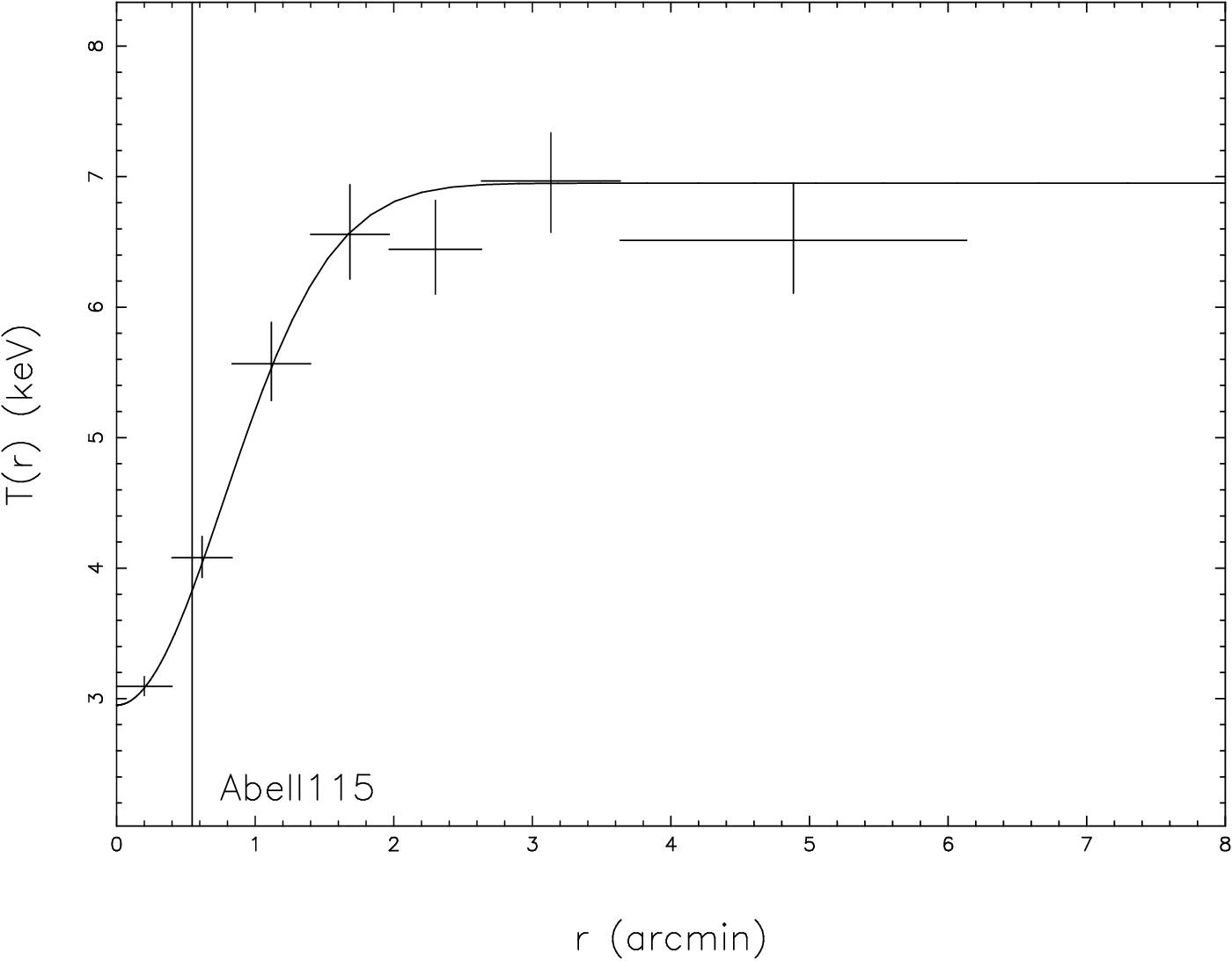}

\includegraphics[angle=270,width=8.5cm]{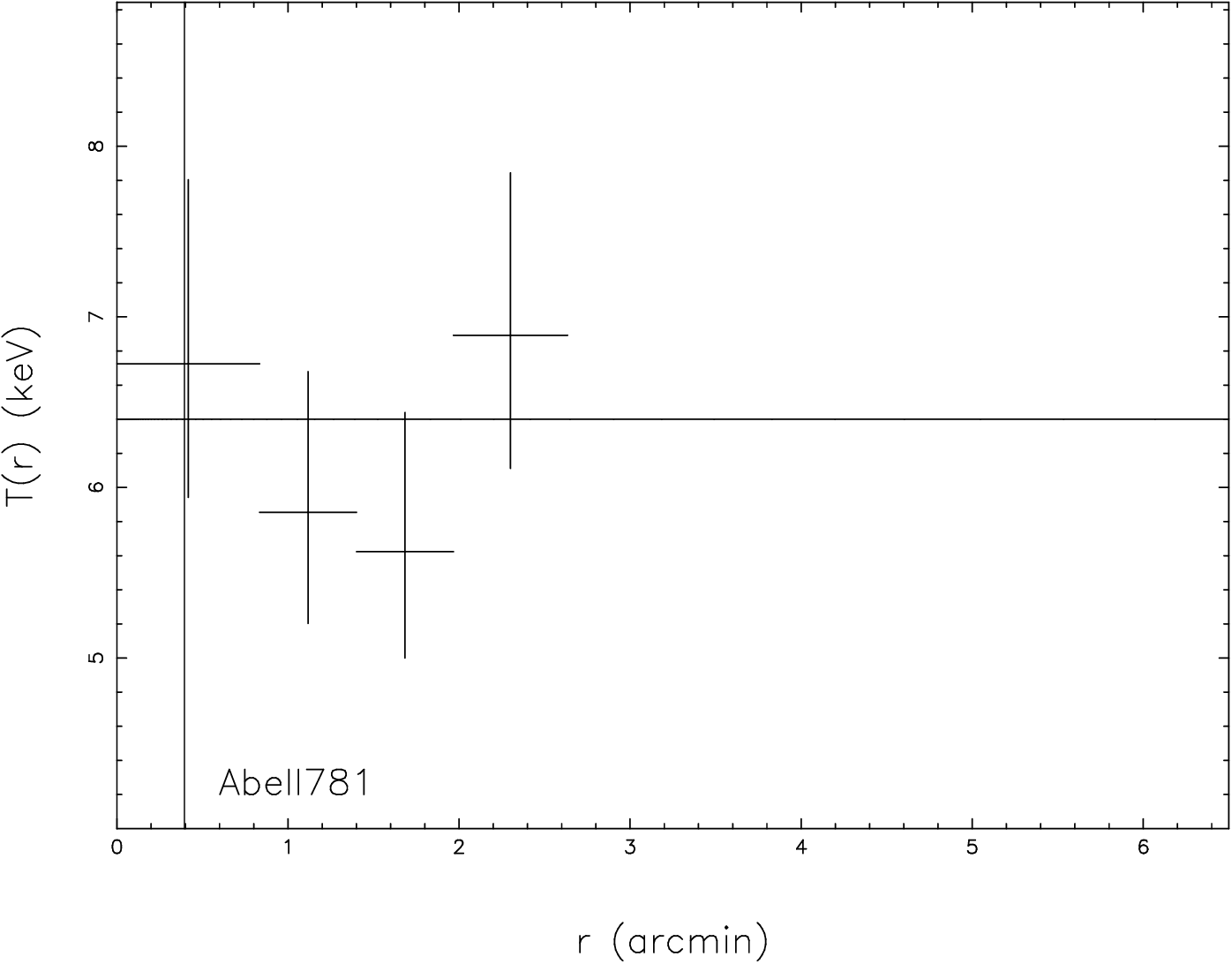}
\includegraphics[angle=270,width=8.5cm]{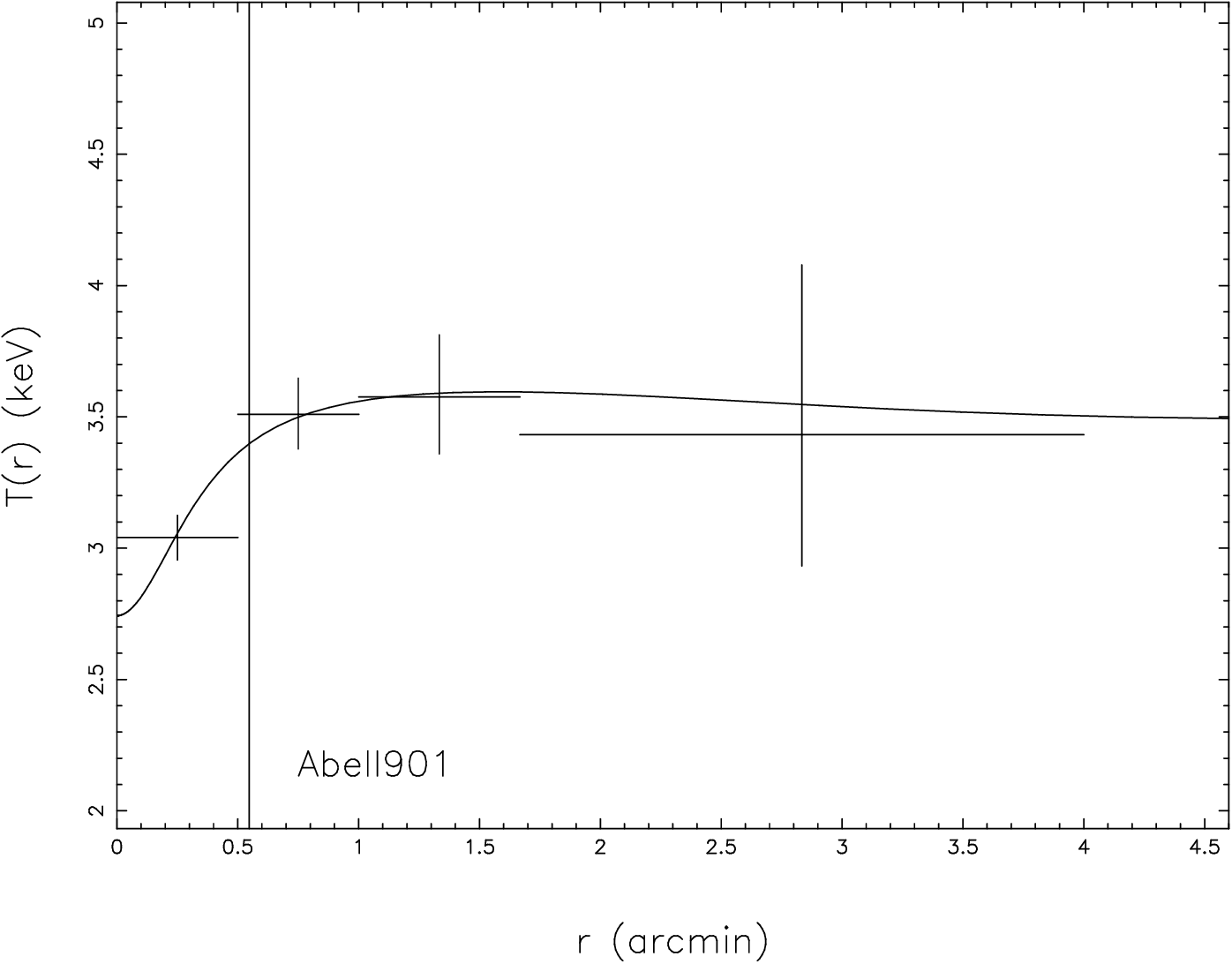}

\includegraphics[angle=270,width=8.5cm]{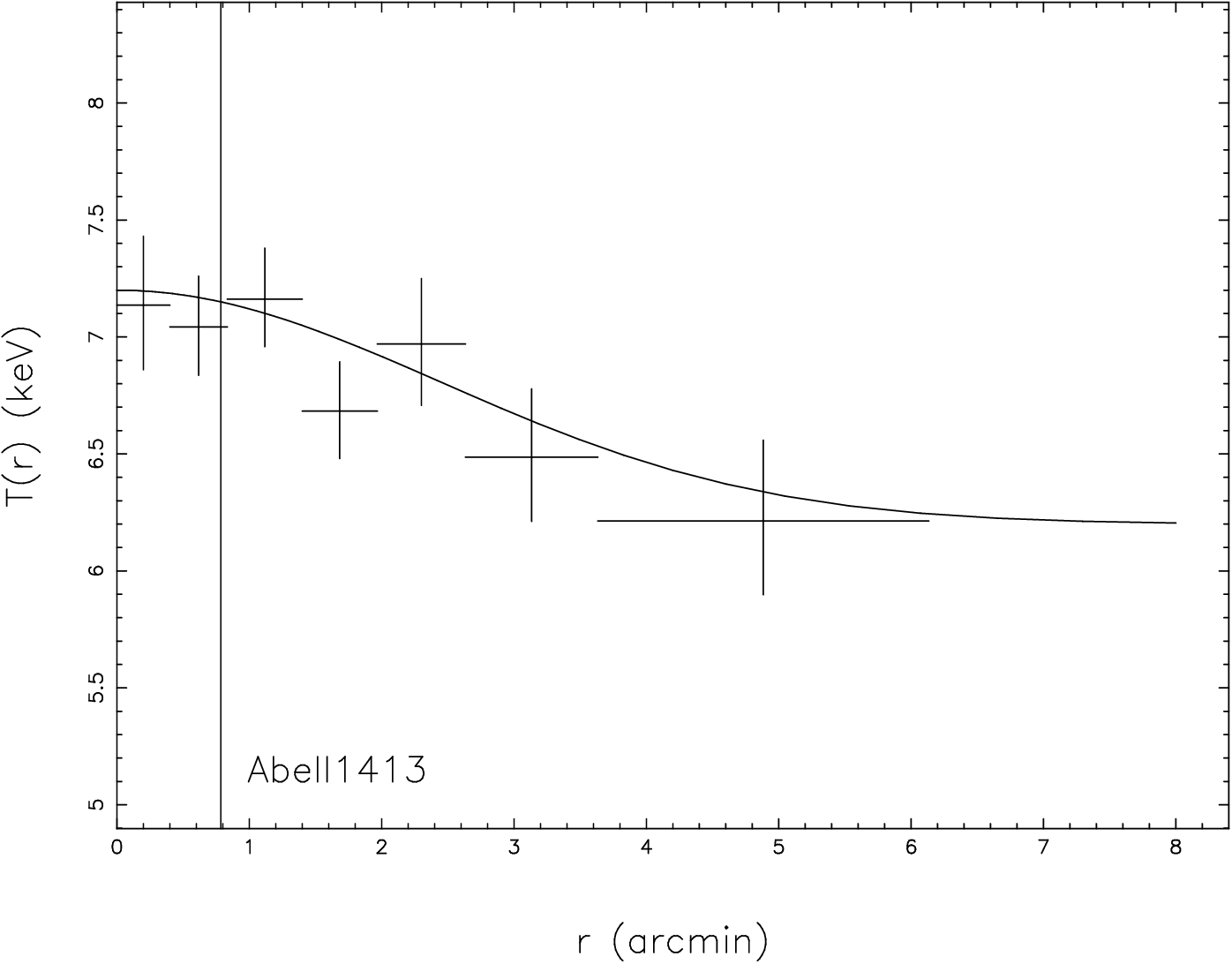}
\includegraphics[angle=270,width=8.5cm]{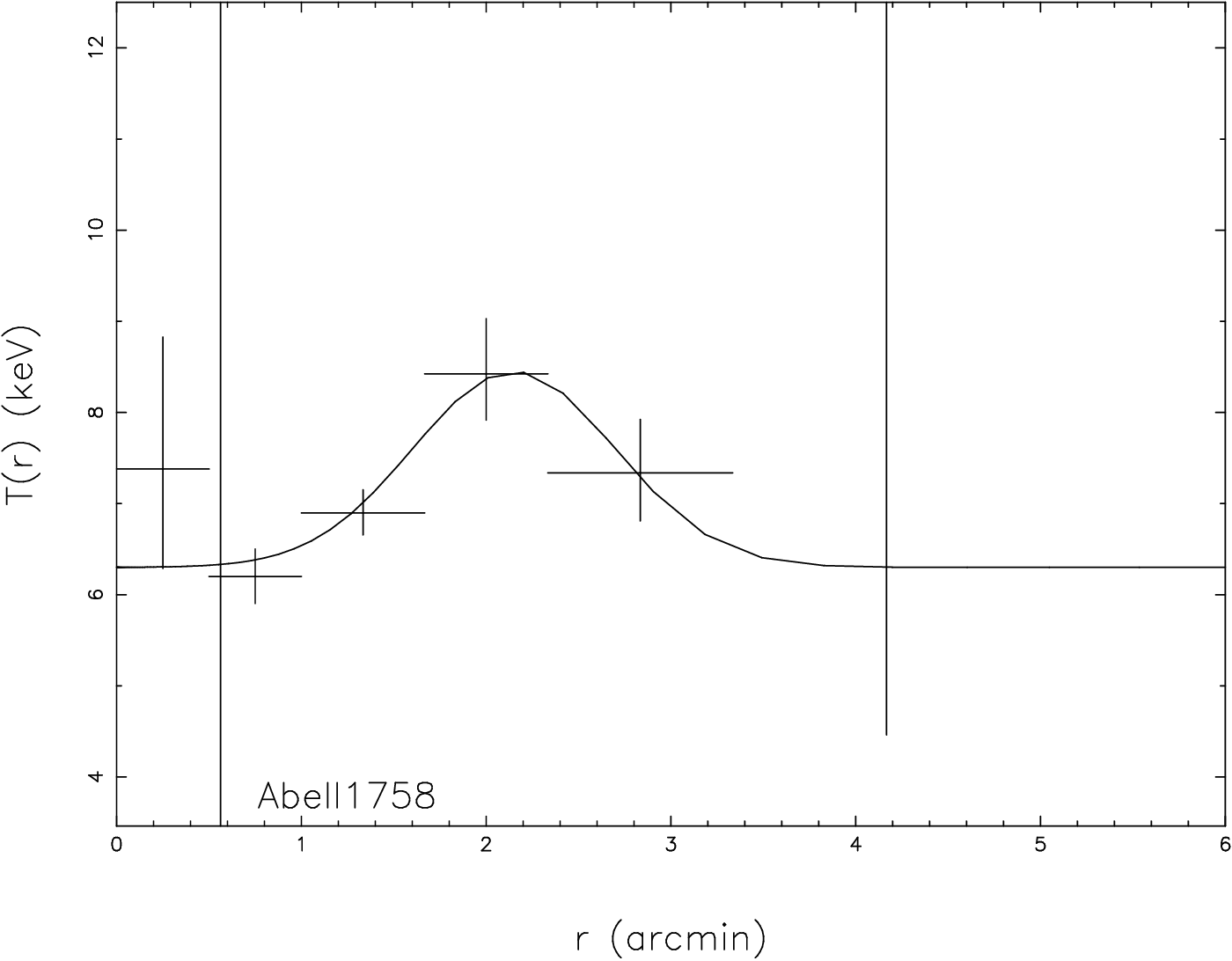}

\end{center}
\caption{See caption in Fig.~\ref{f:ktcom1}.
\label{f:ktcom11}}
\end{figure*}

\begin{figure*}
\begin{center}

\includegraphics[angle=270,width=8.5cm]{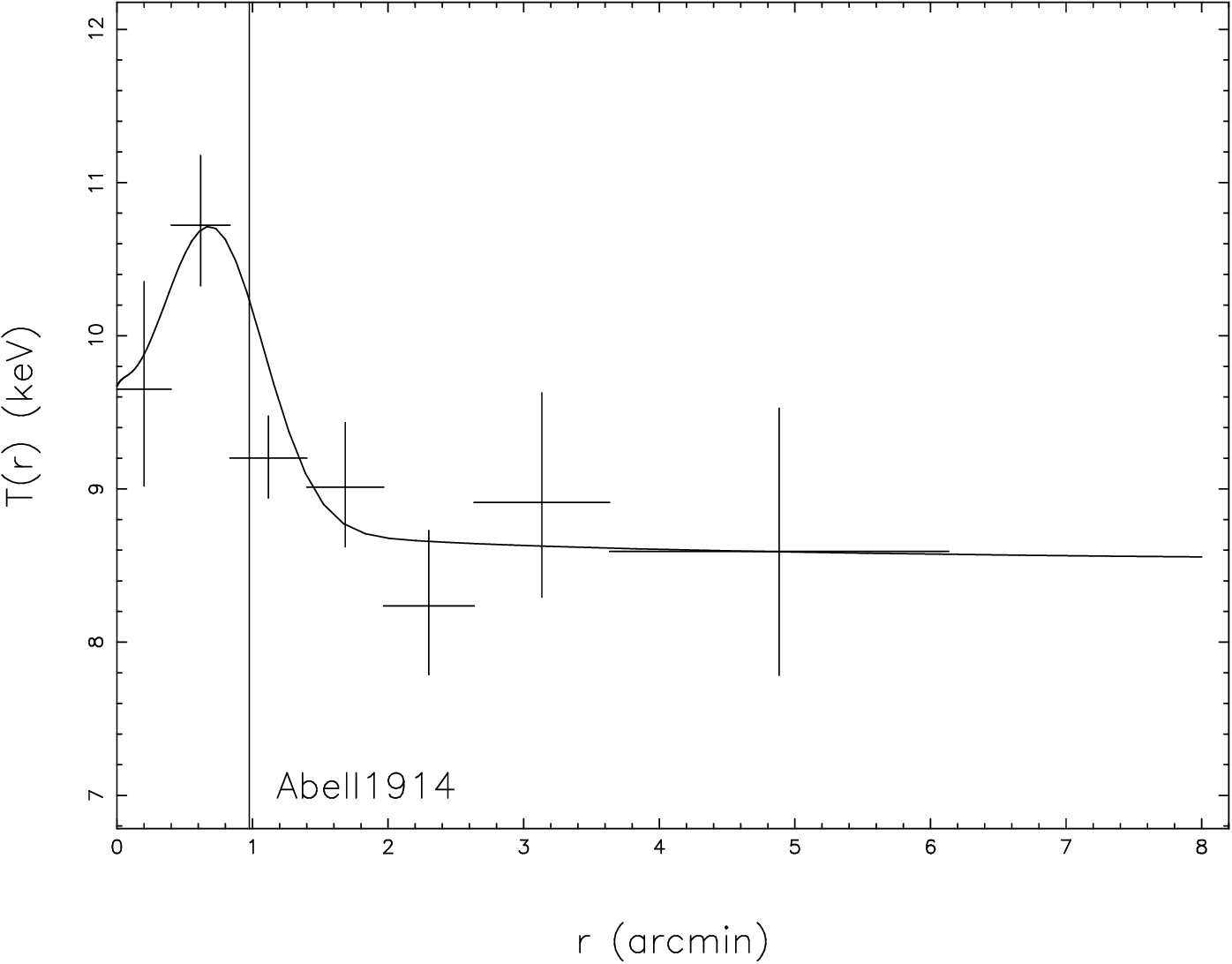}
\includegraphics[angle=270,width=8.5cm]{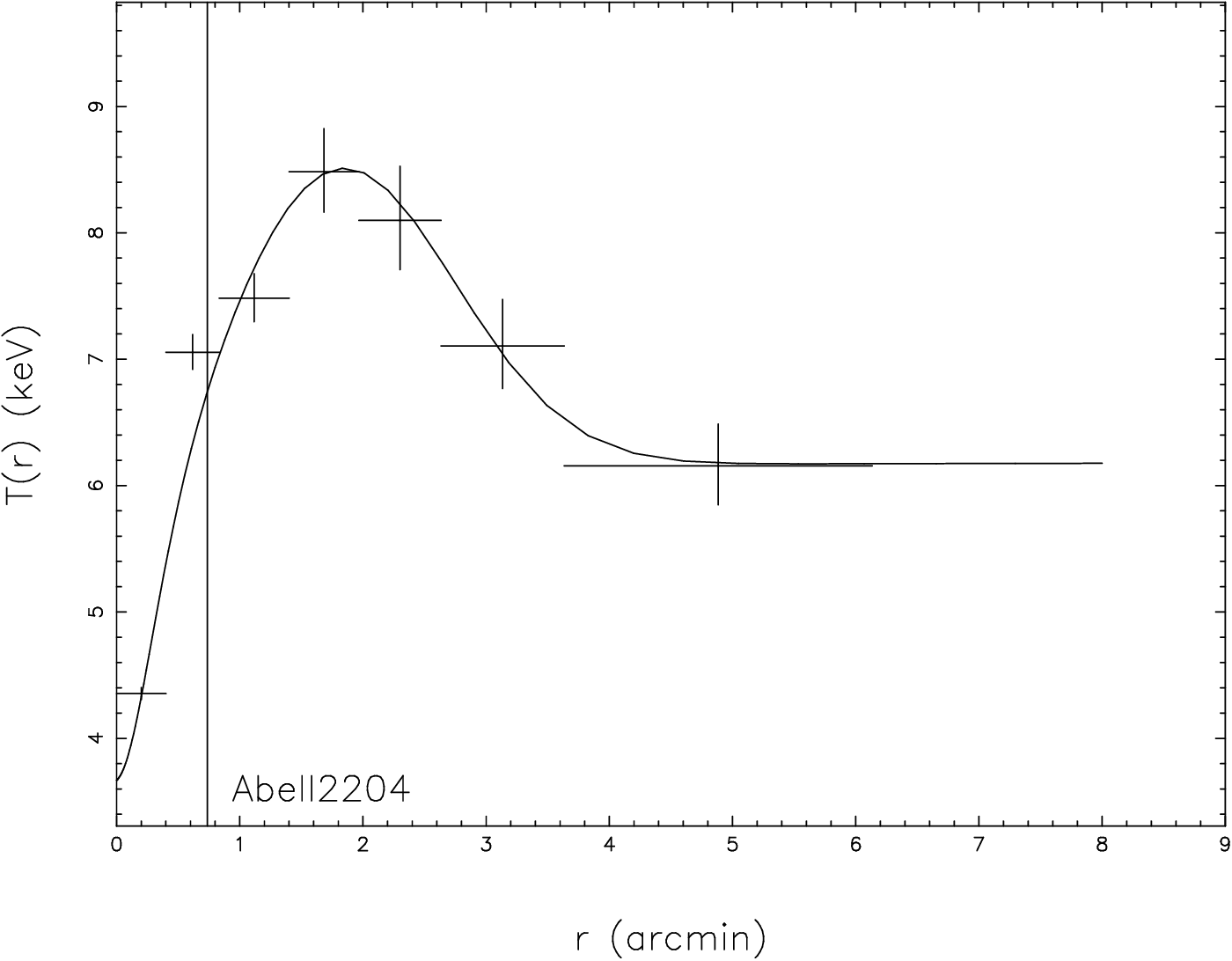}

\includegraphics[angle=270,width=8.5cm]{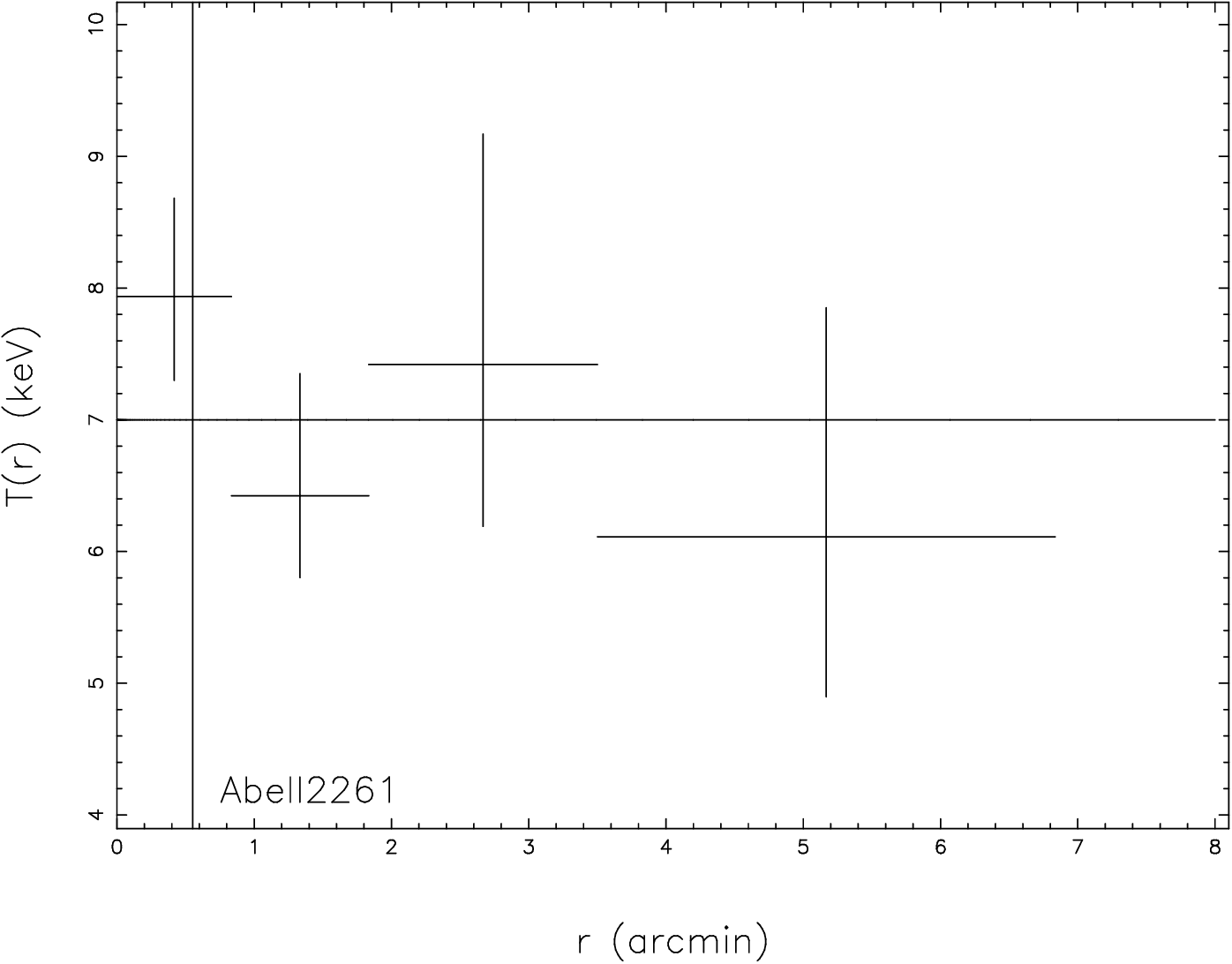}
\includegraphics[angle=270,width=8.5cm]{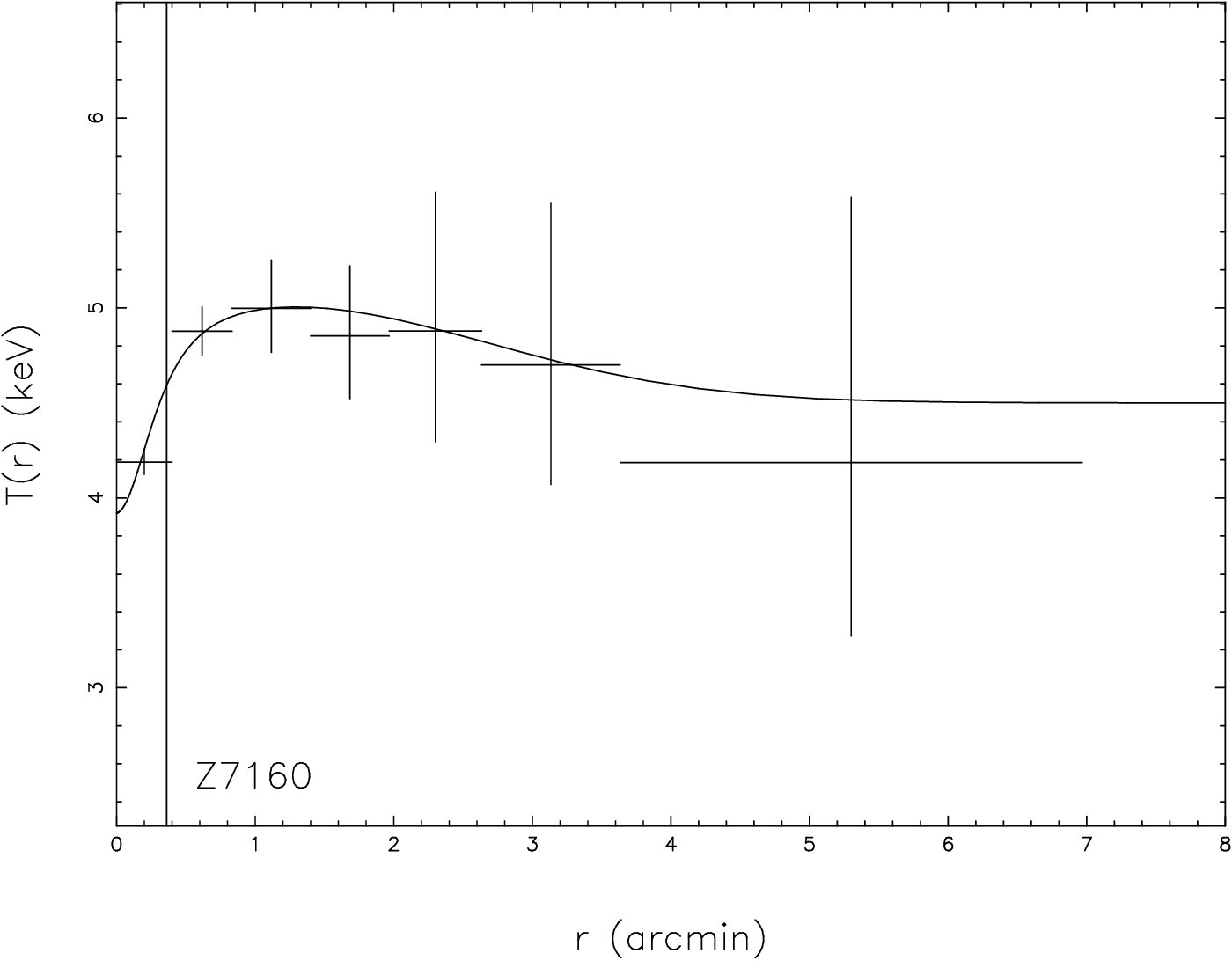}

\end{center}
\caption{See caption in Fig.~\ref{f:ktcom1}.
\label{f:ktcom12}}
\end{figure*}

\emph{XMM-Newton} (as well as \emph{Chandra}), in contrast to earlier
telescopes, has a less energy-dependent, smaller PSF, more reliable to study
cluster temperature profiles. We de-projected the spectra (see the method in
Zhang et al. 2007) and performed the spectral fitting in XSPEC to obtain the
radial profiles of the temperature and metallicity. In
Figs.~\ref{f:ktcom1}--\ref{f:ktcom12} in the electronic edition of the Journal
(together with Fig.~2 in Zhang et al.  2006, and Fig.~1 and Figs.~B.1-B.5 in
Zhang et al. 2007 for the whole sample) we show the radial temperature
profiles. The temperature profiles are approximated by
\begin{equation}
T(r)=P_3
\exp[-(r-P_1)^2/P_2]+P_6(1+r^2/P_4^2)^{-P_5}+P_7\; ,
\label{e:t}
\end{equation}
motivated by the qualitative appearance of the observed temperature
distributions of clusters of galaxies. $P_{i}$, $i=1,...,7$, are used for
parameterization.

A CCC often shows both a peaked surface brightness profile and a steep
temperature drop towards the cluster center. Our sample does not show
pronounced CCC and non-CCC bimodal distribution. For studies of the
segregation of the X-ray scaling relations we empirically define the
CCC as the cluster whose temperature drop is greater than 30 per cent
of the peak temperature towards the cluster center. Nine clusters are
therefore CCCs in this sample. There might be more CCCs in the sample
if the sample would be constructed using the higher resolution
temperature measurements.

The X-ray surface brightness isophotes are also sensitive to cluster
populations. We therefore also used the classification of the cluster
morphologies in Jones \& Forman (1992) and divided the sample to 5 types
(Table~\ref{t:catalog2}): ``single'', ``primary with a small secondary'',
``elliptical'', ``off-center'' and ``complex'' based on the smoothness of its
X-ray surface brightness isophotes in the 0.7--2~keV band.  Therefore
``single'' approximately denotes relaxed clusters, ``primary with a small
secondary'', ``off-center'', ``elliptical'' and ``complex'' denote merging
clusters.

About 80 per cent CCCs are relaxed clusters (``single'' defined in
\S~\ref{s:ylmwlen} using surface brightness isophotes). As we show later,
these 9 CCCs often show shorter central cooling time, larger cooling radii
(scaled to $r_{500}$) and lower central entropies compared to the non-CCCs in
the sample (Table~\ref{t:catalog2}).

\subsection{Surface brightness profiles}
\label{s:sx}

\begin{figure*}
\begin{center}

\includegraphics[angle=270,width=8.5cm]{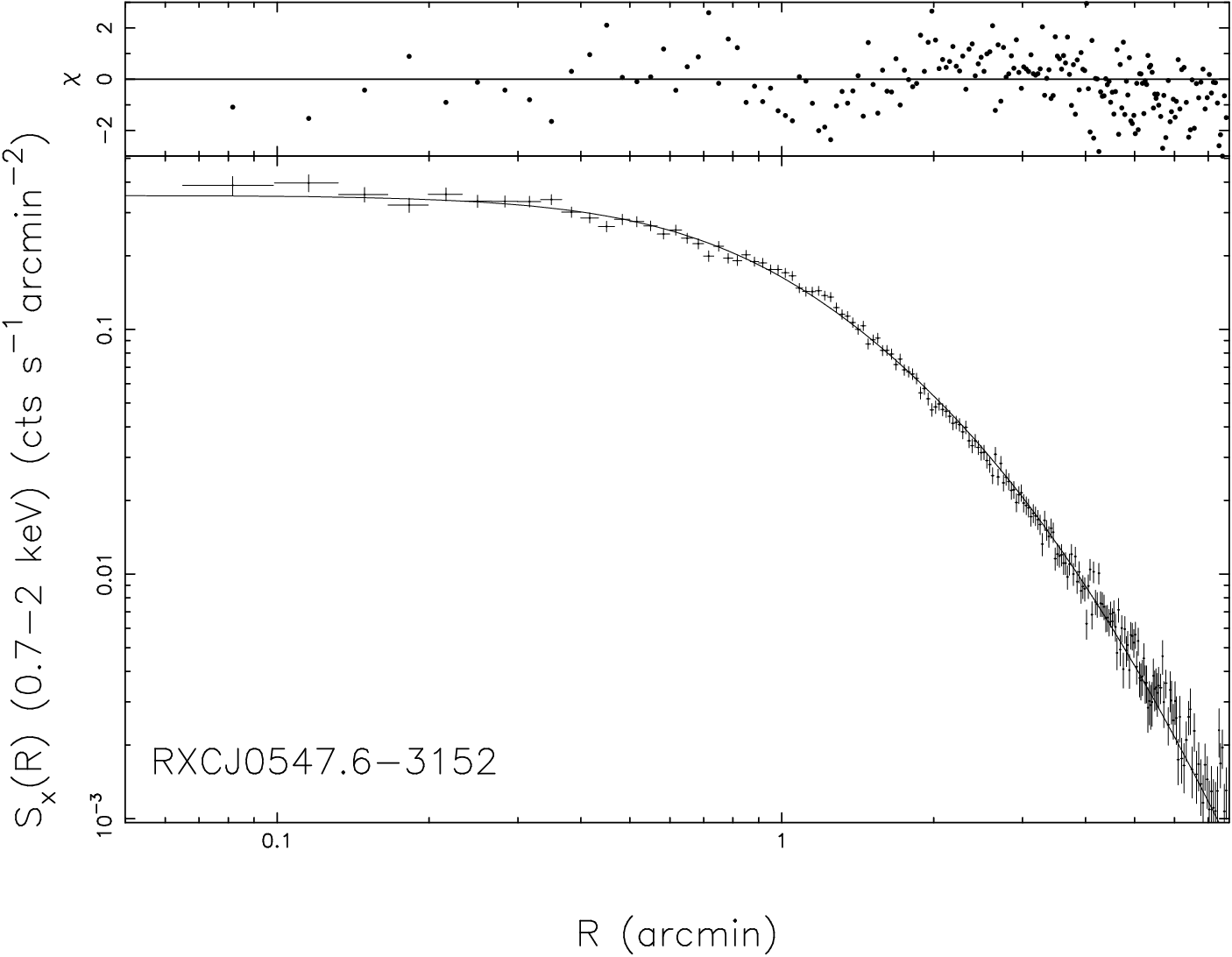}
\includegraphics[angle=270,width=8.5cm]{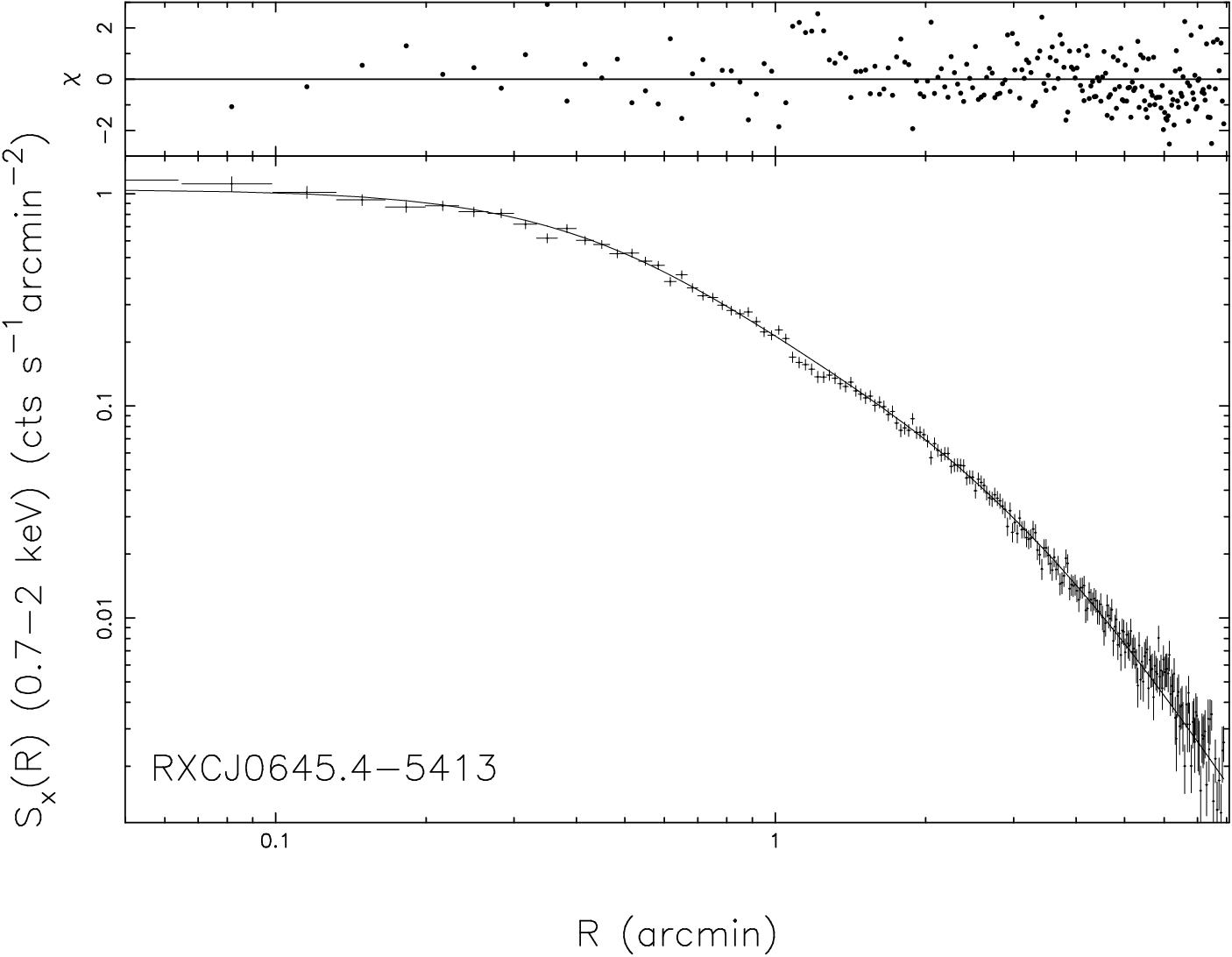}

\includegraphics[angle=270,width=8.5cm]{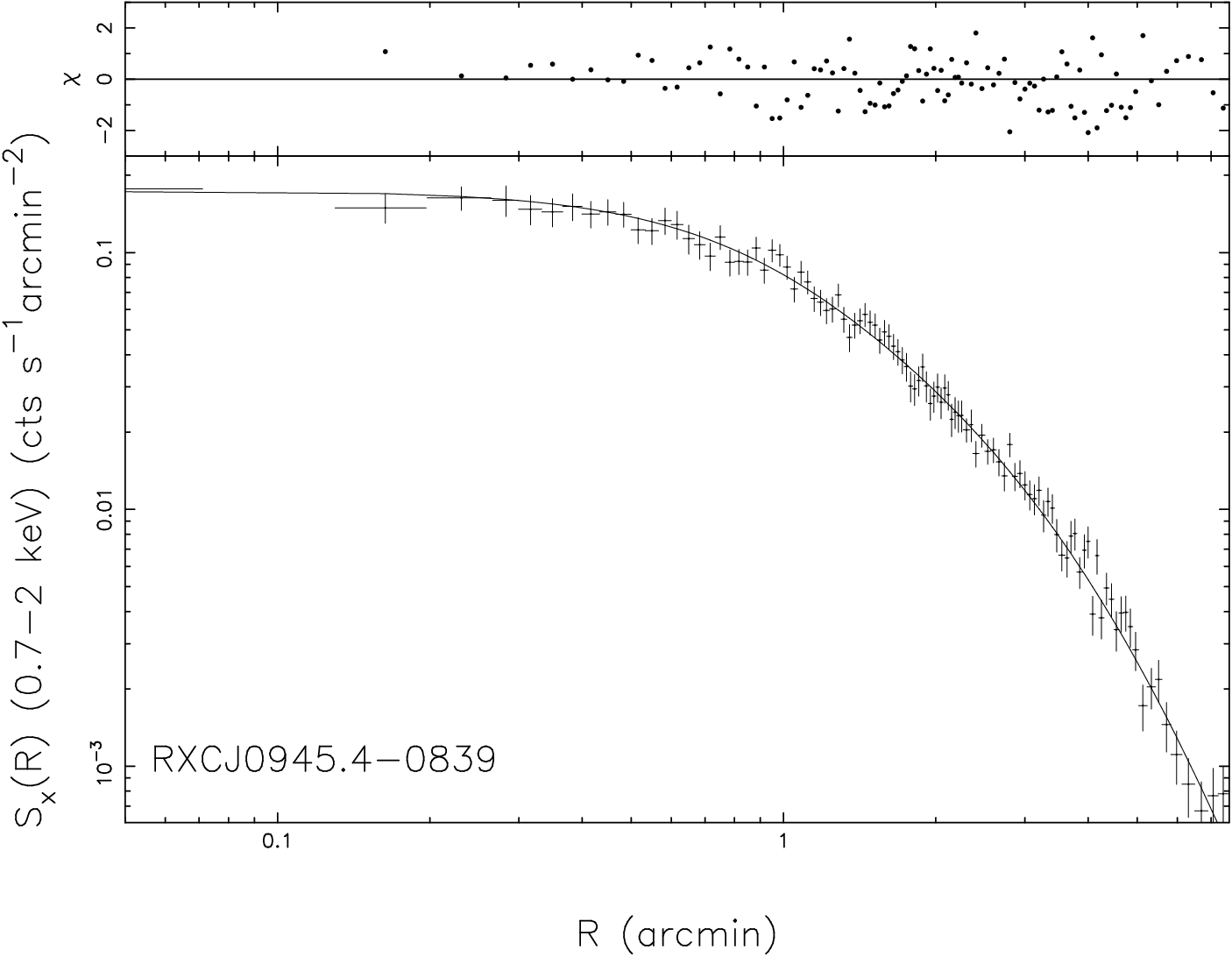}
\includegraphics[angle=270,width=8.5cm]{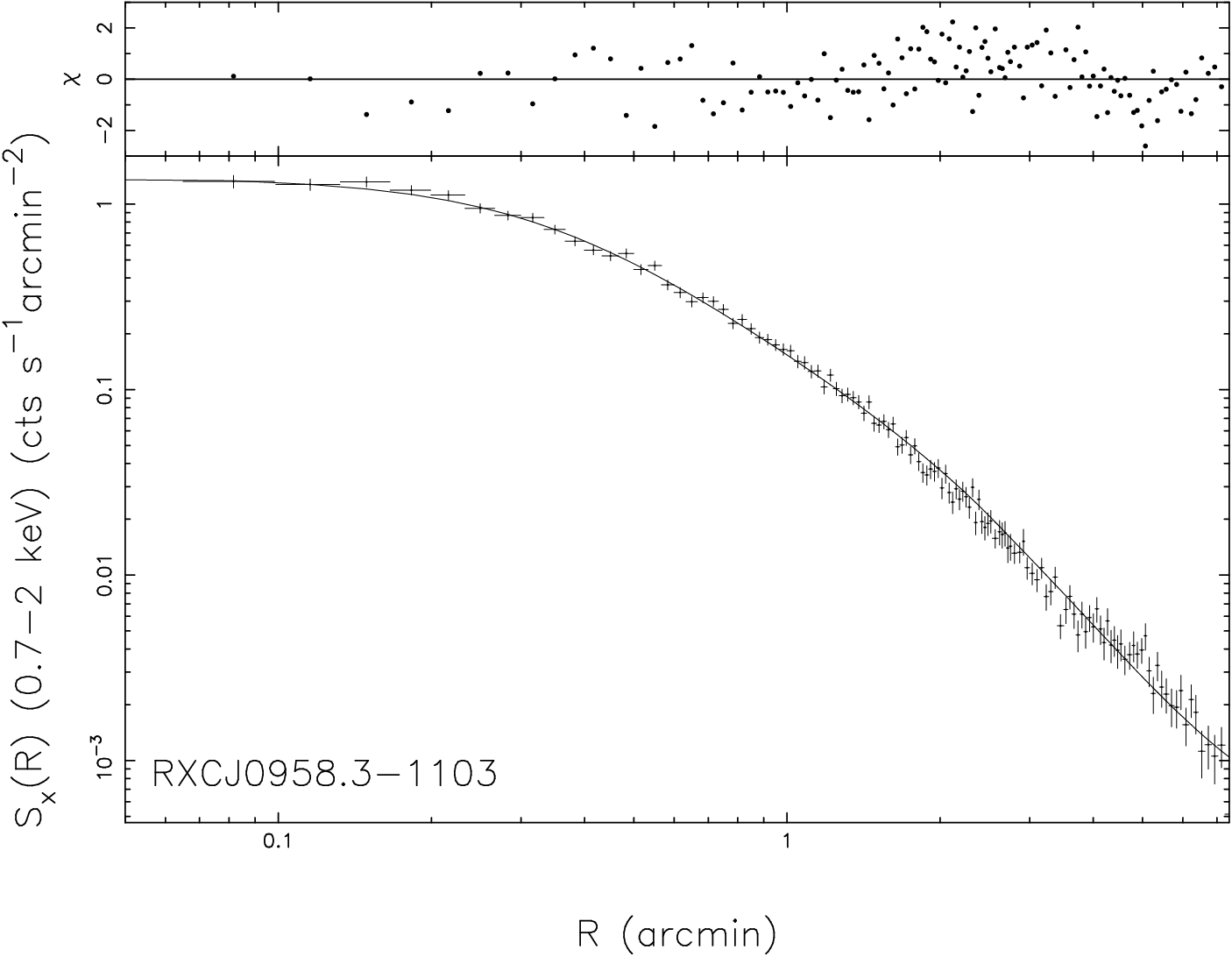}

\includegraphics[angle=270,width=8.5cm]{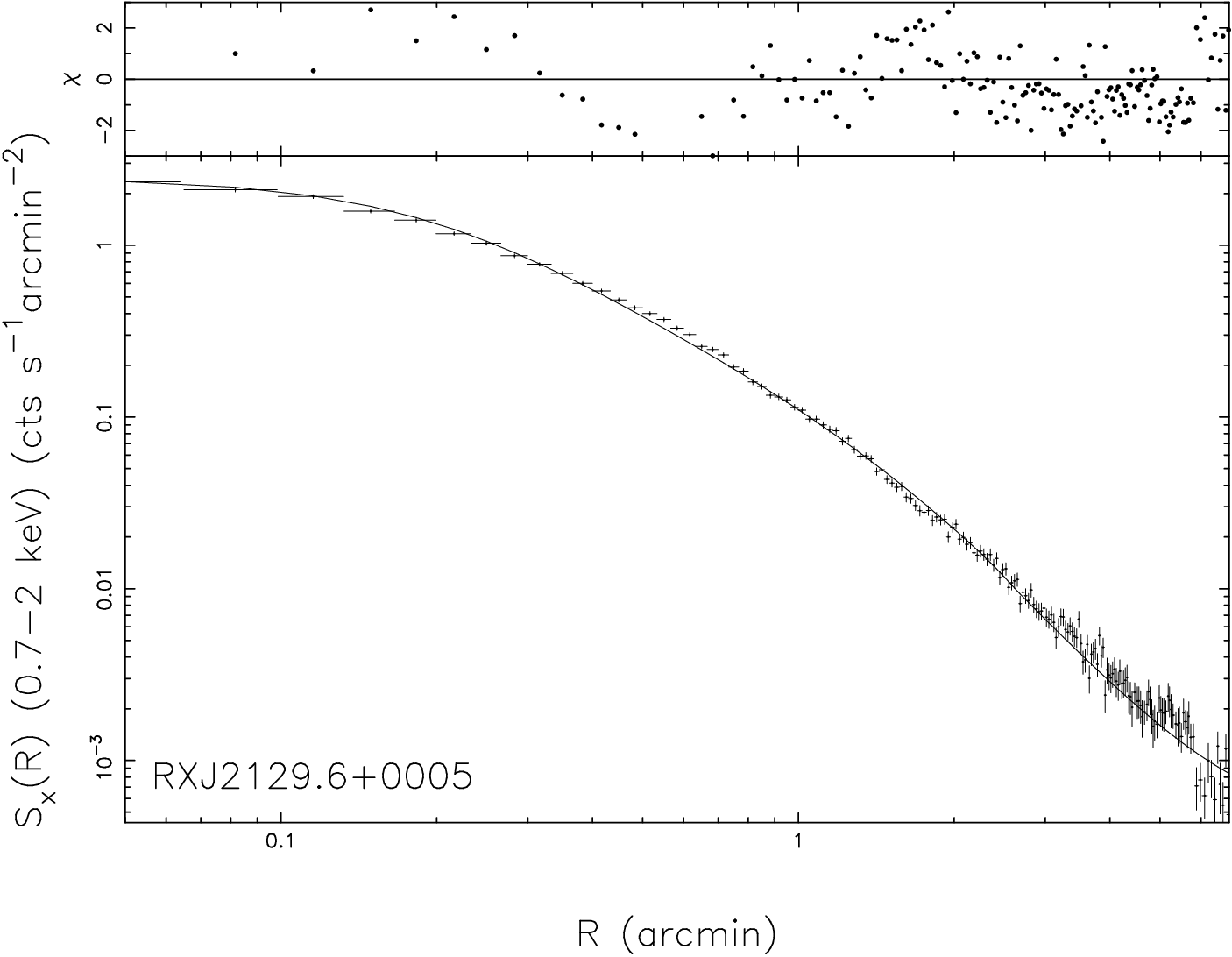}
\includegraphics[angle=270,width=8.5cm]{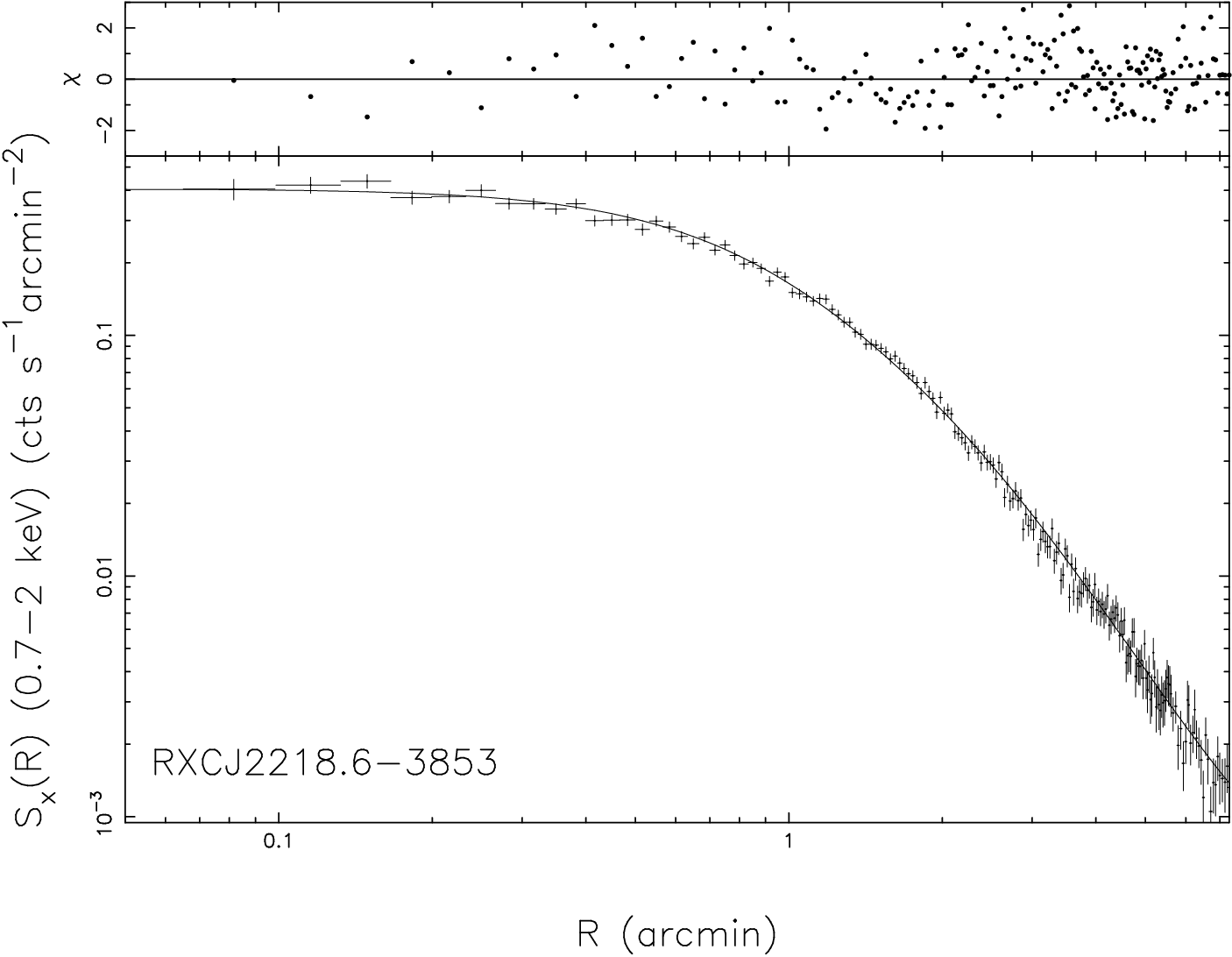}

\end{center}
\caption{Surface
brightness profiles parameterized by double-$\beta$
model.
\label{f:ktcom2}}
\end{figure*}

\begin{figure*}
\begin{center}

\includegraphics[angle=270,width=8.5cm]{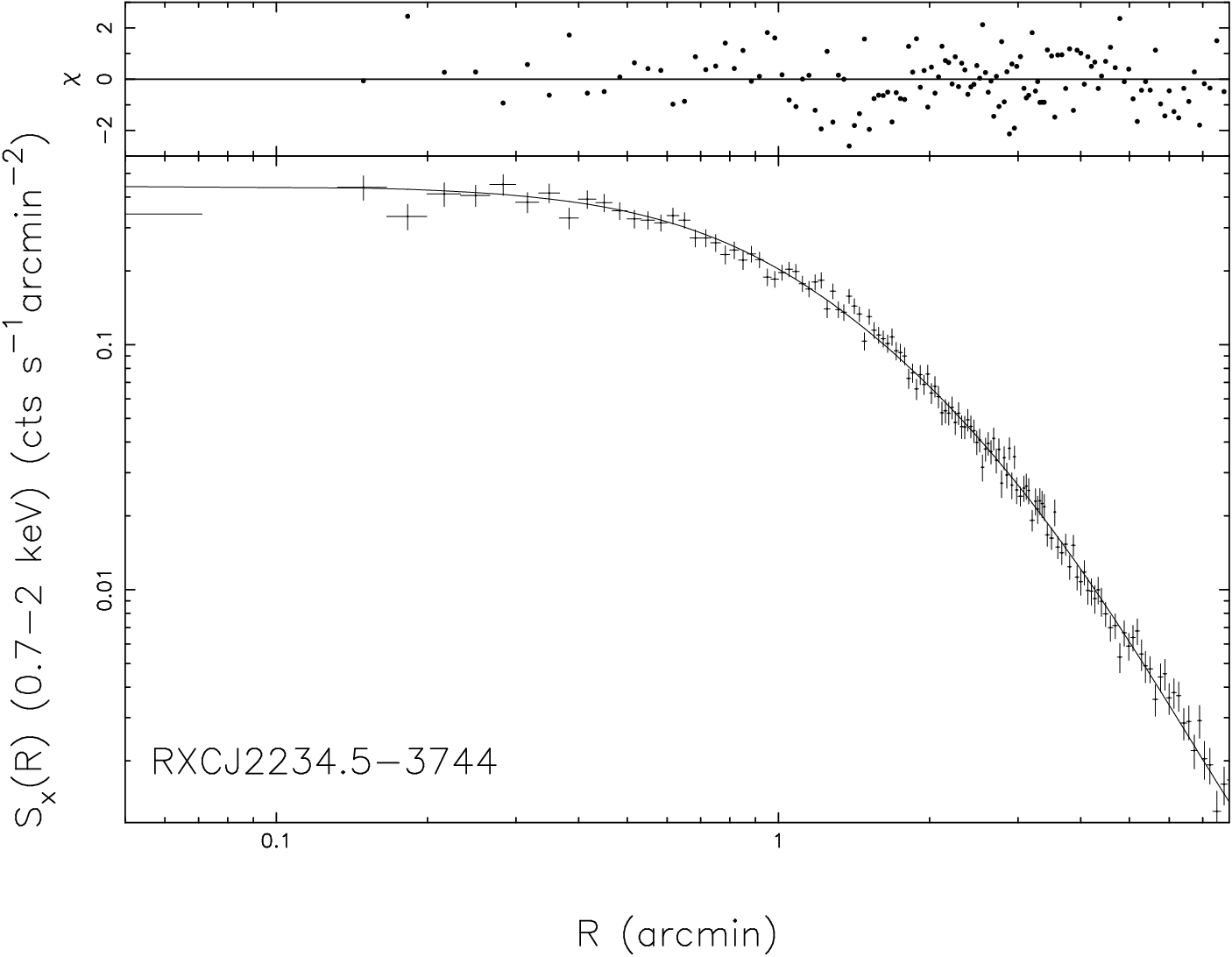}
\includegraphics[angle=270,width=8.5cm]{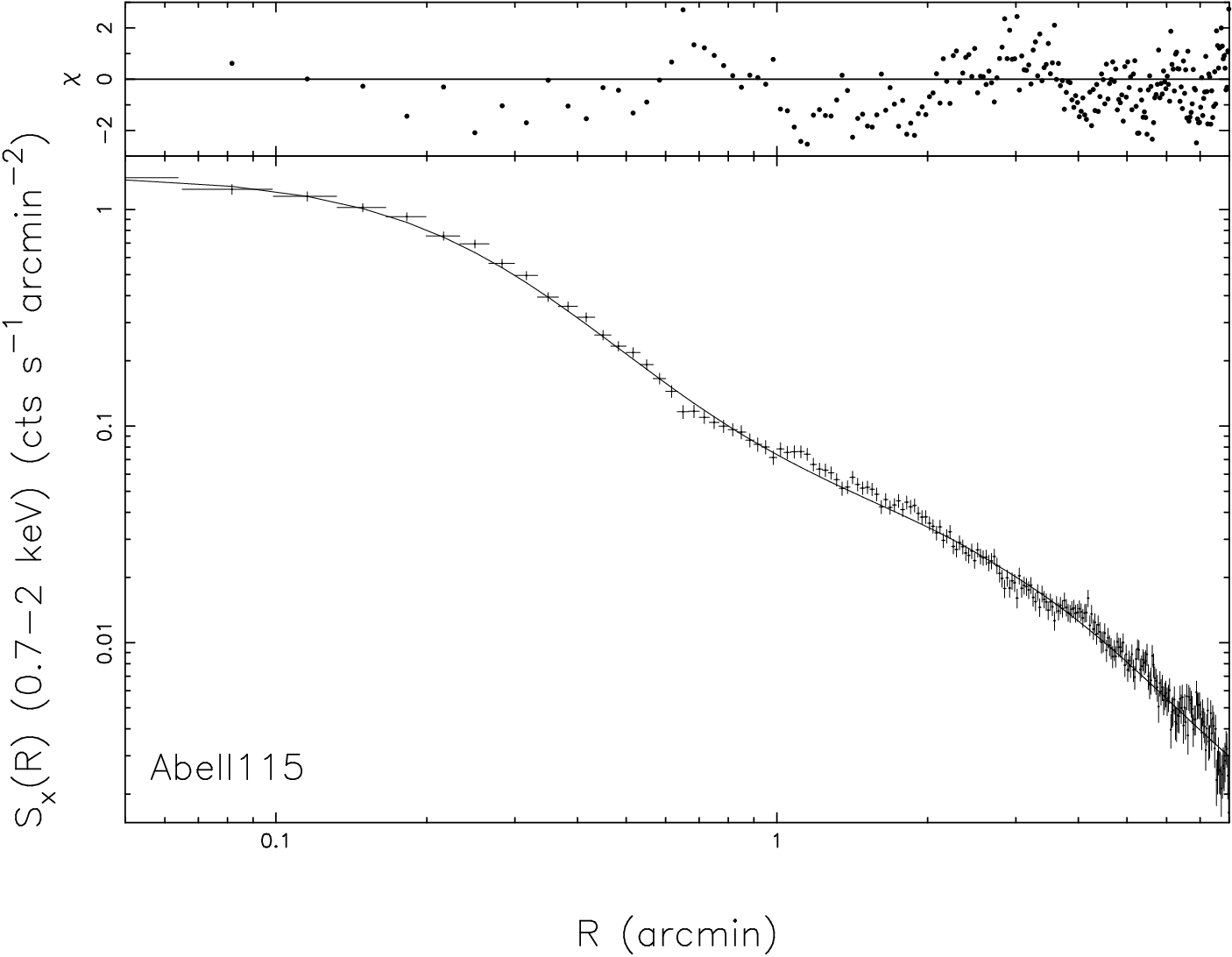}

\includegraphics[angle=270,width=8.5cm]{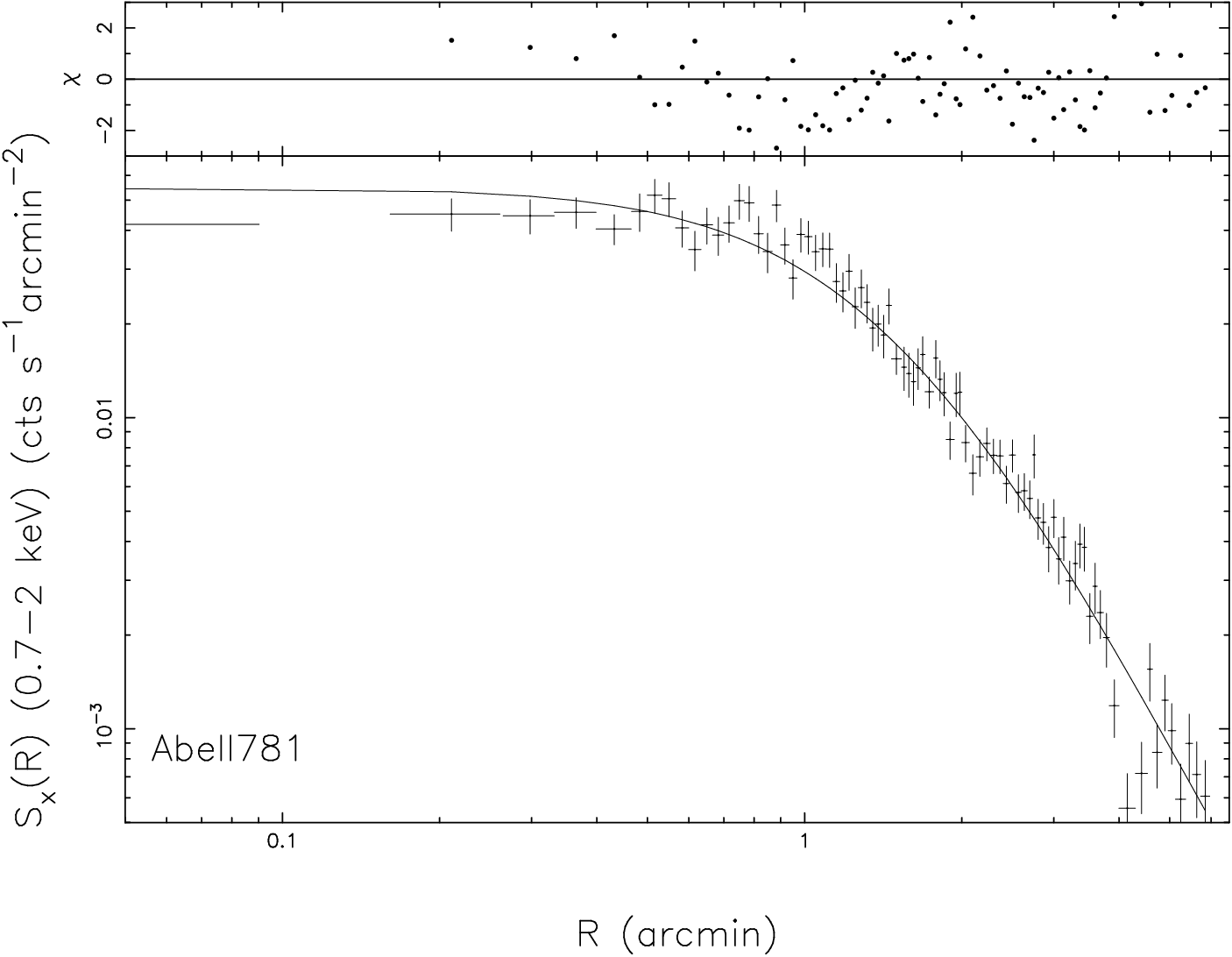}
\includegraphics[angle=270,width=8.5cm]{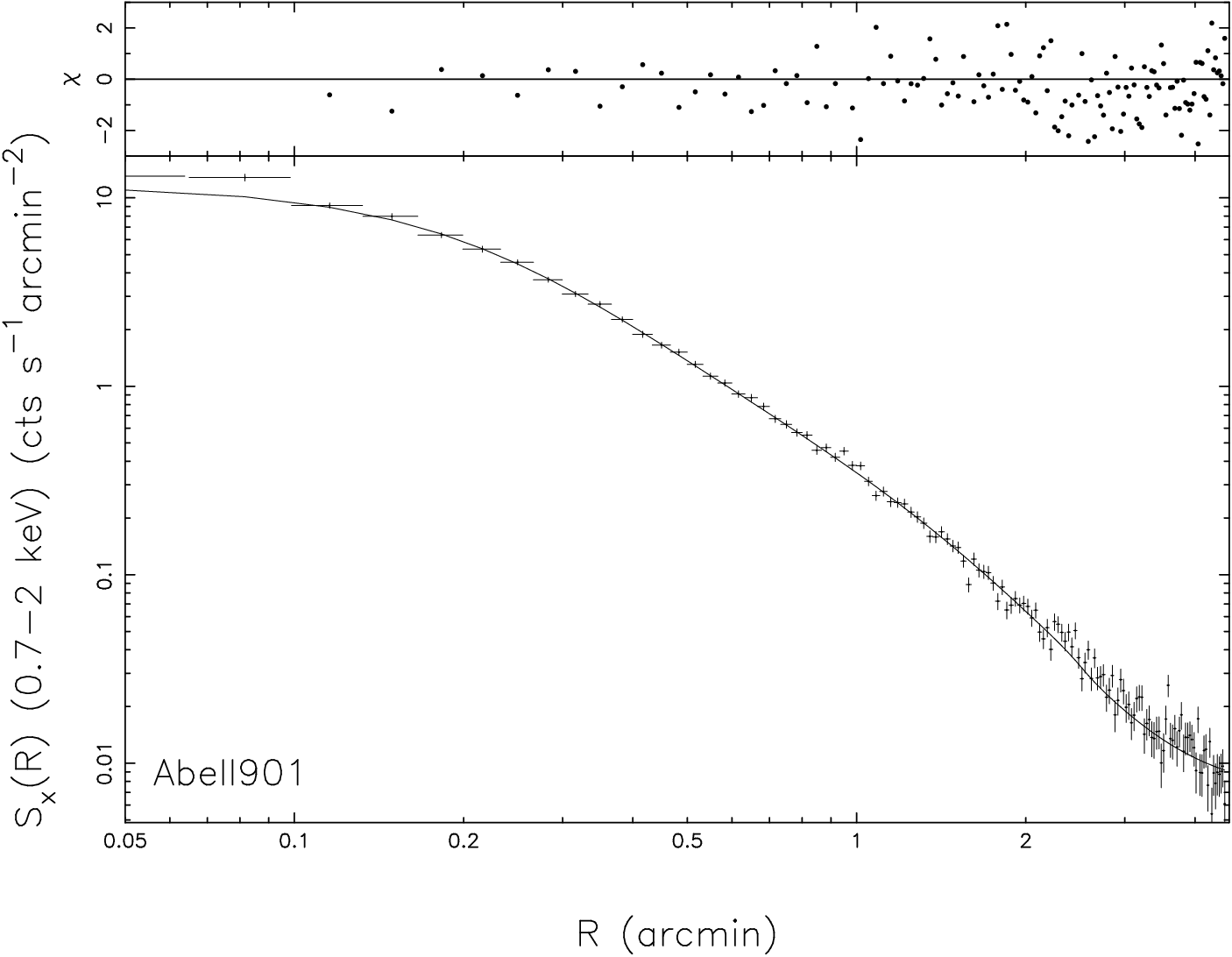}

\includegraphics[angle=270,width=8.5cm]{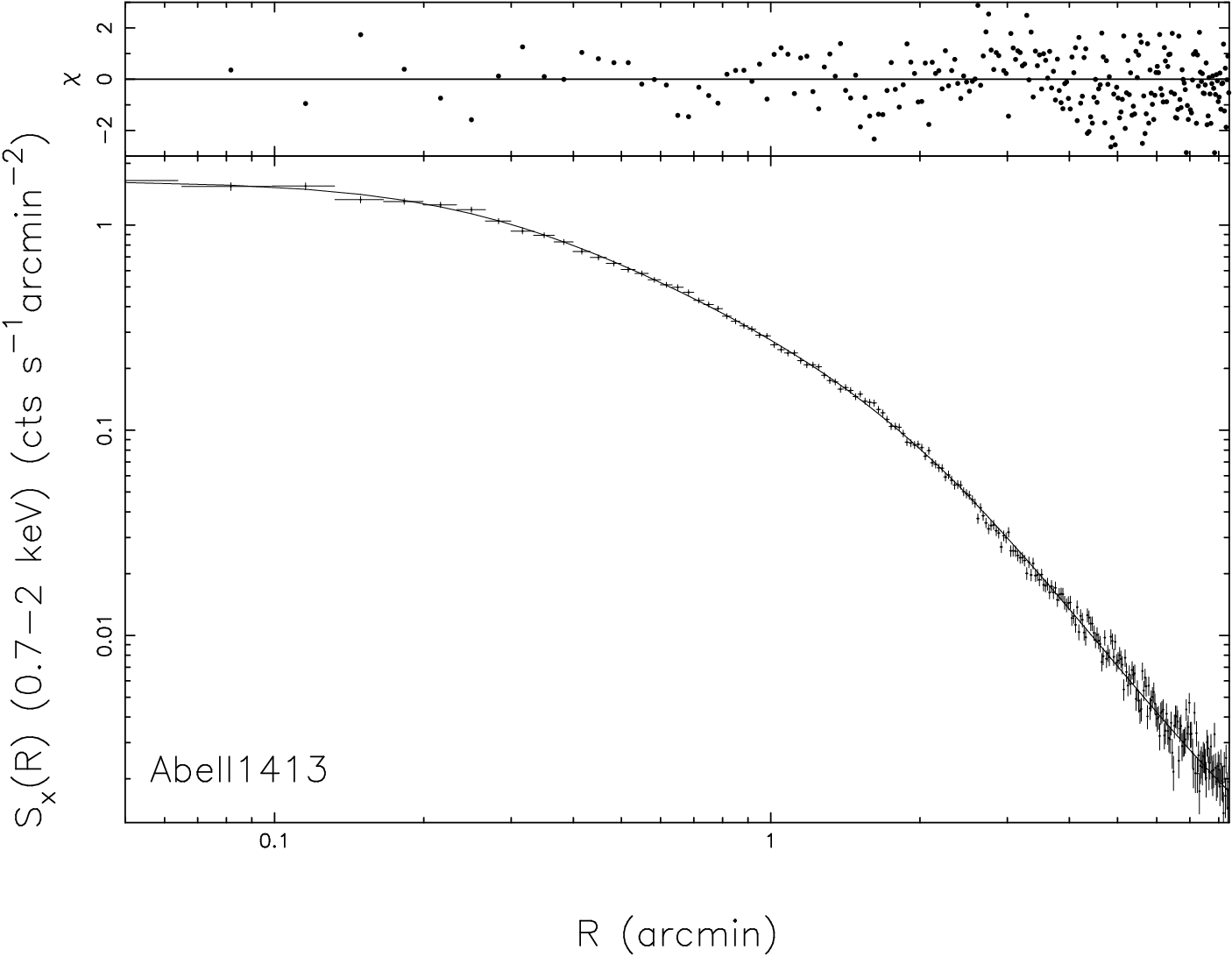}
\includegraphics[angle=270,width=8.5cm]{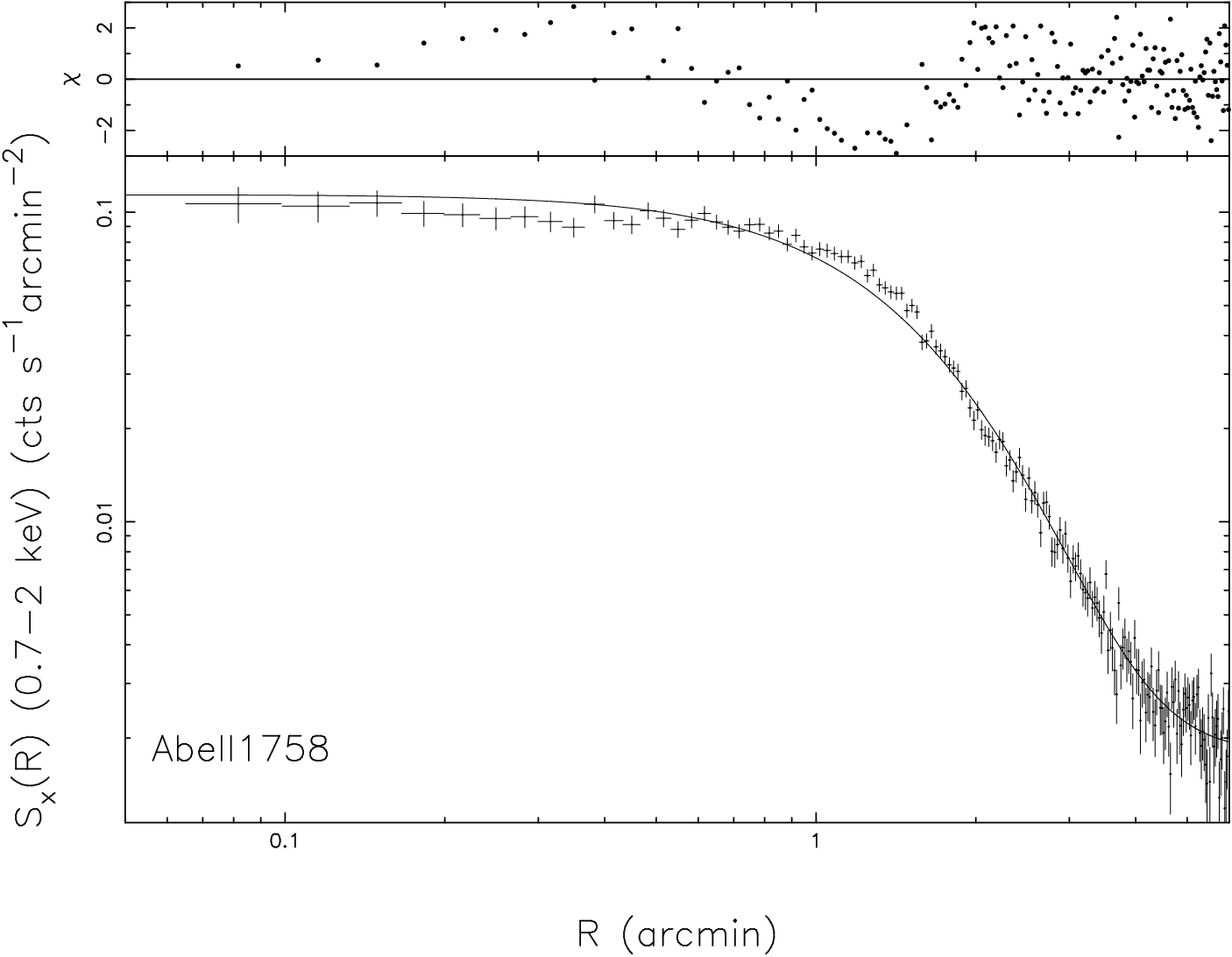}

\end{center}
\caption{See caption in Fig.~\ref{f:ktcom2}.
\label{f:ktcom21}}
\end{figure*}

\begin{figure*}
\begin{center}

\includegraphics[angle=270,width=8.5cm]{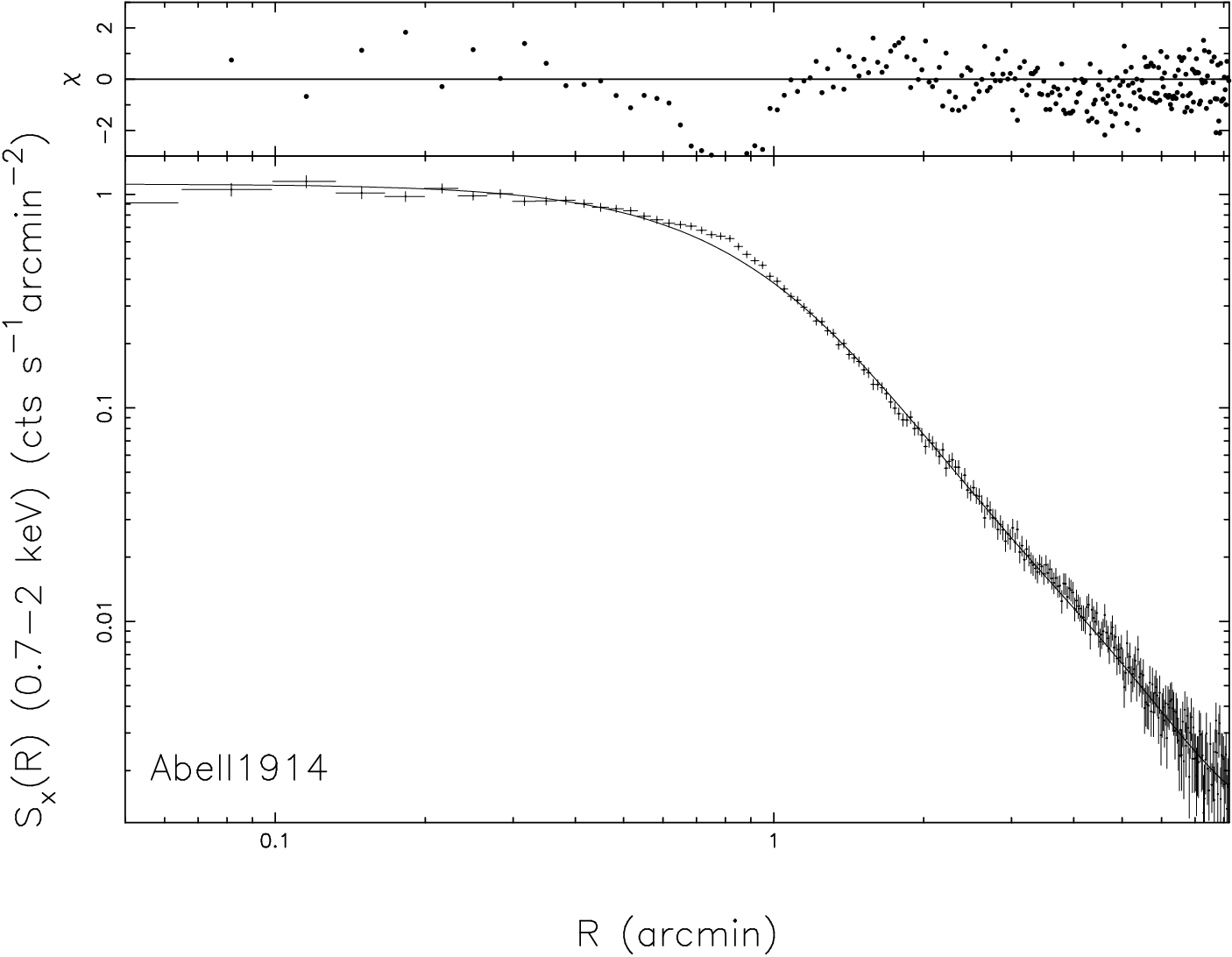}
\includegraphics[angle=270,width=8.5cm]{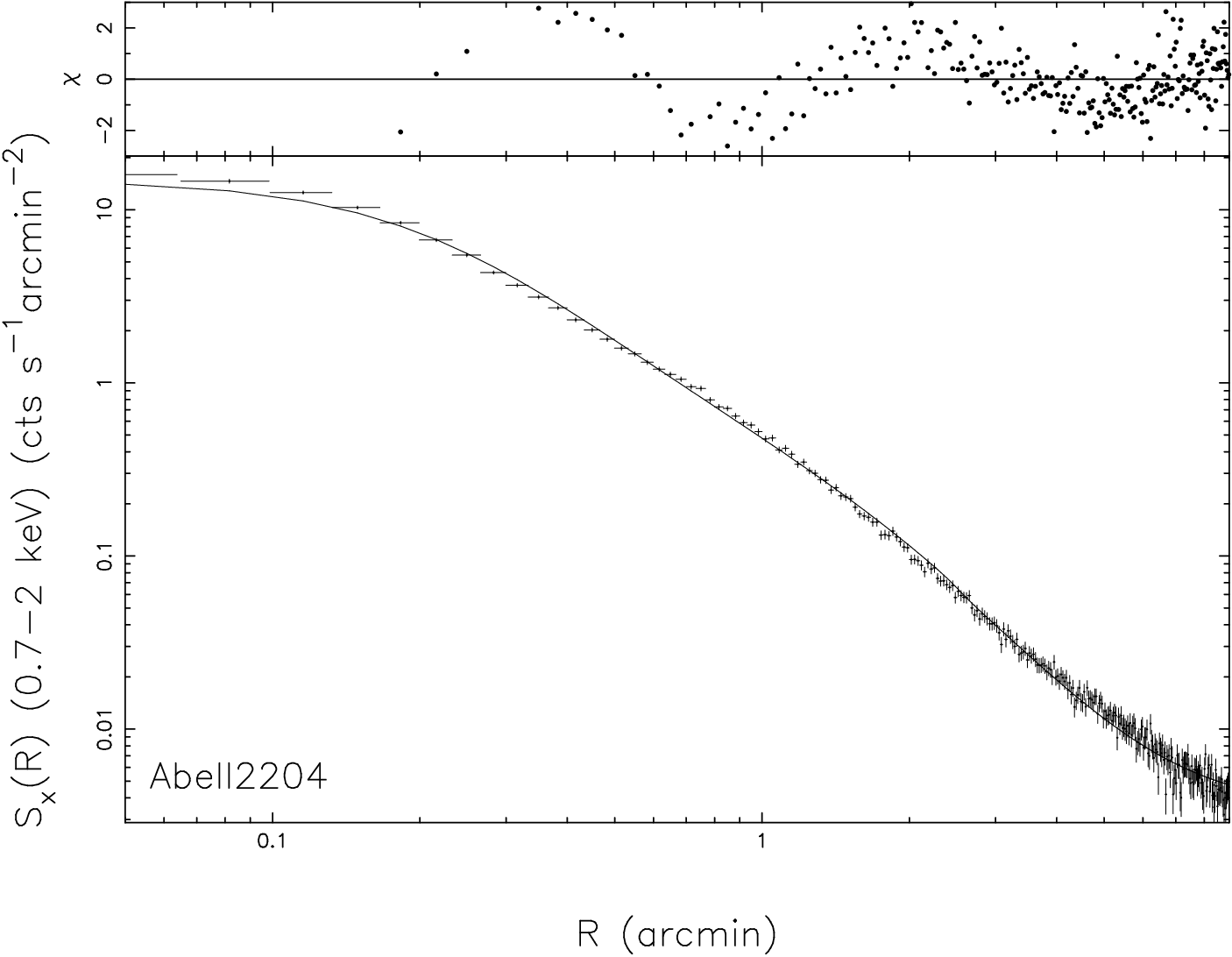}

\includegraphics[angle=270,width=8.5cm]{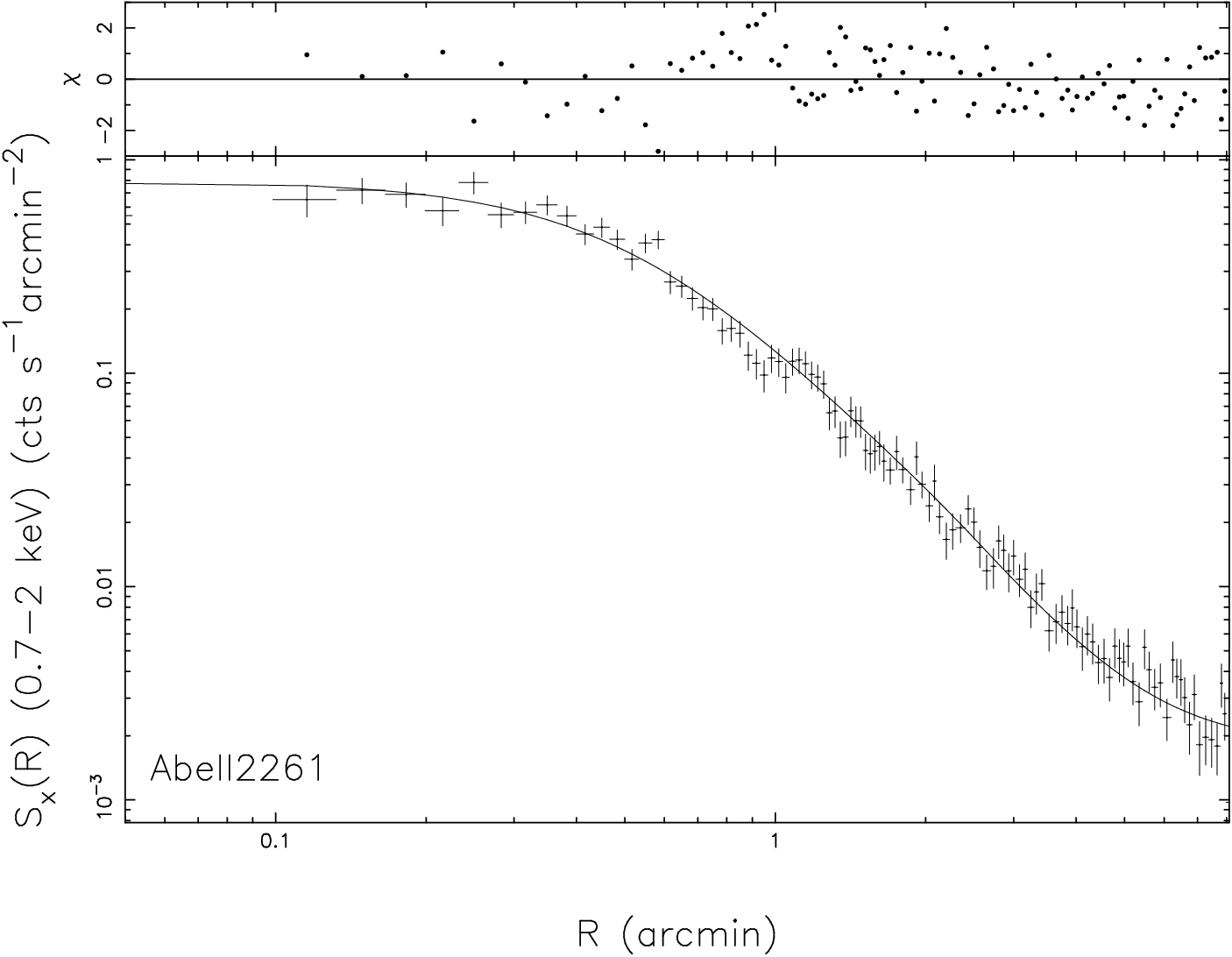}
\includegraphics[angle=270,width=8.5cm]{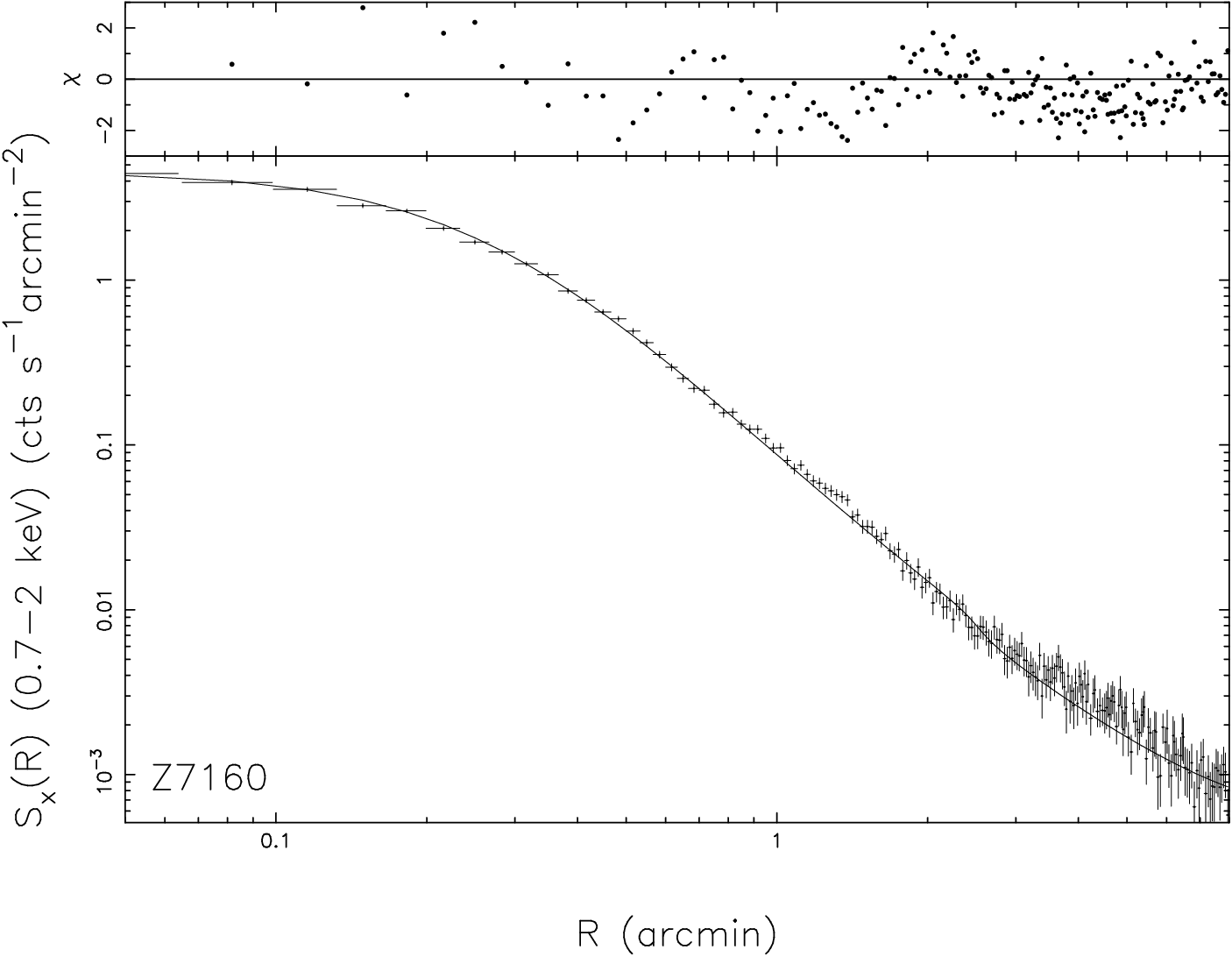}

\end{center}
\caption{See caption in Fig.~\ref{f:ktcom2}.
\label{f:ktcom22}}
\end{figure*}

A $\beta$ model (e.g. Cavaliere \& Fusco-Femiano 1976; Jones \& Forman 1984)
is often used to describe electron density profiles in clusters. To obtain an
acceptable fit for all clusters in the sample, we adopt a double-$\beta$ model
of the electron number density profile, $n_{\rm e}(r)=n_{\rm e01}(1+r^2/r_{\rm
  c1}^2)^{-3\beta/2}+n_{\rm e02}(1+r^2/r_{\rm c2}^2)^{-3\beta/2}$, where
$n_{\rm e0}=n_{\rm e01}+n_{\rm e02}$ is the central electron number density
(see Table~\ref{t:catalog2}), $\beta$ the slope parameter, and $r_{\rm c1}$
and $r_{\rm c2}$ the core radii of the inner and outer components,
respectively. The $\beta$ value is derived by fitting the outer component, and
it is used for the inner component since the inner component is insensitive to
the slope. The soft band (e.g. 0.5--2~keV for 9 REFLEX-DXL clusters and
0.7--2~keV for the remaining 28 clusters) X-ray surface brightness profile
model $S(R)$, in which $R$ is the projected radius, is linked to the radial
profile of the ICM electron number density $n_{\rm e}(r)$ and emissivity
function as an integral performed along the line-of-sight for galaxy clusters.
For massive clusters (i.e. $T >3.5$~keV) the cooling function can be treated
as a temperature independent quantity in the soft band, and the surface
brightness profile is thus described as follows,
\begin{equation}
S_{\rm X}(R)\propto\int_{R}^{\infty} n_{\rm p} n_{\rm e} d\ell.
\end{equation}
For the whole sample, the best-fit ``mekal'' model of the spectra limited to
$0.2r_{500}<R<0.5r_{500}$ is used to convert the surface brightness profile to
derive the density distribution. We fitted the observed surface brightness
profile by this integral convolved with the \emph{XMM-Newton} PSF matrices
(Figs.~\ref{f:ktcom2}--\ref{f:ktcom22} in the electronic edition of the
Journal, Fig.~3 in Zhang et al. 2006, and Fig.~1 and Figs.~B.1-B.5 in Zhang et
al. 2007 for the whole sample) and obtained the parameters of the
double-$\beta$ model of the electron density profile. The fit was performed
within the truncation radius ($r_{\rm t}$, see Table~\ref{t:global})
corresponding to a $S/N$ of 3 of the observational surface brightness profile.
The truncation radii $r_{\rm t}$ are often larger than $r_{500}$. The cluster
cores, referring to the core of the inner component of the double-$\beta$
model, span a broad range up to 0.2~$r_{500}$.

\subsection{Mass distribution}
\label{s:massdis}

The ICM can be used to trace the cluster potential. We assume that,
(1) the ICM is in hydrostatic equilibrium within the gravitational
potential dominated by DM, (2) the DM distribution is spherically
symmetric, and (3) the kinematic term is approximated to zero in
hydrostatic equilibrium. The cluster mass is then calculated from the
X-ray measured ICM density and temperature distributions by,
\begin{equation}
\frac{1}{\mu m_{\rm p} n_{\rm e}(r)}\frac{d[n_{\rm e}(r) T(r)]}{dr}=
  -\frac{GM(<r)}{r^2}~,
\label{e:hyd}
\end{equation}
where $\mu=0.62$ is the mean molecular weight per hydrogen atom. The Boltzmann
constant is included in T as the temperature is in units of keV. Following the
method in Zhang et al. (2007), we use a set of input parameters of the
approximation functions, in which $\beta$, $n_{\rm e0i}$, $r_{\rm ci}$
($i=1,2$) represent the electron number density profile $n_{\rm e}(r)$ and
$P_{\rm i}$ ($i=1,...,7$) represent the temperature profile $T(r)$,
respectively, to compute the mean cluster mass. The mass uncertainties are
propagated using the uncertainties of the electron number density and
temperature measurements by Monte Carlo simulations as follows (see also Zhang
et al. 2007). For each cluster, we simulated the electron density and
temperature profiles of a sample of 100 clusters using observed electron
density and temperature profiles and their errors. The mass profiles and other
properties of the 100 simulated clusters were calculated to derive the errors.

\begin{figure*}
\begin{center}

\includegraphics[angle=270,width=8.5cm]{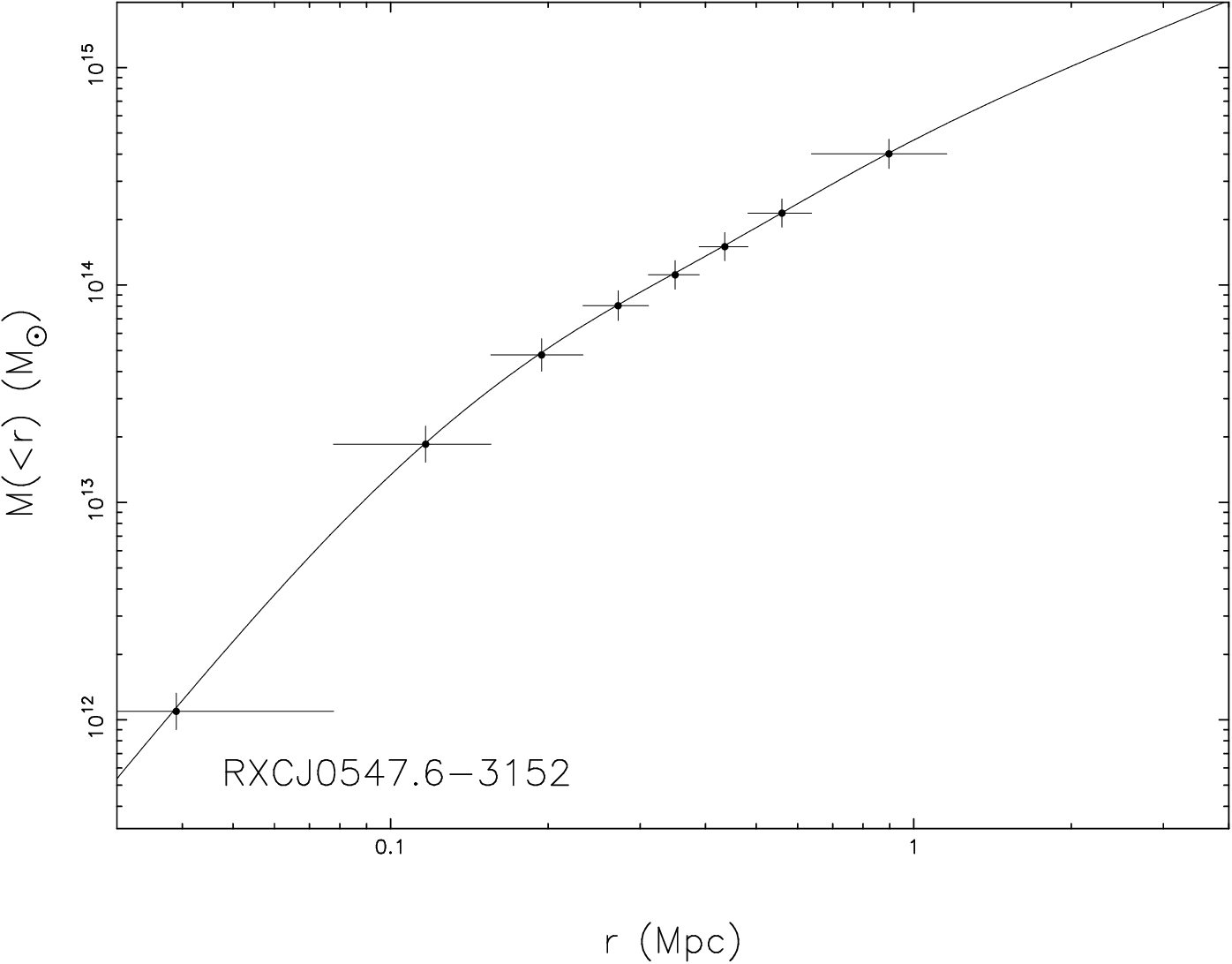}
\includegraphics[angle=270,width=8.5cm]{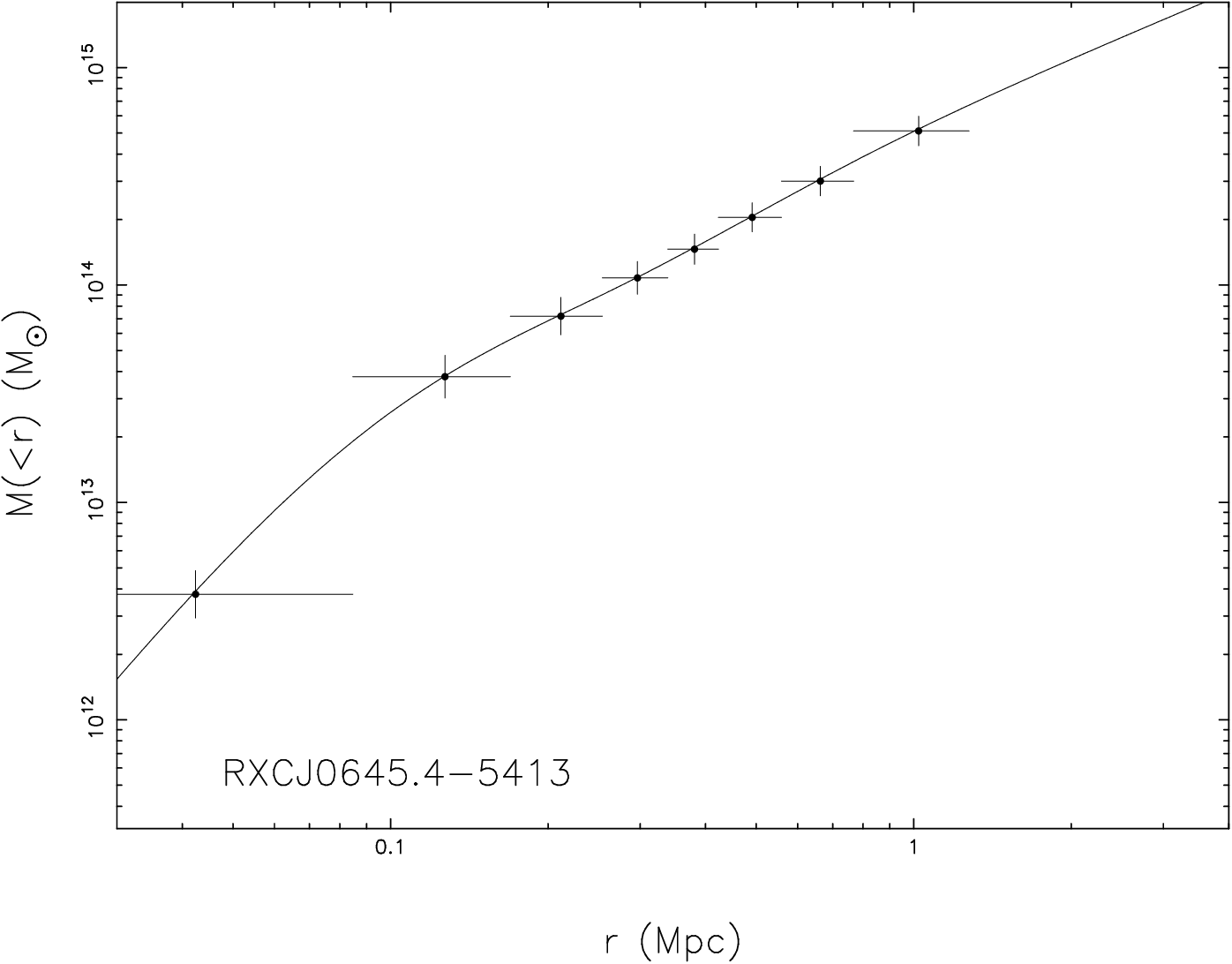}

\includegraphics[angle=270,width=8.5cm]{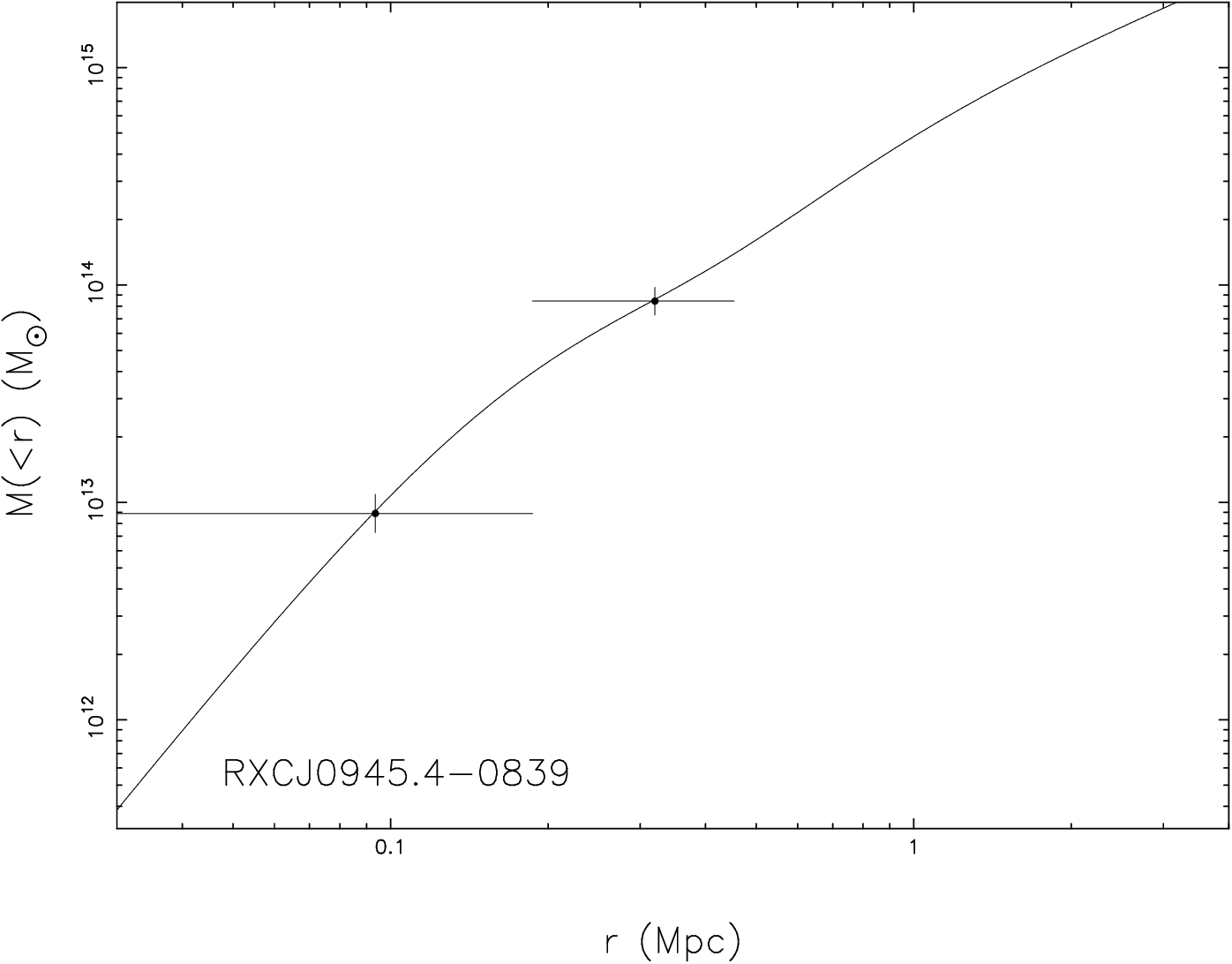}
\includegraphics[angle=270,width=8.5cm]{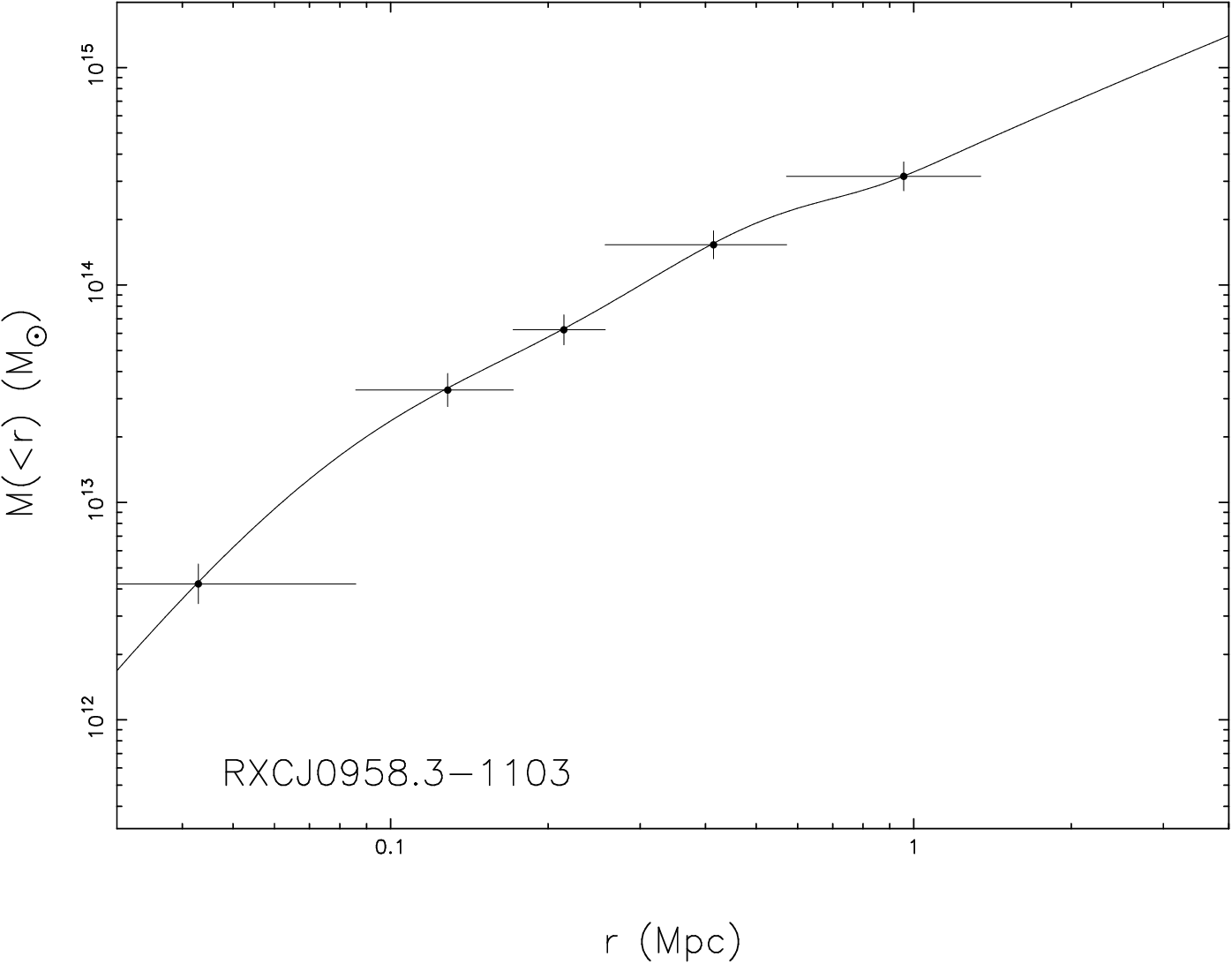}

\includegraphics[angle=270,width=8.5cm]{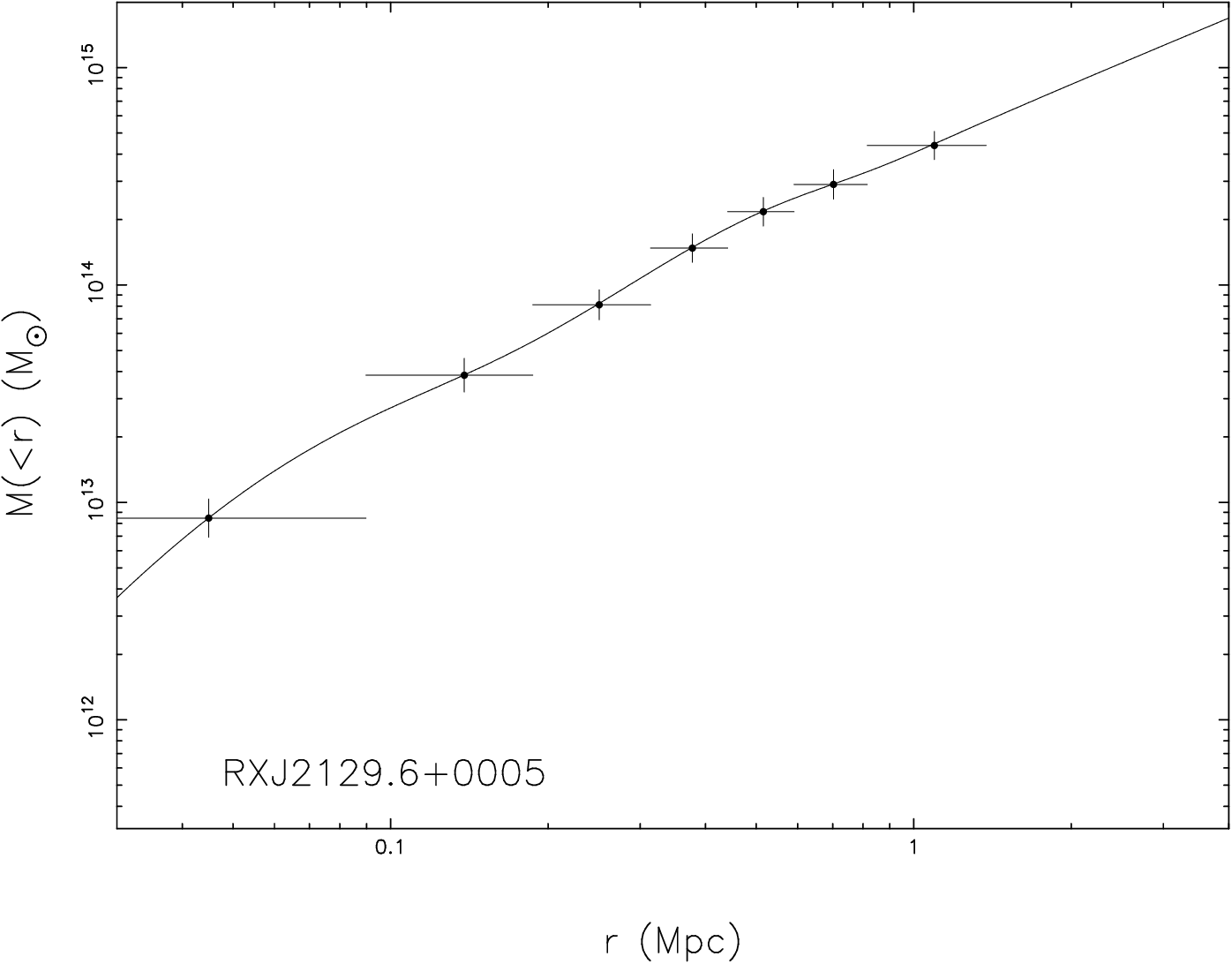}
\includegraphics[angle=270,width=8.5cm]{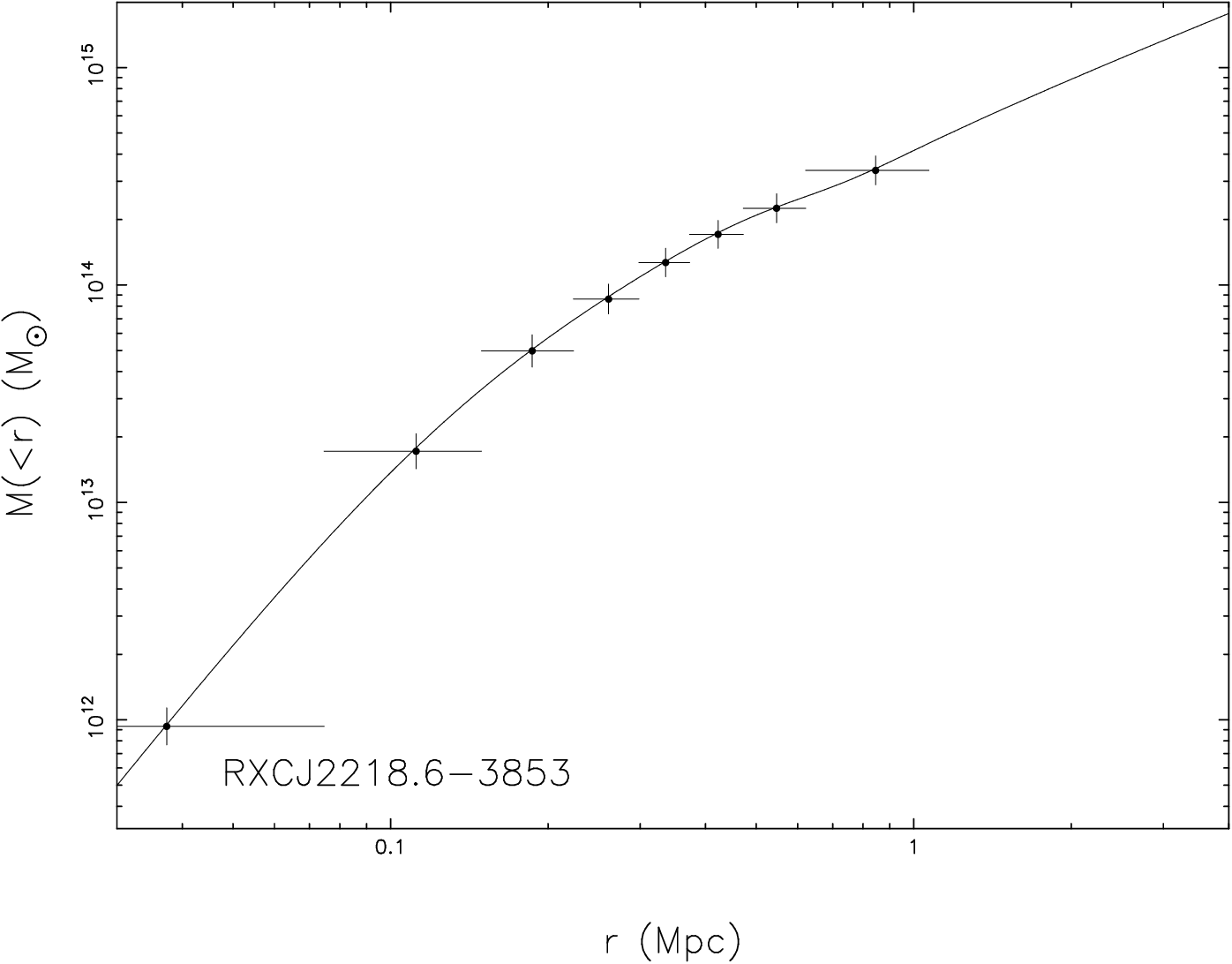}

\end{center}
\caption{Mass profiles with their mass data points binned as the 
temperature measurements to demonstrate the error bars.
\label{f:ktcom3}}
\end{figure*}

\begin{figure*}
\begin{center}

\includegraphics[angle=270,width=8.5cm]{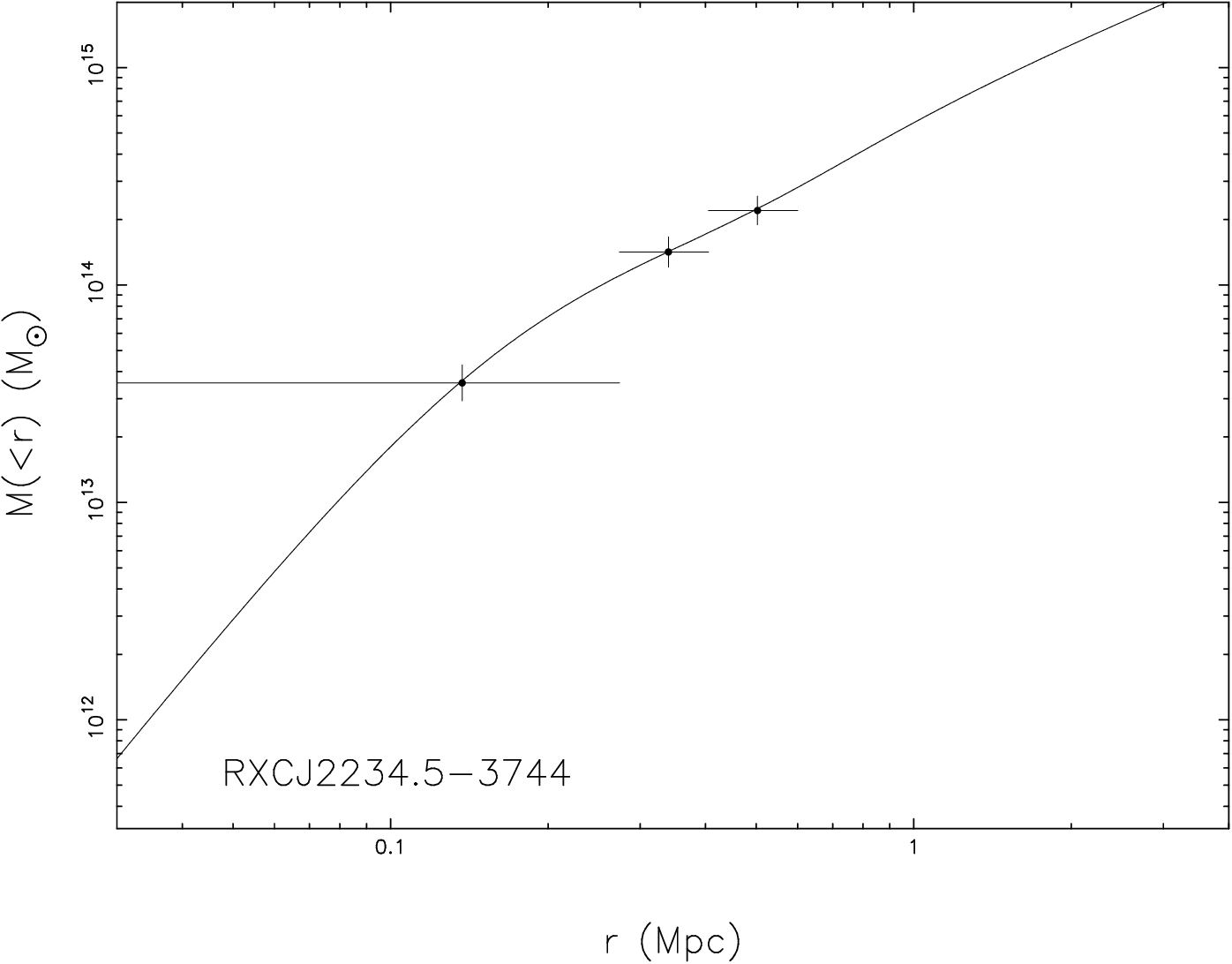}
\includegraphics[angle=270,width=8.5cm]{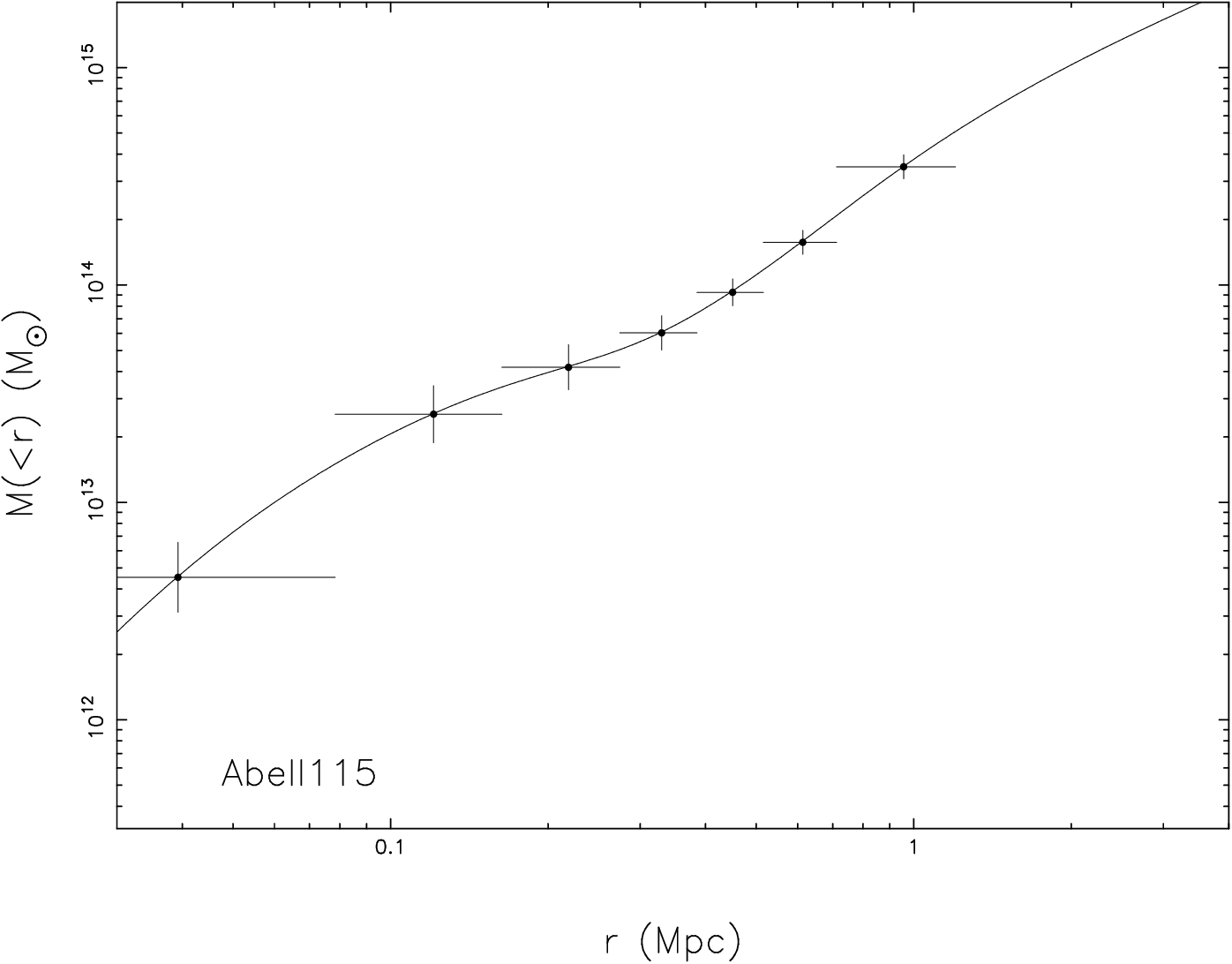}

\includegraphics[angle=270,width=8.5cm]{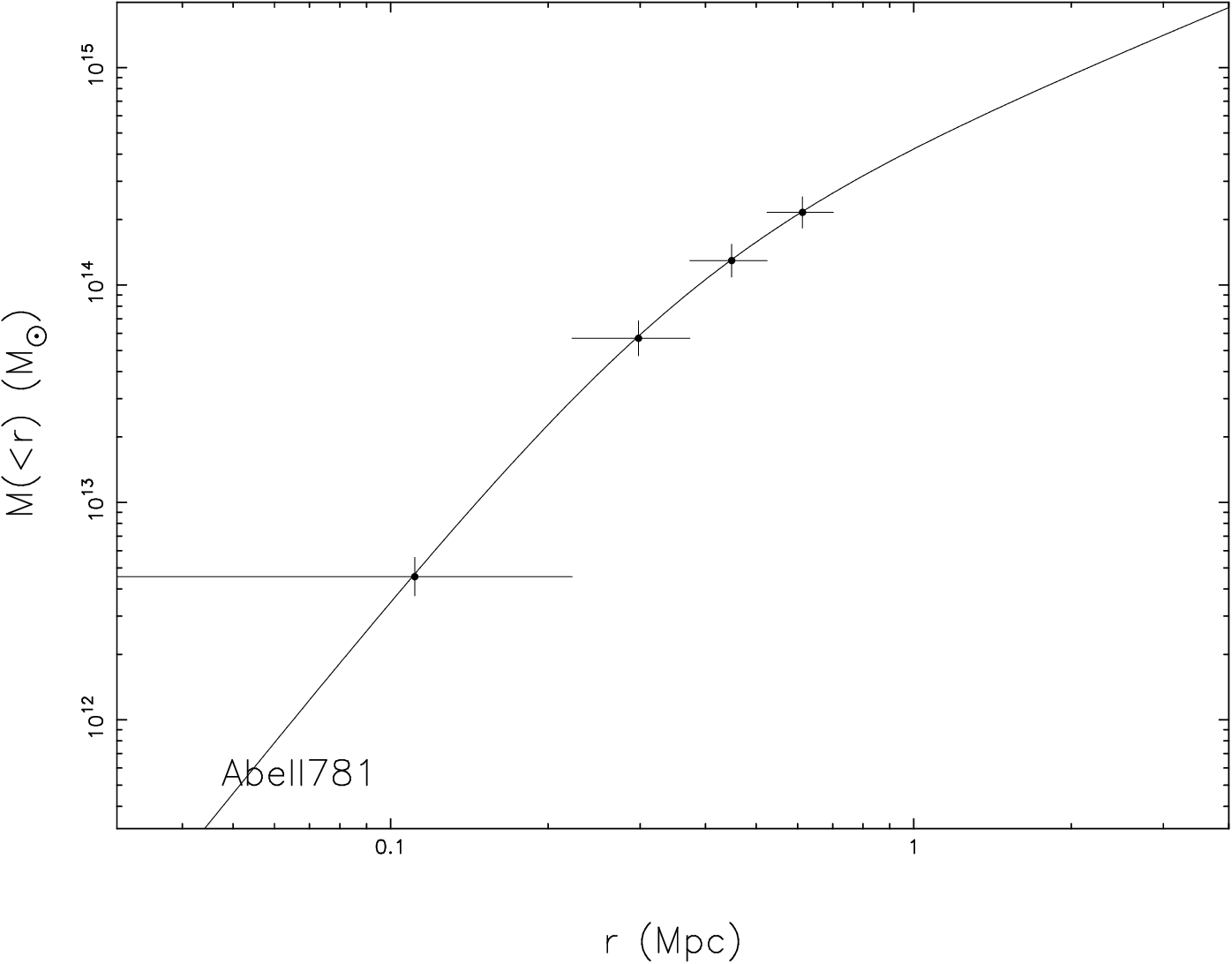}
\includegraphics[angle=270,width=8.5cm]{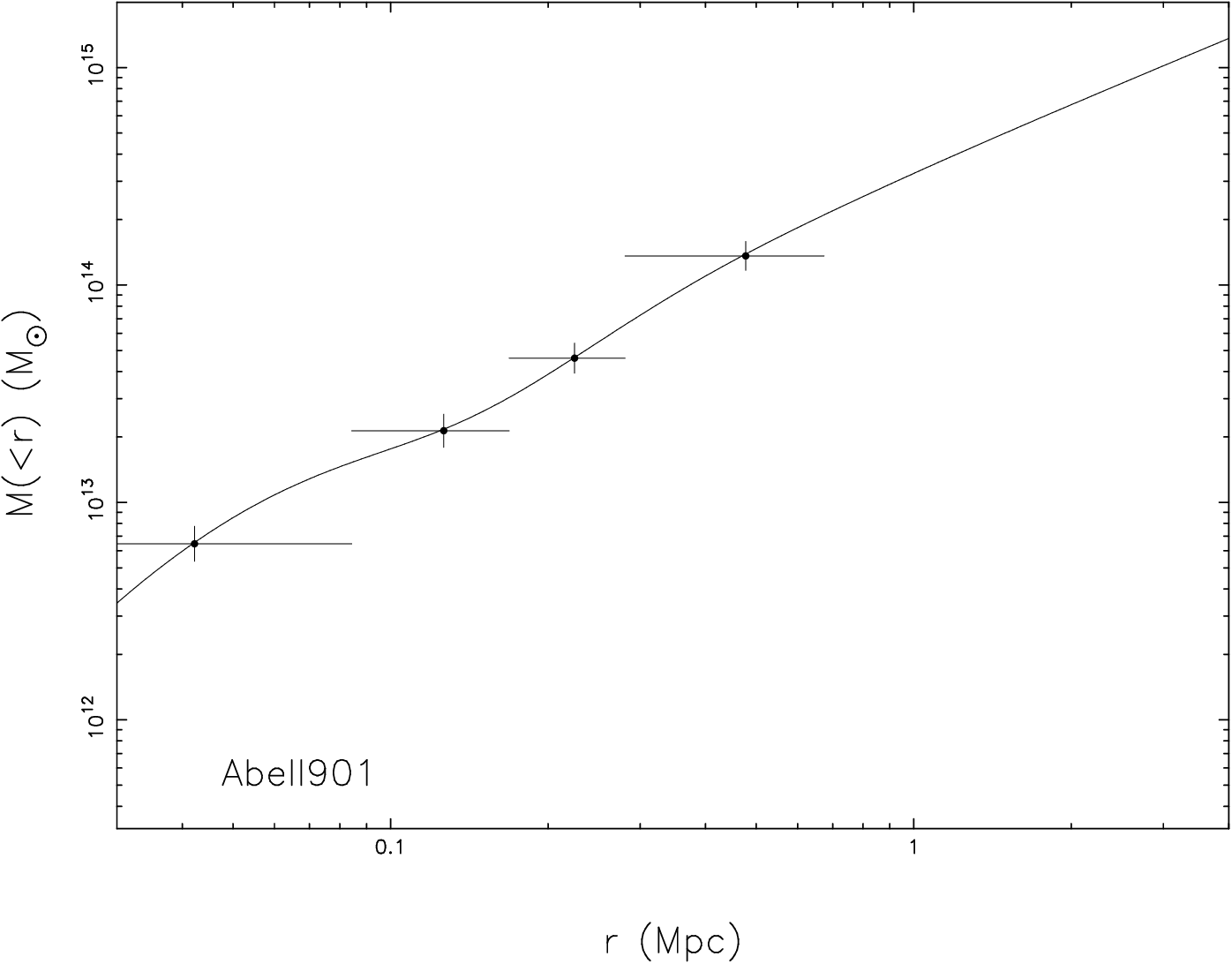}

\includegraphics[angle=270,width=8.5cm]{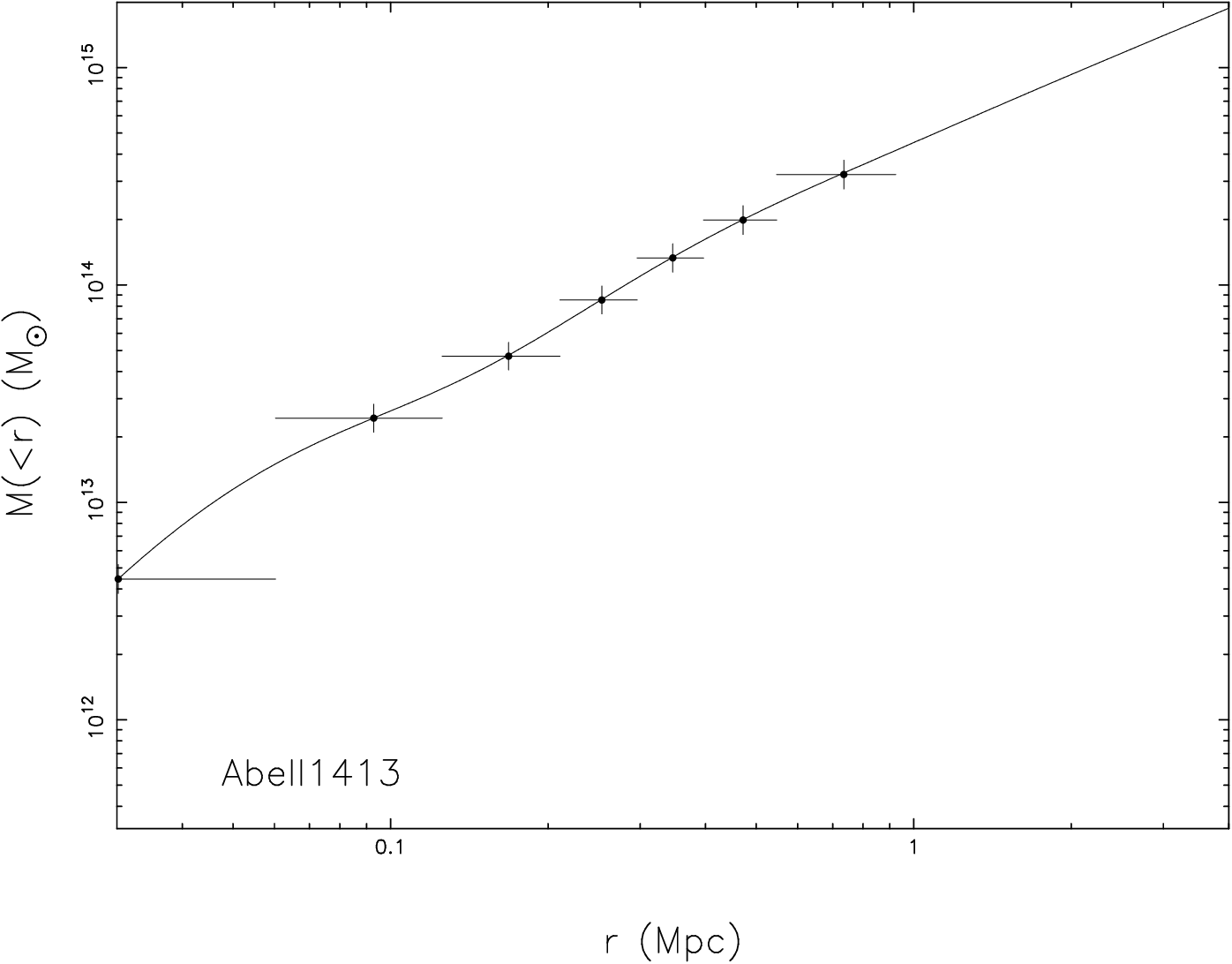}
\includegraphics[angle=270,width=8.5cm]{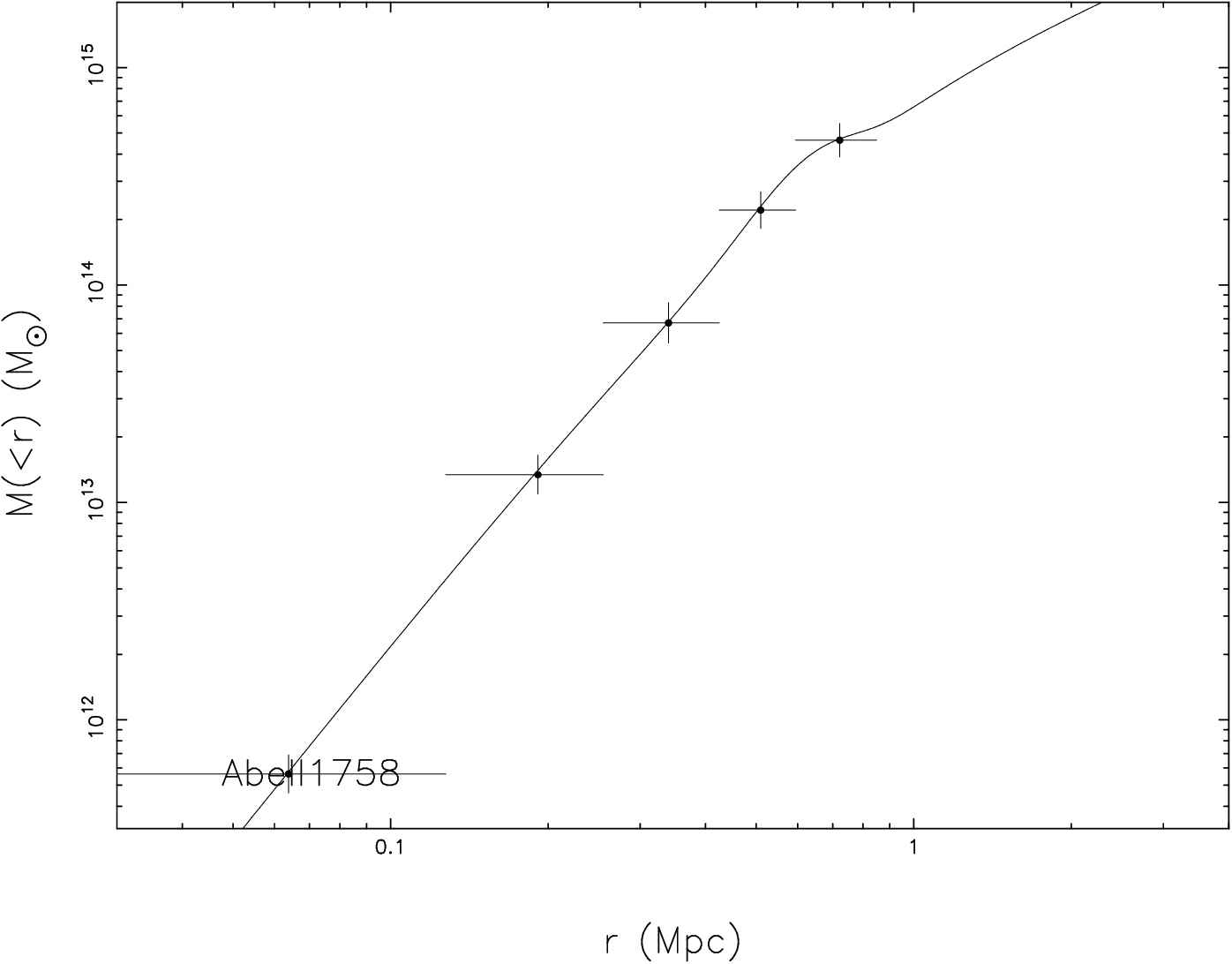}

\end{center}
\caption{See caption in Fig.~\ref{f:ktcom3}.
\label{f:ktcom31}}
\end{figure*}

\begin{figure*}
\begin{center}

\includegraphics[angle=270,width=8.5cm]{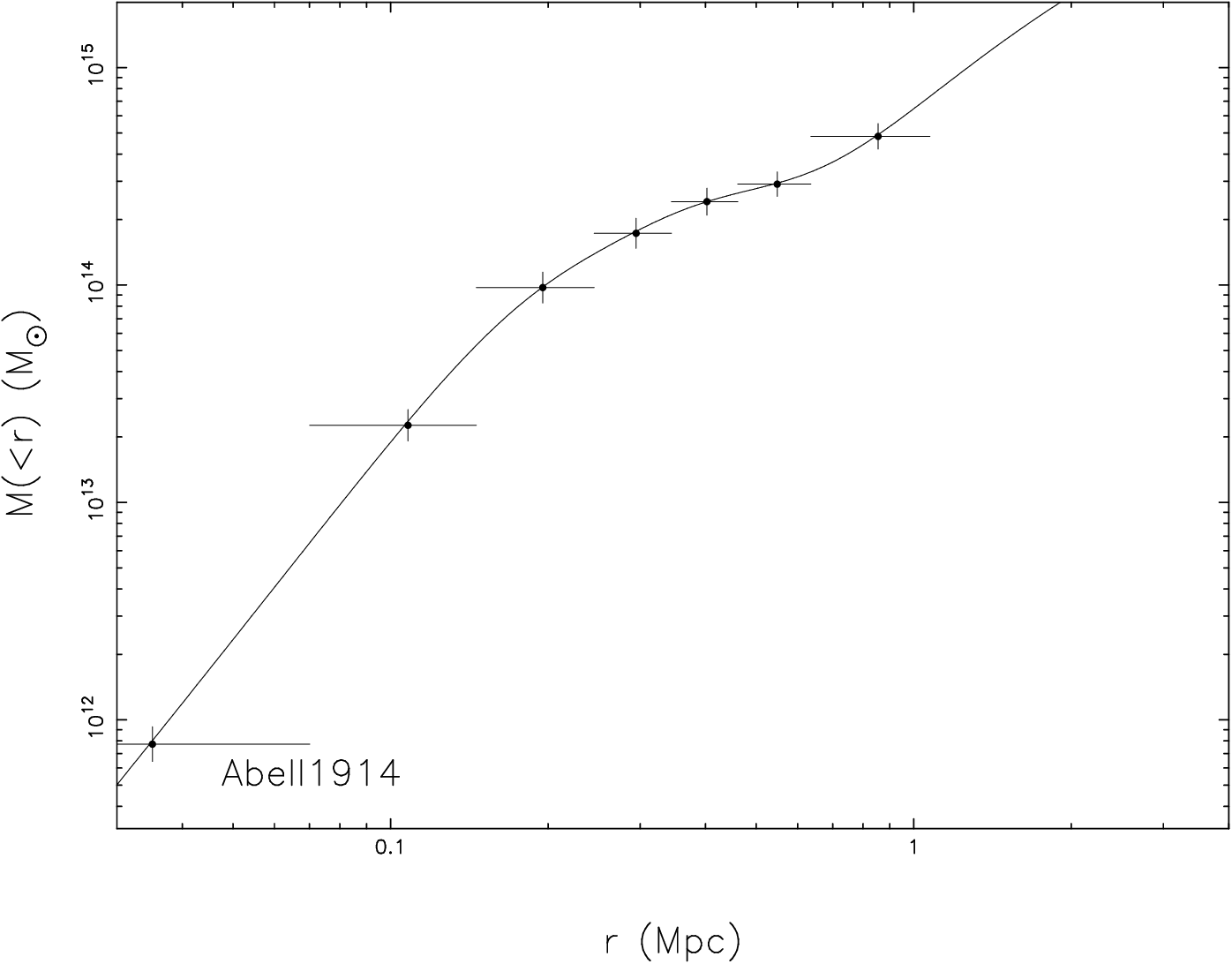}
\includegraphics[angle=270,width=8.5cm]{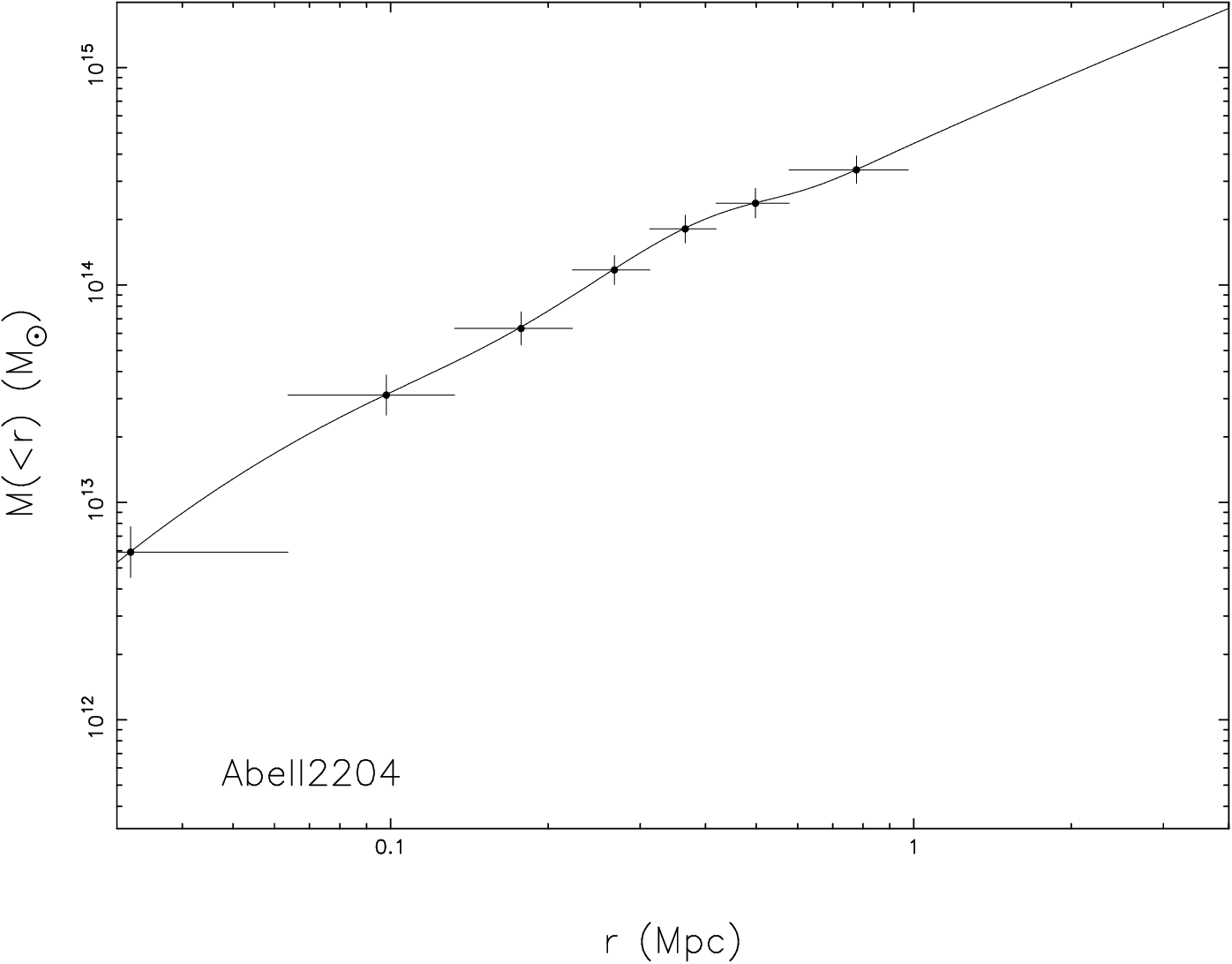}

\includegraphics[angle=270,width=8.5cm]{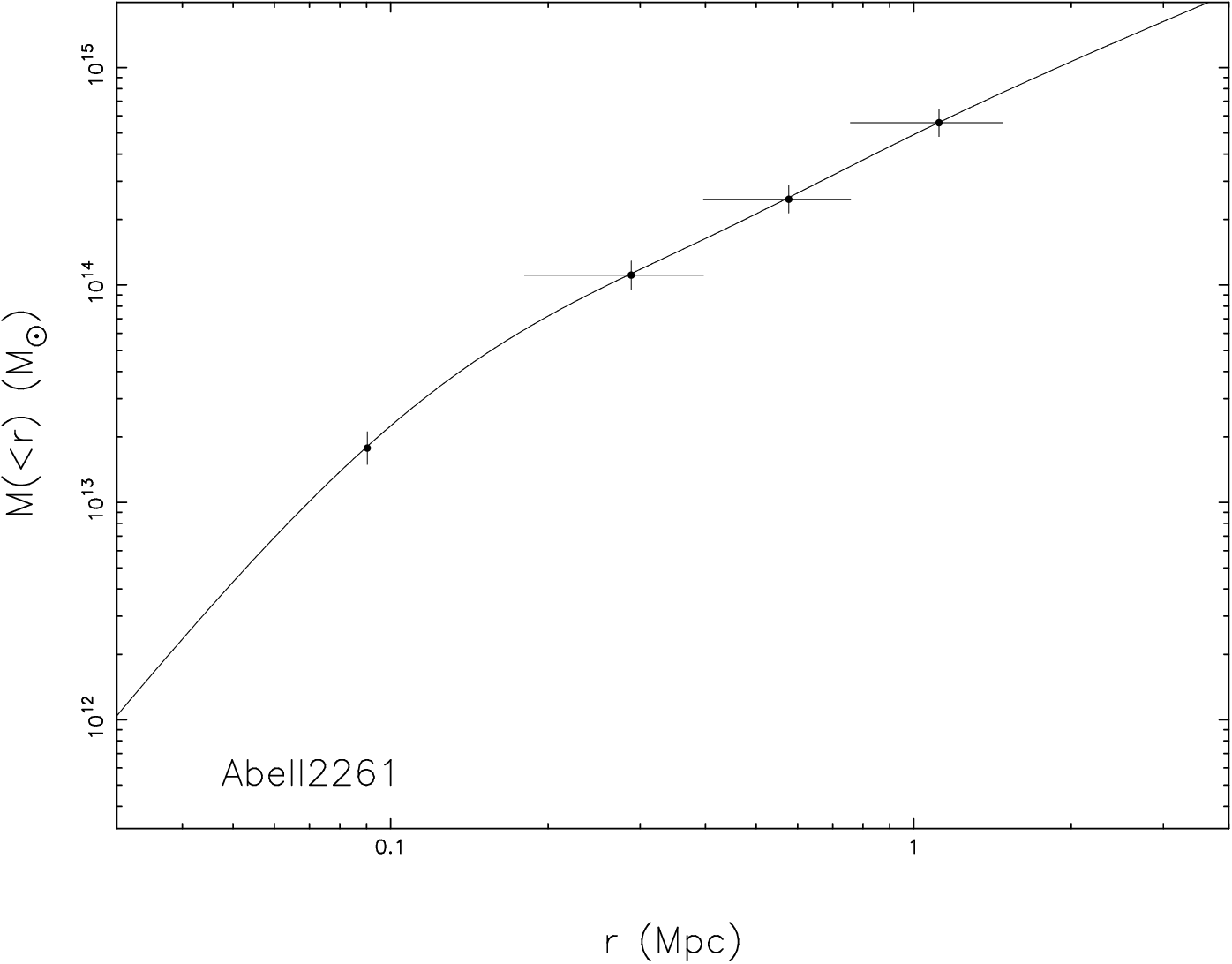}
\includegraphics[angle=270,width=8.5cm]{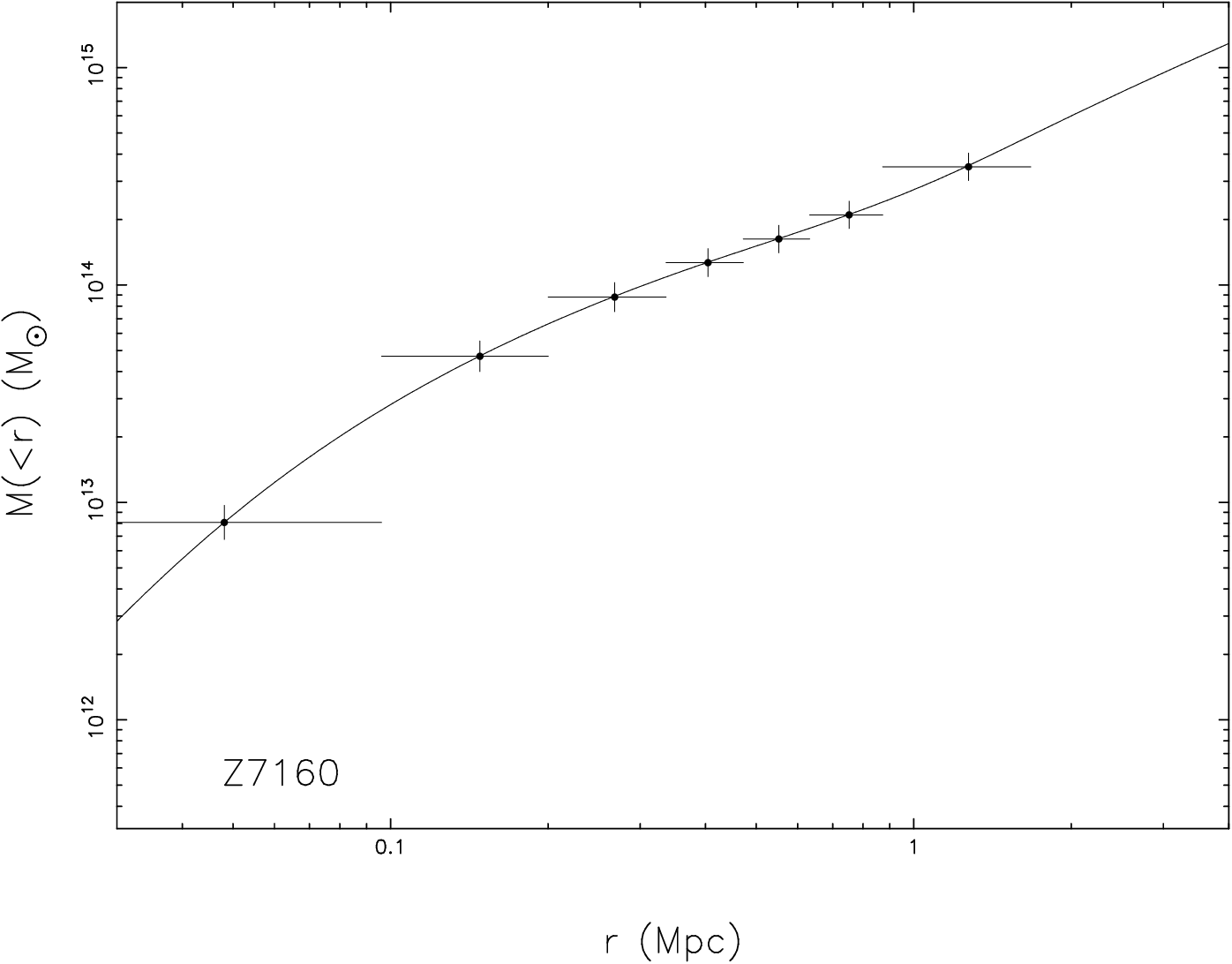}

\end{center}
\caption{See caption in Fig.~\ref{f:ktcom3}.
\label{f:ktcom32}}
\end{figure*}

The derived mass profile was used to calculate the density contrast profile to
derive $M_{500}$ and $r_{500}$ (see Table~\ref{t:catalog2}). In
Figs.~\ref{f:ktcom3}--\ref{f:ktcom32} in the electronic edition of the Journal
(together with Fig.~5 in Zhang et al. 2006, and Fig.~1 and Figs.~B.1-B.5 in
Zhang et al. 2007 for the whole sample), we present the observed mass
profiles.

\subsection{Gas mass fraction distribution}

The gas mass fraction is an important parameter for cluster physics,
e.g. heating and cooling processes, and cosmological applications
using galaxy clusters (e.g. Vikhlinin et al. 2002; Allen et
al. 2004). The gas mass fraction distribution is defined as $f_{\rm
gas}(<r)=M_{\rm gas}(<r)/M(<r)$. The gas mass, total mass and gas mass
fraction at $r_{500}$ are given in Table~\ref{t:catalog2}. We obtain
an average gas mass fraction of $0.121 \pm 0.004$ at $r_{\rm 500}$ for
the sample.

As observed, the gas mass fraction at $r_{\rm 500}$ is generally lower than
the universal baryon fraction, $f_{\rm b}=\Omega_{\rm b}/\Omega_{\rm m}=0.176
\pm 0.019$, based on the WMAP three year results in Spergel et al. (2006).
This is because the baryons in galaxy clusters reside mostly in hot gas
together with a small fraction of stars as implied in observations
(0.003--0.022 in groups, Gastaldello et al. 2007) and simulations (15 per cent
of $f_{\rm gas}$, e.g. Kravtsov et al. 2005). In principle, $\Omega_{\rm m}$
can be determined from the baryon fraction, $f_{\rm b}=f_{\rm gas}+f_{\rm
  gal}$, in which a contribution from stars in galaxies is given by $f_{\rm
  gal}=0.02\pm 0.01 h_{50}^{-1}$ (White et al. 1993). The average gas mass
fraction at $r_{500}$ for the sample supports a low matter density Universe as
also shown in e.g. Allen et al. (2002), Ettori et al. (2003) and Vikhlinin et
al. (2003).

\clearpage 

\begin{table*} { \begin{center} \footnotesize
  {\renewcommand{\arraystretch}{1.3} \caption[]{
 } \label{t:obs}}
  \begin{tabular}{lcccccccccc}
\hline
\hline
  Name           & \multicolumn{2}{c}{X-ray centroid} & Id       & Date           &  \multicolumn{2}{c}{Filter} & Frame & \multicolumn{3}{c}{Net exposure (ks)}   \\
\hline
                 &  R.A. & delc.                      &          &                &     MOS & pn                &    pn  &MOS1 & MOS2 & pn \\
\hline
RXCJ0043.4-2037  & $ 00:43:24.5 $&$ -20:37:31.2 $ &  0042340201  &  2002-01-04    & Thin     & Thin   &       EFF      &   11.4   &   11.4    &    7.1   \\
RXCJ0232.2-4420  & $ 02:32:18.8 $&$ -44:20:51.9 $ &  0042340301  &  2002-07-11    & Thin     & Thin   &       EFF      &   12.7   &   12.1    &    8.7   \\
RXCJ0307.0-2840  & $ 03:07:02.2 $&$ -28:39:55.2 $ &  0042340501  &  2001-02-16    & Thin     & Thin   &       EFF      &   12.4   &   12.7    &    9.3   \\
RXCJ0516.7-5430  & $ 05:16:35.2 $&$ -54:30:36.8 $ &  0205330301  &  2004-01-13    & Thin     & Thin   &       FF       &   10.6   &   11.1    &    8.9   \\
RXCJ0528.9-3927  & $ 05:28:52.5 $&$ -39:28:16.7 $ &  0042340801  &  2001-09-15    & Thin     & Thin   &       EFF      &    7.2   &    6.9    &    3.7   \\
RXCJ0532.9-3701  & $ 05:32:55.9 $&$ -37:01:34.5 $ &  0042341801  &  2002-10-07    & Thin     & Thin   &       EFF      &   10.5   &   11.3    &    7.3   \\
RXCJ0547.6-3152  & $ 05:47:38.3 $&$ -31:52:28.8 $ &  0201900901  &  2004-03-07    & Thin     & Thin   &       EFF      &   21.8   &   22.1    &   17.6   \\
RXCJ0645.4-5413  & $ 06:45:30.0 $&$ -54:13:42.1 $ &  0201901201  &  2004-05-07    & Thin     & Thin   &       EFF      &   11.6   &   12.2    &    9.6   \\
RXCJ0658.5-5556  & $ 06:58:30.2 $&$ -55:56:33.7 $ &  0112980201  &  2000-10-21    & Thin     & Thin   &       EFF      &   25.7   &   23.7    &   21.1   \\
RXCJ0945.4-0839  & $ 09:45:25.1 $&$ -08:39:11.7 $ &  0017540101  &  2001-12-02    & Medium   & Medium &       FF       &    7.1   &    6.4    &    4.4   \\
RXCJ0958.3-1103  & $ 09:58:21.9 $&$ -11:03:48.2 $ &  0201903501  &  2004-06-17    & Thin     & Thin   &       EFF      &    7.9   &    7.9    &    5.3   \\
RXCJ2129.6+0005  & $ 21:29:39.8 $&$ +00:05:18.5 $ &  0093030201  &  2002-10-29    & Medium   & Medium &       EFF      &   38.4   &   39.4    &   25.6   \\
RXCJ2218.6-3853  & $ 22:18:39.9 $&$ -38:53:43.6 $ &  0201903001  &  2004-10-24    & Thin     & Thin   &       EFF      &   20.4   &   20.3    &   12.0   \\
RXCJ2234.5-3744  & $ 22:34:27.1 $&$ -37:44:07.5 $ &  0018741701  &  2001-05-03    & Thin     & Thin   &       FF       &    6.7   &    6.6    &    4.2   \\
RXCJ2308.3-0211  & $ 23:08:22.3 $&$ -02:11:32.1 $ &  0205330501  &  2004-06-05    & Thin     & Thin   &       FF       &    9.6   &   10.0    &    7.8   \\
RXCJ2337.6+0016  & $ 23:37:37.8 $&$ +00:16:15.5 $ &  0042341301  &  2001-12-06    & Thin     & Thin   &       EFF      &   12.4   &   12.0    &    8.3   \\
Abell68          & $ 00:37:06.2 $&$ +09:09:28.7 $ &  0084230201  &  2001-12-14    & Medium   & Medium &       EFF      &   24.9   &   23.8    &   18.2   \\
Abell115         & $ 00:55:50.1 $&$ +26:24:35.7 $ &  0203220101  &  2004-07-16    & Medium   & Medium &       EFF      &   36.0   &   36.8    &   29.5   \\
Abell209         & $ 01:31:52.6 $&$ -13:36:35.5 $ &  0084230301  &  2001-01-15    & Medium   & Medium &       EFF      &   17.3   &   16.1    &   12.8   \\
Abell267         & $ 01:52:42.0 $&$ +01:00:41.2 $ &  0084230401  &  2002-01-02    & Medium   & Medium &       EFF      &   17.2   &   17.1    &   12.4   \\
Abell383         & $ 02:48:03.3 $&$ -03:31:43.6 $ &  0084230501  &  2002-08-17    & Medium   & Medium &       EFF      &   28.1   &   28.0    &   21.5   \\
Abell773         & $ 09:17:52.9 $&$ +51:43:19.4 $ &  0084230601  &  2001-04-26    & Medium   & Medium &       EFF      &   13.0   &   14.7    &   15.9   \\
Abell781         & $ 09:20:24.8 $&$ +30:30:05.7 $ &  0150620201  &  2003-04-22    & Medium   & Medium &       FF       &   14.0   &   14.1    &   11.6   \\
Abell901         & $ 09:55:57.7 $&$ -09:59:06.3 $ &  0148170101  &  2003-05-06    & Thin     & Thin   &       FF       &   19.4   &   18.6    &   53.2   \\
Abell963         & $ 10:17:03.2 $&$ +39:02:56.5 $ &  0084230701  &  2001-11-02    & Medium   & Medium &       EFF      &   23.8   &   24.7    &   17.9   \\
Abell1413        & $ 11:55:18.3 $&$ +23:24:12.7 $ &  0112230501  &  2000-12-06    & Thin     & Thin   &       FF       &   23.6   &   24.4    &   18.8   \\
Abell1689        & $ 13:11:29.3 $&$ -01:20:26.7 $ &  0093030101  &  2001-12-24    & Thin     & Thin   &       EFF      &   36.8   &   37.0    &   32.6   \\
Abell1758        & $ 13:32:44.6 $&$ +50:32:46.5 $ &  0142860201  &  2002-11-12    & Medium   & Medium &       FF       &   38.1   &   40.1    &   19.4   \\
Abell1763        & $ 13:35:18.1 $&$ +41:00:03.9 $ &  0084230901  &  2002-12-13    & Medium   & Medium &       EFF      &   12.3   &   12.0    &    9.3   \\
Abell1835        & $ 14:01:01.9 $&$ +02:52:35.5 $ &  0098010101  &  2000-06-28    & Thin     & Thin   &       FF       &   25.3   &   25.3    &   24.7   \\
Abell1914        & $ 14:26:00.8 $&$ +37:49:38.8 $ &  0112230201  &  2002-12-18    & Thin     & Medium &       EFF      &   20.5   &   21.2    &   14.5   \\
Abell2204        & $ 16:32:47.1 $&$ +05:34:32.3 $ &  0112230301  &  2001-09-12    & Medium   & Medium &       FF       &   17.5   &   18.5    &   14.3   \\
Abell2218        & $ 16:35:53.8 $&$ +66:12:32.4 $ &  0112980101  &  2002-09-28    & Thin     & Thin   &       EFF      &   16.7   &   16.9    &   13.8   \\
Abell2261        & $ 17:22:26.0 $&$ +32:07:47.4 $ &  0093031001  &  2003-08-29    & Thin     & Thin   &       EFF      &    2.7   &    2.9    &    0.9   \\
Abell2390        & $ 21:53:37.1 $&$ +17:41:46.4 $ &  0111270101  &  2001-06-19    & Thin     & Thin   &       FF       &   10.3   &   10.0    &    8.8   \\
Abell2667        & $ 23:51:39.2 $&$ -26:05:03.5 $ &  0148990101  &  2003-06-21    & Medium   & Medium &       FF       &   22.2   &   22.9    &   14.1   \\
Z7160            & $ 14:57:15.2 $&$ +22:20:31.2 $ &  0108670201  &  2002-08-03    & Medium   & Medium &       FF       &   31.2   &   31.6    &   26.3   \\
\hline
\hline
  \end{tabular}
  \end{center}
\hspace*{0.3cm}{\footnotesize The cluster center is in sky
  coordinates in epoch J2000. The MOS data are in EE mode except that
  the MOS1 data of Abell1835 are in window mode, which cannot be used
  for this work.  } }
\end{table*}

\begin{table*} { \begin{center} \footnotesize
      {\renewcommand{\arraystretch}{1.3} \caption[]{ Primary parameters.
          Column~(1): cluster name; Col.~(2): optical redshift (e.g.
          B\"ohringer et al. 2004; Smith et al. 2005; Bardeau et al. 2007);
          Col.~(3): hydrogen column density (Dickey \& Lockman 1990);
          Col.~(4): truncation radius corresponding to a $S/N$ of 3 of the
          observational surface brightness profile; Col.~(5): volume averaged
          radial temperature profile of 0.2--0.5$r_{500}$; Cols.~(6--7):
          Spectroscopic temperature and metallicity in the annulus of
          0.2--0.5$r_{500}$. Cols.~(8--10): bolometric luminosity including
          the $<0.2r_{500}$ region, excluding the $<0.2r_{500}$ region, and
          corrected for the $<0.2r_{500}$ region, respectively.}
  \label{t:global}}
\begin{tabular}{lccccccccccc}
\hline
\hline
Name & $z_{\rm opt}$ & $N_{\rm H}$ & $r_{\rm t}$ & $T_{0.2-0.5r_{500}}$ & $T_{0.2-0.5r_{500}}^{\rm spec}$ & $Z_{0.2-0.5r_{500}}^{\rm spec}$ & $L_{\rm bol}^{\rm incc}$ & $L_{\rm bol}^{\rm excc}$ & $L_{\rm bol}^{\rm corr}$ \\
\hline
    &   & $10^{22}$~cm$^{-2}$ & $\prime$ & keV & keV & $Z_{\odot}$
& $10^{45}$erg/s & $10^{45}$erg/s & $10^{45}$erg/s \\
\hline
RXCJ0043.4-2037  &  0.292   &  0.0154  &   7.0   & $   6.8  \pm   0.4 $ & $   7.0  \pm   0.4 $ & $  0.17  \pm  0.06 $ & $  1.05  \pm  0.10 $ & $  0.59  \pm  0.07 $ & $  0.84  \pm  0.10   $\\
RXCJ0232.2-4420  &  0.284   &  0.0250  &   6.8   & $   6.9  \pm   0.3 $ & $   6.6  \pm   0.3 $ & $  0.23  \pm  0.07 $ & $  1.85  \pm  0.14 $ & $  0.75  \pm  0.08 $ & $  1.10  \pm  0.14   $\\
RXCJ0307.0-2840  &  0.258   &  0.0136  &   4.7   & $   6.9  \pm   0.4 $ & $   7.1  \pm   0.4 $ & $  0.26  \pm  0.08 $ & $  1.31  \pm  0.12 $ & $  0.61  \pm  0.08 $ & $  0.88  \pm  0.12   $\\
RXCJ0516.7-5430  &  0.294   &  0.0686  &   6.4   & $   6.7  \pm   0.5 $ & $   6.7  \pm   0.5 $ & $  0.25  \pm  0.10 $ & $  1.09  \pm  0.15 $ & $  0.80  \pm  0.11 $ & $  1.00  \pm  0.15   $\\
RXCJ0528.9-3927  &  0.284   &  0.0212  &   5.8   & $   6.6  \pm   0.5 $ & $   6.6  \pm   0.5 $ & $  0.20  \pm  0.09 $ & $  1.45  \pm  0.15 $ & $  0.88  \pm  0.10 $ & $  1.10  \pm  0.15   $\\
RXCJ0532.9-3701  &  0.275   &  0.0290  &   7.0   & $   8.2  \pm   0.6 $ & $   7.7  \pm   0.6 $ & $  0.28  \pm  0.09 $ & $  1.27  \pm  0.12 $ & $  0.55  \pm  0.07 $ & $  0.81  \pm  0.12   $\\
RXCJ0547.6-3152  &  0.148   &  0.0205  &   7.7   & $   6.1  \pm   0.2 $ & $   6.0  \pm   0.2 $ & $  0.22  \pm  0.03 $ & $  0.72  \pm  0.06 $ & $  0.39  \pm  0.04 $ & $  0.55  \pm  0.06   $\\
RXCJ0645.4-5413  &  0.164   &  0.0651  &   8.1   & $   8.1  \pm   0.3 $ & $   7.6  \pm   0.3 $ & $  0.11  \pm  0.04 $ & $  1.78  \pm  0.16 $ & $  1.01  \pm  0.12 $ & $  1.30  \pm  0.16   $\\
RXCJ0658.5-5556  &  0.296   &  0.0653  &   8.9   & $  11.1  \pm   0.4 $ & $  10.7  \pm   0.4 $ & $  0.14  \pm  0.04 $ & $  4.95  \pm  0.24 $ & $  2.99  \pm  0.18 $ & $  4.21  \pm  0.24   $\\
RXCJ0945.4-0839  &  0.153   &  0.0359  &   7.6   & $   5.7  \pm   0.5 $ & $   5.3  \pm   0.5 $ & $  0.11  \pm  0.10 $ & $  0.42  \pm  0.06 $ & $  0.24  \pm  0.04 $ & $  0.33  \pm  0.06   $\\
RXCJ0958.3-1103  &  0.167   &  0.0540  &   7.4   & $   6.1  \pm   0.3 $ & $   5.8  \pm   0.3 $ & $  0.27  \pm  0.06 $ & $  0.91  \pm  0.09 $ & $  0.42  \pm  0.06 $ & $  0.57  \pm  0.09   $\\
RXCJ2129.6+0005  &  0.235   &  0.0428  &   6.8   & $   6.4  \pm   0.2 $ & $   6.3  \pm   0.2 $ & $  0.27  \pm  0.03 $ & $  1.43  \pm  0.09 $ & $  0.58  \pm  0.06 $ & $  0.79  \pm  0.09   $\\
RXCJ2218.6-3853  &  0.141   &  0.0138  &   7.3   & $   6.7  \pm   0.2 $ & $   6.2  \pm   0.2 $ & $  0.27  \pm  0.05 $ & $  0.66  \pm  0.05 $ & $  0.34  \pm  0.04 $ & $  0.48  \pm  0.05   $\\
RXCJ2234.5-3744  &  0.151   &  0.0122  &   7.9   & $   7.8  \pm   0.4 $ & $   7.5  \pm   0.4 $ & $  0.22  \pm  0.08 $ & $  1.09  \pm  0.11 $ & $  0.57  \pm  0.07 $ & $  0.80  \pm  0.11   $\\
RXCJ2308.3-0211  &  0.297   &  0.0445  &   7.3   & $   7.7  \pm   0.7 $ & $   7.6  \pm   0.7 $ & $  0.41  \pm  0.16 $ & $  1.21  \pm  0.13 $ & $  0.56  \pm  0.08 $ & $  0.80  \pm  0.13   $\\
RXCJ2337.6+0016  &  0.275   &  0.0382  &   7.0   & $   8.0  \pm   0.5 $ & $   7.5  \pm   0.5 $ & $  0.19  \pm  0.07 $ & $  1.00  \pm  0.09 $ & $  0.54  \pm  0.06 $ & $  0.77  \pm  0.09   $\\
Abell68          &  0.255   &  0.0493  &   6.1   & $   7.7  \pm   0.3 $ & $   7.3  \pm   0.3 $ & $  0.12  \pm  0.05 $ & $  1.14  \pm  0.10 $ & $  0.61  \pm  0.07 $ & $  0.86  \pm  0.10   $\\
Abell115         &  0.197   &  0.0543  &   8.0   & $   6.4  \pm   0.1 $ & $   6.2  \pm   0.1 $ & $  0.20  \pm  0.03 $ & $  1.43  \pm  0.11 $ & $  1.05  \pm  0.09 $ & $  1.16  \pm  0.11   $\\
Abell209         &  0.209   &  0.0164  &   7.4   & $   7.1  \pm   0.3 $ & $   7.1  \pm   0.3 $ & $  0.22  \pm  0.04 $ & $  1.33  \pm  0.11 $ & $  0.84  \pm  0.08 $ & $  1.10  \pm  0.11   $\\
Abell267         &  0.230   &  0.0280  &   5.3   & $   6.5  \pm   0.4 $ & $   6.2  \pm   0.4 $ & $  0.25  \pm  0.08 $ & $  0.77  \pm  0.07 $ & $  0.36  \pm  0.05 $ & $  0.52  \pm  0.07   $\\
Abell383         &  0.187   &  0.0392  &   7.7   & $   5.3  \pm   0.2 $ & $   4.7  \pm   0.2 $ & $  0.18  \pm  0.06 $ & $  0.81  \pm  0.05 $ & $  0.27  \pm  0.04 $ & $  0.36  \pm  0.05   $\\
Abell773         &  0.217   &  0.0144  &   7.7   & $   8.1  \pm   0.4 $ & $   8.3  \pm   0.4 $ & $  0.30  \pm  0.06 $ & $  2.09  \pm  0.16 $ & $  1.06  \pm  0.11 $ & $  1.53  \pm  0.16   $\\
Abell781         &  0.298   &  0.0194  &   6.5   & $   6.4  \pm   0.5 $ & $   6.5  \pm   0.5 $ & $  0.18  \pm  0.10 $ & $  0.63  \pm  0.10 $ & $  0.52  \pm  0.09 $ & $  0.60  \pm  0.10   $\\
Abell901         &  0.163   &  0.0499  &   4.6   & $   3.6  \pm   0.2 $ & $   3.2  \pm   0.2 $ & $  0.12  \pm  0.09 $ & $  1.60  \pm  0.05 $ & $  0.35  \pm  0.03 $ & $  0.57  \pm  0.05   $\\
Abell963         &  0.206   &  0.0140  &   6.1   & $   6.3  \pm   0.2 $ & $   6.5  \pm   0.2 $ & $  0.28  \pm  0.05 $ & $  1.14  \pm  0.09 $ & $  0.55  \pm  0.06 $ & $  0.74  \pm  0.09   $\\
Abell1413        &  0.143   &  0.0219  &   8.4   & $   6.8  \pm   0.1 $ & $   6.6  \pm   0.1 $ & $  0.27  \pm  0.03 $ & $  1.38  \pm  0.08 $ & $  0.68  \pm  0.06 $ & $  0.93  \pm  0.08   $\\
Abell1689        &  0.184   &  0.0182  &   8.0   & $   8.4  \pm   0.2 $ & $   8.5  \pm   0.2 $ & $  0.26  \pm  0.04 $ & $  2.84  \pm  0.10 $ & $  0.93  \pm  0.06 $ & $  1.44  \pm  0.10   $\\
Abell1758        &  0.280   &  0.0106  &   6.0   & $   7.6  \pm   0.2 $ & $   7.9  \pm   0.2 $ & $  0.19  \pm  0.04 $ & $  1.05  \pm  0.07 $ & $  0.66  \pm  0.05 $ & $  0.94  \pm  0.07   $\\
Abell1763        &  0.228   &  0.0936  &   7.5   & $   6.3  \pm   0.3 $ & $   5.8  \pm   0.3 $ & $  0.20  \pm  0.05 $ & $  1.65  \pm  0.15 $ & $  1.14  \pm  0.11 $ & $  1.41  \pm  0.15   $\\
Abell1835        &  0.253   &  0.0232  &   7.0   & $   8.0  \pm   0.3 $ & $   8.4  \pm   0.3 $ & $  0.23  \pm  0.04 $ & $  5.32  \pm  0.17 $ & $  1.51  \pm  0.10 $ & $  2.25  \pm  0.17   $\\
Abell1914        &  0.171   &  0.0095  &   8.2   & $   8.6  \pm   0.3 $ & $   8.8  \pm   0.3 $ & $  0.19  \pm  0.05 $ & $  2.17  \pm  0.11 $ & $  0.62  \pm  0.06 $ & $  0.97  \pm  0.11   $\\
Abell2204        &  0.152   &  0.0567  &   9.0   & $   8.0  \pm   0.2 $ & $   7.6  \pm   0.2 $ & $  0.27  \pm  0.03 $ & $  3.39  \pm  0.12 $ & $  1.09  \pm  0.08 $ & $  1.52  \pm  0.12   $\\
Abell2218        &  0.176   &  0.0324  &   7.7   & $   7.4  \pm   0.3 $ & $   6.6  \pm   0.3 $ & $  0.21  \pm  0.04 $ & $  1.11  \pm  0.08 $ & $  0.68  \pm  0.06 $ & $  0.90  \pm  0.08   $\\
Abell2261        &  0.224   &  0.0328  &   8.1   & $   7.0  \pm   0.6 $ & $   6.6  \pm   0.6 $ & $  0.33  \pm  0.18 $ & $  1.39  \pm  0.22 $ & $  0.67  \pm  0.13 $ & $  0.91  \pm  0.22   $\\
Abell2390        &  0.233   &  0.0680  &   7.7   & $  10.6  \pm   0.6 $ & $  11.6  \pm   0.6 $ & $  0.24  \pm  0.06 $ & $  4.09  \pm  0.27 $ & $  1.95  \pm  0.17 $ & $  2.66  \pm  0.27   $\\
Abell2667        &  0.230   &  0.0165  &   7.6   & $   7.3  \pm   0.3 $ & $   7.0  \pm   0.3 $ & $  0.29  \pm  0.04 $ & $  2.16  \pm  0.13 $ & $  0.83  \pm  0.08 $ & $  1.14  \pm  0.13   $\\
Z7160            &  0.258   &  0.0316  &   8.0   & $   5.0  \pm   0.1 $ & $   4.7  \pm   0.1 $ & $  0.32  \pm  0.04 $ & $  1.70  \pm  0.08 $ & $  0.50  \pm  0.05 $ & $  0.67  \pm  0.08   $\\
\hline
\hline
  \end{tabular}
  \end{center}
\hspace*{0.3cm}{\footnotesize 
The global temperature and luminosity are uniformly calculated as
described in Appendix~\ref{s:resultkt} and Appendix~\ref{s:lumi} for all 37
LoCuSS clusters. In Zhang et al. (2006, 2007), the temperature and
luminosity are defined differently. } }
\end{table*}

\clearpage 

\section{Scaled profiles of the X-ray properties: a structural similarity}
\label{s:prof}

Simulations (e.g. Navarro et al. 1997, 2004) suggest a self-similar
structure for galaxy clusters within hierarchical structure formation
scenarios. The scaled profiles of the X-ray properties and their
scatter can be used to quantify the structural variations, to probe
the regularity of galaxy clusters, and thus to understand the scatter
in the X-ray scaling relations, which is of prime importance for
cosmological applications involving clusters of galaxies.

Because the observational truncation radii ($r_{\rm t}$) in the
surface brightness profiles are above $r_{500}$ but below $r_{200}$
for most clusters in the sample, we use $r_{500}$ for radial scaling.

The following redshift evolution corrections due to LSS growth are
usually used to account for the dependence on the evolution of the
cosmological parameters,

$S_{\rm X} \cdot E^{-3}(z) 
\propto f(T)$,

$S \cdot E^{4/3}(z) 
\propto f(T)$,

$P \cdot E^{-2}(z) 
\propto f(T)$,

$L \cdot E^{-1}(z) 
\propto f(T)$,

$M \cdot E(z) 
\propto f(T)$,

$M_{\rm gas} \cdot E(z) 
\propto f(T)$,

$Y_{\rm X} \cdot E(z) 
\propto f(T)$,

$r \cdot E(z) 
\propto f(T)$,

where $Y_{\rm X}=M_{\rm gas}\cdot T$ is an X-ray analog for the integrated SZ
flux (e.g. Kravtsov et al. 2006; also SZ see Sunyaev \& Zeldovich 1972). The
function $f(T)$ denotes the best-fit power law parameterization.
$\Delta_{c,z}/\Delta_{c,0}$ factor\footnote{$\Delta_{\rm
    c,z}=18\pi^2+82(\Omega_{\rm m,z}-1)-39(\Omega_{\rm m,z}-1)^2$ is the
  analytic approximation derived from the top-hat spherical collapse model for
  a flat Universe (Bryan \& Norman 1998) and $\Omega_{\rm m,z}$ the cosmic
  density parameter at redshift $z$.} is dropped out compared to the redshift
evolution corrections used in Zhang et al. (2006, 2007) for convenience, and
for easier comparison between our sample and published samples.

\subsection{Scaled temperature profiles}

\begin{figure*}
\begin{center}
\includegraphics[width=6.8cm,angle=270]{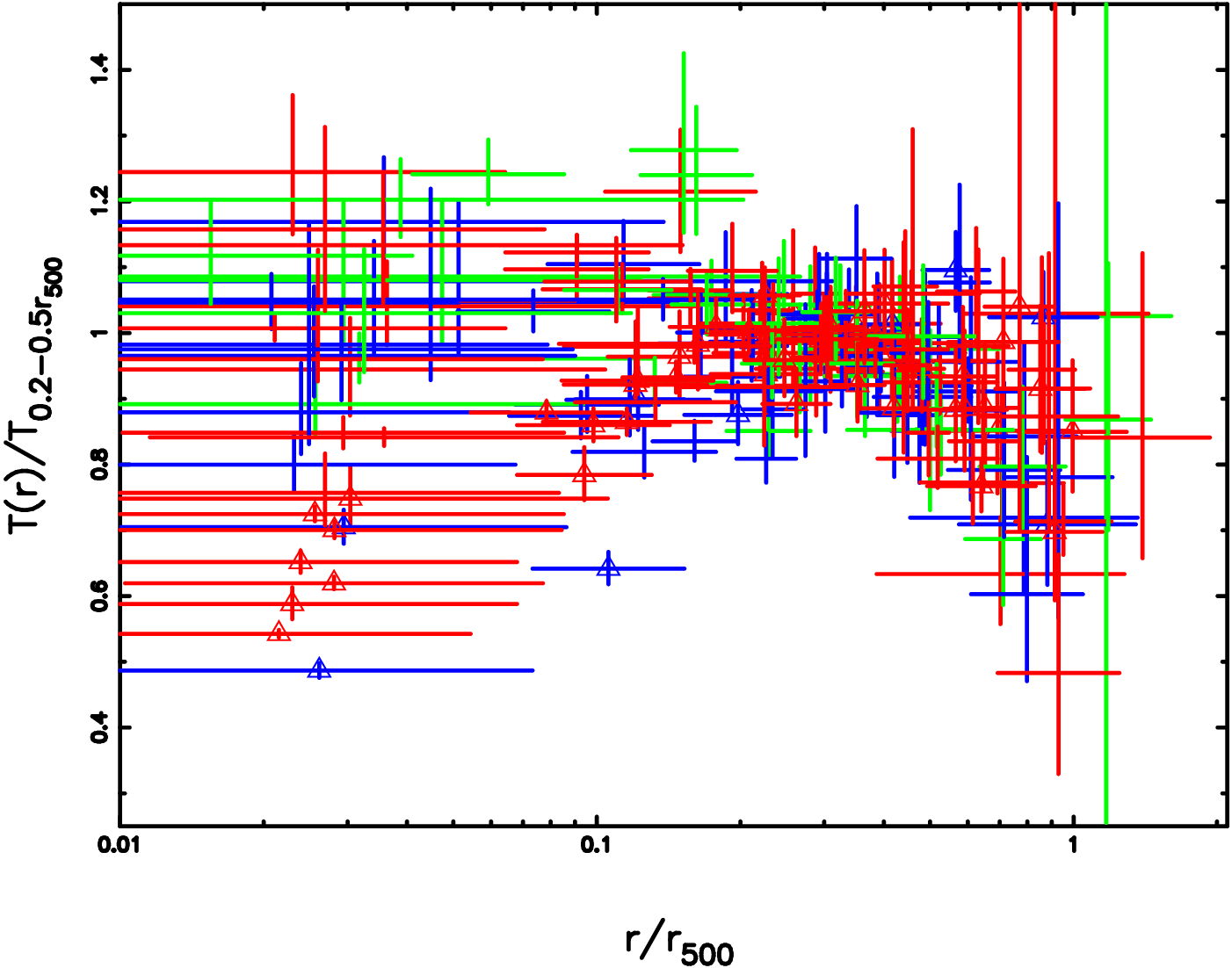}
\includegraphics[width=6.8cm,angle=270]{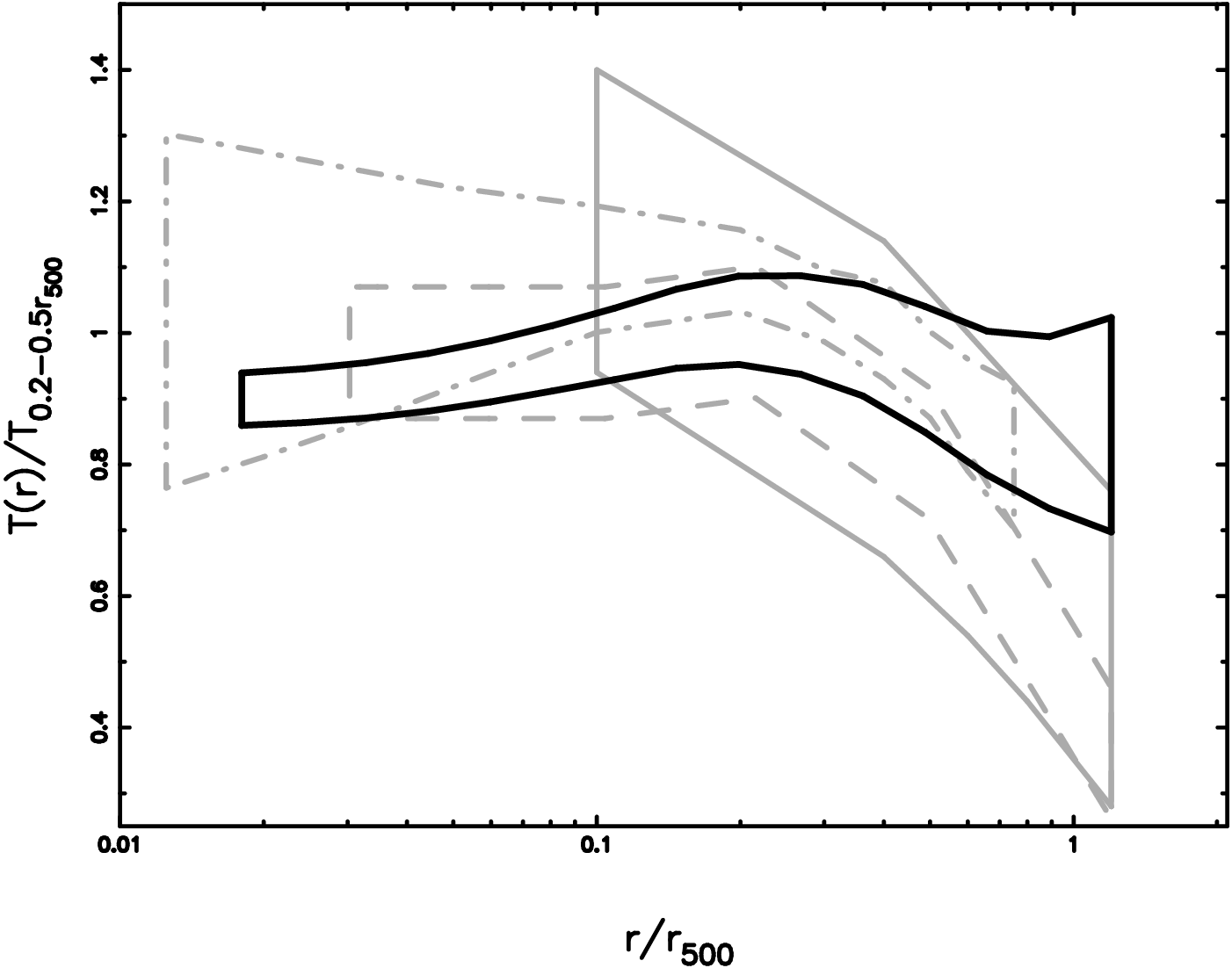}
\end{center}
\caption{{\it Left:} Scaled radial temperature profiles for the sample. The
  CCCs are marked by triangles. The colors have the same meaning as those in
  Fig.~\ref{f:my_1}. {\it Right:} An average temperature profile of the sample
  of 37 LoCuSS clusters (black, solid) compared to the temperature profile
  ranges in Markevitch et al. (1998, gray, solid), Vikhlinin et al. (2005,
  gray, dashed), and Pratt et al.  (2007, gray, dash-dotted). \label{f:scalet}
}
\end{figure*}

We scaled the radial temperature profiles by $T_{0.2-0.5r_{500}}$ and
$r_{500}$ (Fig.~\ref{f:scalet}). Within $0.2r_{\rm 500}$, we observed
a temperature drop to at least 70 per cent of the maximum value
towards the cluster center for 9 clusters (defined as CCCs). Previous
observations have shown that temperature measurements on scales below
$0.2 r_{500}$ tend to show peculiarities linked to the cluster
dynamical history (e.g. Finoguenov et al. 2005; Zhang et al. 2005a,
2005b). For example, the temperatures of merging clusters can be
boosted (Smith et al. 2005, Poole et al. 2006, 2007). However, the
boosting affects mainly the cluster cores. Using global temperatures
excluding the $<0.2 r_{500}$ region can thus (1) tighten the scaled
profiles of the X-ray properties, (2) reduce the scatter in the X-ray
scaling relations, and (3) improve agreement between normalization of
the X-ray scaling relations for the CCCs and non-CCCs, respectively.

We used the 1-$\sigma$ boundary of the scaled radial temperature profiles to
compute the confidence zone for the mean temperature profile. The average
temperature profile for the whole sample gives $T(r) \propto r^{0.13 \pm
  0.04}$, and for the CCC subsample $T(r) \propto r^{0.38 \pm 0.04}$,
respectively, in the $r<0.2 r_{500}$ region. Because the temperature profiles
are not PSF corrected, the significance of the cool cores could be more
pronounced considering the PSF effects. The temperature behavior in the
cluster cores for the CCC subsample is very similar to the behavior for the
nearby CCC sample in Sanderson et al.  (2006) giving $T(r) \propto r^{0.4}$
based on \emph{Chandra} observations. In the outskirts ($0.2
r_{500}<r<r_{500}$), the whole sample shows a structurally similar behavior
giving $T(r) \propto r^{-0.28 \pm 0.19}$ with 30 per cent scatter. The average
temperature profile decreases down to 80 per cent of the maximum value with
intrinsic scatter of $\sim 30$ per cent at $\sim r_{\rm 500}$.

Studies of the cluster temperature distributions indicate a similar
universal temperature profile with a significant decline beyond
cluster cores in observations (e.g. Markevitch et al. 1998; De Grandi
\& Molendi 2002; Vikhlinin et al. 2005; Zhang et al. 2004, 2005a,
2006, 2007; Pratt et al. 2007) and simulations (e.g. Borgani et al.
2004).

Accurate temperature distributions (15 per cent uncertainties) are
measured up to $0.5r{500}$ for our sample. We found steepening of the
average temperature distribution using the parameterization of the
temperature distributions, giving slopes from $0.04 \pm 0.03$ at
0.2$r_{500}$ to $0.16 \pm 0.03$ at 0.5$r_{500}$ (the errors for the
slope are the statistical formal errors). The slope should have been
calculated from the observational data points for the temperature
profile as done for the density profile based on a temperature profile
better resolved spatially. With deep \emph{Chandra} observations,
Vikhlinin et al.  (2005) found more significant steepening of the
temperature distributions in the outskirts.

\subsection{Global temperature}
\label{s:resultkt}

As indicated by the scaled temperature profiles for this sample and also
explained in Zhang et al. (2007), the temperature measurements limited to
0.2--0.5$r_{500}$ are less affected by the cool cores and are thus used as
cluster global temperatures. We present the spectroscopic global temperature
($T^{\rm spec}_{0.2-0.5 r_{500}}$) and metallicity ($Z^{\rm spec}_{0.2-0.5
  r_{500}}$) in the annulus of 0.2--0.5$r_{500}$ together with the volume
average of the radial temperature profile limited to the radial range of
0.2--0.5$r_{500}$ ($T_{0.2-0.5 r_{500}}$) in Table~\ref{t:global}.
$T_{0.2-0.5 r_{500}}$ is used as the cluster global temperature for studies of
the structural similarity of the X-ray properties and the scaling relations.
We present the spectral measurements, temperature versus metallicity, from all
radial bins in Fig.~\ref{f:tc}, and observed no obvious correlation between
cluster temperature and metallicity though the mass range is relatively small.

\subsection{Scaled surface brightness profiles}
\label{s:sxlx}

\begin{figure*}
\begin{center}
\includegraphics[width=6.8cm,angle=270]{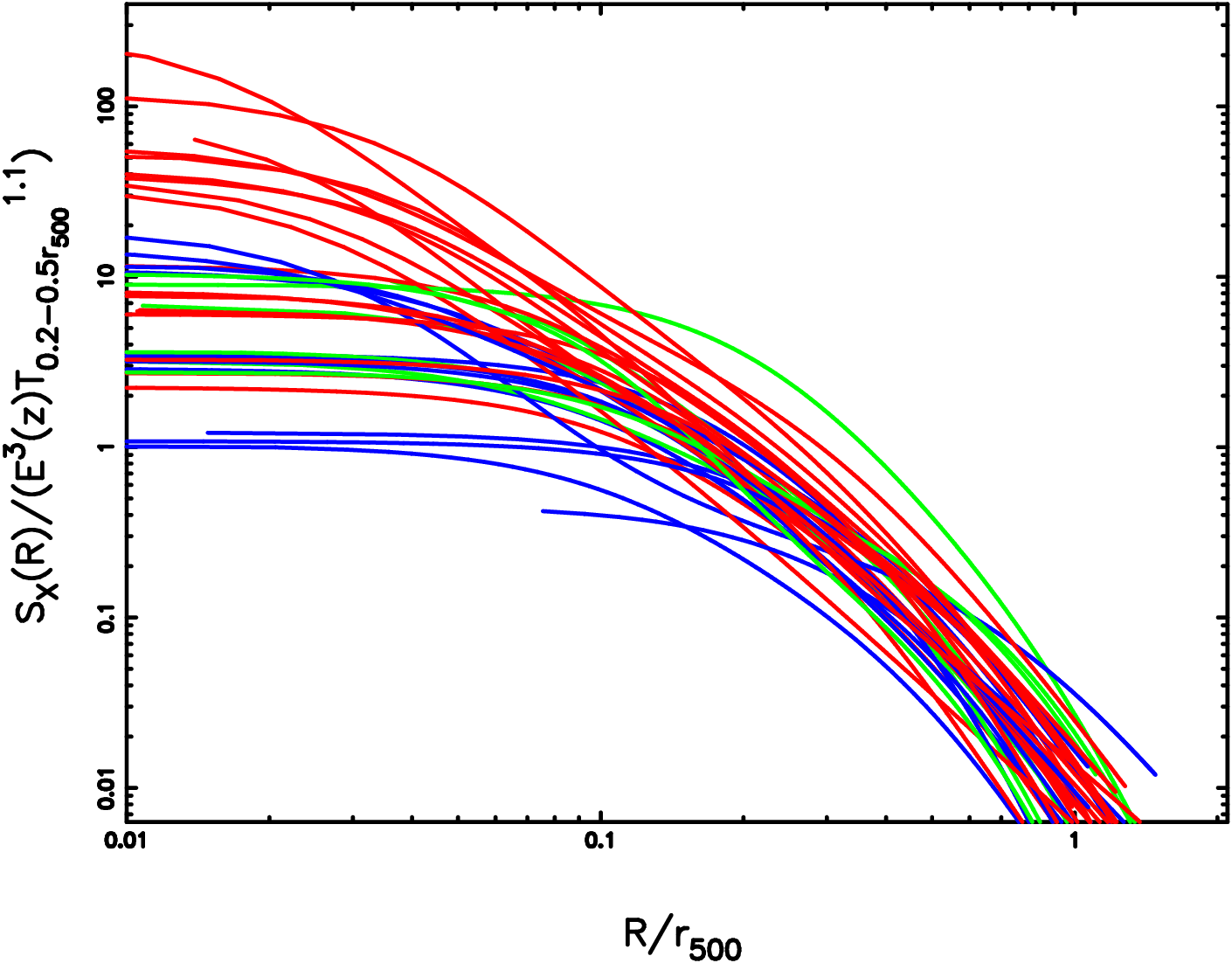}
\includegraphics[width=6.8cm,angle=270]{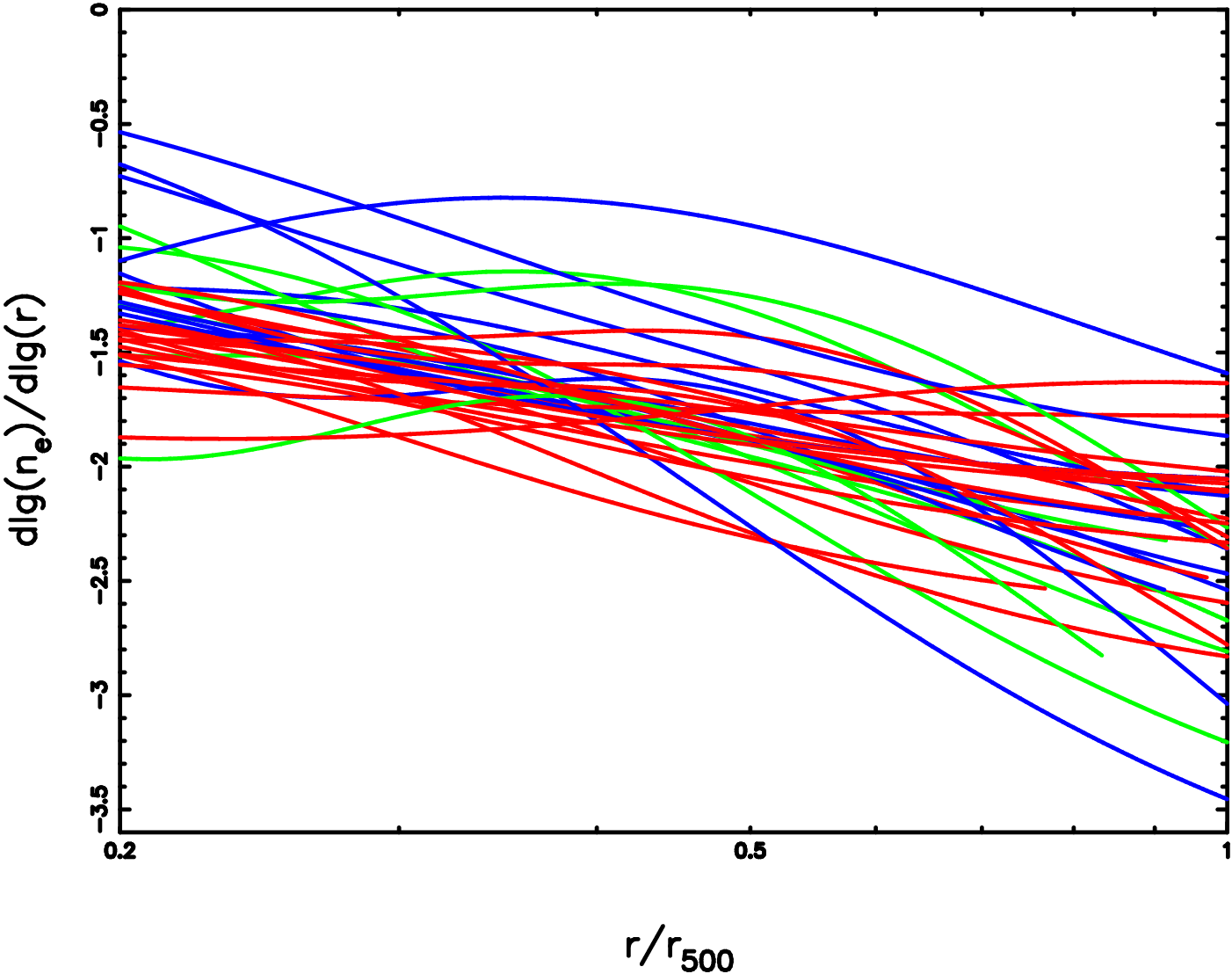}
\end{center}
\caption{{\it Left: }Surface brightness profile fits scaled according
to the empirical scaling, $S_{\rm X} \propto T^{1.1}$. {\it Right:}
Power law slope of the electron number density profiles. The colors
have the same meaning as those in Fig.~\ref{f:my_1}.
\label{f:scalesx} }
\end{figure*}

The standard self-similar model gives $S_{\rm X}
\propto \int_{R}^{\infty} n_{\rm e}^2 d\ell 
\propto M_{\rm gas}^2 \cdot M^{-5/3}
\propto T^{0.5}$ as $M \propto M_{\rm gas} \propto T^{1.5}$. 
The scaling becomes steeper ($S_{\rm X} \propto T^{1.1}$) using the
empirical relations ($M \propto T^{1.5}$ in e.g. Arnaud et al. 2005,
$M_{\rm gas} \propto T^{1.8}$ in e.g. Mohr et al. 1999; Vikhlinin et
al. 1999). We present the empirical scaling $S_{\rm X} \propto
T^{1.1}$ scaled surface brightness profiles in Fig.~\ref{f:scalesx}.
We found a less scattered structurally similar behavior at $R>0.2
r_{500}$ for the scaled surface brightness profiles compared to the
profiles scaled by $S_{\rm X} \propto T^{0.5}$.

\subsection{Global luminosities}
\label{s:lumi}

The bolometric X-ray luminosity (here we use the 0.01--100~keV band) is given
by $L_{\rm bol} \propto \int \widetilde{\Lambda}(r) n^2_{\rm e}(r) dV$,
practically an integral of the X-ray surface brightness to 2.5$r_{500}$
($L_{\rm bol}^{\rm incc}$ in Table\ref{t:global}). The value 2.5$r_{500}$ is
used because the virial radii are about 2.2--2.6$r_{500}$ using our mass
modeling. The luminosity varies by less than 10 per cent when setting the
truncation radius to values between $r_{500}$ and $2.5 r_{500}$. The best-fit
``mekal'' model for the spectra extracted in the annulus limited to
0.2--0.5$r_{500}$ is used to convert the count rate to real flux. Using
different converting ``mekal'' models, e.g. different temperature
($T_{0.2-0.5r_{500}}$ or $T^{\rm spec}_{0.1-0.5r_{500}}$ or $T^{\rm
  spec}_{0.2-0.5r_{500}}$ or $T^{\rm spec}_{<2/3 r_{\rm t}}$) and metallicity
($Z^{\rm spec}_{0.1-0.5r_{500}}$ or $Z^{\rm spec}_{0.2-0.5r_{500}}$ or $Z^{\rm
  spec}_{<2/3 r_{\rm t}}$) causes up to 20 per cent difference in the X-ray
luminosity, in which the difference is more significant for pronounced CCCs.

The core radii populate a broad range of values up to 0.2~$r_{500}$. The X-ray
luminosity is sensitive to the presence of a cool core. It can thus be used to
probe the evolution of a cool cores. We present the bolometric luminosities
including and excluding the $R<0.2 r_{500}$ region ($L_{\rm bol}^{\rm incc}$
and $L_{\rm bol}^{\rm excc}$) in Table~\ref{t:global}. The fractions of the
X-ray luminosity attributed to the $R<0.2 r_{500}$ region span a broad range
up to $\sim 80$ per cent. This introduces large uncertainties in the use of
the luminosity $L_{\rm bol}^{\rm incc}$ as a mass indicator. The X-ray
luminosity excluding the $<0.2 r_{500}$ region ($L_{\rm bol}^{\rm excc}$) is
much less dependent on the presence of the cool core.  However, to better
re-produce the cluster luminosity and the normalization of the
luminosity--temperature/mass relation, we use the X-ray luminosity corrected
for the cool core by assuming\footnote{This assumption was used only in the
  calculation for $L_{\rm bol}^{\rm corr}$, not for any other quantity.}
$S_{\rm X}(R)=S_{\rm X}(0.2 r_{500})$ for the $R<0.2 r_{500}$ region ($L_{\rm
  bol}^{\rm corr}$ in Table~\ref{t:global}). Using $L_{\rm bol}^{\rm corr}$
for the X-ray scaling relations we obtain reduced scatter and consistent
normalization for the CCCs and non-CCCs as shown later. The luminosity is
lower (by up to 10 per cent) assuming a constant luminosity in the cluster
core instead of a $\beta$ model, especially for the pronounced CCCs.  However,
the luminosity corrected for the cool core by the $\beta$ model still
introduces relatively significant scatter dominated by the CCCs due to the
correlation between core radius and slope in the $\beta$ model.

\subsection{Gas profiles in the outskirts}
\label{s:infall}

The generally adopted $\beta$ model ($\beta=2/3$) gives $n_{\rm e}
\propto r^{-2}$. However, Vikhlinin et al. (1999) found a mild trend
for $\beta$ to increase as a function of cluster temperature, which
gives $\beta \sim 0.80$ and $n_{\rm e} \propto r^{-2.4}$ for clusters
around 10~keV. Bahcall (1999) also found that the electron number
density scales as $n_{\rm e} \propto r^{-2.4}$ at large radii. Zhang
et al. (2006, 2007) confirmed their conclusion that $n_{\rm e} \propto
r^{-2.42}$ for the REFLEX-DXL clusters and $n_{\rm e} \propto r^{-2.2
\pm 0.1}$ for the pilot LoCuSS sample. Due to the gradual change in
the slope, one should be cautious when using a single slope
double-$\beta$ model which might introduce a systematic error in the
cluster mass measurements (as also described in e.g. Horner 2001). The
power law fit of the ICM density distributions at $r_{500}$ for the
present sample gives an average of $n_{\rm e}
\propto r^{-2.21\pm 0.05}$.

We made an additional attempt to check the difference in the mass
modeling using a single-$\beta$ model and a double-$\beta$ model for
CCCs caused by the steepening of the slope on large radial scales. We
used a single-$\beta$ model with the value of the core radius fixed to
results for the outer component in the double-$\beta$ model to model
the electron number density distribution. It often gives a higher
value for the slope than the free fitting slope of the double-$\beta$
model by a few per cent, and up to 10 per cent for the most pronounced
CCC. Therefore the cluster total mass $M_{500}^{\rm X}$ can be higher
by up to 20 per cent due to the steepening ICM density in the
outskirts. To suppress this mass bias, we performed the modeling as
follows.  We first fitted the outer region ($>0.5r_{500}$) using a
single-$\beta$ model.  We then fitted the whole radial range with a
single slope, double-$\beta$ model with its slope fixed to the
best-fit value for the $>0.5r_{500}$ region (cf. Appendix~\ref{s:sx}).
As the slope is locally fitted, there should not be a significant bias
due to the beta slope steepening in our X-ray mass modeling. With
deeper observations, the thing that can possibly be improved in the
X-ray mass modeling to reduce the systematics is the slope of the
temperature profile at $r_{500}$. However, the Monte-Carlo error
estimation method accounts for this to some extent (also cf. Croston
et al. 2007).

\subsection{Scaled cooling time profiles}
\label{s:tc}

The cooling time is derived by the total energy of the gas divided by
the energy loss rate
\begin{equation}
t_{\rm c} = \frac{2.5 n_{\rm g}T}{n_{\rm e}^2\widetilde{\Lambda}}
\label{e:tcool}
\end{equation}
where $\widetilde{\Lambda}$, $n_{\rm g}$, $n_{\rm e}$ and $T$ are the
radiative cooling function, gas number density, electron number
density and temperature, respectively. We compute the upper limit of
the age of the cluster as an integral from the cluster redshift $z$ up
to $z=100$. Cooling regions are those showing cooling time less than
the upper limit of the cluster age. The boundary radius of such a
region is called the cooling radius. The cooling radius is zero when
the cooling time in the cluster center is larger than the upper limit
of the cluster age. The cooling time at the resolved inner most radii
of the surface brightness profiles and cooling radii are given in
Table~\ref{t:catalog2}.

\begin{figure*}
\begin{center}
\includegraphics[width=6.8cm,angle=270]{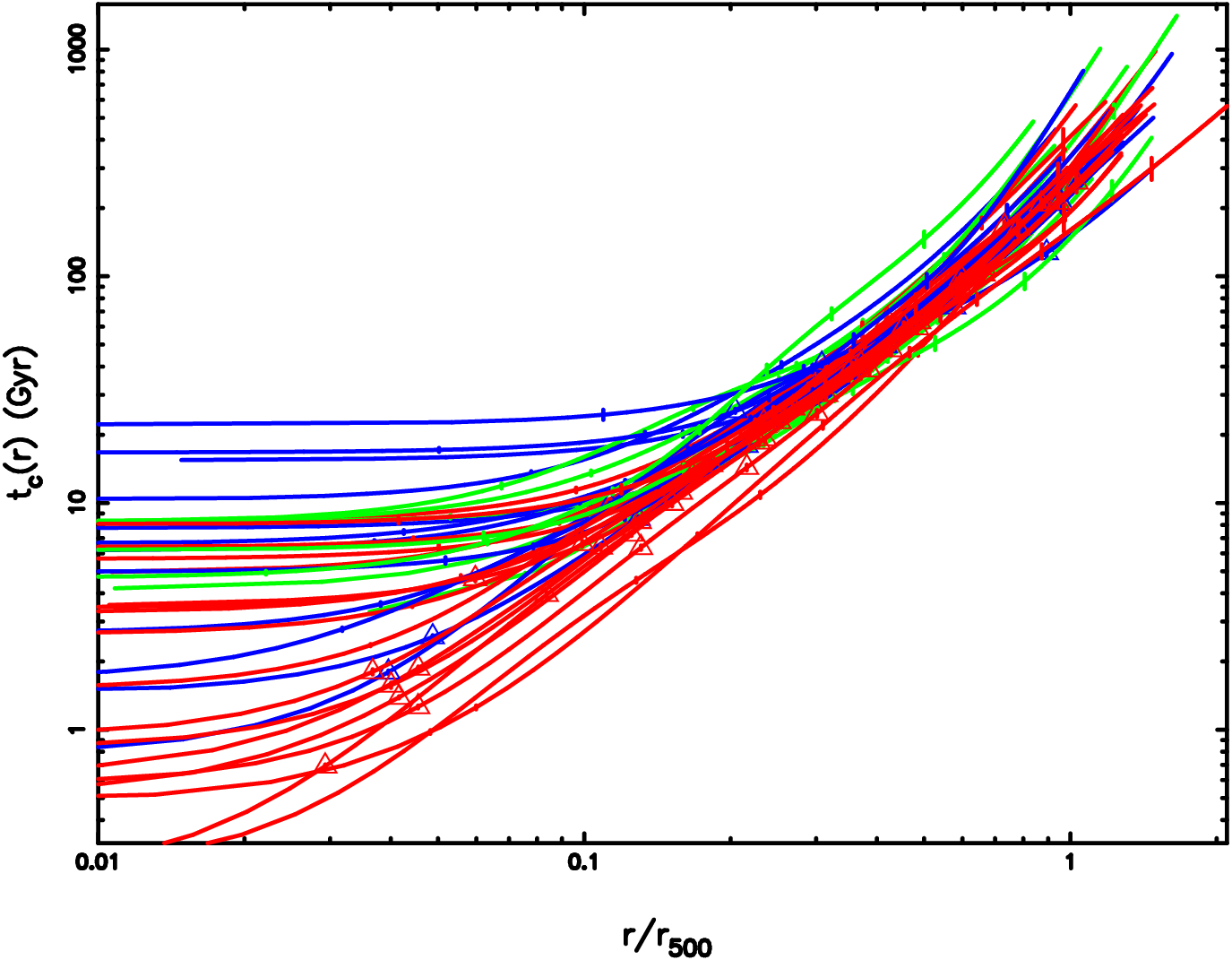}
\includegraphics[width=6.8cm,angle=270]{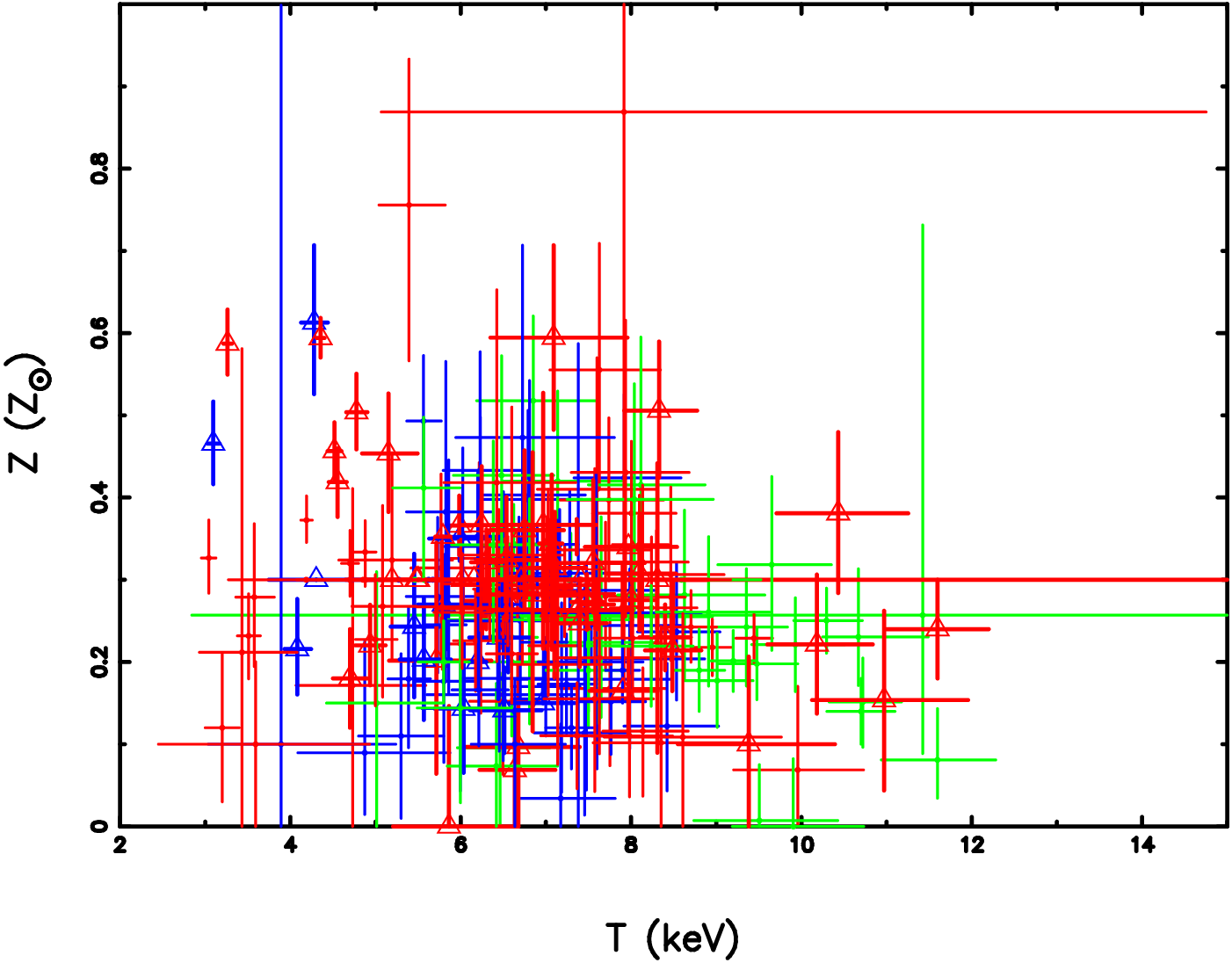}
\end{center}
\caption{Cooling time profiles. The CCCs are marked by triangles. 
The colors have the same meaning as those in Fig.~\ref{f:my_1}.
\label{f:tc} }
\end{figure*}

In Fig.~\ref{f:tc}, we show the cooling time profiles with their radii scaled
to $r_{500}$. The CCCs show larger cooling radii in units of $r_{500}$, and
show much shorter central cooling times than the non-CCCs. The scaled cooling
time profiles show a structurally similar behavior above $0.2 r_{500}$ for the
sample. The best-fit power law above $0.2 r_{500}$ gives $t_{\rm c}(r) \propto
r^{1.58\pm 0.01}$ for the whole sample. The cooling time profiles within
0.2$r_{500}$ show large scatter spreading over 2 magnitudes without showing
pronounced CCC and non-CCC bimodal distribution. Similar studies for nearby
clusters based on \emph{Chandra} data can be found in Sanderson et al. (2006)
and Dunn \& Fabian (2006).

The cooling time is calculated towards the cluster center to the inner most
bin for which the surface brightness profile is resolved. There can be large
uncertainties in the cooling time at the inner most bins where the temperature
measurements are not resolved. The spatial resolution is important to
calculate the proper cooling time in the cluster center. When the cluster core
is not well resolved, the value calculated from the measured temperature and
electron number density gives the upper limit of the cooling time, especially
for the CCCs, which show sharply peaked surface brightness profiles.

\subsection{Scaled entropy profiles}
\label{s:en}

The entropy is the key to the understanding of the thermodynamical
history since it results from shock heating of the gas during cluster
formation. The observed entropy is generally defined as $S=T n_{\rm
  e}^{-2/3}$ for clusters (e.g. Ponman et al. 1999). Radiative cooling
can raise the entropy of the ICM (e.g Pearce et al. 2000) or produce a
deficit below the $S$--$T$ relation from the self-similar prediction
(e.g. Lloyd-Davies et al. 2000).

According to the standard self-similar model the entropy scales as $S
\propto T$ (e.g. Ponman et al. 1999). We investigated the
entropy--temperature relation ($S$--$T$) using $S_{\rm 0.2 r_{500}}$
as the central entropy and $T_{0.2-0.5r_{500}}$ as the cluster
temperature. The CCCs show slightly lower central entropies compared
to the $S$--$T$ scaling law.

\begin{figure*}
\begin{center}
\includegraphics[width=4.5cm,angle=270]{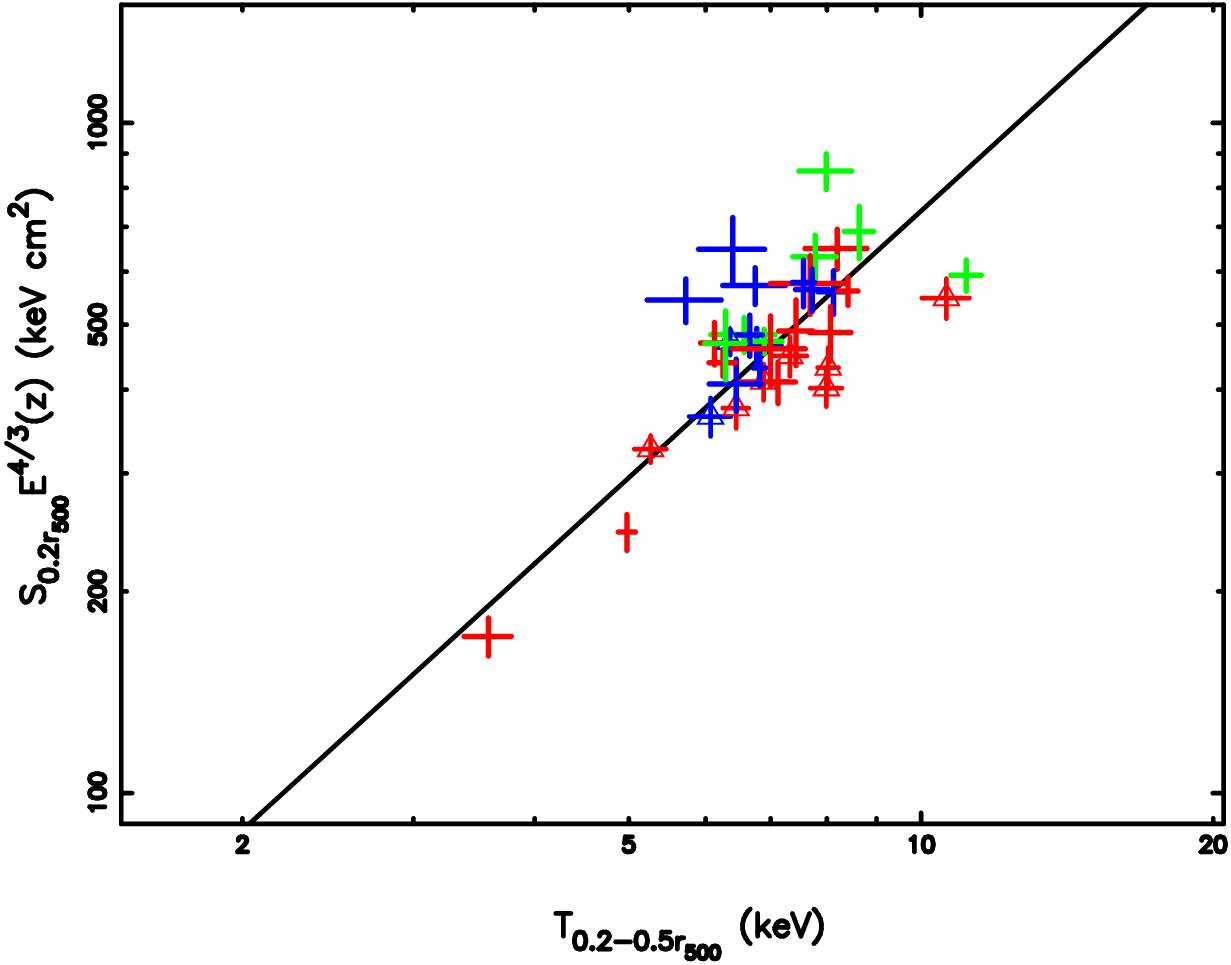}
\includegraphics[width=4.5cm,angle=270]{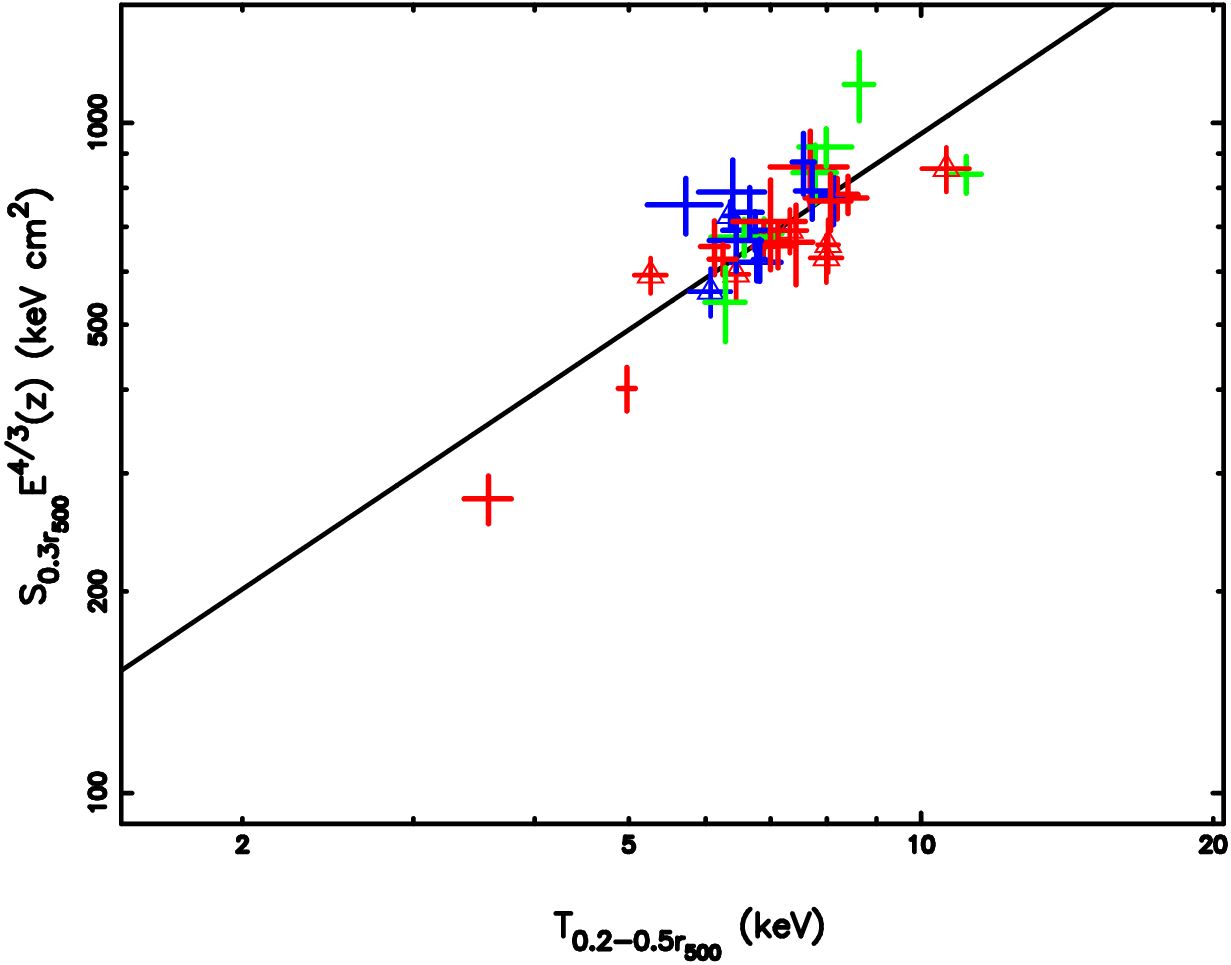}
\includegraphics[width=4.5cm,angle=270]{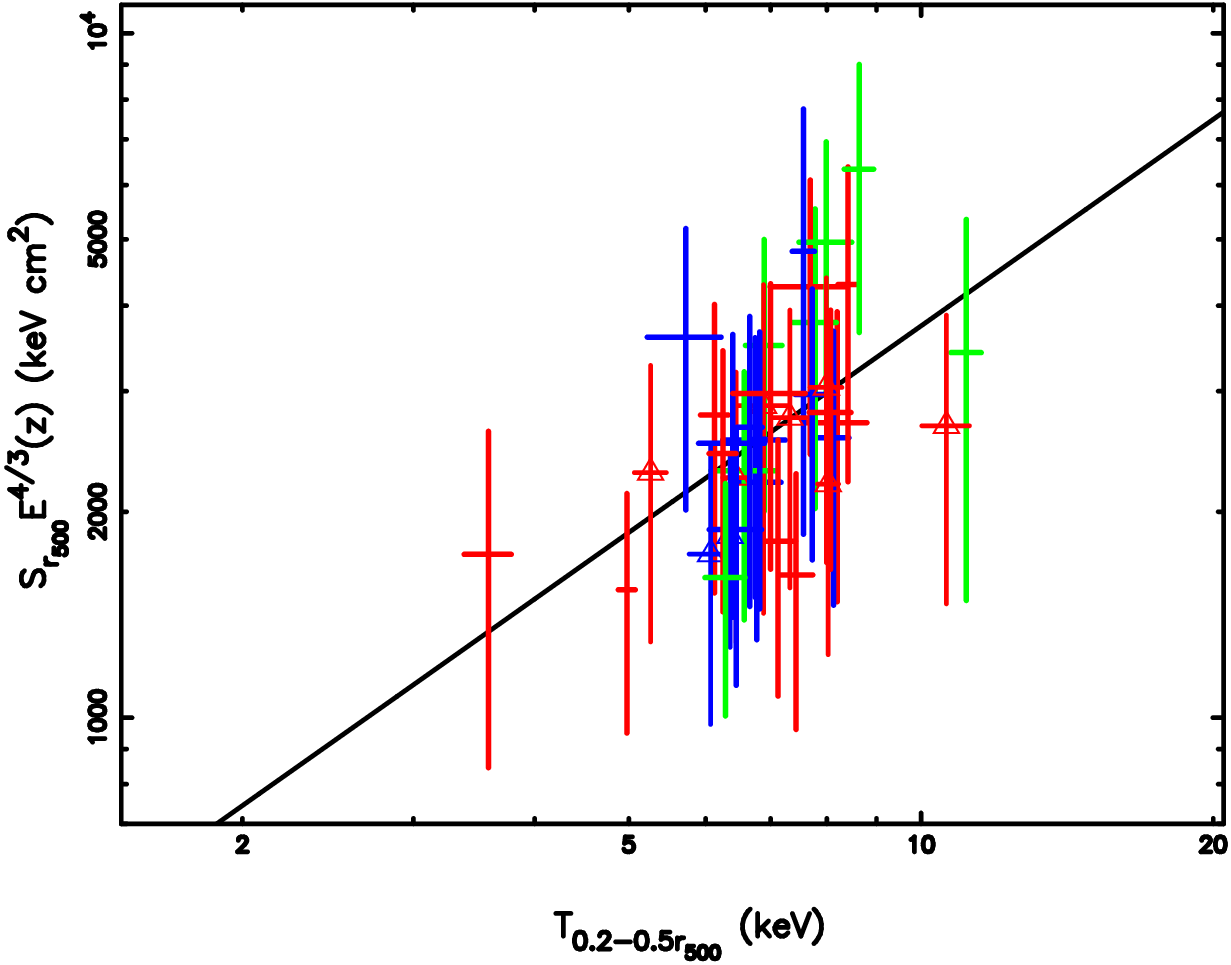}

\includegraphics[width=4.5cm,angle=270]{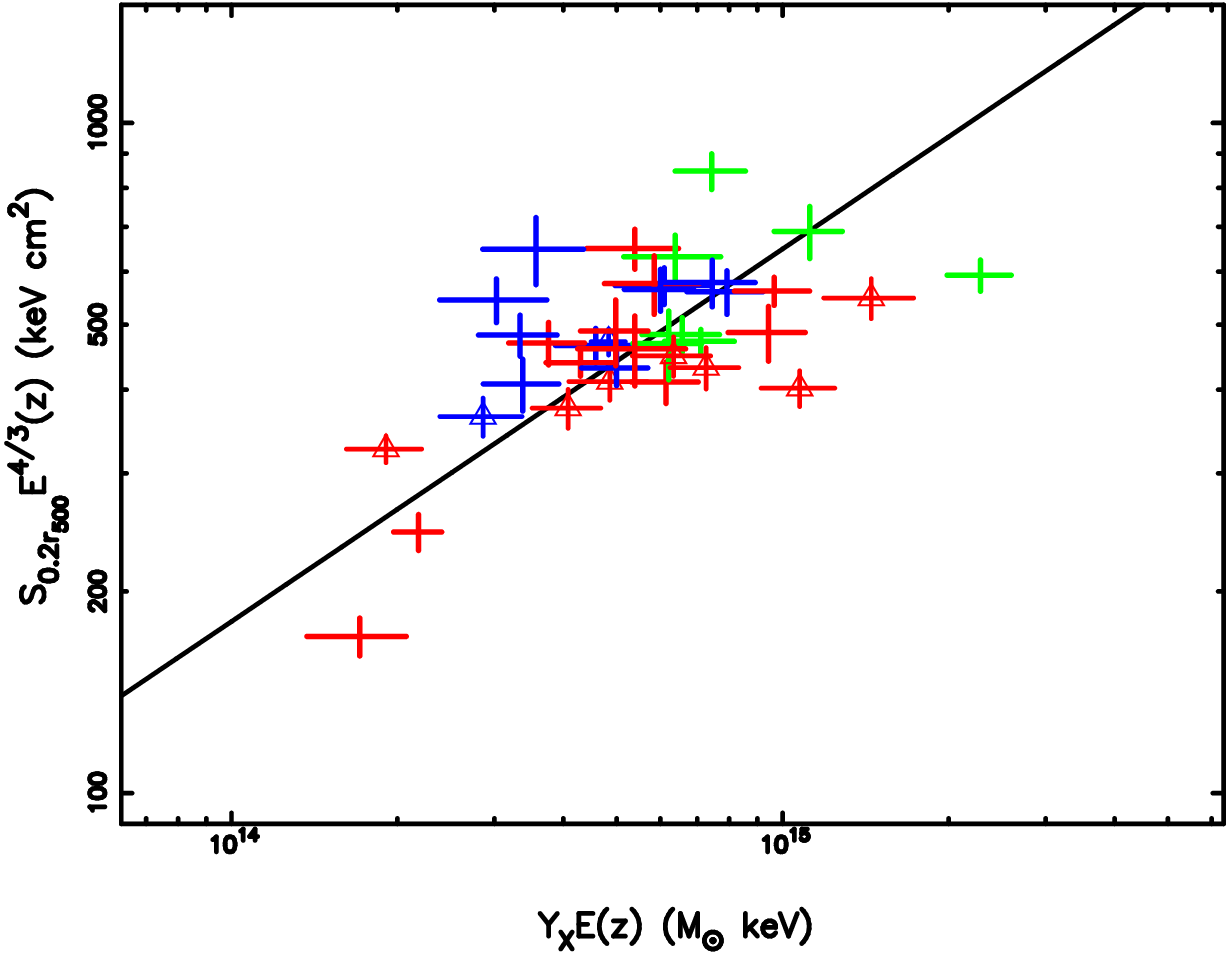}
\includegraphics[width=4.5cm,angle=270]{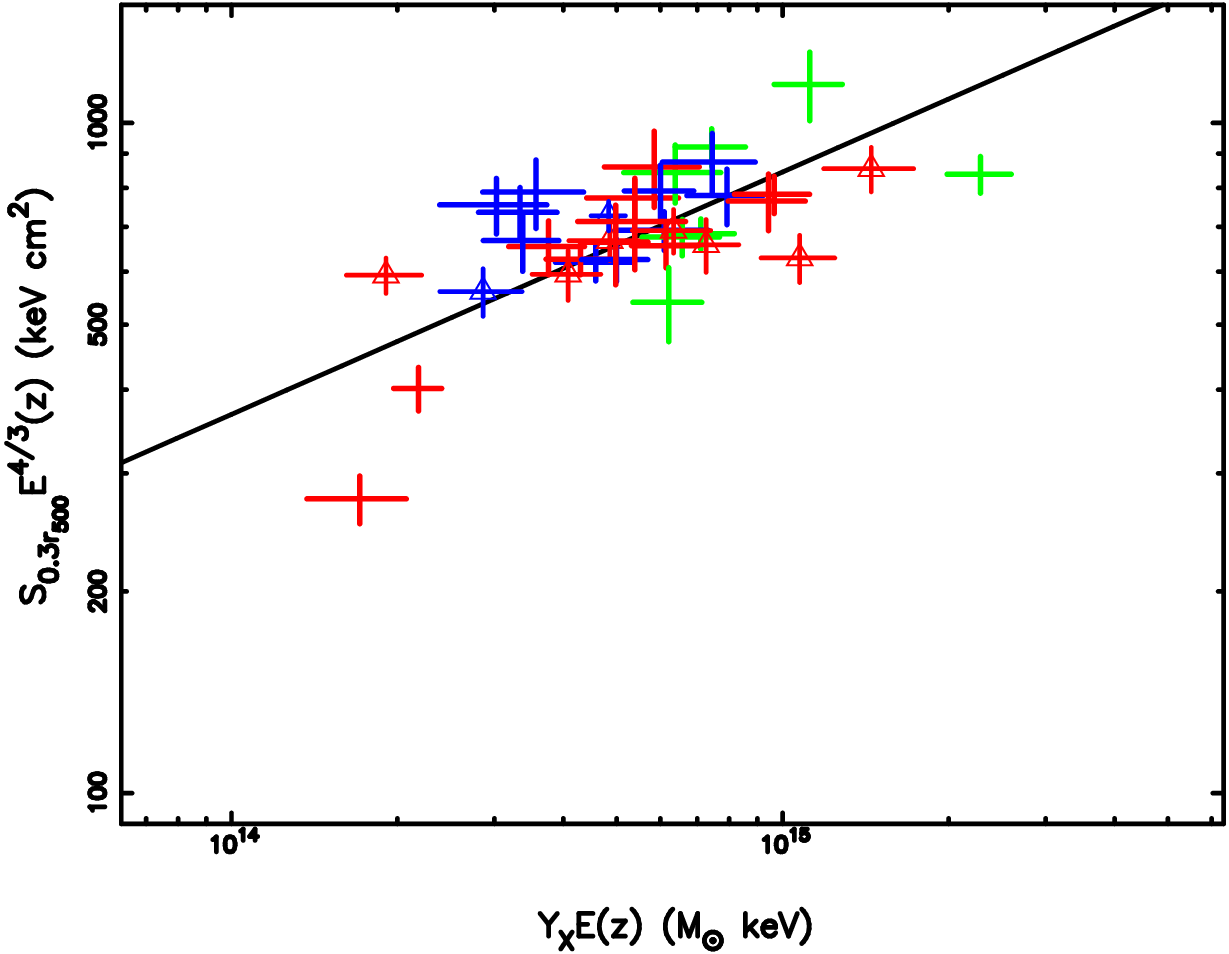}
\includegraphics[width=4.5cm,angle=270]{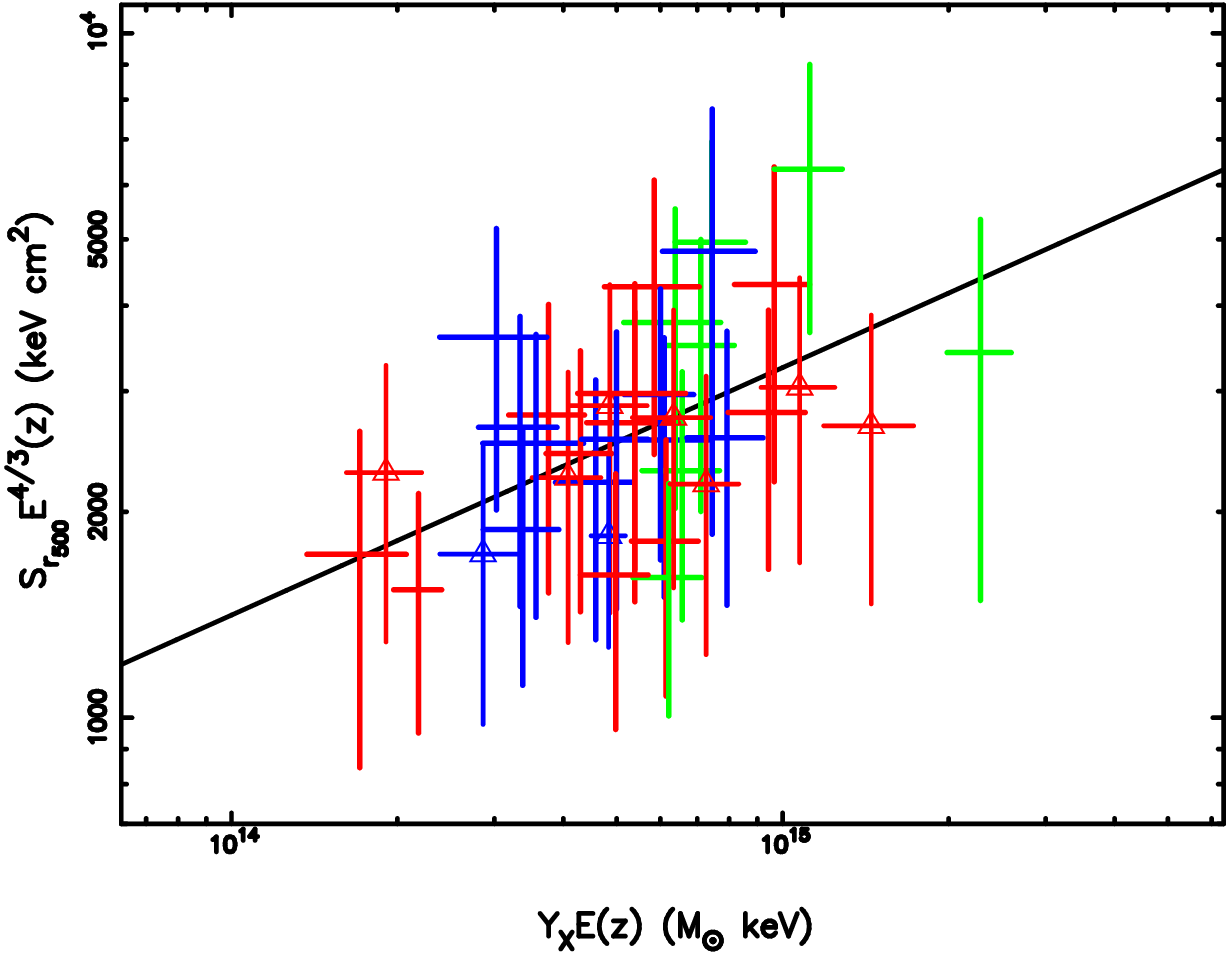}

\includegraphics[width=4.5cm,angle=270]{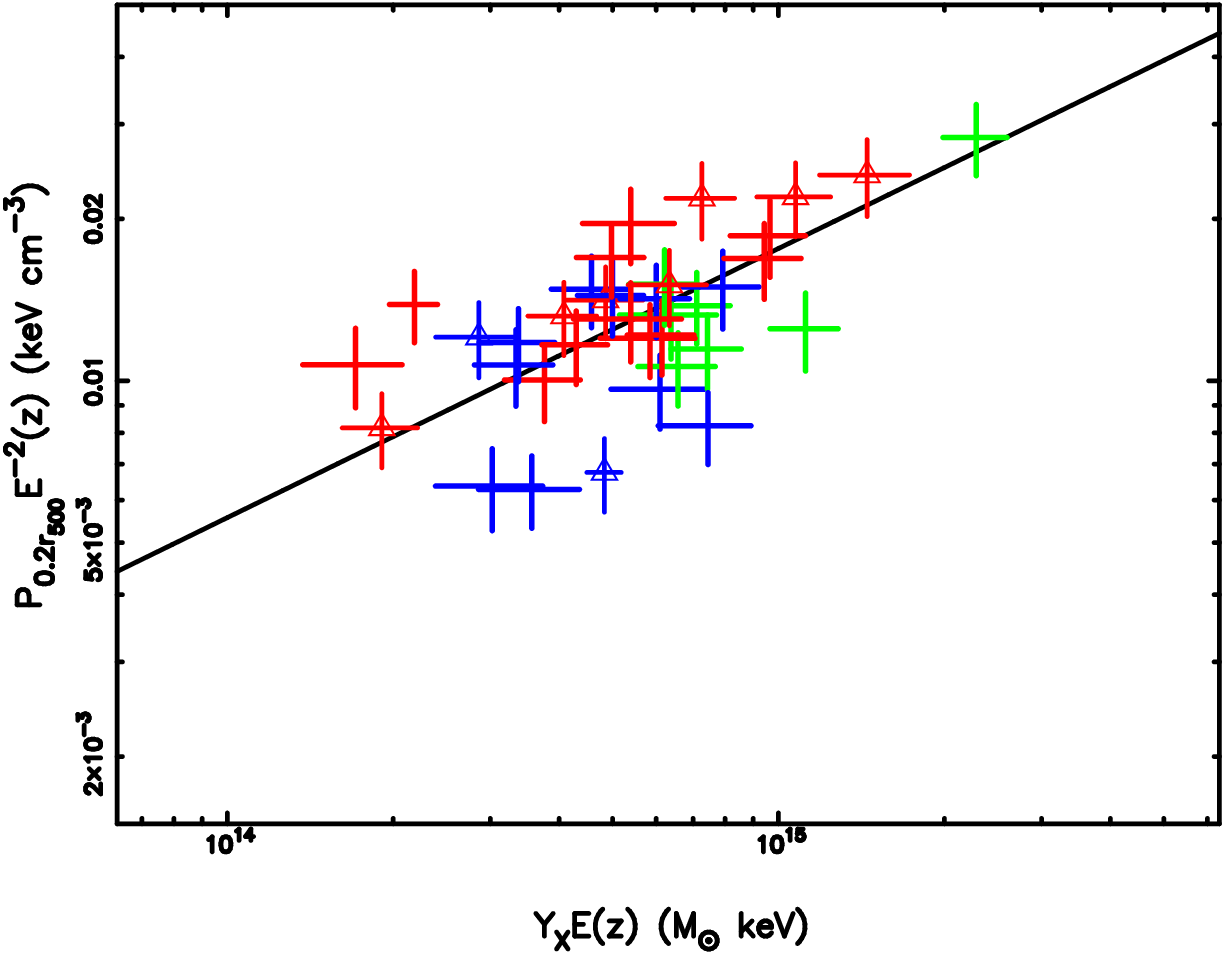}
\includegraphics[width=4.5cm,angle=270]{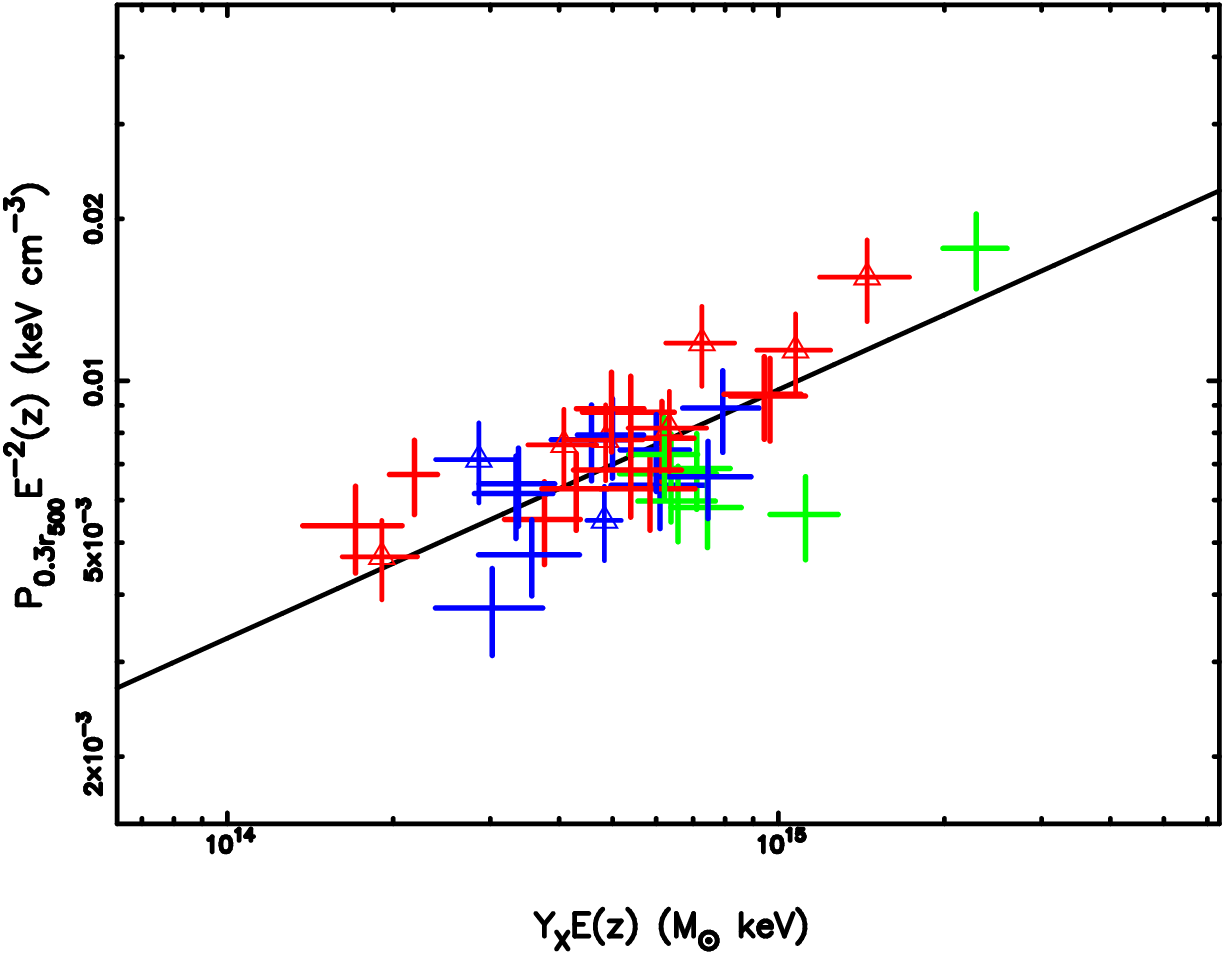}
\includegraphics[width=4.5cm,angle=270]{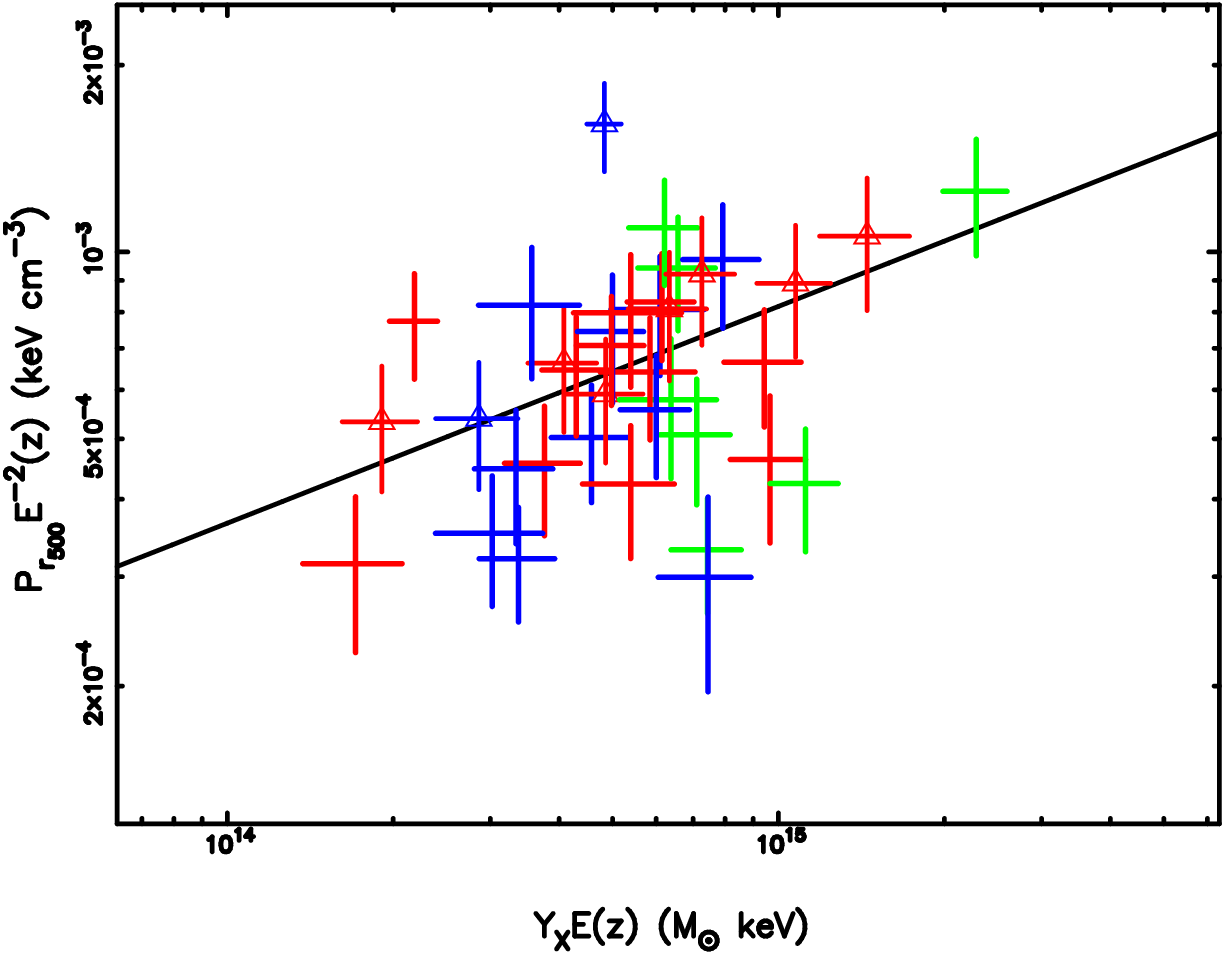}

\end{center}
\caption{Entropy (upper and middle panels) and pressure (lower panels) 
versus temperature (upper panels) and X-ray analog of the SZ flux
(middle and lower panels). The lines denote the best fit for our
sample. The CCCs are marked by triangles. The colors have the same
meaning as those in Fig.~\ref{f:my_1}.
\label{f:cores}
}
\end{figure*}

Ponman et al. (2003) suggested to scale the entropy as $S \propto T^{0.65}$
based on observations for nearby clusters. We show the $S$--$T$ relations in
Fig.~\ref{f:cores} and Table~\ref{t:psty}. The $S_{\rm 0.2 r_{500}}$--$T$
relation ($S_{0.2 r_{500}} \propto T^{1.32\pm 0.16}$) for our sample agrees
better with the standard self-similar model $S \propto T$. The CCCs ($S_{\rm
  0.2 r_{500}} \propto T^{0.63 \pm 0.20}$) show better agreement with the
empirically determined scaling (Ponman et al. 2003) $S \propto T^{0.65}$. The
$S$--$T$ relation becomes less steep using the entropies at $0.3 r_{500}$
(Fig.~\ref{f:cores}). Pratt et al. (2006) found that further away from the
cluster center, the scatter in the entropy of the $S$--$T$ relation decreases.
This is not evident for our sample due to the limited radial temperature bins.
There is no noticeable evolution in the $S$--$T$ relation (e.g. using $S_{\rm
  0.3 r_{500}}$) comparing our sample to the nearby relaxed cluster sample in
Pratt et al. (2006) after the redshift evolution correction due to LSS growth
has been applied.

As $S \propto T$ and $M_{\rm gas} \propto T^{1.5}$ according to the standard
self-similarity, one derives $S \propto (T M_{gas})^{0.5} \propto Y_{\rm
  X}^{0.5}$. In Fig.~\ref{f:cores}, we present both the entropy and the
pressure at $0.2 r_{500}$, $0.3 r_{500}$ and $r_{500}$ versus X-ray analog of
the SZ flux. The best-fit power law is given in Table~\ref{t:psty}. The
CCCs show large scatter in $S$--$T$, $S$--$Y_{\rm X}$ and $P$--$Y_{\rm X}$
relations. The $P$--$Y_{\rm X}$ relations show least segregation between CCCs
and non-CCCs, especially at $r_{500}$.

\begin{figure*}
\begin{center}
\includegraphics[width=6.8cm,angle=270]{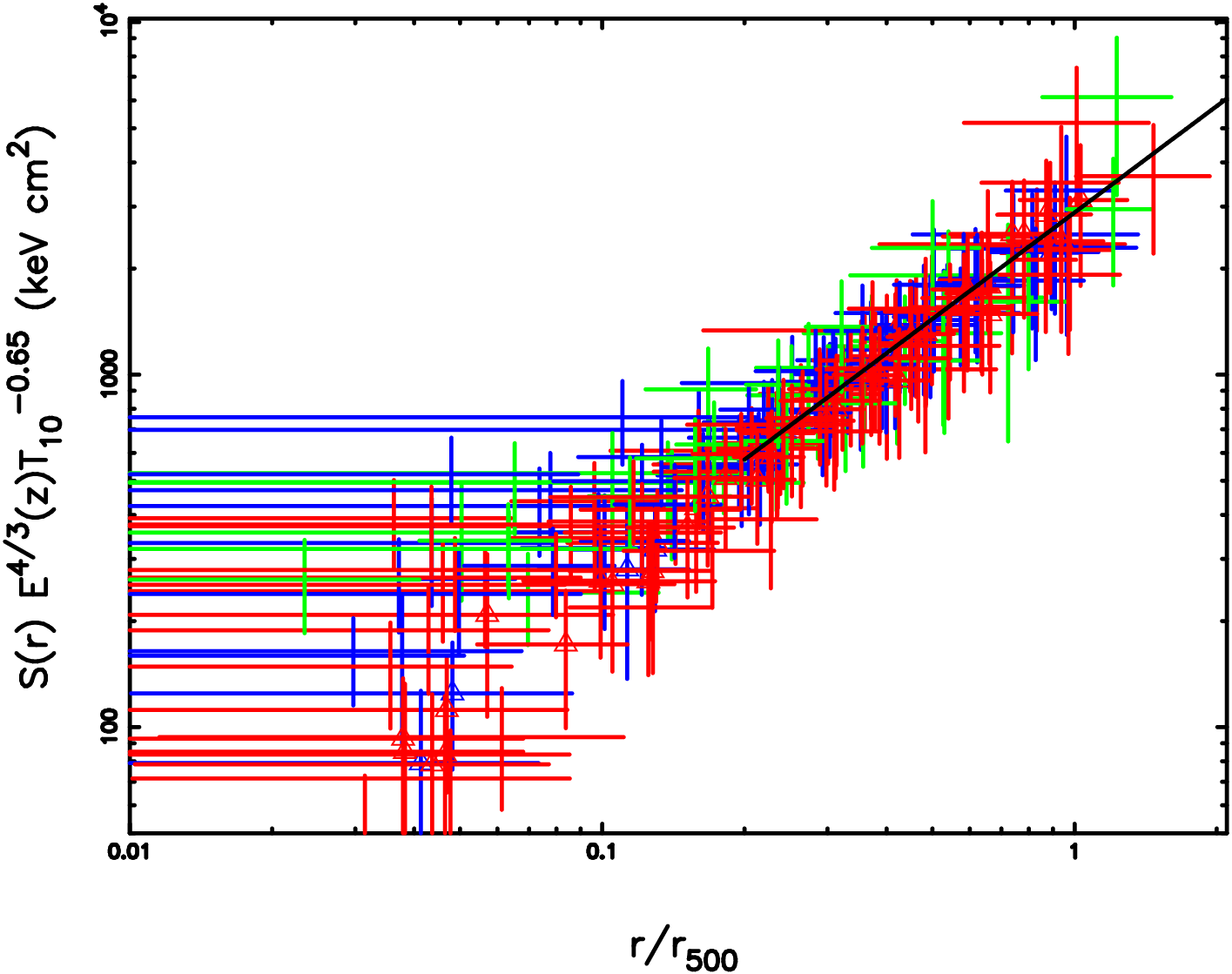}
\includegraphics[width=6.8cm,angle=270]{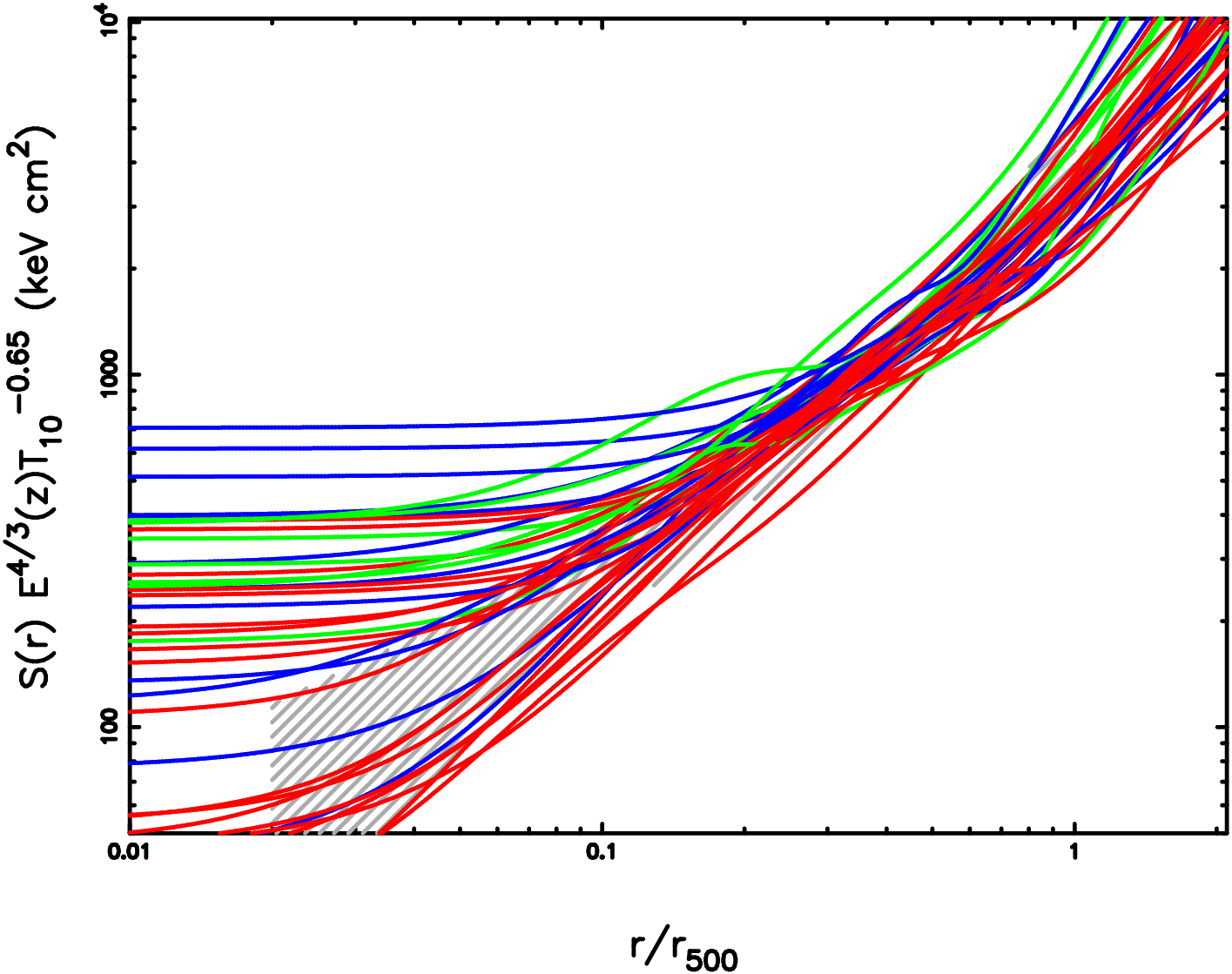}
\end{center}
\caption{{\it Left:} Scaled entropy profiles for the sample and the
combined best fit. {\it Right:} Scaled entropy profile fits for the
sample compared to the sample in Pratt et al. (2006, gray,
hatched). The CCCs are marked by triangles in the left panel. The
colors have the same meaning as those in Fig.~\ref{f:my_1}. $T_{10}$
denotes $T_{0.2-0.5r_{500}}/10$~keV.
\label{f:en} }
\end{figure*}

We scaled the radial entropy profiles using the empirically scaling (Ponman et
al. 2003) $S\propto T^{0.65}$ and $r_{500}$. As shown in Fig.~\ref{f:en}, the
scaled entropy profiles for the sample are structurally similar above $0.2
r_{500}$ and show least scatter around 0.2--0.5$r_{500}$. After the redshift
correction, we obtained an agreement between the scaled entropy profiles for
this sample, the Birmingham-CfA sample (Ponman et al. 2003), and the cluster
sample in Pratt et al. (2006). There is no evident evolution in the scaled
entropy profiles after the redshift evolution correction due to LSS growth has
been applied.

The scaled entropy profiles for our sample give the best fit
$S(r)\propto r^{1.00\pm 0.03}$ above $0.2r_{\rm 500}$. A similar power
law was also found as $S\propto r^{0.97}$ by Ettori et al. (2002), and
$S\propto r^{0.95}$ by Piffaretti et al. (2005), respectively. The
spherical accretion shock model predicts $S\propto r^{1.1}$
(e.g. Tozzi \& Norman 2001; Kay 2004). The combined fit of the entropy
profiles for the 9 CCCs in our sample gives $S\propto r^{1.05\pm
0.04}$, close to the prediction of the spherical accretion shock model
and consistent with the relaxed nearby clusters in Pratt et al. (2006)
giving $S\propto r^{1.08}$. 

\subsection{Scaled total mass and gas mass profiles}

\begin{figure*}
\begin{center}
\includegraphics[width=6.8cm,angle=270]{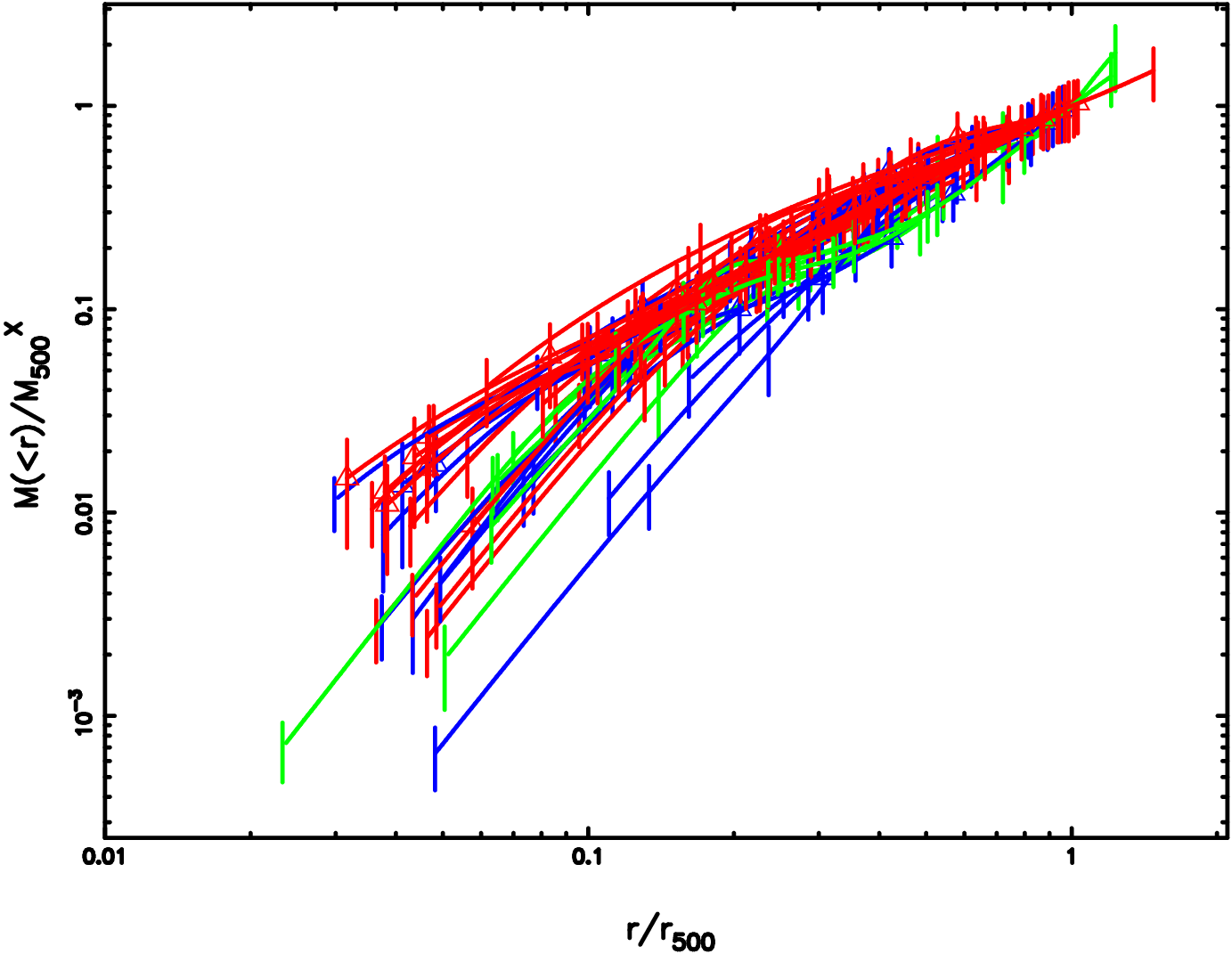}
\includegraphics[width=6.8cm,angle=270]{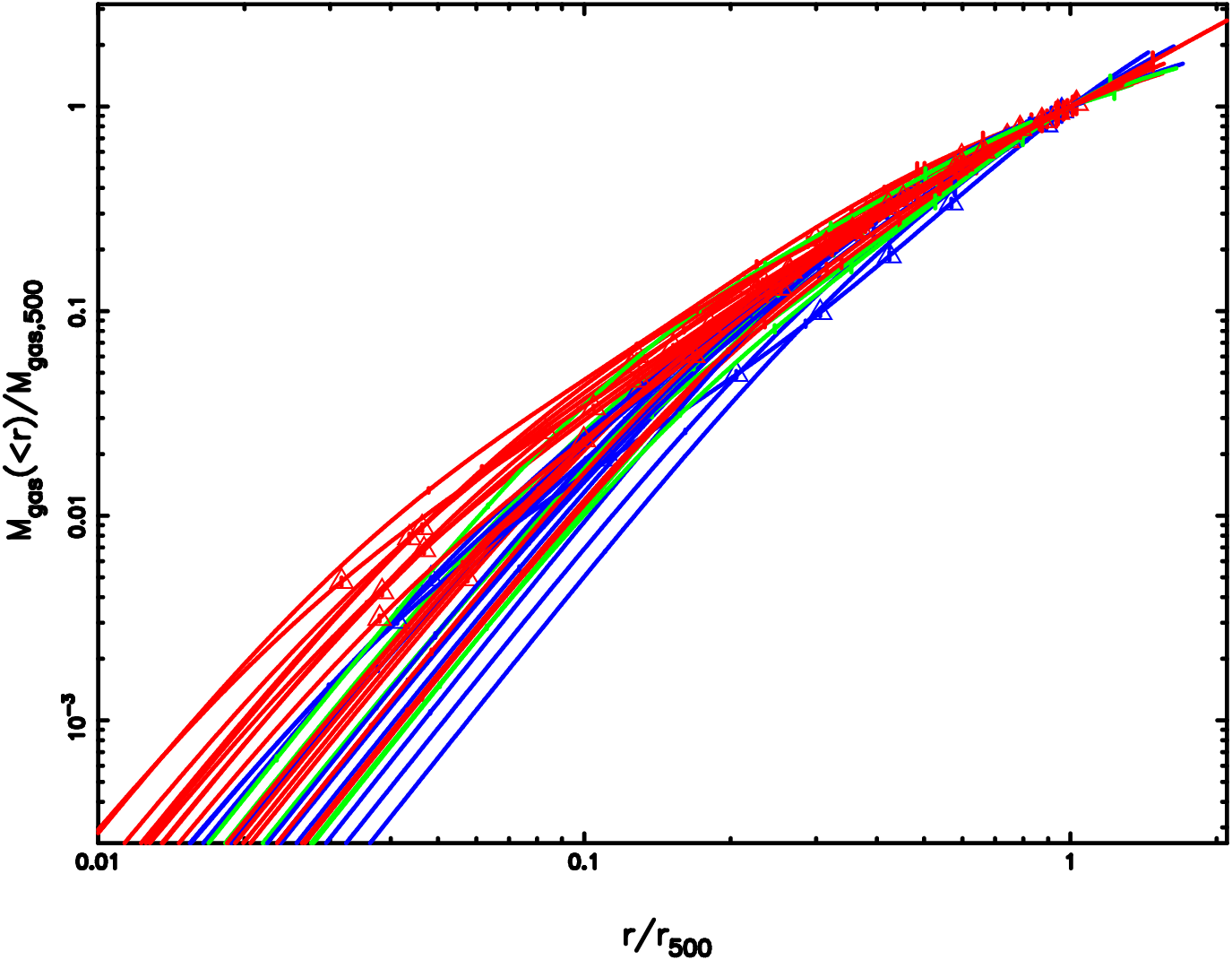}
\end{center}
\caption{Scaled total mass profiles (left) and gas mass profiles
(right). The CCCs are marked by triangles. The colors have the same
meaning as those in Fig.~\ref{f:my_1}. The error bars (a few per
cent) of the gas mass profiles are too small to be seen in the right
panel.
\label{f:scaleym} }
\end{figure*}

The mass profiles were scaled with respect to $M_{\rm 500}^{\rm X}$
and $r_{\rm 500}$, respectively (Fig.~\ref{f:scaleym}). We found least
scatter at radii above $0.2 r_{\rm 500}$. In the inner regions ($<0.2
r_{\rm 500}$), the mass profiles show large scatter spreading one
magnitude. This might be due to the complex dynamics near the cluster
center. As found in the strong lensing studies based on high quality
HST data (Smith et al. 2005), the cluster cores can be very
complicated with multi-clump DM halos.

Similar to the mass profiles, the scaled gas mass profiles appear
structurally similar at radii above $0.2 r_{\rm 500}$ but show 
smaller scatter (a few per cent) than the scaled mass profiles.

\subsection{Summary of the structural similarity}

In summary, we observed the structural similarity of the scaled profiles of
the X-ray properties outside the cluster cores. It supports simulations that
major mergers are not common recently (e.g. Cohn \& White 2005), while the
structurally similar behavior shows an empirical structural similarity, which
is in agreement with simulations (e.g. Poole et al. 07) that the signatures of
merger impact are retained by the cluster for a substantial fraction of cosmic
time.

There is large scatter in the cluster cores for the scaled profiles
of the X-ray properties, but no pronounced CCC and non-CCC bimodal
distributions. The peculiarities in the cluster cores for the
individual clusters may explain the large scatter in the scaled
profiles within cluster cores where the ICM does not always trace DM.

As found in Majerowicz et al. (2002), the cluster total mass for the
pronounced CCC Abell~1835 is lower by less than 10 per cent than the mass
derived ignoring the temperature drop in the cluster center. Therefore
the cluster total masses on large radial scales are relatively
insensitive to the cluster cores. The scaling relations can be less
sensitive to the peculiarities in the cluster cores by (1) measuring
the global temperatures excluding the cluster cores, (2) and
calculating the luminosities corrected for the cluster cores.

\clearpage

\begin{table*} { \begin{center} \footnotesize
  {\renewcommand{\arraystretch}{1.3} \caption[]{ Deduced properties of
  37 LoCuSS galaxy clusters. Column~(1): cluster name; Col.~(2):
  electron number density of the inner most radial bin; Col.~(3):
  entropy at $0.2 r_{500}$; Col.~(4): inner most radial bin of the
  surface brightness profile; Col.~(5): cooling time measured at $
  r_{\rm cen}$; Col.~(6): cooling radius; Col.~(7): $r_{500}$;
  Cols.~(8--10): gas mass, total mass and gas mass fraction at
  $r_{500}$, Col.~(11): cluster morphology using the cluster
  classification in Jones \& Forman (1992), 0 denotes ``single'', 1
  denotes ``primary with small secondary'', 2 denotes ``elliptical'',
  3 denotes ``off-center'' and 4 denotes ``complex''.}
  \label{t:catalog2}} \begin{tabular}{lrrrrrrrrrl}
\hline
\hline
Name & $n_{e0}$        & $S_{0.2 r_{500}} $    & $ r_{\rm cen}$  & $t_{\rm c}
$ & $r_{\rm cool}$ & $r_{500}$ & $M_{\rm gas, 500}$ & $M_{500}^{\rm X}$
& $f_{\rm gas, 500}$ & Mor.\\
\hline
\multicolumn{2}{r}{$10^{-3}$~cm$^{-3}$} & keV~cm$^{2}$ & kpc &
Gyr & $r_{500}$ &  Mpc & $10^{14} M_{\odot}$ & $10^{14} M_{\odot}$
& & \\
\hline
RXCJ0043.4-2037    & $    5.8  \pm     0.4 $ & $  391  \pm    23 $ & $     8.76 $ & $      7.8  \pm       0.3 $ & $     0.11 $ & $     1.08 $ & $     0.58  \pm      0.06 $ & $     4.83  \pm      1.43 $ & $    0.121  \pm     0.066 $ &   2                \\
RXCJ0232.2-4420    & $   14.1  \pm     0.1 $ & $  402  \pm    16 $ & $    47.14 $ & $      3.3  \pm       0.2 $ & $     0.14 $ & $     1.30 $ & $     0.89  \pm      0.09 $ & $     8.43  \pm      2.48 $ & $    0.106  \pm     0.057 $ &   1                \\
RXCJ0307.0-2840    & $   10.2  \pm     0.3 $ & $  359  \pm    22 $ & $    12.01 $ & $      3.6  \pm       0.3 $ & $     0.15 $ & $     1.14 $ & $     0.62  \pm      0.06 $ & $     5.47  \pm      2.01 $ & $    0.113  \pm     0.083 $ & $ 0    ^{\rm \;c}$ \\
RXCJ0516.7-5430    & $    2.9  \pm     0.2 $ & $  474  \pm    29 $ & $    17.59 $ & $     15.5  \pm       0.4 $ & $     0.00 $ & $     1.19 $ & $     0.78  \pm      0.09 $ & $     6.40  \pm      1.83 $ & $    0.121  \pm     0.068 $ &   2                \\
RXCJ0528.9-3927    & $   10.6  \pm     0.4 $ & $  423  \pm    25 $ & $    12.87 $ & $      4.2  \pm       0.3 $ & $     0.12 $ & $     1.19 $ & $     0.87  \pm      0.07 $ & $     6.40  \pm      1.87 $ & $    0.135  \pm     0.071 $ &   3                \\
RXCJ0532.9-3701    & $   13.2  \pm     0.5 $ & $  555  \pm    38 $ & $     8.38 $ & $      3.3  \pm       0.3 $ & $     0.12 $ & $     1.13 $ & $     0.57  \pm      0.07 $ & $     5.40  \pm      1.59 $ & $    0.106  \pm     0.062 $ &   0                \\
RXCJ0547.6-3152    & $    5.3  \pm     0.2 $ & $  433  \pm    31 $ & $     2.32 $ & $      8.3  \pm       0.3 $ & $     0.10 $ & $     1.21 $ & $     0.57  \pm      0.07 $ & $     5.84  \pm      1.73 $ & $    0.098  \pm     0.058 $ &   0                \\
RXCJ0645.4-5413    & $   14.9  \pm     1.0 $ & $  513  \pm    38 $ & $     1.78 $ & $      2.7  \pm       0.3 $ & $     0.12 $ & $     1.26 $ & $     0.90  \pm      0.11 $ & $     6.60  \pm      1.95 $ & $    0.136  \pm     0.080 $ &   2                \\
RXCJ0658.5-5556    & $    6.1  \pm     0.2 $ & $  495  \pm    26 $ & $     8.84 $ & $      8.2  \pm       0.2 $ & $     0.11 $ & $     1.42 $ & $     1.76  \pm      0.17 $ & $    10.99  \pm      5.31 $ & $    0.160  \pm     0.178 $ &   3                \\
RXCJ0945.4-0839    & $    3.8  \pm     0.2 $ & $  500  \pm    37 $ & $     3.37 $ & $     10.3  \pm       0.2 $ & $     0.05 $ & $     1.27 $ & $     0.49  \pm      0.07 $ & $     6.73  \pm      1.95 $ & $    0.073  \pm     0.043 $ &   2                \\
RXCJ0958.3-1103    & $   21.6  \pm     0.9 $ & $  337  \pm    22 $ & $     2.56 $ & $      1.4  \pm       0.3 $ & $     0.16 $ & $     0.99 $ & $     0.43  \pm      0.05 $ & $     3.29  \pm      0.97 $ & $    0.132  \pm     0.077 $ & $ 2    ^{\rm \;c}$ \\
RXCJ2129.6+0005    & $   50.4  \pm     1.4 $ & $  332  \pm    22 $ & $     2.36 $ & $      0.6  \pm       0.3 $ & $     0.16 $ & $     1.06 $ & $     0.56  \pm      0.06 $ & $     4.33  \pm      1.27 $ & $    0.130  \pm     0.073 $ & $ 0    ^{\rm \;c}$ \\
RXCJ2218.6-3853    & $    6.2  \pm     0.3 $ & $  450  \pm    32 $ & $     2.22 $ & $      6.1  \pm       0.3 $ & $     0.12 $ & $     1.14 $ & $     0.47  \pm      0.06 $ & $     4.83  \pm      1.42 $ & $    0.097  \pm     0.058 $ &   2                \\
RXCJ2234.5-3744    & $    6.0  \pm     0.3 $ & $  580  \pm    45 $ & $     3.33 $ & $      8.2  \pm       0.3 $ & $     0.08 $ & $     1.35 $ & $     0.76  \pm      0.11 $ & $     8.11  \pm      2.36 $ & $    0.094  \pm     0.057 $ &   1                \\
RXCJ2308.3-0211    & $    9.2  \pm     0.3 $ & $  486  \pm    49 $ & $    13.27 $ & $      5.0  \pm       0.3 $ & $     0.11 $ & $     1.24 $ & $     0.65  \pm      0.07 $ & $     7.42  \pm      2.28 $ & $    0.088  \pm     0.051 $ &   0                \\
RXCJ2337.6+0016    & $    5.7  \pm     0.3 $ & $  710  \pm    43 $ & $    12.59 $ & $      8.2  \pm       0.3 $ & $     0.05 $ & $     1.43 $ & $     0.81  \pm      0.07 $ & $    10.95  \pm      3.22 $ & $    0.074  \pm     0.039 $ &   3                \\
Abell68            & $    6.4  \pm     0.3 $ & $  487  \pm    35 $ & $     3.55 $ & $      6.6  \pm       0.2 $ & $     0.10 $ & $     1.21 $ & $     0.68  \pm      0.07 $ & $     6.51  \pm      1.93 $ & $    0.105  \pm     0.060 $ &   2                \\
Abell115           & $   35.4  \pm     1.4 $ & $  430  \pm    19 $ & $     2.06 $ & $      0.8  \pm       0.2 $ & $     0.12 $ & $     1.07 $ & $     0.69  \pm      0.04 $ & $     4.23  \pm      1.14 $ & $    0.163  \pm     0.072 $ & $ 4    ^{\rm \;c}$ \\
Abell209           & $    5.7  \pm     0.2 $ & $  369  \pm    26 $ & $     3.74 $ & $      6.3  \pm       0.3 $ & $     0.14 $ & $     1.15 $ & $     0.78  \pm      0.08 $ & $     5.33  \pm      1.71 $ & $    0.146  \pm     0.089 $ &   0                \\
Abell267           & $    7.4  \pm     0.4 $ & $  364  \pm    32 $ & $     3.29 $ & $      4.9  \pm       0.4 $ & $     0.15 $ & $     1.06 $ & $     0.47  \pm      0.04 $ & $     4.29  \pm      1.30 $ & $    0.109  \pm     0.060 $ &   2                \\
Abell383           & $   54.8  \pm     1.5 $ & $  302  \pm    14 $ & $     1.98 $ & $      0.5  \pm       0.2 $ & $     0.16 $ & $     0.98 $ & $     0.33  \pm      0.04 $ & $     3.17  \pm      0.94 $ & $    0.104  \pm     0.061 $ & $ 0    ^{\rm \;c}$ \\
Abell773           & $    7.4  \pm     0.4 $ & $  433  \pm    40 $ & $     3.15 $ & $      5.7  \pm       0.5 $ & $     0.14 $ & $     1.33 $ & $     1.05  \pm      0.12 $ & $     8.30  \pm      2.45 $ & $    0.126  \pm     0.070 $ &   0                \\
Abell781           & $    1.8  \pm     0.1 $ & $  536  \pm    61 $ & $     6.28 $ & $     22.1  \pm       1.1 $ & $     0.00 $ & $     1.05 $ & $     0.48  \pm      0.06 $ & $     4.49  \pm      1.33 $ & $    0.107  \pm     0.064 $ &   4                \\
Abell901           & $  113.6  \pm     1.7 $ & $  158  \pm    10 $ & $     1.78 $ & $      0.2  \pm       0.4 $ & $     0.24 $ & $     0.99 $ & $     0.44  \pm      0.07 $ & $     3.22  \pm      0.96 $ & $    0.137  \pm     0.085 $ &   0                \\
Abell963           & $   14.3  \pm     0.8 $ & $  393  \pm    18 $ & $     2.14 $ & $      2.6  \pm       0.2 $ & $     0.14 $ & $     1.14 $ & $     0.62  \pm      0.06 $ & $     5.19  \pm      1.52 $ & $    0.120  \pm     0.066 $ &   0                \\
Abell1413          & $   25.3  \pm     1.0 $ & $  400  \pm    23 $ & $     1.59 $ & $      1.7  \pm       0.3 $ & $     0.14 $ & $     1.18 $ & $     0.68  \pm      0.08 $ & $     5.38  \pm      1.57 $ & $    0.127  \pm     0.073 $ &   2                \\
Abell1689          & $   30.0  \pm     0.9 $ & $  505  \pm    24 $ & $     1.95 $ & $      1.5  \pm       0.2 $ & $     0.13 $ & $     1.44 $ & $     1.05  \pm      0.14 $ & $    10.26  \pm      2.96 $ & $    0.102  \pm     0.060 $ &   0                \\
Abell1758          & $    2.4  \pm     0.2 $ & $  487  \pm    39 $ & $     3.80 $ & $     16.7  \pm       0.3 $ & $     0.00 $ & $     1.43 $ & $     0.85  \pm      0.14 $ & $    11.15  \pm      3.37 $ & $    0.077  \pm     0.050 $ &   2                \\
Abell1763          & $    6.5  \pm     0.3 $ & $  402  \pm    47 $ & $     4.00 $ & $      6.1  \pm       0.5 $ & $     0.13 $ & $     1.12 $ & $     0.88  \pm      0.08 $ & $     4.96  \pm      1.46 $ & $    0.178  \pm     0.094 $ &   1                \\
Abell1835          & $   63.9  \pm     1.3 $ & $  351  \pm    22 $ & $     2.50 $ & $      0.6  \pm       0.3 $ & $     0.18 $ & $     1.30 $ & $     1.18  \pm      0.14 $ & $     8.01  \pm      2.32 $ & $    0.148  \pm     0.084 $ & $ 0    ^{\rm \;c}$ \\
Abell1914          & $   10.5  \pm     0.5 $ & $  629  \pm    56 $ & $     1.84 $ & $      4.7  \pm       0.3 $ & $     0.11 $ & $     1.71 $ & $     1.19  \pm      0.13 $ & $    16.76  \pm      4.87 $ & $    0.071  \pm     0.039 $ &   3                \\
Abell2204          & $  143.7  \pm     2.3 $ & $  401  \pm    28 $ & $     1.68 $ & $      0.2  \pm       0.3 $ & $     0.17 $ & $     1.17 $ & $     0.84  \pm      0.10 $ & $     5.31  \pm      1.52 $ & $    0.158  \pm     0.086 $ & $ 0    ^{\rm \;c}$ \\
Abell2218          & $    5.8  \pm     0.2 $ & $  444  \pm    50 $ & $     3.26 $ & $      8.0  \pm       0.5 $ & $     0.12 $ & $     1.07 $ & $     0.61  \pm      0.06 $ & $     4.18  \pm      1.27 $ & $    0.147  \pm     0.085 $ &   0                \\
Abell2261          & $   12.4  \pm     0.9 $ & $  408  \pm    49 $ & $     3.95 $ & $      3.4  \pm       0.5 $ & $     0.14 $ & $     1.19 $ & $     0.69  \pm      0.10 $ & $     6.02  \pm      1.73 $ & $    0.114  \pm     0.068 $ &   0                \\
Abell2390          & $   40.3  \pm     1.6 $ & $  484  \pm    33 $ & $     2.31 $ & $      0.9  \pm       0.3 $ & $     0.14 $ & $     1.29 $ & $     1.22  \pm      0.16 $ & $     7.67  \pm      2.28 $ & $    0.159  \pm     0.095 $ & $ 0    ^{\rm \;c}$ \\
Abell2667          & $   40.8  \pm     1.4 $ & $  398  \pm    27 $ & $     2.33 $ & $      0.8  \pm       0.3 $ & $     0.15 $ & $     1.19 $ & $     0.77  \pm      0.09 $ & $     6.02  \pm      1.74 $ & $    0.128  \pm     0.073 $ & $ 0    ^{\rm \;c}$ \\
Z7160              & $   65.1  \pm     1.5 $ & $  218  \pm    13 $ & $     2.53 $ & $      0.5  \pm       0.3 $ & $     0.21 $ & $     0.87 $ & $     0.39  \pm      0.03 $ & $     2.39  \pm      0.70 $ & $    0.161  \pm     0.084 $ &   0\\
\hline
Average             &  ---                   & ---           & ---      & ---                     & ---           & ---       & ---                     & ---     & $0.121 \pm 0.004$    & \\
\hline
\hline
  \end{tabular}
  \end{center}
\hspace*{0.3cm}{\footnotesize 

  1. The errors are calculated using the uncertainties of the
  electron number density and temperature measurements by Monte Carlo
  simulations. For each cluster, we simulated electron density and
  temperature profiles of a sample of 100 clusters using observed
  electron density and temperature profiles. The mass profiles and other
  properties of the 100 simulated clusters were calculated to derive the
  errors. In Zhang et al. (2006), the errors are simply
  propagated for the REFLEX-DXL clusters.

  2. $^{\rm \;c}$ in Col.~(11) denotes CCCs.} }
\end{table*}

\begin{table*} { \begin{center} \footnotesize
  {\renewcommand{\arraystretch}{1.3} \caption[]{ Power law,
  $Y=Y_0\;X^{\gamma}$, parameterized entropy--temperature,
  entropy--$Y_{\rm X}$ and pressure--$Y_{\rm X}$ relations.}
  \label{t:psty}}
\begin{tabular}{lllllllll}
\hline
\hline
$X$ & $Y$ & $Y_0$ & $\gamma$ & \multicolumn{4}{c}{Scatter (dex)} & Sample \\
\hline
    &     &       &           &  \multicolumn{2}{c}{mean} & \multicolumn{2}{c}{average} & \\
    &     &       &           &  $X$ & $Y$ &  $X$ & $Y$ & \\
\hline
$\frac{T_{0.2-0.5r_{500}}}{\rm keV}$ &
$\frac{S_{0.2r_{500}}}{\rm keV\;cm^{2}} \; E(z)^{4/3} $
   & $10^{1.55\pm0.13}$ & $1.32 \pm 0.16$ & 0.03 & 0.03 & 0.05 & 0.06 & whole sample \\
&  & $10^{1.45\pm0.15}$ & $1.45 \pm 0.17$ & 0.03 & 0.03 & 0.04 & 0.06 & non-CCC subsample \\
&  & $10^{2.09\pm0.17}$ & $0.63 \pm 0.20$ & 0.31 & 0.18 & 0.15 & 0.09 & CCC subsample \\
\hline
$\frac{T_{0.2-0.5r_{500}}}{\rm keV}$ &
$\frac{S_{0.3r_{500}}}{\rm keV\;cm^{2}} \; E(z)^{4/3} $
   & $10^{2.01\pm0.10}$ & $0.97 \pm 0.11$ & 0.03 & 0.03 & 0.05 & 0.05 & whole sample \\
&  & $10^{1.87\pm0.11}$ & $1.14 \pm 0.13$ & 0.03 & 0.03 & 0.04 & 0.04 & non-CCC subsample \\
&  & $10^{2.46\pm0.14}$ & $0.44 \pm 0.17$ & 0.13 & 0.16 & 0.13 & 0.06 & CCC subsample \\
\hline
$\frac{T_{0.2-0.5r_{500}}}{\rm keV}$ &
$\frac{S_{r_{500}}}{\rm keV\;cm^{2}} \; E(z)^{4/3} $
   & $10^{2.57\pm0.22}$ & $1.00 \pm 0.26$ & 0.21 & 0.23 & 0.10 & 0.22 & whole sample \\
&  & $10^{2.44\pm0.27}$ & $1.18 \pm 0.32$ & 0.04 & 0.23 & 0.09 & 0.22 & non-CCC subsample \\
&  & $10^{2.90\pm0.27}$ & $0.54 \pm 0.32$ & 0.38 & 0.41 & 0.23 & 0.22 & CCC subsample \\
\hline
$\frac{Y_{\rm X}}{\rm M_{\odot}keV}  \; E(z) $ &
$\frac{S_{0.2r_{500}}}{\rm keV\;cm^{2}} \; E(z)^{4/3} $
   & $10^{-5.53\pm1.27}$ & $0.56 \pm 0.09$ & 0.07 & 0.10 & 0.15 & 0.08 & whole sample \\
&  & $10^{-6.69\pm1.61}$ & $0.64 \pm 0.11$ & 0.18 & 0.03 & 0.12 & 0.08 & non-CCC subsample \\
&  & $10^{-5.63\pm0.90}$ & $0.22 \pm 0.06$ & 0.29 & 0.10 & 0.45 & 0.10 & CCC subsample \\
\hline
$\frac{Y_{\rm X}}{\rm M_{\odot}keV}  \; E(z) $ &
$\frac{S_{0.3r_{500}}}{\rm keV\;cm^{2}} \; E(z)^{4/3} $
   & $10^{-2.50\pm0.82}$ & $0.36 \pm 0.06$ & 0.09 & 0.03 & 0.17 & 0.06 & whole sample \\
&  & $10^{-3.80\pm1.07}$ & $0.45 \pm 0.07$ & 0.08 & 0.03 & 0.15 & 0.06 & non-CCC subsample \\
&  & $10^{ 6.25\pm0.79}$ & $1.50 \pm 0.05$ & 0.32 & 0.14 & 0.42 & 0.07 & CCC subsample \\
\hline
$\frac{Y_{\rm X}}{\rm M_{\odot}keV}  \; E(z) $ &
$\frac{S_{r_{500}}}{\rm keV\;cm^{2}} \; E(z)^{4/3} $
   & $10^{-1.91\pm1.44}$ & $0.36 \pm 0.10$ & 0.09 & 0.23 & 0.68 & 0.22 & whole sample \\
&  & $10^{-3.15\pm1.87}$ & $0.54 \pm 0.32$ & 0.11 & 0.23 & 0.23 & 0.21 & non-CCC subsample \\
&  & $10^{0.67\pm1.53}$ & $0.18 \pm 0.10$ & 0.22 & 0.42 & 0.27 & 0.22 & CCC subsample \\
\hline
$\frac{Y_{\rm X}}{\rm M_{\odot}keV}  \; E(z) $ &
$\frac{P_{0.2r_{500}}}{\rm keV\;cm^{-3}} \; E(z)^{-2} $
   & $10^{-9.26\pm1.31}$ & $0.50 \pm 0.09$ & 0.05 & 0.08 & 0.18 & 0.10 & whole sample \\
&  & $10^{-8.43\pm1.50}$ & $0.44 \pm 0.10$ & 0.08 & 0.08 & 0.20 & 0.10 & non-CCC subsample \\
&  & $10^{-10.90\pm2.83}$ & $0.61 \pm 0.19$ & 0.22 & 0.20 & 0.17 & 0.10 & CCC subsample \\
\hline
$\frac{Y_{\rm X}}{\rm M_{\odot}keV}  \; E(z) $ &
$\frac{P_{0.3r_{500}}}{\rm keV\;cm^{-3}} \; E(z)^{-2} $
   & $10^{-8.95\pm1.02}$ & $0.46 \pm 0.07$ & 0.10 & 0.08 & 0.16 & 0.08 & whole sample \\
&  & $10^{-8.15\pm1.18}$ & $0.41 \pm 0.08$ & 0.08 & 0.08 & 0.17 & 0.08 & non-CCC subsample \\
&  & $10^{-10.41\pm1.67}$ & $0.56 \pm 0.11$ & 0.24 & 0.16 & 0.17 & 0.09 & CCC subsample \\
\hline
$\frac{Y_{\rm X}}{\rm M_{\odot}keV}  \; E(z) $ &
$\frac{P_{r_{500}}}{\rm keV\;cm^{-3}} \; E(z)^{-2} $
   & $10^{-8.32\pm1.88}$ & $0.35 \pm 0.13$ & 0.19 & 0.08 & 0.36 & 0.13 & whole sample \\
&  & $10^{-8.47\pm2.10}$ & $0.36 \pm 0.14$ & 0.25 & 0.08 & 0.38 & 0.12 & non-CCC subsample \\
&  & $10^{-8.47\pm3.45}$ & $0.37 \pm 0.23$ & 0.37 & 0.27 & 0.50 & 0.18 & CCC subsample \\
\hline
 $\frac{r}{r_{500}}$
   & $\frac{S}{\rm keV\;cm^{2}} \; E(z)^{4/3} \left(\frac{T_{0.2-0.5r_{500}}}{10keV}\right)^{-0.65}$ 
   & $10^{3.46\pm0.01}$ & $1.00 \pm 0.03$ & --- & --- & --- & --- & whole sample \\
&  & $10^{3.46\pm0.02}$ & $0.99 \pm 0.04$ & --- & --- & --- & --- & non-CCC subsample \\
&  & $10^{3.46\pm0.02}$ & $1.05 \pm 0.04$ & --- & --- & --- & --- & CCC subsample \\
\hline
\hline
  \end{tabular}
  \end{center}
  \hspace*{0.3cm}{\footnotesize Here we provide both the 
    mean scatter derived by a Gaussian fit to the histogram and the
    average scatter. } }
\end{table*}

\begin{figure*}
\begin{center}
\includegraphics[width=8.5cm,angle=270]{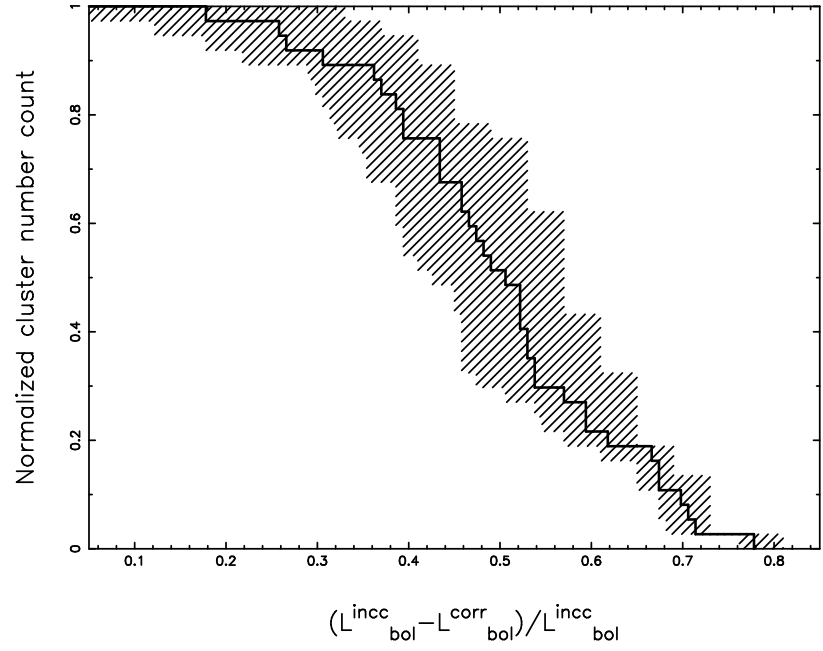}
\end{center}
\caption{Normalized cumulative cluster number count as a function of
the fraction of the total bolometric luminosity attributed to the
$<0.2 r_{500}$ region.
\label{f:fraclcc}}
\end{figure*}

\section{Mass--observable relations at different radius}
\label{apx:simuyx2}

\begin{table*} { \begin{center} \footnotesize
      {\renewcommand{\arraystretch}{1.3} \caption[]{ Deduced properties of 37
          LoCuSS galaxy clusters at the radii ($r^{\rm Y_{\rm X},X}_{500}$)
          determined by the method described in \S~\ref{s:simuyx}.
          Column~(1): cluster name; Col.~(2): $r^{\rm Y_{\rm X},X}_{500}$;
          Cols.~(3): X-ray mass at $r^{\rm Y_{\rm X},X}_{500}$. Cols.~(4--5):
          weak lensing masses at $r^{\rm Y_{\rm X},X}_{500}$ for the D06 and
          B07 clusters.}
  \label{t:yx500}} \begin{tabular}{lrrrr}
\hline
\hline
Name & $r^{\rm Y_{\rm X},X}_{500}$ & $M^{\rm X}(r^{\rm Y_{\rm X},X}_{500})$  & \multicolumn{2}{c}{$M^{\rm wl}(r^{\rm Y_{\rm X},X}_{500})$}  \\
\hline
     &  Mpc      & $10^{14} M_{\odot}$ & \multicolumn{2}{c}{$10^{14} M_{\odot}$} \\
     &           &                   &  D06                &  B07              \\
\hline
RXCJ0043.4-2037    &     1.07  &$     4.78  \pm     1.40  $& --- & --- \\
RXCJ0232.2-4420    &     1.22  &$     7.66  \pm     2.20  $& --- & --- \\
RXCJ0307.0-2840    &     1.15  &$     5.51  \pm     2.01  $& --- & --- \\
RXCJ0516.7-5430    &     1.20  &$     6.49  \pm     1.83  $& --- & --- \\
RXCJ0528.9-3927    &     1.23  &$     6.80  \pm     1.98  $& --- & --- \\
RXCJ0532.9-3701    &     1.08  &$     5.16  \pm     1.51  $& --- & --- \\
RXCJ0547.6-3152    &     1.13  &$     5.39  \pm     1.60  $& --- & --- \\
RXCJ0645.4-5413    &     1.32  &$     6.97  \pm     2.07  $& --- & --- \\
RXCJ0658.5-5556    &     1.55  &$    14.38  \pm     6.07  $& --- & --- \\
RXCJ0945.4-0839    &     1.09  &$     5.44  \pm     1.57  $& --- & --- \\
RXCJ0958.3-1103    &     1.05  &$     3.48  \pm     1.02  $& --- & --- \\
RXCJ2129.6+0005    &     1.13  &$     4.60  \pm     1.34  $& $5.51 \pm 3.87$ & --- \\
RXCJ2218.6-3853    &     1.09  &$     4.59  \pm     1.34  $& --- & --- \\
RXCJ2234.5-3744    &     1.26  &$     7.44  \pm     2.13  $& --- & --- \\
RXCJ2308.3-0211    &     1.19  &$     7.05  \pm     2.16  $& --- & --- \\
RXCJ2337.6+0016    &     1.18  &$     8.41  \pm     2.41  $& --- & --- \\
Abell68            &     1.16  &$     6.20  \pm     1.82  $&$11.41 \pm 8.49$ & $4.68 \pm 1.25$ \\
Abell115           &     1.31  &$     5.79  \pm     1.53  $& $6.30 \pm 3.99$ & --- \\
Abell209           &     1.19  &$     5.52  \pm     1.74  $& $8.41 \pm 5.09$ & $5.08 \pm 1.29$ \\
Abell267           &     0.99  &$     3.83  \pm     1.16  $& $7.53 \pm 4.13$ & $2.27 \pm 0.92$ \\
Abell383           &     1.02  &$     3.30  \pm     0.97  $& ---             & $2.95 \pm 0.92$ \\
Abell773           &     1.28  &$     7.84  \pm     2.25  $&$10.01 \pm 5.14$ & --- \\
Abell781           &     1.10  &$     4.75  \pm     1.39  $& $7.69 \pm 4.27$ & --- \\
Abell901           &     1.00  &$     3.27  \pm     0.96  $& --- & ---\\
Abell963           &     1.16  &$     5.28  \pm     1.52  $& ---             & $3.31 \pm 0.57$ \\
Abell1413          &     1.23  &$     5.63  \pm     1.63  $& --- & --- \\
Abell1689          &     1.32  &$     9.24  \pm     2.69  $& ---             &$11.54 \pm 2.13$ \\
Abell1758          &     1.18  &$     8.46  \pm     2.55  $&$13.34 \pm 6.91$ & --- \\
Abell1763          &     1.25  &$     6.11  \pm     1.77  $& $7.64 \pm 3.81$ & $8.03 \pm 1.66$ \\
Abell1835          &     1.36  &$     8.43  \pm     2.45  $& $8.33 \pm 4.17$ &$13.41 \pm 2.62$ \\
Abell1914          &     1.37  &$    11.56  \pm     3.30  $& $7.40 \pm 4.23$ & --- \\ 
Abell2204          &     1.29  &$     5.90  \pm     1.69  $& $7.76 \pm 5.00$ & ---\\
Abell2218          &     1.13  &$     4.47  \pm     1.36  $& ---             & $6.63 \pm 1.36$ \\
Abell2261          &     1.22  &$     6.21  \pm     1.76  $& $6.09 \pm 3.43$ & ---\\
Abell2390          &     1.39  &$     8.25  \pm     2.41  $& $9.56 \pm 4.39$ & $7.52 \pm 1.56$ \\
Abell2667          &     1.24  &$     6.31  \pm     1.81  $& --- & --- \\
Z7160              &     1.04  &$     2.85  \pm     0.83  $& $7.29 \pm 4.46$ & --- \\ 
\hline
\hline
  \end{tabular}
  \end{center}
  \hspace*{0.3cm}{\footnotesize ``D06'' denotes the clusters in the D06 subsample, 
    and ``B07'' the B07 subsample, respectively. We used the error of $M^{\rm
      wl}_{500}$ as the error for the weak
    lensing masses here.} }
\end{table*}

\begin{table*} { \begin{center} \footnotesize
      {\renewcommand{\arraystretch}{1.3} \caption[]{ Deduced properties of 37
          LoCuSS galaxy clusters at the radii ($r^{\rm Y_{\rm X},wl}_{500}$)
          determined by the method described in \S~\ref{s:ylmwlen_simuyx1}.
          Column~(1): cluster name; Col.~(2): $r^{\rm Y_{\rm X},wl}_{500}$;
          Cols.~(3): X-ray mass at $r^{\rm Y_{\rm X},wl}_{500}$. Cols.~(4--5):
          weak lensing masses at $r^{\rm Y_{\rm X},wl}_{500}$ for the D06 and
          B07 clusters.}
  \label{t:ywl500}} \begin{tabular}{lrrrr}
\hline
\hline
Name & $r^{\rm Y_{\rm X},wl}_{500}$ & $M^{\rm X}(r^{\rm Y_{\rm X},wl}_{500})$  & \multicolumn{2}{c}{$M^{\rm wl}(r^{\rm Y_{\rm X},wl}_{500})$}  \\
\hline
     &  Mpc      & $10^{14} M_{\odot}$ & \multicolumn{2}{c}{$10^{14} M_{\odot}$} \\
     &           &                   &  D06                &  B07              \\
\hline
RXCJ0043.4-2037    &     1.06  &$     4.72  \pm     1.40  $& --- & --- \\
RXCJ0232.2-4420    &     1.20  &$     7.54  \pm     2.20  $& --- & --- \\
RXCJ0307.0-2840    &     1.14  &$     5.44  \pm     2.01  $& --- & --- \\
RXCJ0516.7-5430    &     1.18  &$     6.37  \pm     1.83  $& --- & --- \\
RXCJ0528.9-3927    &     1.22  &$     6.68  \pm     1.98  $& --- & --- \\
RXCJ0532.9-3701    &     1.07  &$     5.10  \pm     1.51  $& --- & --- \\
RXCJ0547.6-3152    &     1.12  &$     5.32  \pm     1.60  $& --- & --- \\
RXCJ0645.4-5413    &     1.30  &$     6.89  \pm     2.07  $& --- & --- \\
RXCJ0658.5-5556    &     1.53  &$    13.77  \pm     6.07  $& --- & --- \\
RXCJ0945.4-0839    &     1.08  &$     5.36  \pm     1.57  $& --- & --- \\
RXCJ0958.3-1103    &     1.04  &$     3.44  \pm     1.02  $& --- & --- \\
RXCJ2129.6+0005    &     1.12  &$     4.55  \pm     1.34  $& $5.45 \pm 3.87$ & --- \\
RXCJ2218.6-3853    &     1.08  &$     4.53  \pm     1.34  $& --- & --- \\
RXCJ2234.5-3744    &     1.24  &$     7.34  \pm     2.13  $& --- & --- \\
RXCJ2308.3-0211    &     1.18  &$     6.96  \pm     2.16  $& --- & --- \\
RXCJ2337.6+0016    &     1.17  &$     8.28  \pm     2.41  $& --- & --- \\
Abell68            &     1.15  &$     6.13  \pm     1.82  $&$11.28 \pm 8.49$ & $4.64 \pm 1.25$ \\
Abell115           &     1.29  &$     5.68  \pm     1.53  $& $6.23 \pm 3.99$ & --- \\
Abell209           &     1.18  &$     5.46  \pm     1.74  $& $8.33 \pm 5.09$ & $5.03 \pm 1.29$ \\
Abell267           &     0.98  &$     3.77  \pm     1.16  $& $7.45 \pm 4.13$ & $2.26 \pm 0.92$ \\
Abell383           &     1.01  &$     3.27  \pm     0.97  $& ---             & $2.92 \pm 0.92$ \\
Abell773           &     1.27  &$     7.72  \pm     2.25  $& $9.92 \pm 5.14$ & --- \\
Abell781           &     1.09  &$     4.67  \pm     1.39  $& $7.59 \pm 4.27$ & --- \\
Abell901           &     0.99  &$     3.24  \pm     0.96  $& --- & ---\\
Abell963           &     1.15  &$     5.21  \pm     1.52  $& ---             & $3.29 \pm 0.57$ \\
Abell1413          &     1.22  &$     5.57  \pm     1.63  $& --- & --- \\
Abell1689          &     1.31  &$     9.14  \pm     2.69  $& ---             &$11.43 \pm 2.13$ \\
Abell1758          &     1.17  &$     8.34  \pm     2.55  $&$13.19 \pm 6.91$ & --- \\
Abell1763          &     1.24  &$     5.99  \pm     1.77  $& $7.56 \pm 3.81$ & $7.93 \pm 1.66$ \\
Abell1835          &     1.35  &$     8.34  \pm     2.45  $& $8.25 \pm 4.17$ &$13.25 \pm 2.62$ \\
Abell1914          &     1.36  &$    11.35  \pm     3.30  $& $7.34 \pm 4.23$ & --- \\ 
Abell2204          &     1.28  &$     5.82  \pm     1.69  $& $7.67 \pm 5.00$ & ---\\
Abell2218          &     1.12  &$     4.41  \pm     1.36  $& ---             & $6.58 \pm 1.36$ \\
Abell2261          &     1.21  &$     6.14  \pm     1.76  $& $6.04 \pm 3.43$ & ---\\
Abell2390          &     1.37  &$     8.15  \pm     2.41  $& $9.46 \pm 4.39$ & $7.45 \pm 1.56$ \\
Abell2667          &     1.23  &$     6.23  \pm     1.81  $& --- & --- \\
Z7160              &     1.03  &$     2.82  \pm     0.83  $& $7.20 \pm 4.46$ & --- \\ 
\hline
\hline
  \end{tabular}
  \end{center}
  \hspace*{0.3cm}{\footnotesize ``D06'' denotes the clusters in the D06 subsample, 
    and ``B07'' the B07 subsample, respectively. We used the error of $M^{\rm
      wl}_{500}$ as the error for the weak
    lensing masses here.} }
\end{table*}

\begin{table*} { \begin{center} \footnotesize
      {\renewcommand{\arraystretch}{1.3} \caption[]{ Deduced properties of 37
          LoCuSS galaxy clusters at the radii ($r^{\rm Y_{\rm X},si}_{500}$)
          determined by the method described in \S~\ref{s:ylmwlen_simuyx2}.
          Column~(1): cluster name; Col.~(2): $r^{\rm Y_{\rm X},si}_{500}$;
          Cols.~(3): X-ray mass at $r^{\rm Y_{\rm X},si}_{500}$. Cols.~(4--5):
          weak lensing masses at $r^{\rm Y_{\rm X},si}_{500}$ for the D06 and
          B07 clusters.}
  \label{t:ywl500si}} \begin{tabular}{lrrrr}
\hline
\hline
Name & $r^{\rm Y_{\rm X},si}_{500}$ & $M^{\rm X}(r^{\rm Y_{\rm X},si}_{500})$  & \multicolumn{2}{c}{$M^{\rm wl}(r^{\rm Y_{\rm X},si}_{500})$}  \\
\hline
     &  Mpc      & $10^{14} M_{\odot}$ & \multicolumn{2}{c}{$10^{14} M_{\odot}$} \\
     &           &                   &  D06                &  B07              \\
\hline
RXCJ0043.4-2037    &     1.15  &$     5.21  \pm     1.52  $& --- & --- \\
RXCJ0232.2-4420    &     1.30  &$     8.39  \pm     2.42  $& --- & --- \\
RXCJ0307.0-2840    &     1.23  &$     6.02  \pm     2.16  $& --- & --- \\
RXCJ0516.7-5430    &     1.29  &$     7.24  \pm     2.04  $& --- & --- \\
RXCJ0528.9-3927    &     1.33  &$     7.67  \pm     2.21  $& --- & --- \\
RXCJ0532.9-3701    &     1.15  &$     5.53  \pm     1.62  $& --- & --- \\
RXCJ0547.6-3152    &     1.21  &$     5.84  \pm     1.73  $& --- & --- \\
RXCJ0645.4-5413    &     1.42  &$     7.54  \pm     2.23  $& --- & --- \\
RXCJ0658.5-5556    &     1.69  &$    18.47  \pm     6.80  $& --- & --- \\
RXCJ0945.4-0839    &     1.17  &$     6.02  \pm     1.73  $& --- & --- \\
RXCJ0958.3-1103    &     1.13  &$     3.75  \pm     1.09  $& --- & --- \\
RXCJ2129.6+0005    &     1.21  &$     4.97  \pm     1.44  $& $5.89 \pm 3.87$ & --- \\
RXCJ2218.6-3853    &     1.17  &$     4.97  \pm     1.45  $& --- & --- \\
RXCJ2234.5-3744    &     1.35  &$     8.11  \pm     2.31  $& --- & --- \\
RXCJ2308.3-0211    &     1.28  &$     7.65  \pm     2.32  $& --- & --- \\
RXCJ2337.6+0016    &     1.26  &$     9.26  \pm     2.67  $& --- & --- \\
Abell68            &     1.24  &$     6.71  \pm     1.97  $&$12.27 \pm 8.49$ & $4.98 \pm 1.25$ \\
Abell115           &     1.42  &$     6.51  \pm     1.73  $& $6.77 \pm 3.99$ & --- \\
Abell209           &     1.27  &$     5.97  \pm     1.82  $& $8.98 \pm 5.09$ & $5.43 \pm 1.29$ \\
Abell267           &     1.06  &$     4.25  \pm     1.27  $& $8.08 \pm 4.13$ & $2.40 \pm 0.92$ \\
Abell383           &     1.10  &$     3.53  \pm     1.03  $& ---             & $3.18 \pm 0.92$ \\
Abell773           &     1.38  &$     8.76  \pm     2.52  $&$10.71 \pm 5.14$ & --- \\
Abell781           &     1.19  &$     5.19  \pm     1.53  $& $8.30 \pm 4.27$ & --- \\
Abell901           &     1.07  &$     3.51  \pm     1.03  $& --- & ---\\
Abell963           &     1.24  &$     5.75  \pm     1.65  $& ---             & $3.47 \pm 0.57$ \\
Abell1413          &     1.33  &$     6.07  \pm     1.75  $& --- & --- \\
Abell1689          &     1.41  &$     9.97  \pm     2.91  $& ---             &$12.29 \pm 2.13$ \\
Abell1758          &     1.26  &$     9.33  \pm     2.80  $&$14.35 \pm 6.91$ & --- \\
Abell1763          &     1.35  &$     6.93  \pm     1.99  $& $8.18 \pm 3.81$ & $8.71 \pm 1.66$ \\
Abell1835          &     1.46  &$     9.08  \pm     2.59  $& $8.86 \pm 4.17$ &$14.53 \pm 2.62$ \\
Abell1914          &     1.47  &$    13.04  \pm     3.74  $& $7.87 \pm 4.23$ & --- \\ 
Abell2204          &     1.39  &$     6.35  \pm     1.81  $& $8.27 \pm 5.00$ & ---\\
Abell2218          &     1.21  &$     4.88  \pm     1.48  $& ---             & $6.99 \pm 1.36$ \\
Abell2261          &     1.32  &$     6.74  \pm     1.95  $& $6.51 \pm 3.43$ & ---\\
Abell2390          &     1.49  &$     8.88  \pm     2.58  $&$10.20 \pm 4.39$ & $7.95 \pm 1.56$ \\
Abell2667          &     1.33  &$     6.82  \pm     1.95  $& --- & --- \\
Z7160              &     1.12  &$     3.08  \pm     0.88  $& $7.85 \pm 4.46$ & --- \\ 
\hline
\hline
  \end{tabular}
  \end{center}
  \hspace*{0.3cm}{\footnotesize ``D06'' denotes the clusters in the D06 subsample, 
    and ``B07'' the B07 subsample, respectively. We used the error of $M^{\rm
      wl}_{500}$ as the error for the weak
    lensing masses here.} }
\end{table*}

\begin{figure*}
\begin{center}
\includegraphics[angle=270,width=8.5cm]{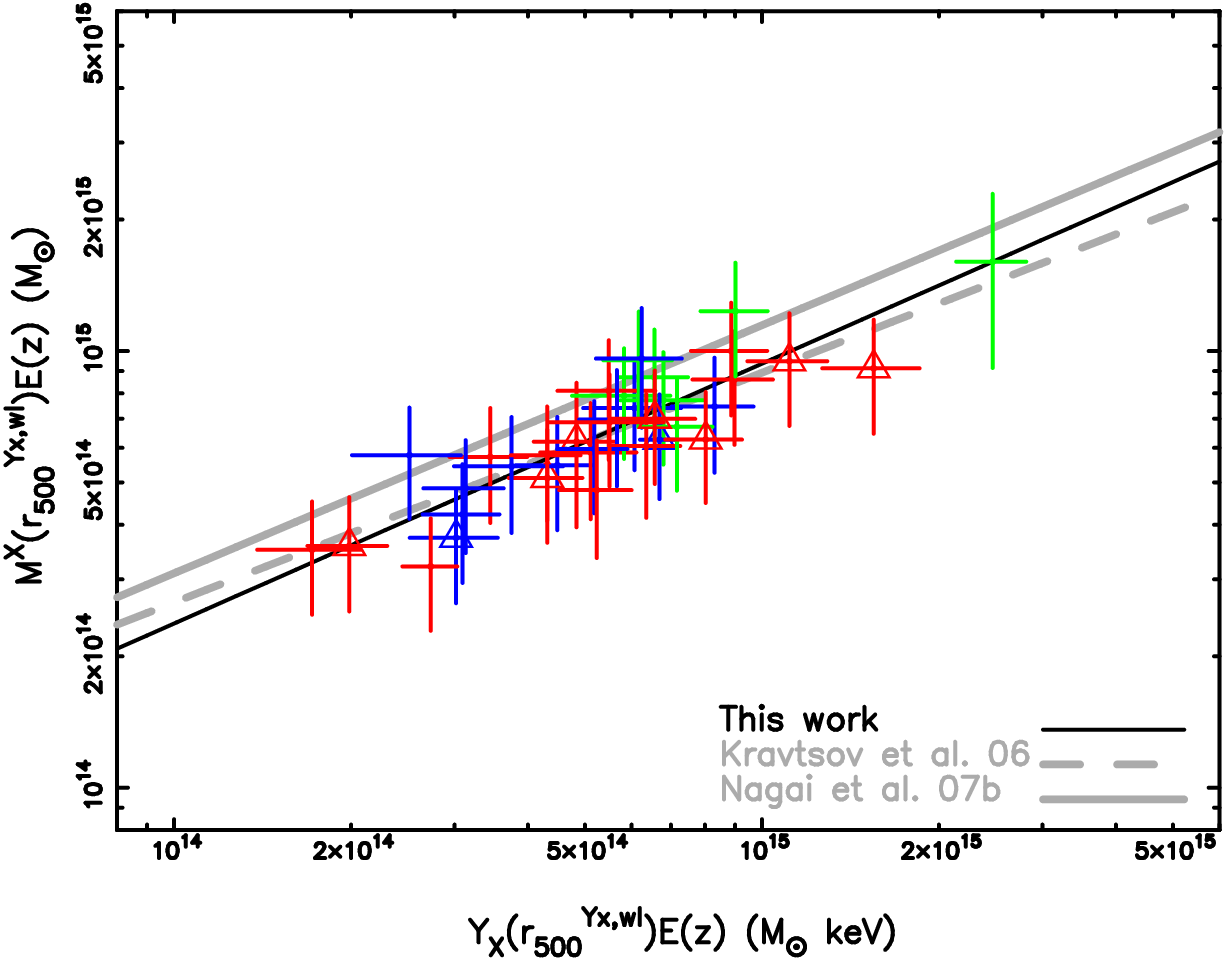}
\includegraphics[angle=270,width=8.5cm]{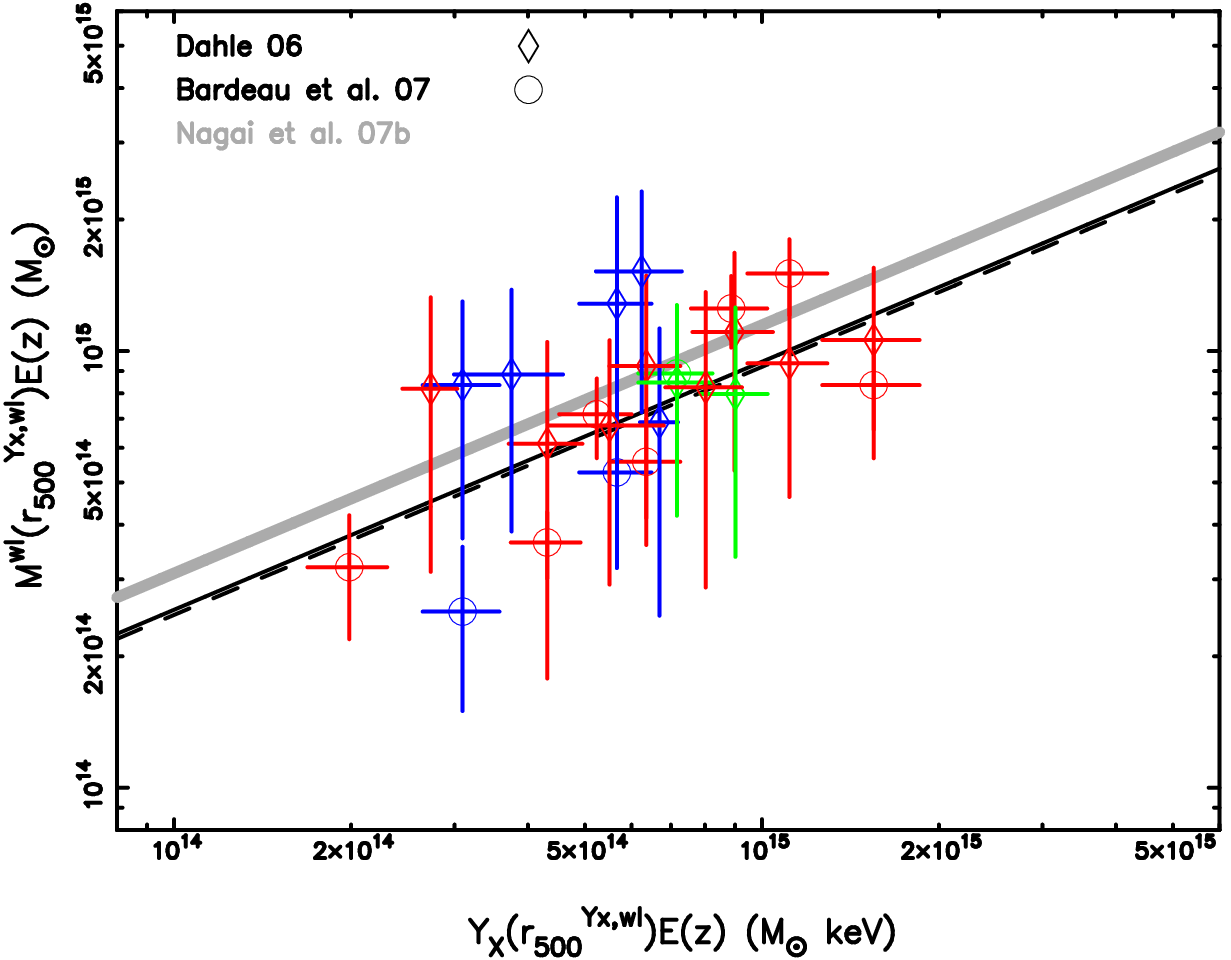}

\includegraphics[angle=270,width=8.5cm]{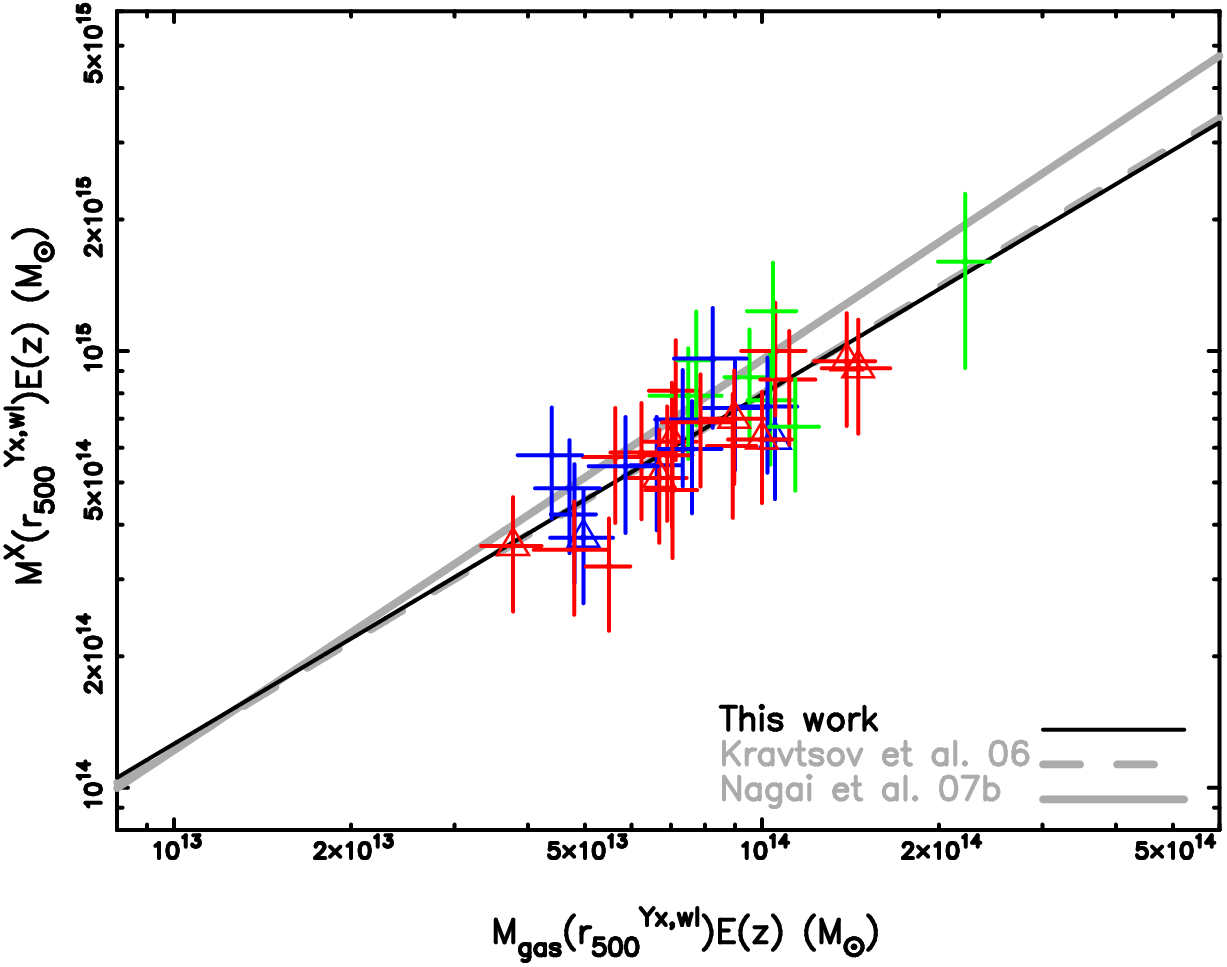}
\includegraphics[angle=270,width=8.5cm]{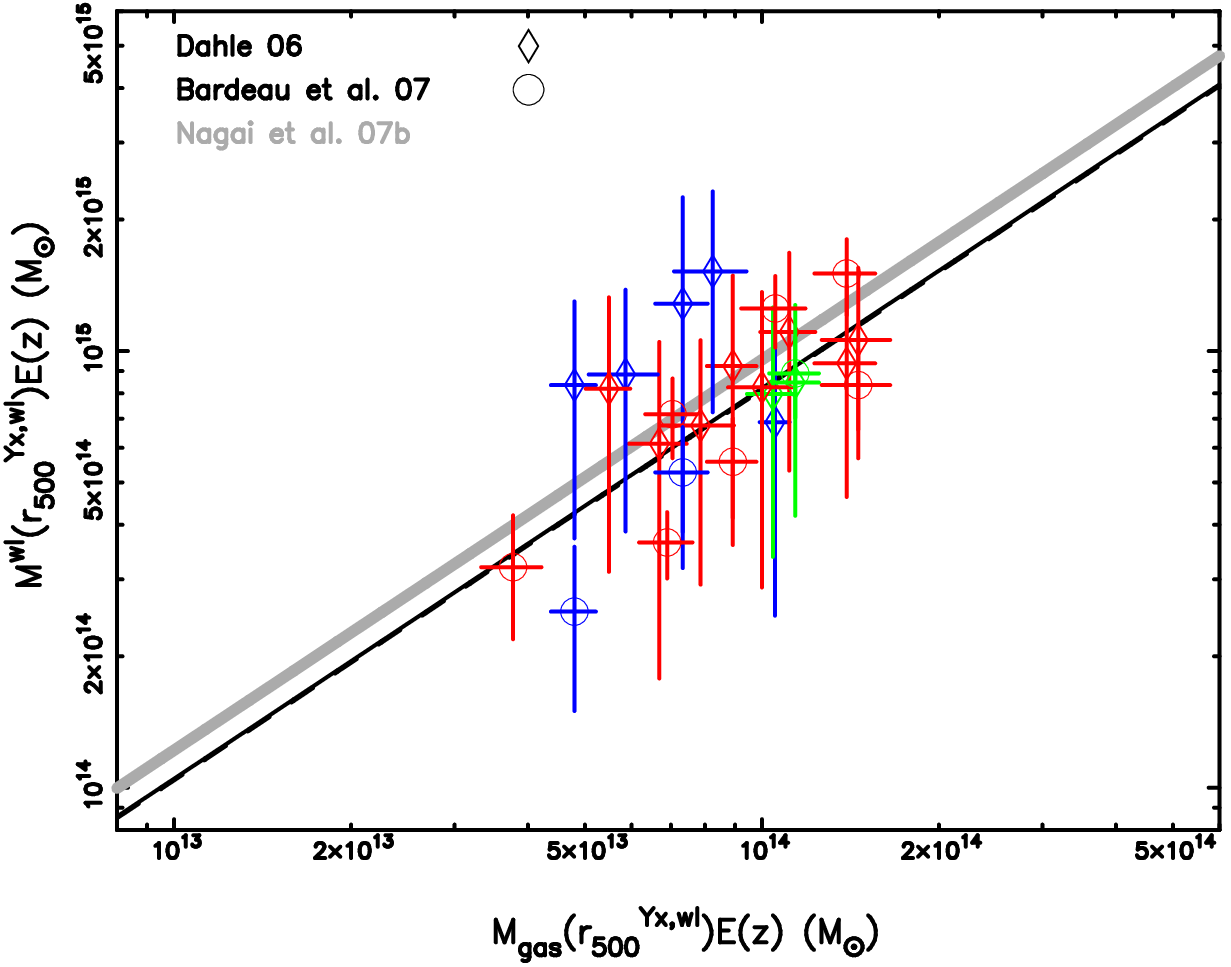}
\end{center}
\caption{See caption in Fig.~\ref{f:my_simuyx_1} except that the radius
  ($r_{500}^{\rm Y_{\rm X},wl}$) is determined by the method described in
  \S~\ref{s:ylmwlen_simuyx1}.
  \label{f:my_simuyx1_1}}
\end{figure*}

\begin{figure*}
\begin{center}
\includegraphics[angle=270,width=8.5cm]{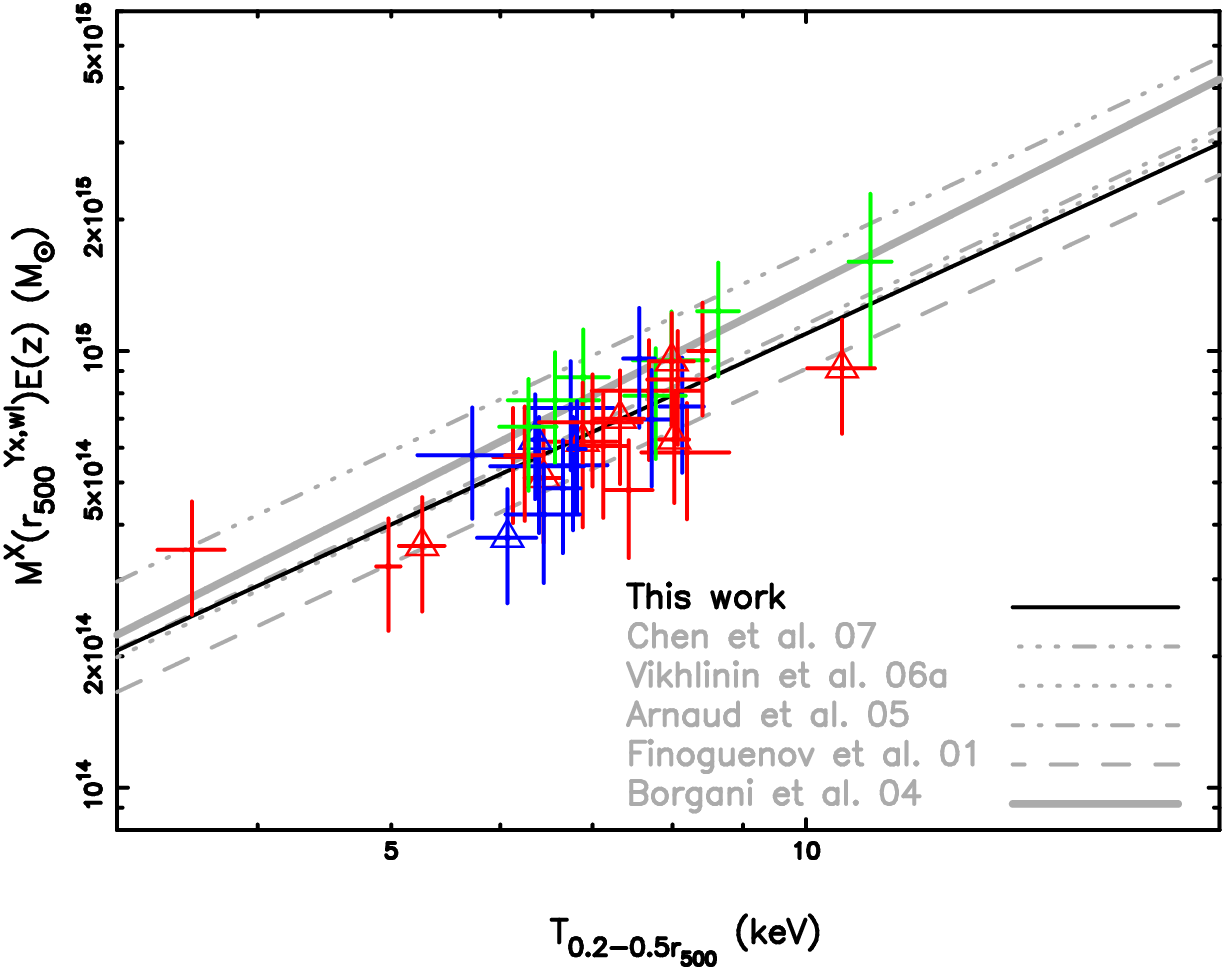}
\includegraphics[angle=270,width=8.5cm]{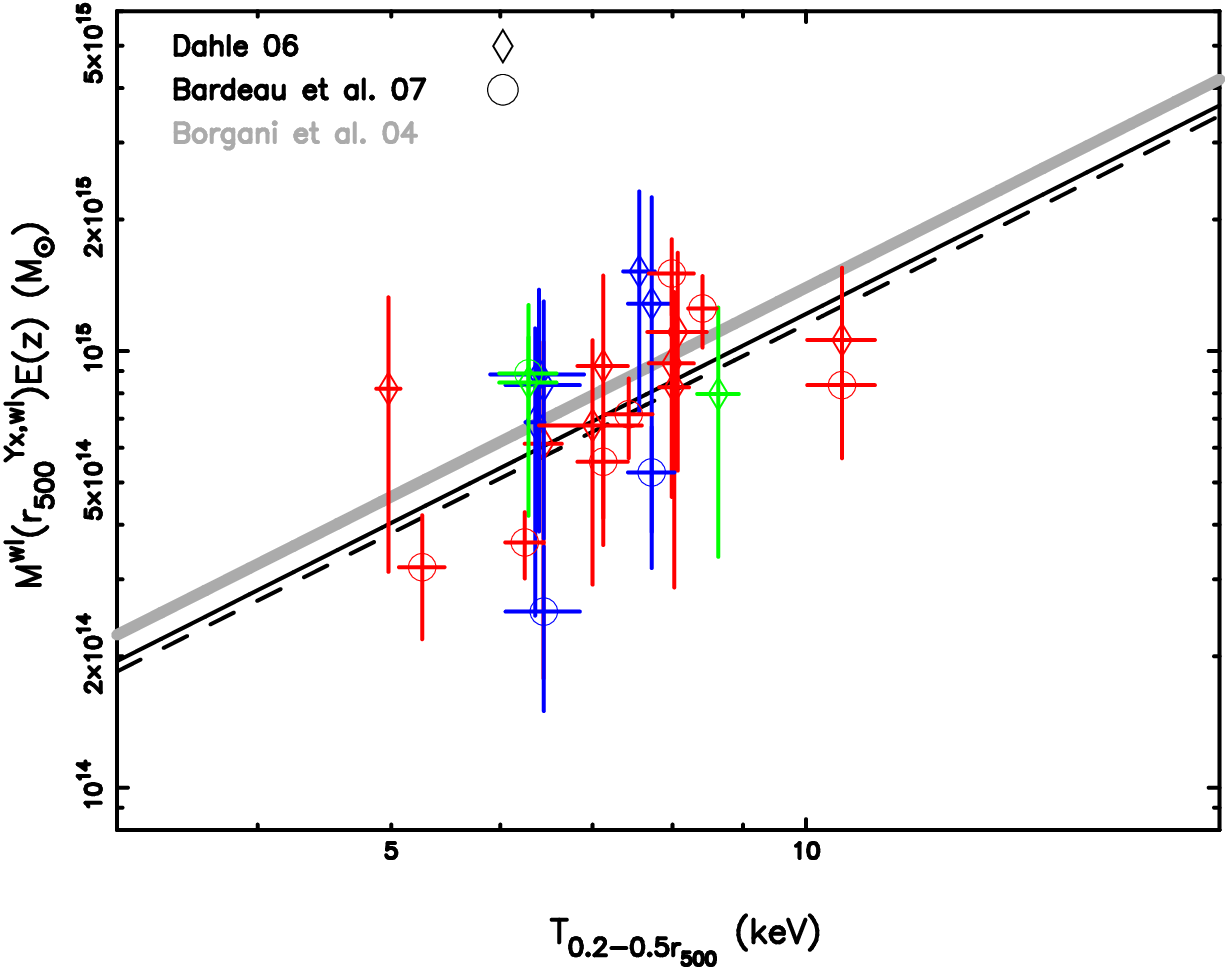}

\includegraphics[angle=270,width=8.5cm]{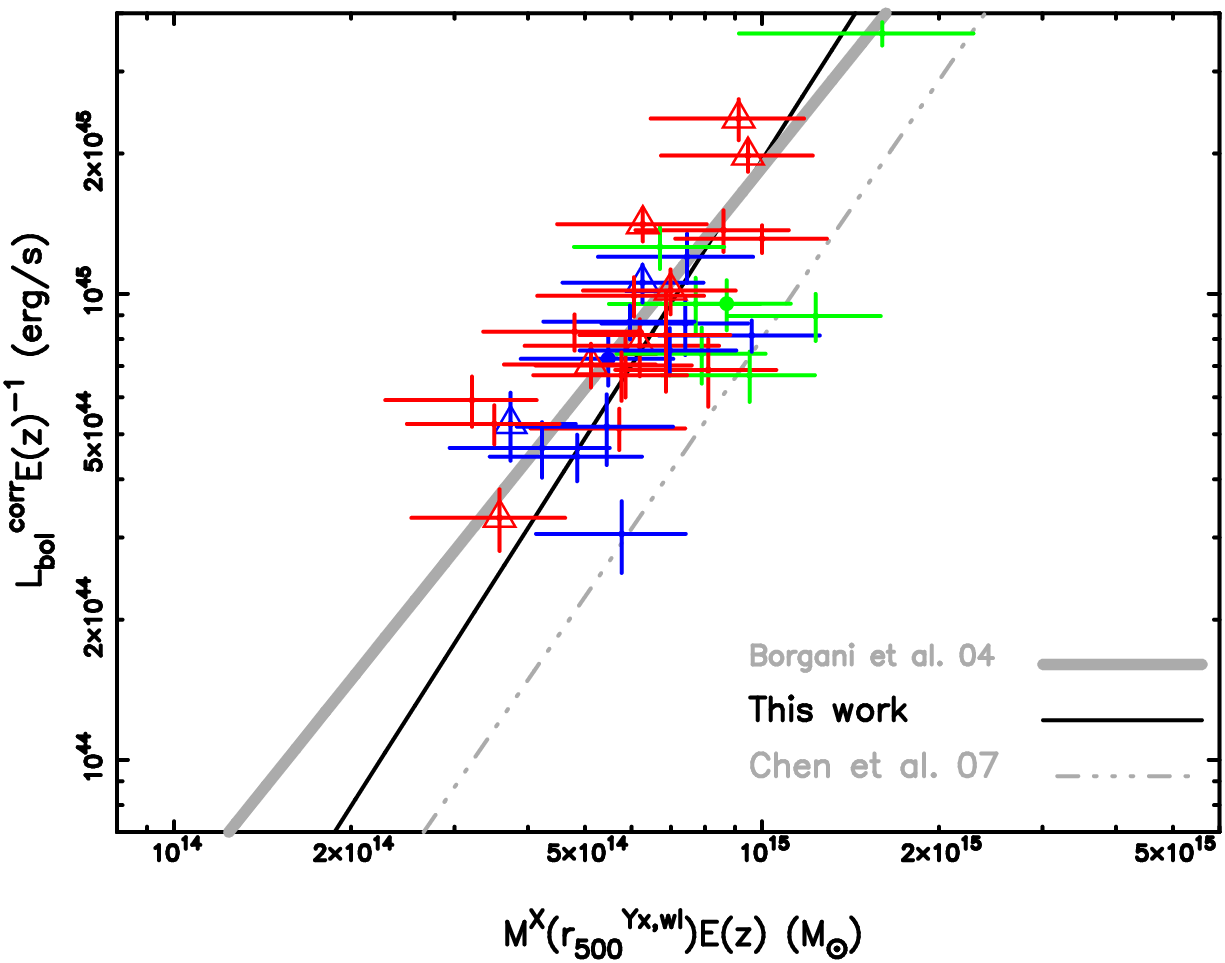}
\includegraphics[angle=270,width=8.5cm]{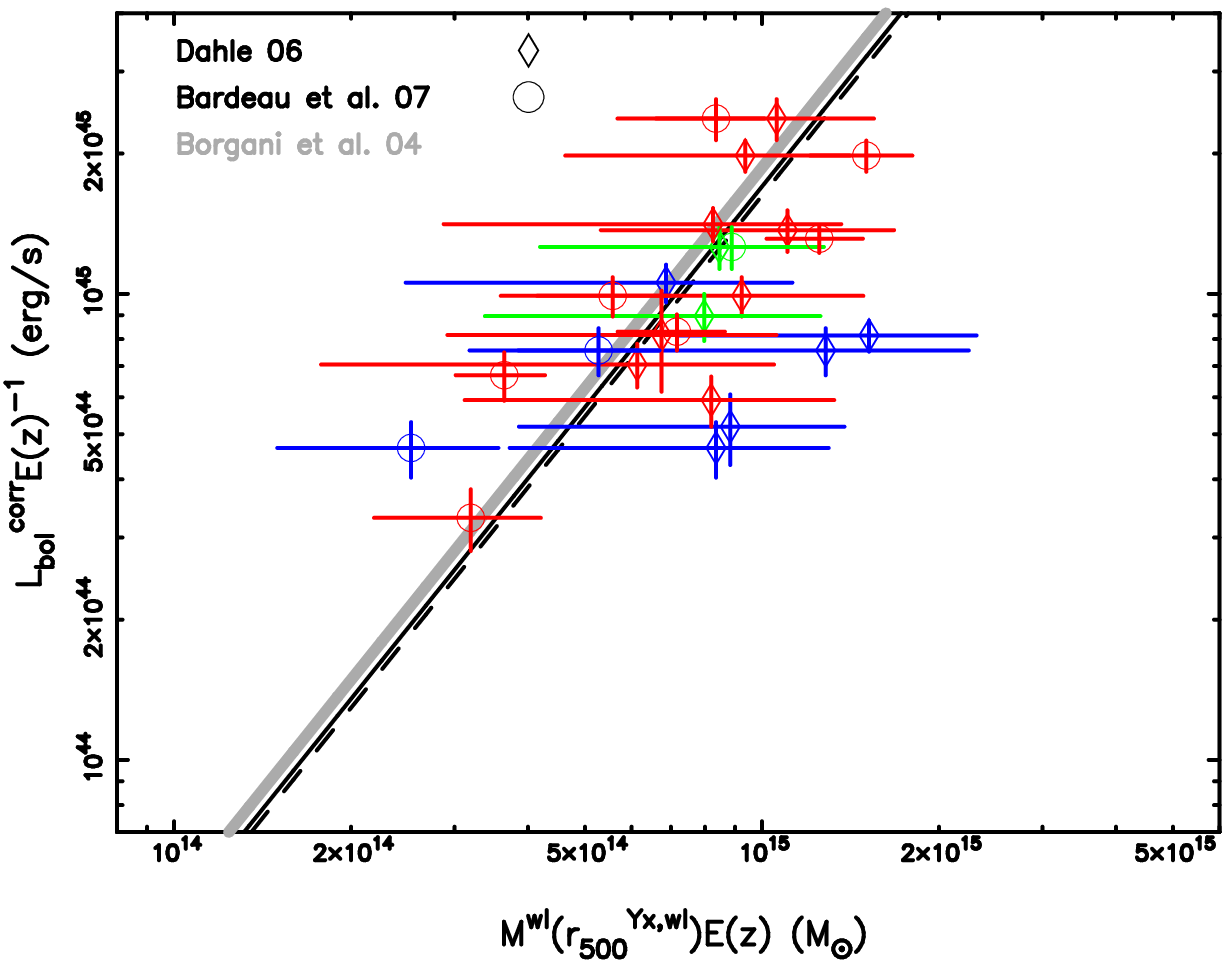}
\end{center}
\caption{See caption in Fig.~\ref{f:my_simuyx_2} except that the radius
  ($r_{500}^{\rm Y_{\rm X},wl}$) is determined by the method described in
  \S~\ref{s:ylmwlen_simuyx1}.
  \label{f:my_simuyx1_2}}
\end{figure*}

\begin{figure*}
\begin{center}
\includegraphics[angle=270,width=8.5cm]{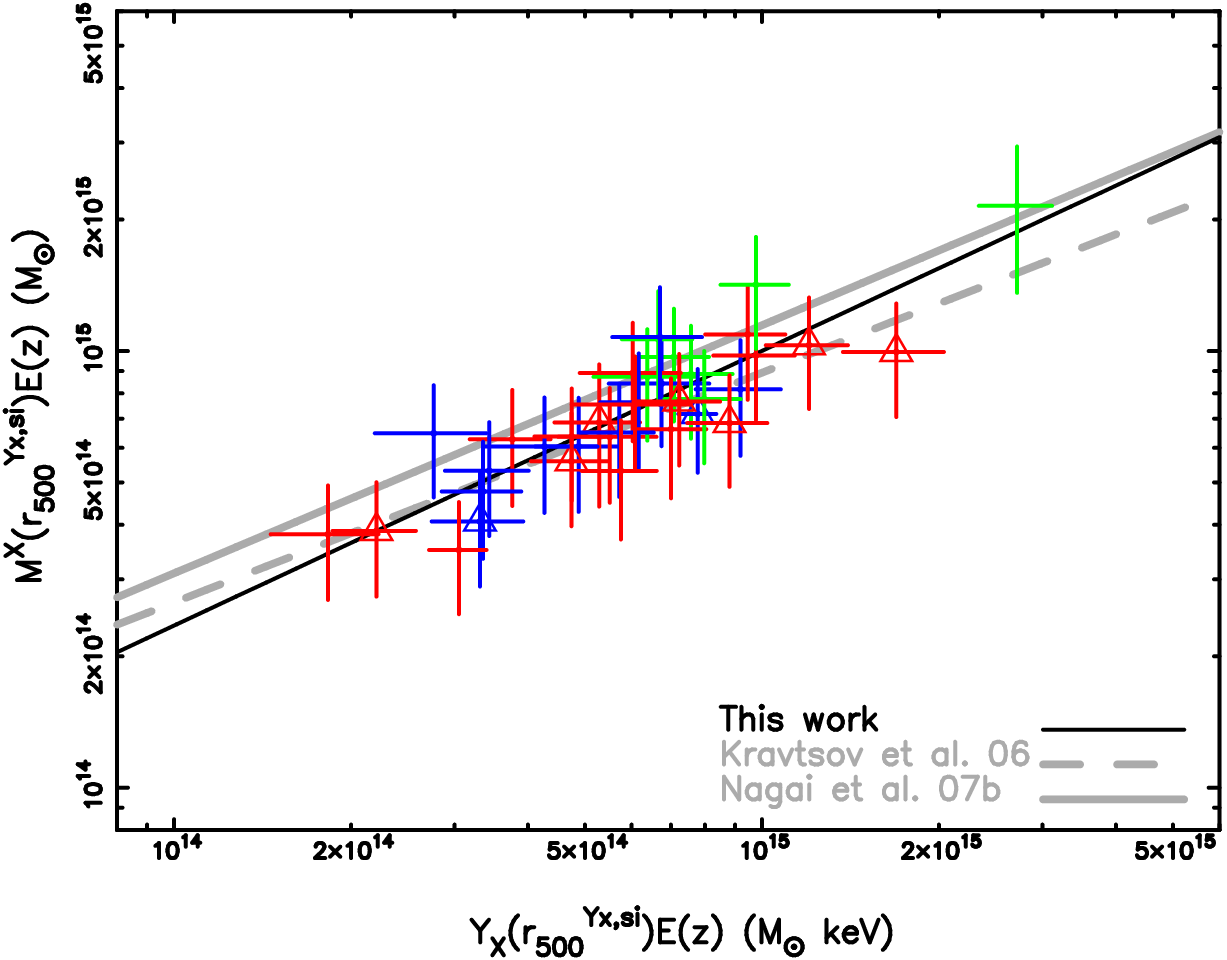}
\includegraphics[angle=270,width=8.5cm]{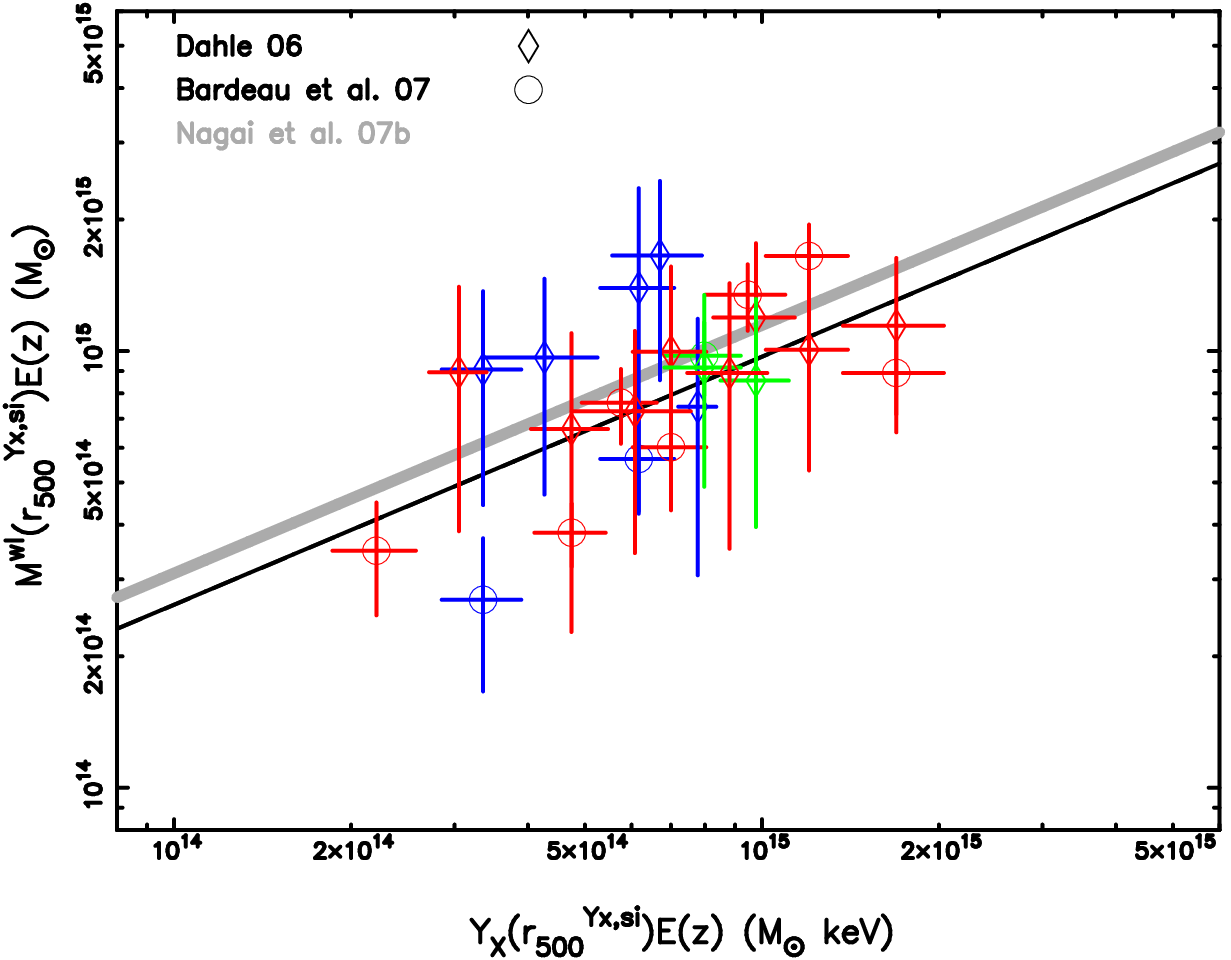}

\includegraphics[angle=270,width=8.5cm]{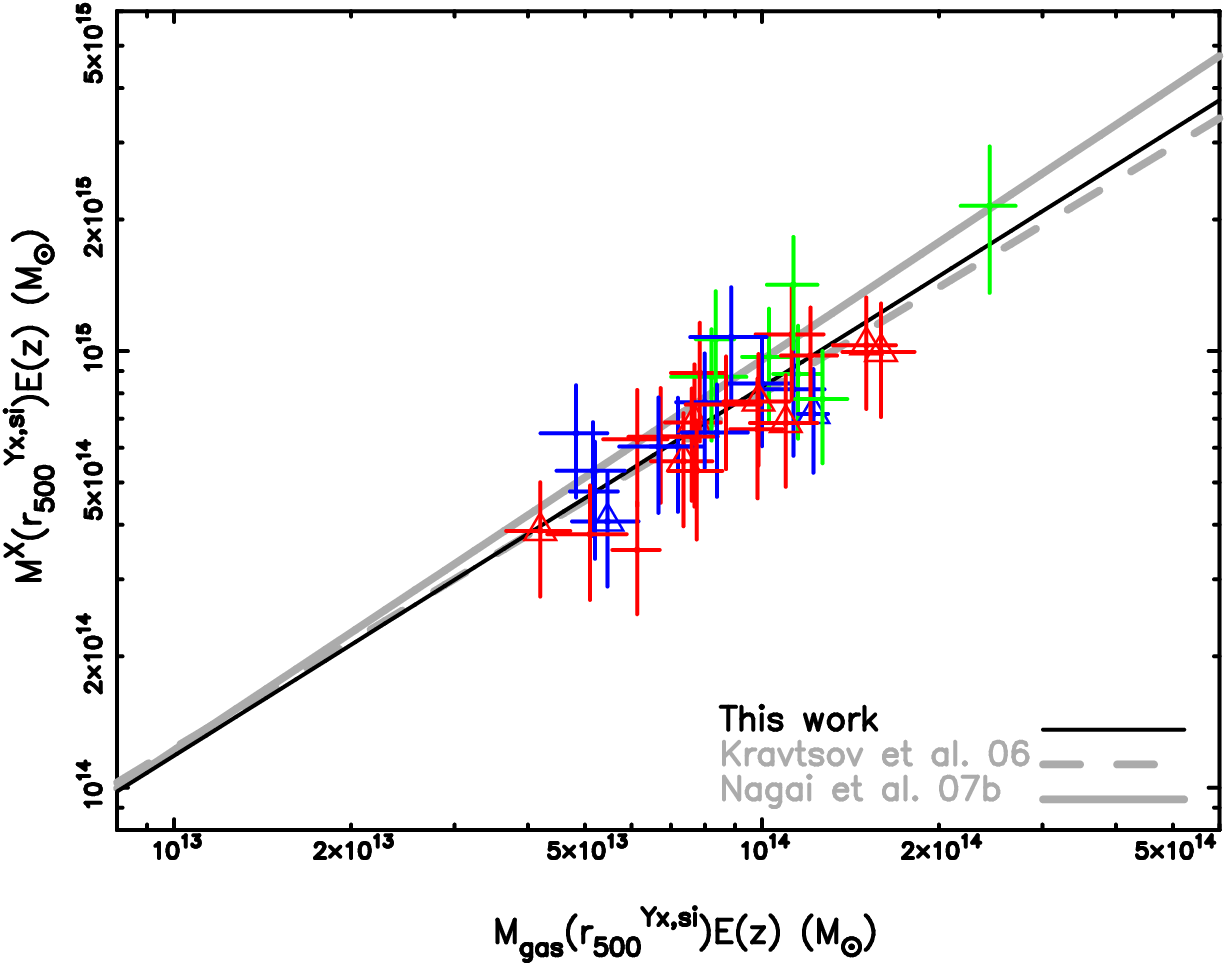}
\includegraphics[angle=270,width=8.5cm]{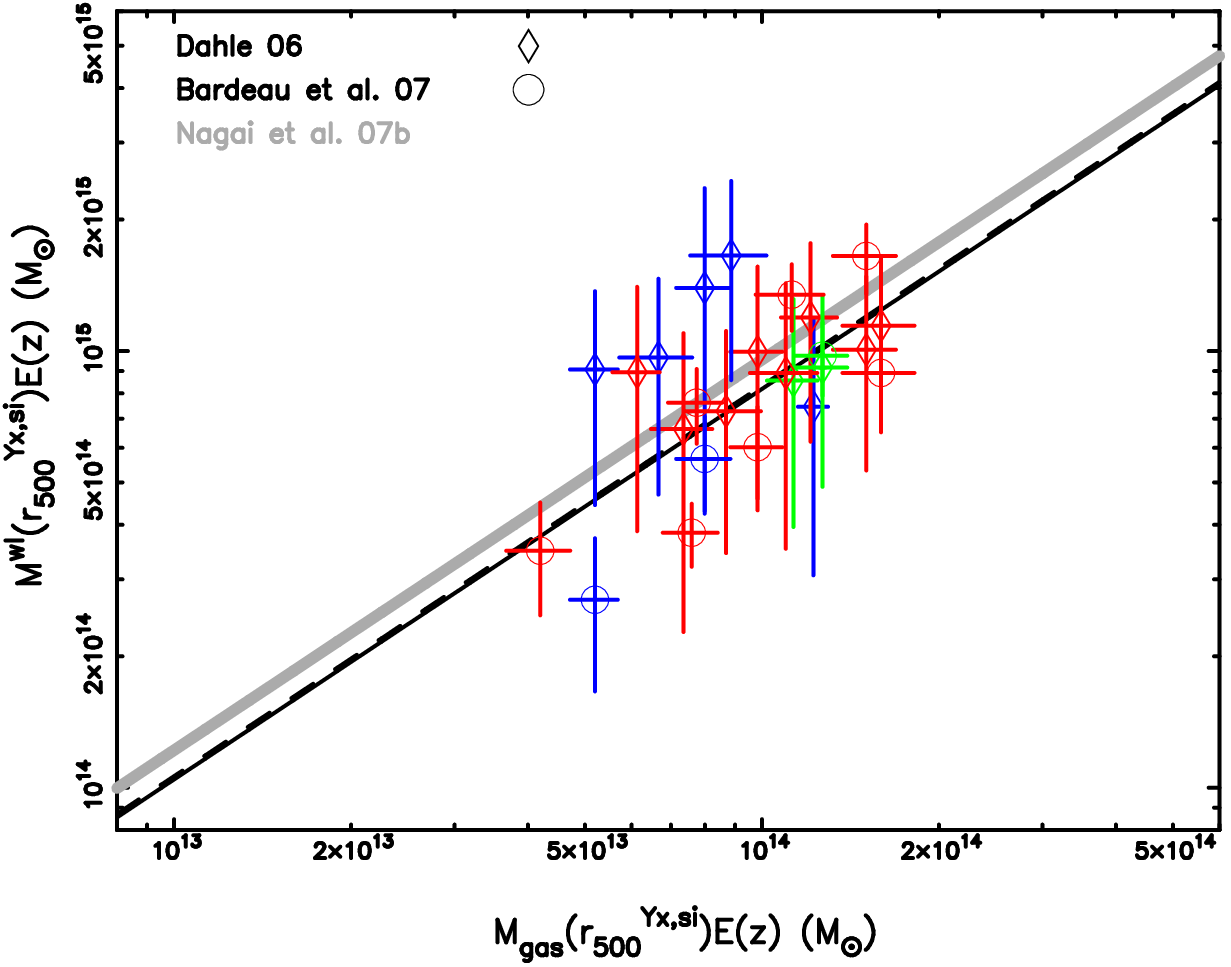}
\end{center}
\caption{See caption in Fig.~\ref{f:my_simuyx_1} except that the
  radius ($r_{500}^{\rm Y_{\rm X},si}$) is determined by the method
  described in \S~\ref{s:ylmwlen_simuyx2} but combining the $Y_{\rm
    X}(r)$ profile and the $M$--$Y_{\rm X}$ relation from simulations
  in Nagai et al.  (2007b).
\label{f:my_simuyx2_1}}
\end{figure*}

\begin{figure*}
\begin{center}
\includegraphics[angle=270,width=8.5cm]{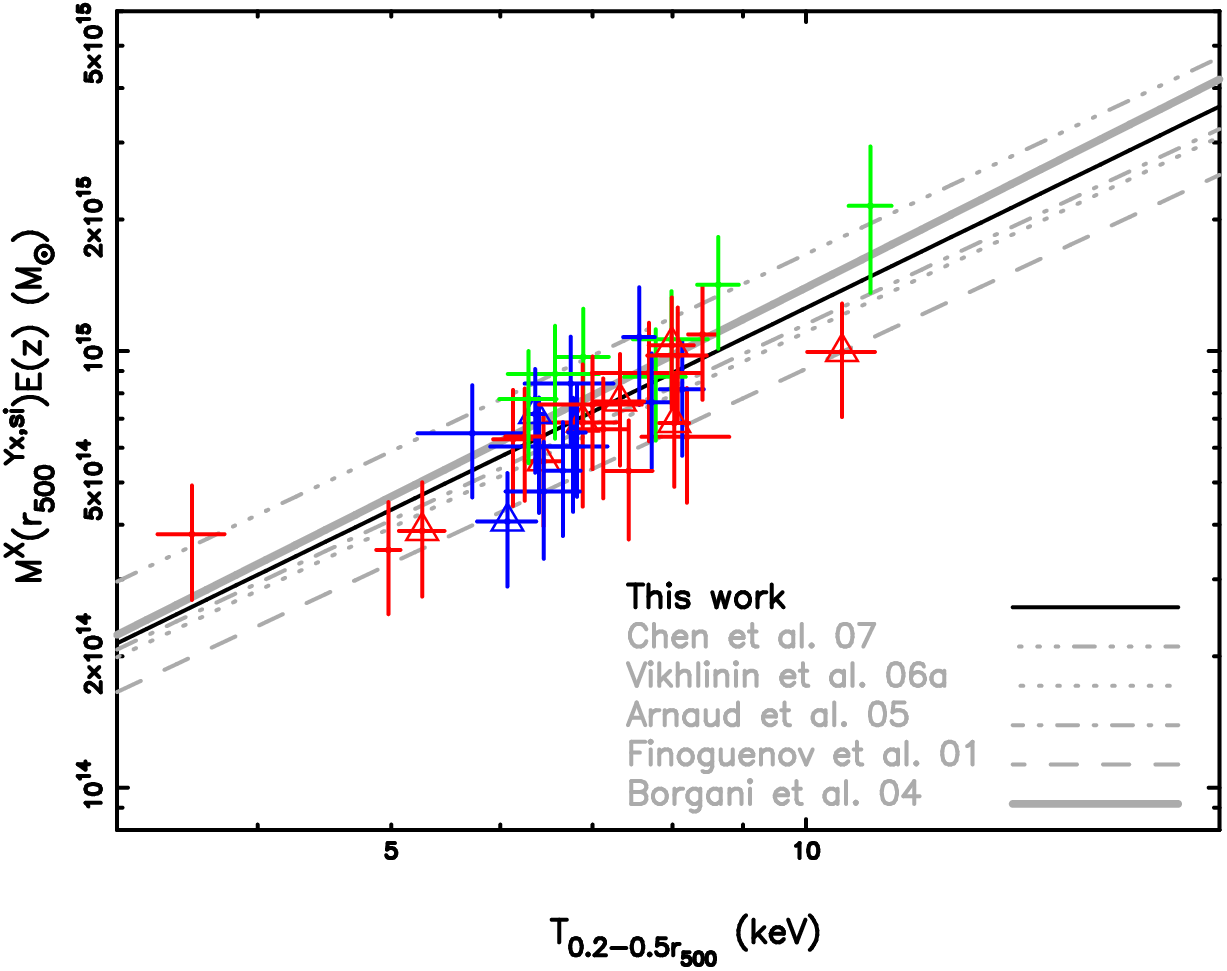}
\includegraphics[angle=270,width=8.5cm]{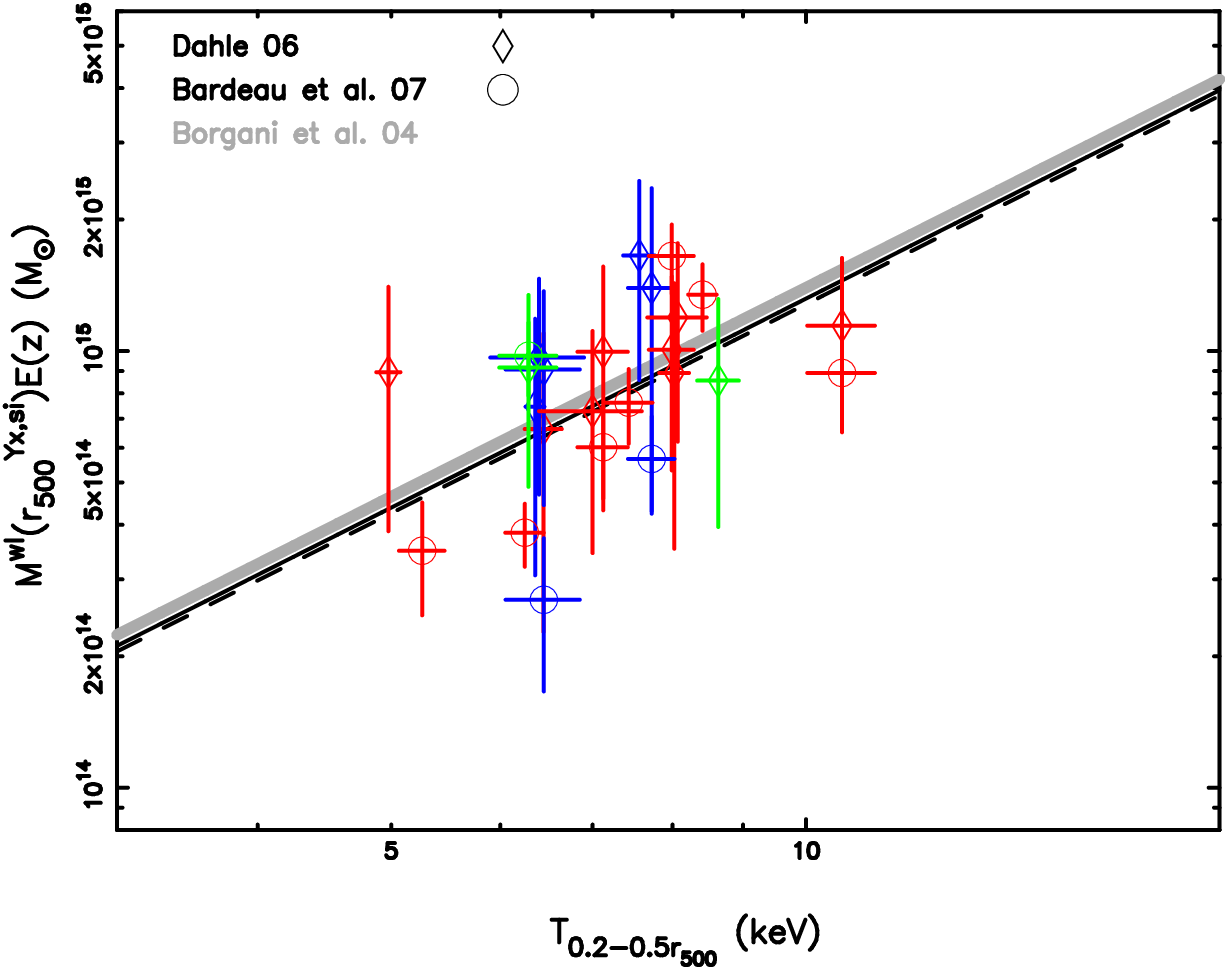}

\includegraphics[angle=270,width=8.5cm]{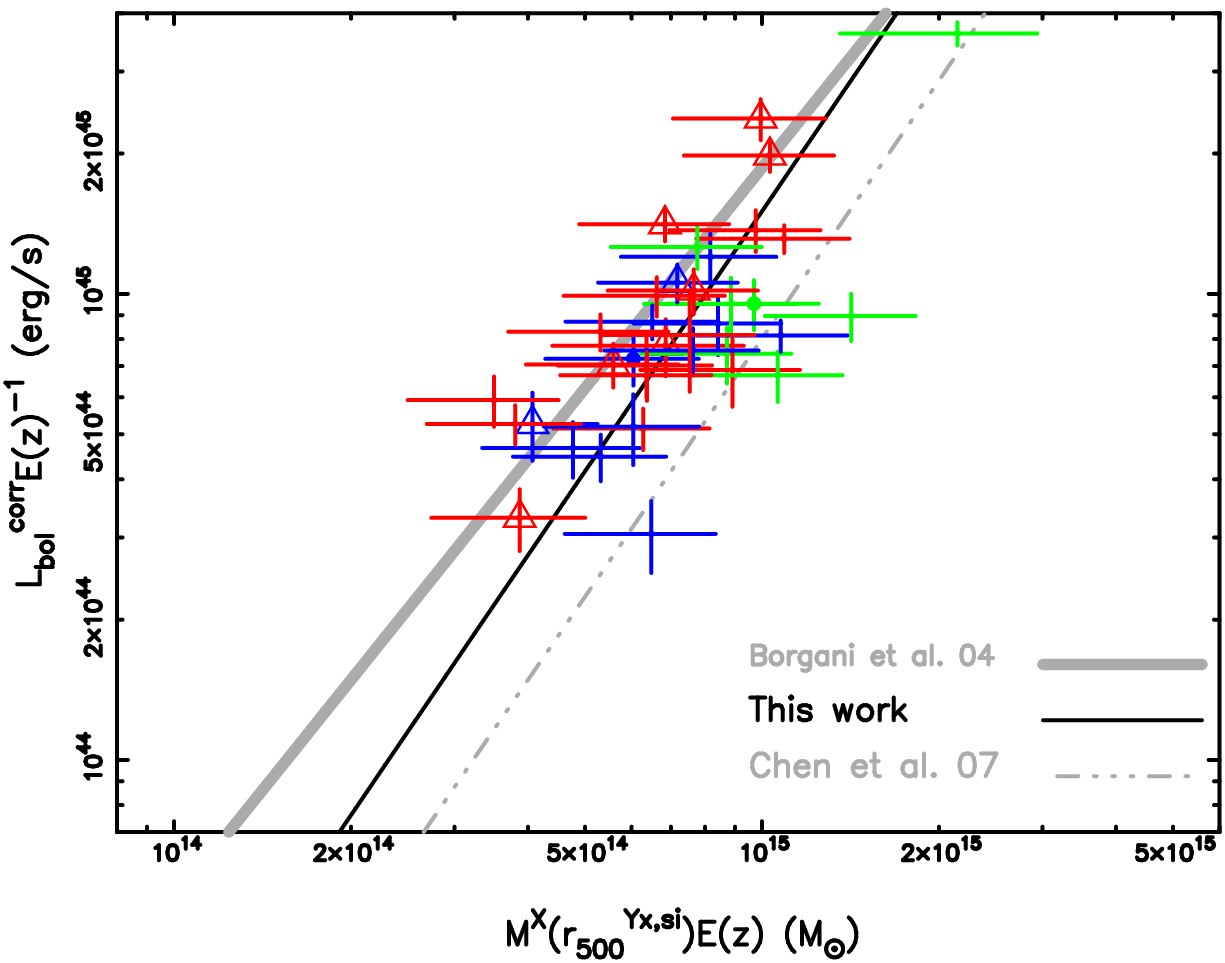}
\includegraphics[angle=270,width=8.5cm]{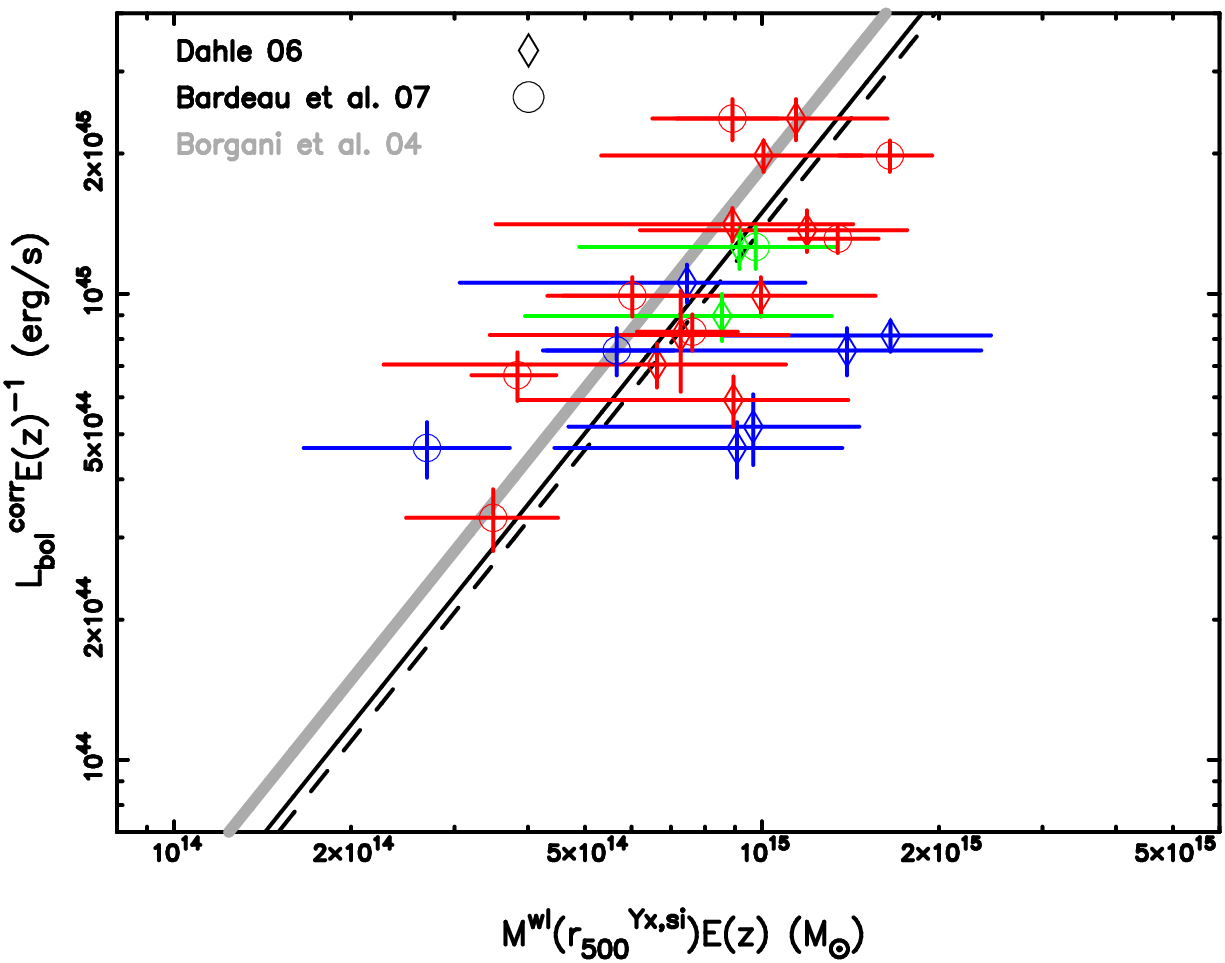}
\end{center}
\caption{See caption in Fig.~\ref{f:my_simuyx_2} except that the
  radius ($r_{500}^{\rm Y_{\rm X},si}$) is determined by the method
  described in \S~\ref{s:ylmwlen_simuyx2} but combining the $Y_{\rm
    X}(r)$ profile and the $M$--$Y_{\rm X}$ relation from simulations
  in Nagai et al.  (2007b).
\label{f:my_simuyx2_2}}
\end{figure*}

\begin{figure*}
\begin{center}
\includegraphics[angle=270,width=8.5cm]{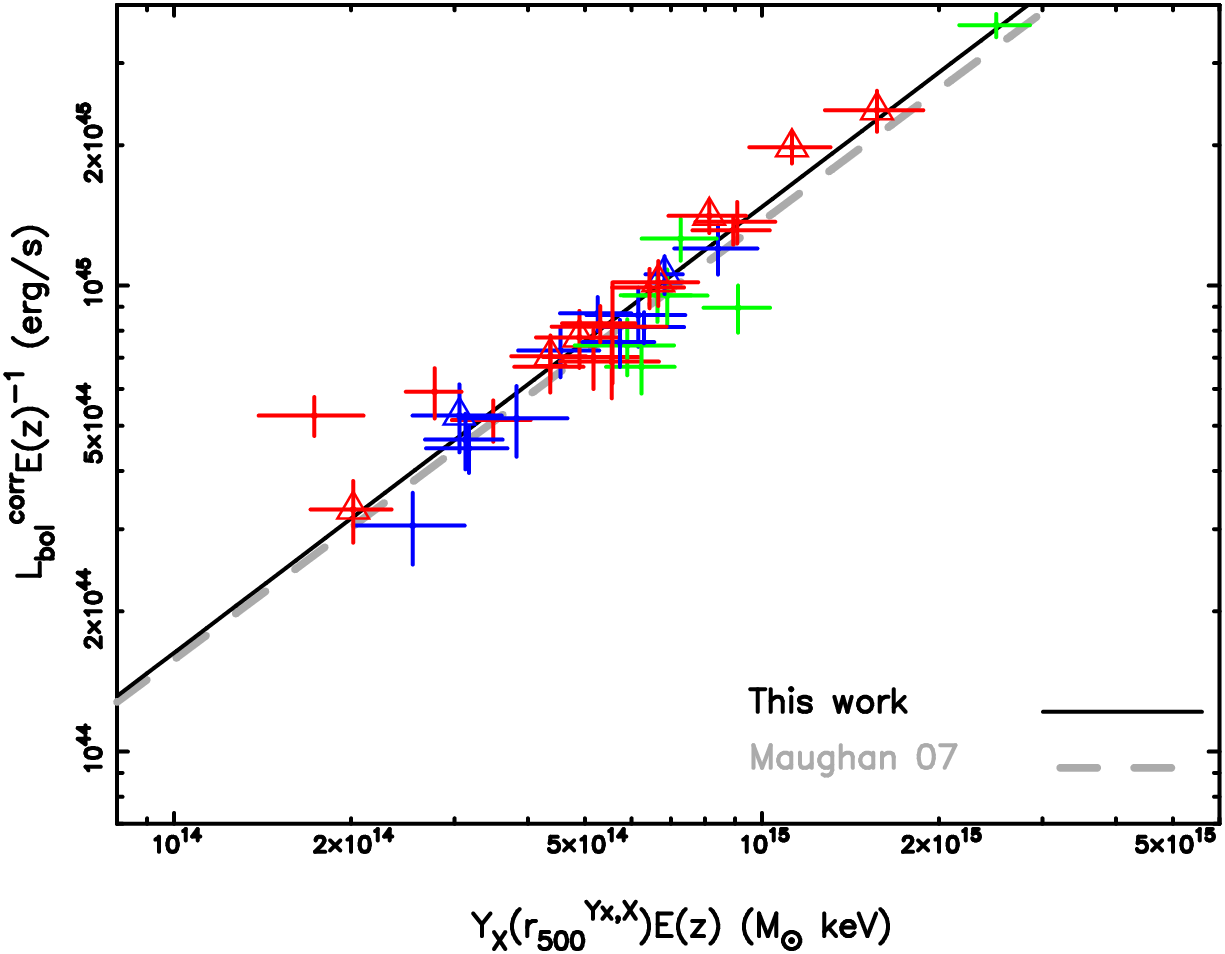}
\includegraphics[angle=270,width=8.5cm]{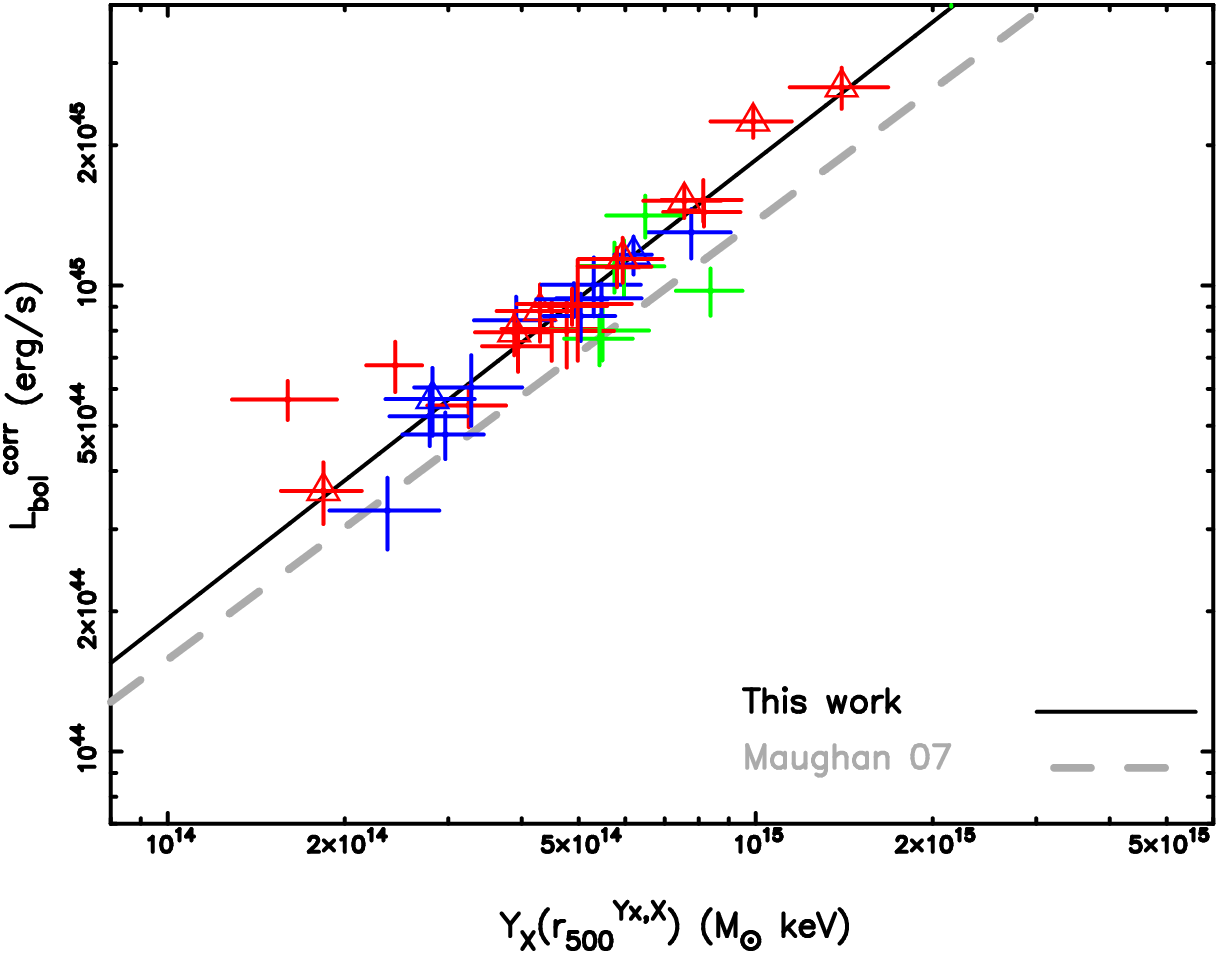}

\includegraphics[angle=270,width=8.5cm]{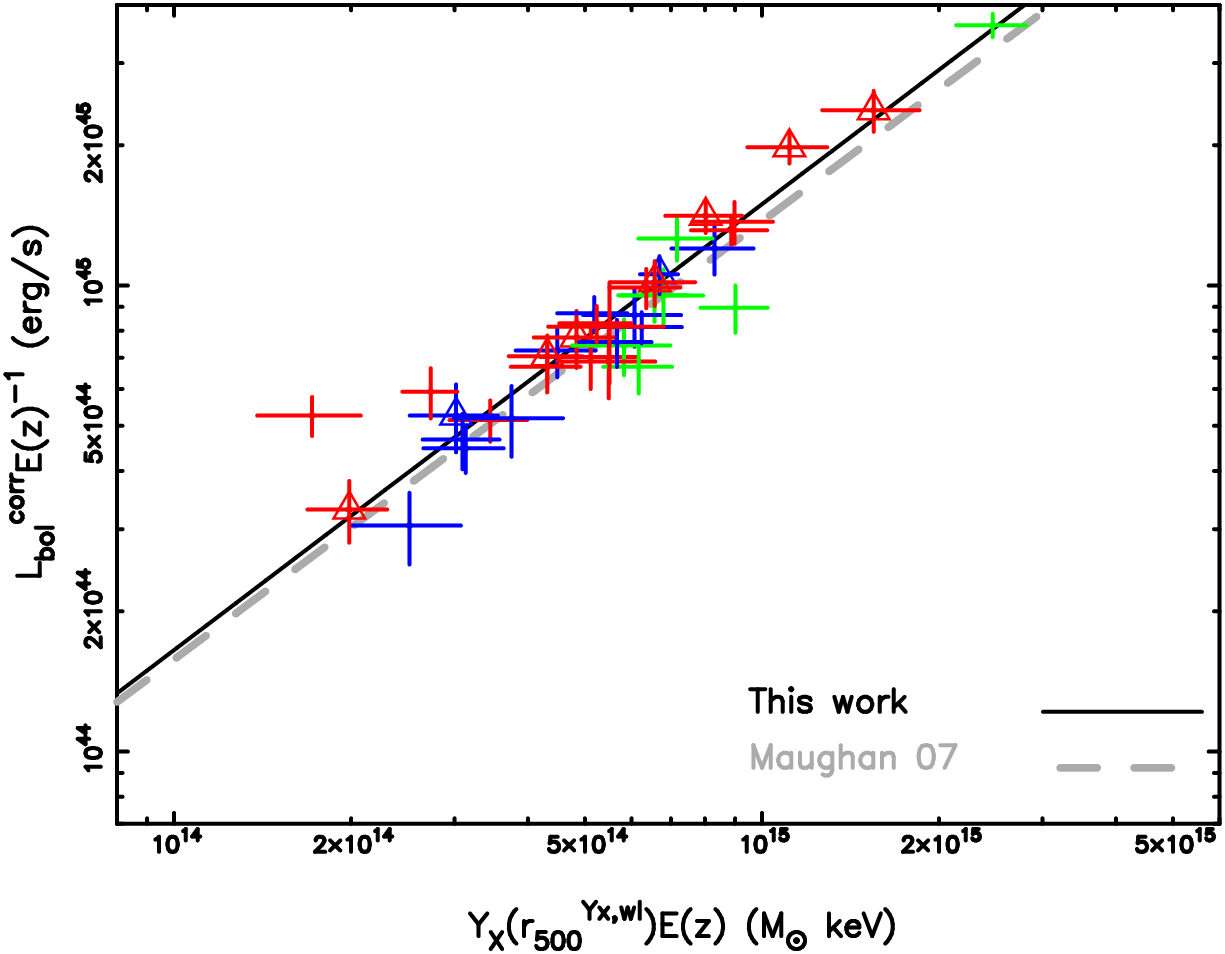}
\includegraphics[angle=270,width=8.5cm]{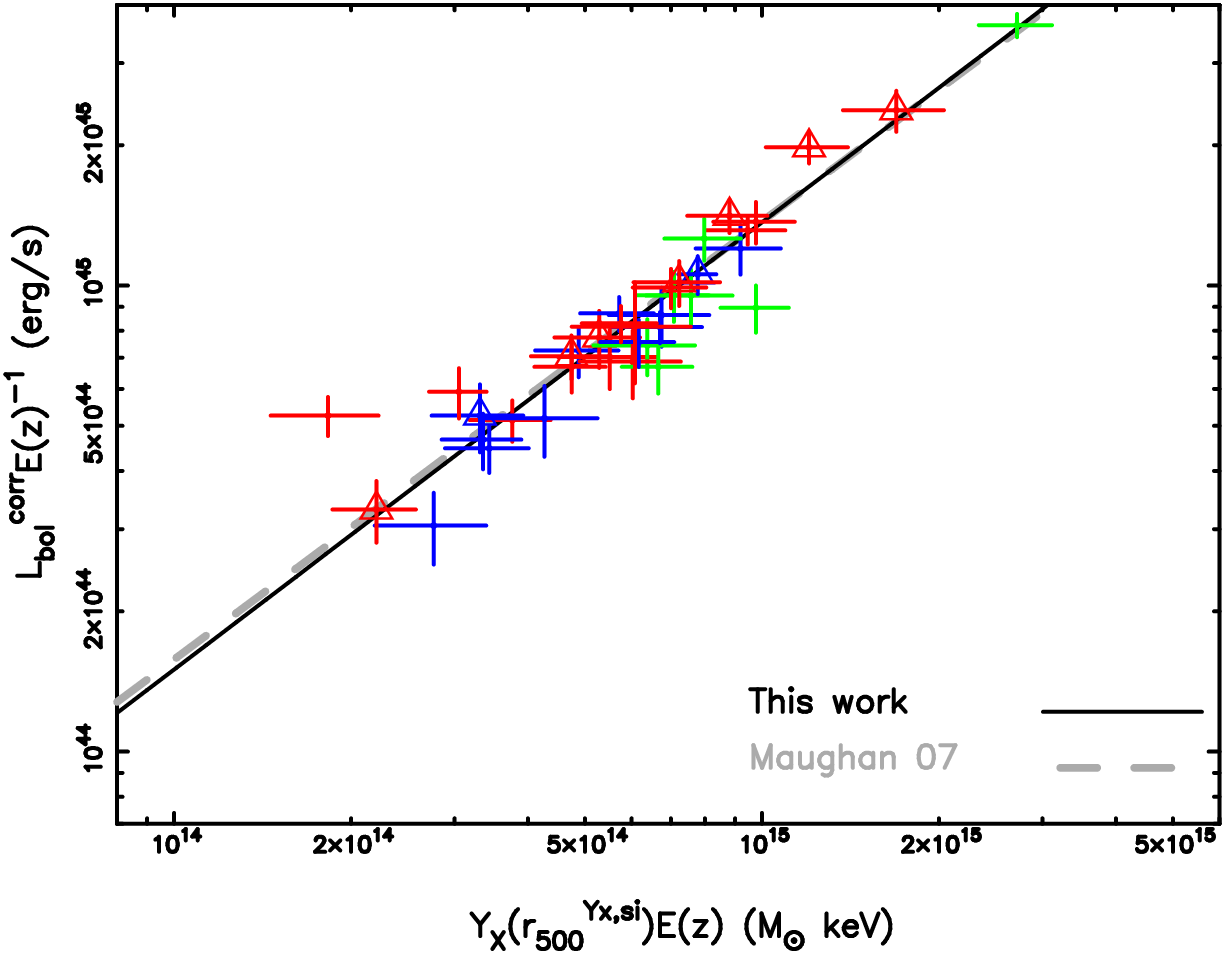}
\end{center}
\caption{Luminosity--$Y_{\rm X}$ relation.  The CCCs are marked by
  triangles. The clusters appearing ``primary with small secondary''
  or ``off-center'' morphology are in green, appearing ``elliptical''
  or ``complex'' morphology are in blue, and appearing ``single''
  morphology are in red using the classification in Jones \& Forman
  (1992). The $Y_{\rm X}$ parameters are measured at $r_{500}^{\rm
    Y_{\rm X},X}$ in the upper panels, at $r_{500}^{\rm Y_{\rm X,wl}}$
  in the left lower panel, and at $r_{500}^{\rm Y_{\rm X,si}}$ in the
  right lower panel. In \S~\ref{s:simuyx}, \S~\ref{s:ylmwlen_simuyx1}
  and \S~\ref{s:ylmwlen_simuyx2}, we gave the description of the
  derivation of the radius by combining the $Y_{\rm X}(r)$ profile and
  the $M$--$Y_{\rm X}$ relation.
  \label{f:lboly_simuyx2}}
\end{figure*}

\end{document}